\documentclass[defaultstyle,11pt]{cuthesis}

\input dissertation_preamble.tex

\numberwithin{equation}{chapter}

\makeatletter
\g@addto@macro\bfseries{\boldmath}
\makeatother

\title{A $C^*$-Algebraic Approach to Parametrized Quantum Spin Systems and Their Phases in One Spatial Dimension}

\author{Daniel D.}{Spiegel}

\otherdegrees{B.A., University of Chicago, 2017}

\degree{Doctor of Philosophy}		%  #1 {long descr.}
	{Ph.D., Physics}		%  #2 {short descr.}

\dept{Department of}			%  #1 {designation}
	{Physics}		%  #2 {name}

\advisor{Prof.}				%  #1 {title}
	{Markus Pflaum}			%  #2 {name}

\reader{Oliver DeWolfe}		%  2nd person to sign thesis
\readerThree{Agn\`es Beaudry}		%  3rd person to sign thesis
\readerFour{Michael Hermele} 
\readerFive{Victor Gurarie}

\abstract{  %\OnePageChapter	% because it is very short

This thesis aims to lay a foundation for a $C^*$-algebraic approach to parametrized quantum spin systems in the thermodynamic limit. Our main physical result is the construction of a phase invariant for one-dimensional quantum spin chains parametrized by a topological space $X$. This invariant is constructed using $C^*$-algebraic techniques and takes values in degree one \v{C}ech cohomology $\check{H}^1(X;\PU(\hilbH))$, where $\PU(\hilbH)$ is the projective unitary group of an infinite-dimensional Hilbert space $\hilbH$, endowed with the quotient of the strong operator topology on $\Unitary(\hilbH)$. One may equivalently view this invariant as a fiber bundle whose fibers are projective Hilbert spaces \cite{Lawson,SchottenloherUnitaryStrongTopology,Steenrod} or, using Dixmier--Douady theory, as an element of $\check{H}^2(X;\Unitary(1))$ or $\check{H}^3(X;\bbZ)$ \cite{Brylinski,DixmierDouady,Husemoller_Schottenloher}. An exactly solvable model of a one-dimensional spin system parametrized by the 3-sphere $X = \bbS^3$ is presented and it is shown that its invariant is nontrivial. This model has appeared in our collaboration's preprint \textit{Flow of (higher) Berry curvature and bulk boundary correspondence in parametrized quantum systems} \cite{Xueda}.

We also prove several mathematical results, concerning topological aspects of the pure state space $\pstate(\fA)$ of a $C^*$-algebra $\fA$, that have been inspired by our study of parametrized quantum systems. Our first new result is that $\pstate(\fA)$, endowed with the weak* topology, has trivial fundamental group for every UHF algebra $\fA$. This result has been submitted for publication in a paper by our collaboration entitled \textit{Homotopical foundations of parametrized quantum spin systems} \cite{beaudry2023homotopical}. 

With respect to the norm topology on $\pstate(\fA)$, we show how one can construct a fiber bundle over $\pstate(\fA)$ whose fiber over a pure state is the Hilbert space coming from the corresponding Gelfand--Naimark--Segal construction. This result was published in our collaboration's paper \textit{Continuous dependence on the initial data in the Kadison transitivity theorem and GNS construction} \cite{Spiegel}. This line of thinking leads further to the construction of a distinguished class in $\check{H}^1(\pstate(\fA);\Unitary(1))$ that generalizes the principal $\Unitary(1)$-bundle $\sphere \hilbH \rightarrow \bbP \hilbH$ in a representation independent way, where $\sphere \hilbH$ and $\bbP \hilbH$ are the unit sphere and projective Hilbert space of $\hilbH$, respectively. 

As the article title above suggests, we also show how the operators obtained from the Kadison transitivity theorem can be chosen in a continuous way. Using the Michael selection theorem, we prove that for any nonzero irreducible representation $(\hilbH, \pi)$ of a $C^*$-algebra $\fA$ and any $n \in \bbN$, there exists a continuous function $A(\x, \y)$ with values in $\fA$ such that $\pi(A(\x,\y))x_i = y_i$ for all $i \in \qty{1,...,n}$, where $\x, \y \in \hilbH^n$ and the components $\x$ are linearly independent. We also prove versions of this result where $A(\x, \y)$ is self-adjoint or unitary. With this result we prove that the unitary group $\Unitary(\fA)$ of a unital $C^*$-algebra is a fiber bundle over every path component of $\pstate(\fA)$ with the norm topology. These results were also published in the aforementioned paper \cite{Spiegel}.
	}

\dedication[Dedication]{	% NEVER use \OnePageChapter here.
	To David, Lisa, Isaac, and Breanna.
	}

\acknowledgements{	\OnePageChapter	% *MUST* BE ONLY ONE PAGE!
	It took me some time to find my academic niche and I am grateful to many people for cultivating my interests and making my graduate career a productive one. Oliver DeWolfe is half the reason I came to Boulder in the first place (the other half being the mountains) and I would like to thank him for his support and understanding as I transitioned to mathematical physics. 
	Of course, I owe it to Markus Pflaum for showing me that mathematical physics was even an option at all. 
	Thank you, Markus, for finding me a home in the math department and especially for all your guidance in finding interesting problems and for your help in writing this thesis and our articles.

	I would like to thank Michael Hermele, Marvin Qi, and Xueda Wen for keeping us grounded in physics. 
	Mike's model is the key example of this thesis, Marvin aided in the construction of the $\check{H}^1(X, \PU(\hilbH))$ class, and the idea for the parametrized phase invariant suddenly clicked when Xueda was explaining how projective representations arise in SPT phases. 
	I am also grateful to Juan Moreno for teaching me some things about fiber bundles. I would like to thank Victor Gurarie for taking the time to serve on my committee. Finally, a special thank you goes to Agn\`es Beaudry for her countless hours of incredibly hard work, for her sense of humor, and for her friendship.  

	Outside the academic sphere, I want to thank my parents, David and Lisa, for raising me right and my brother, Isaac, for blazing the trail I followed throughout my early years.  I want to thank all my high school and college friends for making those chapters of my life so awesome. You guys are the best. I would like to thank Mango for his constant companionship. Most of all, thank you, Breanna, for your endless love, devotion, and support, and for nodding patiently while I expounded on the abstract beauty of ``sea star algebras.''
	}

\LoFisShort	% use this only for 1-page Table of Figures
 \emptyLoT	% use this if there is no List of Tables

\begin{document}

%!TEX root = dissertation.tex

\newpage
\chapter{Introduction}

While everyone is familiar with the everyday phases of matter---solid, liquid, and gas---condensed matter physicists have long been interested in more subtle notions of phase and phase transition. For example, a bar magnet is in the ferromagnetic phase, in which the magnetic moments of its atoms are aligned, until it is heated above its Curie temperature, at which point it enters the paramagnetic phase of disordered magnetic moments. When cooled to very low temperatures, certain materials may take on a \textit{quantum} phase of matter in which exotic macroscropic quantum phenomena are exhibited, such as topological superconductivity and the quantum Hall effect. These phenomena characterize phases of matter in the sense that these behaviors are shared among microscopically disparate materials and can be created or destroyed in abrupt phase transitions. In the past fifteen years, interest has centered around \textit{symmetry protected topological (SPT) phases}. These are quantum phases at zero temperature that categorize materials possessing a given symmetry. The presence of the symmetry gives rise to a wider variety of phases than exists if no symmetry is required.

This dissertation is about a newly recognized type of quantum phase, where the object of study is not a single quantum system but an entire family of them, depending continuously on a parameter in a topological space. For such \textit{parametrized systems}, the global topology of the parameter space can give rise to nontrivial phases even when the system is trivial for any given choice of parameters. \textit{Parametrized phases} are a relatively unexplored but fast growing frontier in condensed matter physics. Indeed, there has been a flurry of activity on the topic in recent years \cite{BachmannThouless,beaudry2023homotopical,CordovaFreedLamSeiberg1,CordovaFreedLamSeiberg2,GaiottoJF,HsinKapustinThorngren,Kitaev_differential_forms,KSp_higher_berry_curvature,KSp_thouless,KSo_local_Noether,ShiozakiGeneralizedThouless,ohyama2023discrete,ShiozakiAdiabaticCycles,Spiegel,Xueda}.  Our approach will utilize topological and especially $C^*$-algebraic techniques to take a  step towards a mathematically rigorous classification of parametrized phases for one-dimensional systems of quantum spins.  In fact, this dissertation is just as much about parametrized phases as it is about a number of abstract $C^*$-algebraic results that have sprung from our investigations. Such is the nature of mathematical physics.

\section{Motivation and Historical Background}

Very roughly, a \textit{parametrized quantum system} is a continuous function from a topological space $X$, dubbed the \textit{parameter space}, into either a space of Hamiltonians or a space of states. The types of systems we will consider in this dissertation are many-body systems of quantum spins on a crystalline lattice. Near zero temperature, quantum systems tend to lie in their ground state. Therefore when we model a parametrized quantum system as a continuous function into a space of states, we typically think of the values of this function as the ground states of some parametrized family of Hamiltonians, living in the background. To ensure that the macroscopic quantum properties of the system are stable under deformation, one typically requires that the Hamiltonian have a unique ground state and a gap in the spectrum above the ground state energy, with a positive lower bound on the size of the gap that is independent of the parameter. One says a parametrized quantum system is \textit{trivial} if the system is constant, i.e., independent of the parameter, and there is no entanglement between the different particles (or degrees of freedom more generally) that comprise the system.

Two gapped parametrized quantum systems are in the same \textit{phase} if there exists a homotopy between them that maintains a uniform positive lower bound on the gap. It is permissible that this homotopy could exist only after combining each system with a trivial system. The action of adding on auxiliary non-interacting degrees of freedom is known as \textit{stacking}; it enlarges the space of Hamiltonians/states, allowing more room for the homotopy to take place. More precisely, the stack of two (not-necessarily-trivial) systems is described by the sum of their Hamiltonians and the tensor product of their ground states.

Thus, we can say a \textit{parametrized phase} is an equivalence class of parametrized quantum systems with respect to the equivalence relation generated by homotopy and stacking with trivial systems. We say a phase is \textit{trivial} if it contains a trivial parametrized system. The goal is then to classify phases over a given parameter space $X$ and find interesting examples of nontrivial phases. It is widely believed that the map from gapped Hamiltonians to ground states should be a weak homotopy equivalence and therefore for the purposes of classifying phases it should not matter whether one works with Hamiltonians or ground states \cite{KapustinSopenkoYang,kitaevIAS,Xiong}. 
%Indeed, one often thinks of a phase as a pattern of entanglement of the (parametrized) ground state.

The goal of classifying parametrized phases stems from a conjecture of  Kitaev that he described in a number of talks \cite{KitaevSRE,KitaevHomotopy,kitaevIAS} and that continues to be developed by others \cite{beaudry2023homotopical,GaiottoJF,shiozaki2018generalized,Xiong2018,Xiong}. Kitaev defined a (non-parametrized) lattice system to be \textit{invertible} if it admits an inverse system, i.e., a system such that the stack of the system with its inverse can be connected by a path to a trivial system. He then conjectured that gapped invertible systems should be described by a \textit{loop spectrum} (``spectrum'' in the sense of homotopy theory), indexed by the spatial dimension $d$ of the system, with different spectrums for bosonic and fermionic systems. A loop spectrum is a sequence of pointed topological spaces $Y_0,Y_1, Y_2, \ldots$ and weak homotopy equivalences 
\[
Y_d \xrightarrow{\simeq} \Omega Y_{d+1}
\]
for all $d$, where $\Omega Y_{d+1}$ is the space of loops in $Y_{d+1}$ based at the basepoint of $Y_{d+1}$. The basepoint corresponds to a trivial system. A loop spectrum gives rise to a \textit{generalized cohomology theory} defined as
\[
E^d(X) \defeq [X, Y_d],
\]
where the right hand side is the set of homotopy classes of maps $X \rightarrow Y_d$. We interpret $E^d(X)$ as classifying $d$-dimensional systems parametrized by $X$, with the group operation on $E^d(X)$ corresponding to the stacking of phases---a perspective expanded upon by our collaboration in the paper \textit{Flow of (higher) Berry curvature and bulk-boundary correspondence in parametrized quantum systems} \cite{Xueda}. The homotopy type of the spaces $Y_d$ is therefore of great importance for the purpose of classifying parametrized phases. 

For example, for $0$-dimensional bosonic systems (by $d = 0$ we mean a ``blob'' of particles with no significant spatial extent) it is well-understood that $Y_0$ should be an Eilenberg--Maclane space of type $K(\bbZ, 2)$. More explicitly, one can think of $Y_0$ as $\bbC \bbP^\infty$. Phases of 0-dimensional bosonic systems parametrized by $X$ are then classified by
\[
[X, K(\bbZ, 2)] \cong H^2(X, \bbZ).
\]
The $H^2(X, \bbZ)$ phase invariant can be thought of as the isomorphism class or first Chern class of the line bundle of ground states of the system. Alternatively, it can be thought of as the de Rham cohomology class of the Berry curvature $\Omega$. The latter is a priori an element of $H^2_\tn{dR}(X, \bbR)$, but can be thought of as an element of singular cohomology $H^2_\tn{sing}(X, \bbZ)$ after noting that $\Omega$ has quantized integrals. See, e.g., \cite{beaudry2023homotopical,KSp_higher_berry_curvature} for a review.

Another widely held belief is that invertible gapped phases of quantum matter are described by topological quantum field theories (TQFT). Freed and Hopkins have classified the latter and identified the relevant spectra \cite{FreedHopkinsSym,freed_hopkins}. This leads to predictions for the classification of phases of lattice systems. In this dissertation, we are particularly interested in bosonic systems in one spatial dimension. For these systems, the corresponding space in the loop spectrum is expected to be an Eilenberg--Maclane space of type $K(\bbZ, 3)$, hence one-dimensional bosonic parametrized phases over the parameter space $X$ are expected to be classified by the third cohomology group:
\[
[X, K(\bbZ, 3)] \cong H^3(X, \bbZ).
\]
There is a desire to understand and prove these predictions directly from considerations of lattice models---from the ground up, so to speak. No proof has been given that the space of lattice systems should be a $K(\bbZ, 3)$ and we will not discuss this point further, however the derivation of an $H^3(X, \bbZ)$ phase invariant for lattice models is the subject of Chapter \ref{chp:cocycle_construction}.

Very significant work on the derivation of phase invariants for parametrized lattice systems has already been done by  Kapustin,  Spodyneiko, and Sopenko, partially inspired by ideas of Kitaev \cite{Kitaev_differential_forms,KSp_higher_berry_curvature,KSp_thouless,KSo_local_Noether}. Given a family of Hamiltonians in $d$ spatial dimensions parametrized by a smooth manifold $X$, Kapustin and Spodyneiko define a closed $(d+2)$-form $\Omega^{(d+2)}$ called \textit{higher Berry curvature}, giving a class in de Rham cohomology ${H}_\tn{dR}^{d+2}(X, \bbR)$. When $d = 0$ this is the usual Berry curvature 2-form. They show that the cohomology class of $\Omega^{(d+2)}$ is a topological invariant of the family and argue that $\Omega^{(d+2)}$ should have quantized integrals over $(d+2)$-dimensional spheres in $X$. This is very close to a full proof that $\Omega^{(d+2)}$ gives the desired $H^3(X,\bbZ)$ invariant for $d = 1$. However, this approach leaves open a few mysteries. For one, it is not clear that the cohomology class of $\Omega^{(d+2)}$ depends only on the parametrized family of ground states. Furthermore, de Rham cohomology cannot detect invariants in the torsion subgroup of $H^3(X,\bbZ)$. Finally, computing $\Omega^{(d+2)}$ can be rather difficult!

The first issue---that of obtaining an invariant that depends only on the family of ground states---was rectified in a later article by Kapustin and Sopenko \cite{KSo_local_Noether}. In that article they construct a class in $H^{d+2}_\tn{dR}(X,\bbR)$ directly from a smooth family of states. It is also claimed that this class is quantized, although no proof is provided. Very recently, Ohyama, Terashima, and Shiozaki have shown how to derive a Deligne cohomology class in $H^3(X, \bbZ)$ (and more generally a \v{C}ech cohomology class in $\check{H}^3(X,\bbZ)$) from a parametrized family of injective matrix product states in a way that captures elements of the torsion subgroup \cite{ohyama2023discrete}. Their method appears to only capture elements of the torsion subgroup of $H^3(X, \bbZ)$. 

In this dissertation we will provide a new perspective on the $H^3(X, \bbZ)$ invariant for one-dimensional bosonic systems. Our approach is designed to treat parametrized families of pure states in the thermodynamic limit of infinitely many particles. We use $C^*$-algebraic techniques to derive a \v{C}ech 1-cocycle in $\check{H}^1(X, \PU(\hilbH))$, where $\PU(\hilbH)$ is the projective unitary group of a Hilbert space $\hilbH$ (which is infinite-dimensional in the thermodynamic limit), then we use Dixmier-Douady theory \cite{Brylinski,Husemoller_Schottenloher} to map 
\[
\check{H}^1(X; \PU(\hilbH)) \to \check{H}^2(X;\Unitary(1)) \to \check{H}^3(X; \bbZ).
\]
A precise statement will be given in the following section. One advantage of our approach is its generality. In particular, we do not require our states to be matrix product states and we can obtain a class in $\check{H}^1(X, \PU(\hilbH))$ for arbitrary topological spaces; mapping to $\check{H}^3(X,\bbZ)$ requires only that $X$ is paracompact Hausdorff.

It is important to work in the thermodynamic limit if one wants to be mathematically precise about the notion of phase for $d > 0$. As Kapustin--Sopenko put it, ``\ldots the distinction between phases becomes sharp only in infinite volume\ldots'' \cite{KapustinSopenkoHallConductance}. Indices for classifying phases are sometimes most clearly defined in the thermodynamic limit and working with infinite systems often reveals new perspectives and mathematical phenomena. In an online talk at the One World IAMP seminar, Ogata analogized the indices for SPT phases in the thermodynamic limit to the Fredholm index of an operator, which is zero for an operator on any finite-dimensional space but may be nonzero in infinite-dimensions \cite{Ogata_IAMP}. One of the most significant infinite-dimensional phenomena affecting this thesis is Kuiper's theorem, which states that $\Unitary(\hilbH)$ is contractible for an infinite-dimensional Hilbert space $\hilbH$ in the norm topology \cite{Kuiper}. In fact, it is also contractible in the strong operator topology \cite{SchottenloherUnitaryStrongTopology,AtiyahSegalTwistedKTheory}. One consequence of this is that the map $\check{H}^1(X; \PU(\hilbH)) \to \check{H}^2(X; \Unitary(1))$ is a bijection \cite{Brylinski}.

Of course, working in the thermodynamic limit brings many subtleties to contend with as well. One might want to describe a system consisting of infinitely many particles by using a tensor product of infinitely many on-site Hilbert spaces, but this is a mathematically ill-advised object (one can construct such a thing, but not in a very useful way). On the other hand, it is possible to take an infinite tensor product of on-site observable algebras to obtain a well-defined (and useful) \textit{algebra of quasi-local observables}. This is an example of a \textit{uniformly hyperfinite (UHF) $C^*$-algebra}. Thus, $C^*$-algebras are a foundational necessity for a mathematically rigorous treatment of the thermodynamic limit.

Mathematical physicists have developed a wealth of tools for working with systems in the $C^*$-algebraic framework (\cite{BMNS_automorphic_equivalence,OM_automorphic_equivalence,Matsui_split_property,Matsui13,NOS2006_LR_Bounds,NS_LR_Bounds_Exp_Clustering,NSY_quasilocality_bounds}, to name a few). Notably, Ogata has used $C^*$-algebraic techniques to derive indices for SPT phases in one and two spatial dimensions \cite{ogata_Z2_timereversal,ogata_Z2_reflection,ogata_2d,ogata_spin_chains_split} using the split property proven by Matsui \cite{Matsui_split_property,Matsui13}. Our derivation of the $\check{H}^3(X, \bbZ)$ class for parametrized systems is heavily inspired by her work. 

I believe $C^*$-algebras constitute some of the most beautiful objects in mathematics. We will often study these from an abstract perspective. Not all of our $C^*$-algebraic results have found physical applications, but all have in some way sprouted from considerations of parametrized systems. Perhaps our most significant abstract $C^*$-algebraic results arose from a desire to understand how one can perform classical $C^*$-algebraic maneuvers, like the Gelfand--Naimark--Segal (GNS) construction and Kadison transitivity theorem, in a way that depends continuously on the initial data, i.e., for a parametrized family of pure states. These problems are considered in Chapter \ref{chp:fiber_bundles} and are resolved with the construction of a number of fiber bundles built from $C^*$-algebraic structures. The precise results will be outlined in the next section. Most of the results of this chapter were published in our article \textit{Continuous dependence on the initial data in the Kadison transitivity theorem and GNS construction}, published in Reviews in Mathematical Physics \cite{Spiegel}. 

Finally, we acknowledge that there is still much work to be done on the subject of parametrized phases. The ultimate goal to construct Kitaev's loop spectra has yet to be realized. Much is still unknown about two-dimensional parametrized systems. For example, there does not yet exist a method for calculating invariants for two-dimensional parametrized systems that can detect torsion. Closer to home, we describe the precise data required to do our $\check{H}^3(X, \bbZ)$ construction, but these data are rather abstract and formulated in terms of states; it would be desirable to provide conditions on a family of gapped Hamiltonians such that their ground states fulfill our hypotheses. These are all difficult problems, but we have hope that none are insurmountable.

\section{Outline and Main Results}

We begin in Chapter \ref{chp:background} by introducing the main mathematical tools used in this dissertation. Our hope is for this dissertation to be of use to mathematicians and physicists alike, so we have included a generous amount of exposition in this chapter and beyond. Section \ref{sec:PH} introduces the topology of projective Hilbert space and the projective unitary group and constructs several important principal fiber bundles. Most of the material of section \ref{subsec:metrics} and \ref{subsec:fiber_bundle_PH} and comes from Markus Pflaum's online notes \cite{PflaumGeometryCQF}, which we polished and expanded upon for publication in \cite{Spiegel}. Small bits and pieces of this material can be found in the literature and some of it might be considered folklore, but there does not seem to exist a complete exposition like the one we give here elsewhere in the literature. Section \ref{subsec:PU(H)} gives a brief review of the topology of the projective unitary group, following D.\ J.\ Simms \cite{Simms_1970} with a slight reorganization and slightly different methods of proof.

Section \ref{sec:crash_course} consists of mostly standard textbook fare on $C^*$-algebras, which we have included for the benefit of physicists who might not be familiar with that subject. Almost all proofs are omitted in that section, with the exception of Section \ref{subsec:Aut(A)_Inn(A)}. That section consists of a few elementary results on the topologies of the automorphism and inner automorphism groups of a $C^*$-algebra. The proofs there are quite basic, but we are unaware of where to find them in the literature. In Section \ref{subsec:noninteracting} we present a noninteracting parametrized 1d system to illustrate some of the various topologies and $C^*$-algebraic concepts. Section \ref{subsec:noninteracting} is adapted from \cite{Spiegel}.

Chapter \ref{chp:pure_state_topology} begins with a review of more advanced algebraic and topological properties of the pure state space of a $C^*$-algebras, most importantly the $C^*$-algebraic notion of superselection sector. Good references for this material exist and have been provided, but the results are scattered and we have collected them in a more or less comprehensive exposition. We have also provided new proofs in several cases and filled in proofs that the original authors left as exercises. Section \ref{sec:UHF_simply_connected} is then devoted to the proof of the following original result. To clarify notation, we let $\pstate(\fA)$ denote the set of pure states of a $C^*$-algebra $\fA$. 

\begin{theorem*}[\ref{thm:fundaUHFfinal}]
Let $\fA$ be a UHF algebra. Endowed with the weak* topology, the fundamental group of $\pstate(\fA)$ is trivial.
\end{theorem*}

This result is not used elsewhere in the thesis, but fits well with the material of Chapter \ref{chp:pure_state_topology}. Excluding introductory remarks, section \ref{sec:UHF_simply_connected} is a facsimile of \cite[Sec.~4]{beaudry2023homotopical}, which has been submitted for publication in Reviews in Mathematical Physics, while the rest of Chapter \ref{chp:pure_state_topology} was published more or less verbatim in \cite{Spiegel}.

Chapter \ref{chp:cocycle_construction} contains our derivation of the $\check{H}^3(X; \bbZ)$ class for parametrized one-dimensional spin systems. The invariant is  constructed as an element of $\check{H}^1(X; \PU(\hilbH))$ for an infinite-dimensional separable Hilbert space $\hilbH$, where $\PU(\hilbH)$ is the projective unitary group of $\hilbH$ endowed with the quotient of the strong operator topology. Using Dixmier-Douady theory, one has bijections $\check{H}^1(X; \PU(\hilbH)) \rightarrow \check{H}^2(X; \Unitary(1)) \rightarrow \check{H}^3(X;\bbZ)$, as mentioned previously. We prove two important new auxiliary results, Theorem \ref{thm:cont_family_PU_from_aut} and Theorem \ref{thm:equivalence_product_state}, on our way to the construction. To state the auxiliary results and the main result in full generality would require some technical development that we will not take up just yet. A more specialized version of the construction is presented below.

\begin{theorem*}
Consider a spin system on the lattice $\bbZ$, i.e., for each $v \in \bbZ$ let $n_v \in \bbN$ be the dimension of the on-site Hilbert space at site $v$ and let
\[
\fA = \bigotimes_{v \in \bbZ} M_{n_v}(\bbC)
\]
be the associated quasi-local algebra. Let $X$ be a paracompact Hausdorff space and let $\omega:X \rightarrow \pstate(\fA)$ be a weak*-continuous family of states. Assume the following data can be obtained:
\begin{itemize}
	\item a point $v_0 \in \bbZ$ where we ``cut'' the lattice into left and right halves, denoting
	\[
	\fA_L = \bigotimes_{v \leq v_0} M_{n_v}(\bbC) \qqtext{and} \fA_R = \bigotimes_{v > v_0} M_{n_v}(\bbC),
	\]
	\item fixed pure states $\omega_L \in \pstate(\fA_L)$ and $\omega_R \in \pstate(\fA_R)$ with GNS representations denoted \newline $(\hilbH_L, \pi_L, \Omega_L)$ and $(\hilbH_R, \pi_R, \Omega_R)$,
	\item an indexed open cover $\cO = (O_i)_{i \in I}$ of $X$,
	\item for each index $i \in I$, two strongly continuous families of automorphisms $\alpha_i:O_i \rightarrow \Aut(\fA_L)$ and $\beta_i:O_i \rightarrow \Aut(\fA_R)$,
	\item for each $i \in I$, a norm-continuous family of inner automorphisms $\eta_i:O_i \rightarrow \Inn(\fA)$,
\end{itemize}
such that for all $i \in I$ and $x \in O_i$, we have
\[
\omega_x = (\omega_L \otimes \omega_R) \circ (\alpha_{i,x} \otimes \beta_{i,x}) \circ \eta_{i,x}.
\]
Then for all $i,j \in I$ such that $O_i \cap O_j \neq \varnothing$, there exists a unique strongly continuous family $\bbU_{ij}:O_i \cap O_j \rightarrow \PU(\hilbH_R)$ of projective unitaries such that
\[
\bbU_{ij,x}\pi_{R}(A)\bbU_{ij,x}^{-1} = (\pi_{R} \circ \beta_{i,x} \circ \beta_{j,x}^{-1})(A)
\]
for all $A \in \fA_R$ and $x \in O_i \cap O_j$. Moreover, the collection of maps $\bbU_{ij}$ is a \v{C}ech 1-cocycle and the image of its cohomology class under the bijection $\check{H}^1(X; \PU(\hilbH_R)) \rightarrow \check{H}^2(X; \Unitary(1))$ is independent of all choices for the bulleted data above.
\end{theorem*}

In Section \ref{sec:1d_example} we introduce a toy model of a gapped parametrized one-dimensional quantum spin system with parameter space $X = \sphere^3$. This model was invented by Michael Hermele; he first presented it in a talk at Harvard's CMSA \cite{Hermele2021} and it later appeared in our collaboration's preprint \cite{Xueda}. At any point in parameter space, the system is in a state of decoupled dimers, with the configuration of dimers differing between the two hemispheres of $\sphere^3$. The system is therefore exactly solvable and we work out the ground state explicitly. We show that the bulleted data above can be found and the invariant is nontrivial for this model. Our construction of the invariant for this model has not yet been published elsewhere.

Finally, Chapter \ref{chp:fiber_bundles} contains a number of new abstract $C^*$-algebraic results that have been inspired by our investigations of parametrized quantum systems. This chapter is adapted from our publication \cite{Spiegel} and contains some additional results as well. In Section \ref{sec:fiberwise_GNS_construction} we consider how the GNS construction can be performed in families, i.e., what happens to the GNS construction if we vary the input state. Our answer to this question is the construction of a Hilbert bundle over pure state space whose fibers are the GNS Hilbert spaces. Kuiper's theorem implies that infinite-dimensional Hilbert bundles with structure group $\Unitary(\hilbH)$ are guaranteed to be trivial, but in this case we can essentially reduce the structure group to $\Unitary(1)$ to get a nontrivial bundle. We are led to a generalization of the principal $\Unitary(1)$-bundle $\sphere \hilbH \rightarrow \bbP\hilbH$ to the setting of abstract $C^*$-algebras. Here, $\sphere \hilbH$ and $\bbP\hilbH$ are the unit sphere and projective Hilbert space of $\hilbH$, respectively. The following is the culmination of a number of results spread out across Section \ref{sec:fiberwise_GNS_construction}. We use the fact that degree one \v{C}ech cohomology $\check{H}^1(X;G)$ over a space $X$ with values in a topological group $G$ is in bijective correspondence with the isomorphism classes of principal $G$-bundles over $X$ \cite{Lawson}.

\begin{theorem*}%[\ref{thm:SH->PH_generalization}]
Let $\fA$ be a nonzero $C^*$-algebra and equip its pure state space $\pstate(\fA)$ with the norm topology.  There exists a unique \v{C}ech cohomology class $\xi \in \check{H}^1(\pstate(\fA);\Unitary(1))$ such that for any nonzero irreducible representation $(\hilbH, \pi)$ of $\fA$, the class $f^*\xi \in \check{H}^1(\bbP \hilbH;\Unitary(1))$ corresponds to the isomorphism class of the bundle $\Unitary(1) \rightarrow \sphere \hilbH \rightarrow \bbP \hilbH$, where $f:\bbP \hilbH \rightarrow \pstate(\fA)$ is the map that associates to each ray the pure state it represents.
\end{theorem*}

We conclude in Section \ref{sec:continuous_kadison} by showing how the operators obtained from the Kadison transitivity theorem can be chosen to depend continuously on the initial data, that is, a set of vectors in the Hilbert space of an irreducible representation of a $C^*$-algebra $\fA$. We call this the \textit{continuous Kadison transitivity theorem}. We let $\Unitary(\fA)$ denote the unitary group of a unital $C^*$-algebra $\fA$.

\begin{theorem*}[Continuous Kadison Transitivity]
Let $\fA$ be a $C^*$-algebra and let $(\hilbH, \pi)$ be an irreducible representation. Let $n$ be a positive integer and let
\[
X = \qty{(\x, \y) \in \hilbH^{2n}: \tn{$x_1,\ldots, x_n$ are linearly independent}},
\]
equipped with the subspace topology inherited from $\hilbH^{2n}$, where $\x = (x_1,\ldots, x_n)$ and $\y = (y_1,\ldots, y_n)$. There exists a continuous map $A:X \rightarrow \fA$ such that
\[
\pi(A(\x, \y)) x_i = y_i
\]
for all $(\x, \y) \in X$ and $i = 1,\ldots, n$.

On the subspace $X_\tn{sa} \subset X$ of pairs $(\x, \y)$ such that there exists a self-adjoint $T \in \cB(\hilbH)$ with $Tx_i = y_i$ for all $i = 1,\ldots, n$, there exists a continuous map $A:X_\tn{sa} \rightarrow \fA_\tn{sa}$ satisfying the same property. If $\fA$ is unital, then on the subspace $X_\tn{u} \subset X$ of pairs $(\x, \y)$ such that there exists a unitary $T \in \Unitary(\hilbH)$ with $Tx_i = y_i$ for all $i = 1,\ldots, n$, every point $(\x_0, \y_0)$ has a neighborhood $O \subset X_\tn{u}$ for which there exists a continuous map $A:O \rightarrow \Unitary(\fA)$ which agains satisfies the same property.
\end{theorem*}

The continuous Kadison transitivity theorem, especially the unitary clause, is perhaps our deepest and most difficult result. The theorem states the existence of a continuous selection of operators as obtained from the Kadison transitivity theorem. The key ingredient of the proof is thus the Michael selection theorem \cite[Thm.~$3.2''$]{MichaelSelection} as it provides general conditions under which continuous selections may be found. Applying the Michael selection theorem for the unitary version is by no means straightforward, however. The continuous Kadison transitivity theorem has the corollary below. First, we introduce one final piece of notation.  If $\omega$ is a state on a unital $C^*$-algebra $\fA$ and $U \in \Unitary(\fA)$, then we define a new state $U \cdot \omega$ on $\fA$ as
\[
(U \cdot \omega)(A) = \omega(U^*AU)
\]
for all $A \in \fA$.

\begin{corollary*}
Let $\fA$ be a unital $C^*$-algebra, let $\omega \in \pstate(\fA)$, and let $\Unitary(\fA)$ be the unitary group of $\fA$. There exists a norm-continuous map $U:\qty{\psi \in \pstate(\fA): \norm{\psi - \omega} < 2} \rightarrow \Unitary(\fA)$ such that
\[
\psi = U_\psi \cdot \omega
\]
for all $\psi \in \pstate(\fA)$ with $\norm{\psi - \omega} < 2$. In particular, defining $\pstate_\omega(\fA) = \qty{U \cdot \omega: U \in \Unitary(\fA)}$ and $\Unitary_\omega(\fA) = \qty{U \in \Unitary(\fA): U \cdot \omega = \omega}$, the map
\[
\Unitary(\fA) \rightarrow \pstate_\omega(\fA), \quad U \mapsto U \cdot \omega
\]
is a principal $\Unitary_\omega(\fA)$-bundle. Here, $\pstate_\omega(\fA)$ is given the norm topology.
\end{corollary*}

We prove that the bundle $\Unitary(\fA) \rightarrow \pstate_\omega(\fA)$ is nontrivial in a few cases, the most interesting case being when $\fA$ is a UHF algebra. These results were first published in our collaboration's paper \cite{Spiegel}. While we have not yet found immediate physical consequences for the results of Chapter \ref{chp:fiber_bundles}, we believe they are rather significant in their own right and may find physical applications, potentially in the near future as the $C^*$-algebraic theory of parametrized phases is further developed.

%!TEX root = dissertation.tex

\newpage
\chapter{Mathematical Background}\label{chp:background}

In this chapter we review the background necessary for the study of parametrized phases. In Section \ref{sec:fiber_bundles_and_cech_cohomology} we review the basic theory of fiber bundles and \v{C}ech cohomology, including the relationship between principal fiber bundles and non-abelian \v{C}ech cohomology. In Section \ref{sec:PH} we review the topology of projective Hilbert space and finally in Section \ref{sec:crash_course} we review the basics of $C^*$-algebra theory and its applications to quantum spin systems.

\section{Fiber Bundles and \v{C}ech Cohomology}
\label{sec:fiber_bundles_and_cech_cohomology}

We review the basics of the theory of fiber bundles and principal fiber bundles and use this to motivate our review of \v{C}ech cohomology. The material on fiber bundles is based on the classic textbook by Steenrod \cite{Steenrod} and the lecture notes by Cohen \cite{CohenNotes}, while our review of \v{C}ech cohomology is derived from a variety of sources including \cite{Brylinski,DixmierDouady,EilenbergSteenrod,Forster,Hirzebruch,Lawson,Steenrod}.

\subsection{Fiber Bundles and Principal Fiber Bundles}
\label{subsec:fiber_bundles_and_principal_fiber_bundles}

\begin{definition}
Let $E$ and $X$ be topological spaces and let $p:E \rightarrow X$ be a continuous map. A \textdef{local trivialization} of $p$  consists of an open subset $U \subset X$, a topological space $F$, and a homeomorphism $\phi:p^{-1}(U) \rightarrow U \times F$ such that the diagram below commutes
\begin{equation}\label{cd:loc_triv}
\begin{tikzcd}
p^{-1}(U)\arrow[rr,"\phi"]\arrow[dr,"p"'] && U \times F \arrow[dl,"\proj_U"] \\
& U & 
\end{tikzcd}
\end{equation}
where $\proj_U:U \times F \rightarrow U$ is the projection onto $U$. Let us call $U$ the \textdef{base} and $F$ the \textdef{typical fiber} of the local trivialization. 

We say $(E, X, p)$  is a \textdef{fiber bundle} if there exists a set $\cA$ of local trivializations whose bases cover $X$. We call such a set $\cA$ a \textdef{trivializing atlas} for the fiber bundle. We call $E$ the \textdef{total space}, $X$ the \textdef{base space}, and $p$ the \textdef{projection} of the fiber bundle. If there exists a trivializing atlas for the fiber bundle with the same typical fiber $F$ across all local trivializations, then we say $(E, X, p)$ is a fiber bundle with typical fiber $F$. We often denote a fiber bundle $(E, X, p)$ with typical fiber $F$ as $F \rightarrow E \xrightarrow{p} X$ or $F \rightarrow E \rightarrow X$ if the projection is understood. In this form, the arrow $F \rightarrow E$ is just a piece of notation; we are not actually specifying a function $F \rightarrow E$.
\end{definition}

If $E$ is just a set rather than a topological space, then one can topologize $E$ with a collection of bijections $\phi:p^{-1}(U) \rightarrow U \times F$ satisfying the commutative diagram \eqref{cd:loc_triv} and that are pairwise compatible in the sense below. Our presentation of this fact is a slight variation on a  standard result. Similar results are \cite[Thm.~3.2]{Steenrod} and \cite[Ch.~3, Prop.~2]{Lang_Manifolds}. However, Steenrod requires fixed model fibers (i.e, $F_\phi = F_\chi$ in the notation below), while Lang works with vector bundles and requires the base space $X$ to be manifold. Since we are unsure where to find the exact statement below in the literature, we have given a full proof.

\begin{theorem}\label{thm:topologize_total_space}
Let $E$ be a set, let $X$ be a topological space, and let $p:E \rightarrow X$ be a surjection. Assume to be given a set $\cA$ of bijections $\phi:p^{-1}(U_\phi) \rightarrow U_\phi \times F_\phi$, where $U_\phi \subset X$ is open and $F_\phi$ is a topological space,  such that the diagram below commutes
\[
\begin{tikzcd}
p^{-1}(U_\phi) \arrow[dr,"p"'] \arrow[rr,"\phi"]& & U_\phi \times F_\phi\arrow[dl,"\proj_{U_\phi}"]\\
&U_\phi & 
\end{tikzcd}
\]
where $\proj_{U_\phi}$ is the projection onto $U_\phi$. Suppose further that $\qty{U_\phi}_{\phi \in \cA}$ covers $X$ and that for any $\phi, \chi \in \cA$, the function 
\begin{equation}\label{eq:transition_function_def}
\chi \circ \phi^{-1} : (U_\phi \cap U_\chi) \times F_\phi \rightarrow (U_\phi \cap U_\chi) \times F_\chi
\end{equation}
is continuous. Then there exists a unique topology on $E$ such that for each $\phi \in \cA$, the set $p^{-1}(U_\phi)$ is open and $\phi$ is a homeomorphism. With respect to this topology, $p$ is continuous, hence $(E, X, p)$ is a fiber bundle.
\end{theorem}

Note that the hypotheses imply that \eqref{eq:transition_function_def} is a homeomorphism by symmetry in $\phi$ and $\chi$.

\begin{proof}
Define
\[
\cT = \qty{U \subset E: \phi(p^{-1}(U_\phi) \cap U) \tn{ is open in $U_\phi \times F_\phi$ for all $\phi \in \cA$.}}
\]
We claim this is a topology on $E$. It is clear that $\varnothing, E \in \cT$. Suppose $\cU$ is an arbitrary collection of elements of $\cT$. Then for any $\phi \in \cA$, we have
\begin{align*}
\phi\qty(p^{-1}(U_\phi) \cap \bigcup_{U \in \cU} U) = \bigcup_{U \in \cU} \phi(p^{-1}(U_\phi) \cap U),
\end{align*}
which is open in $U_\phi \times F$ since each $\phi(p^{-1}(U_\phi) \cap U)$ is open.  If $U_1,\ldots, U_n \in \cT$, then
\[
\phi\qty(p^{-1}(U_\phi) \cap \bigcap_{i=1}^n U_i) = \bigcap_{i=1}^n \phi(p^{-1}(U_\phi) \cap U_i)
\]
is open in $U_\phi \times F$ since each $\phi(p^{-1}(U_\phi) \cap U_i)$ is open. Note that we have used the fact that $\phi$ is injective. This proves that $\cT$ is a topology on $E$.

We show that $p$ is continuous. Suppose $U \subset X$ is open and let $(U_\phi, \phi) \in \cA$. Observe that
\[
\phi(p^{-1}(U_\phi) \cap p^{-1}(U)) = (U_\phi \cap U) \times F,
\]
which is open in $U \times F$. Therefore $p^{-1}(U)$ is open, so $p$ is continuous.

We show that for each $\phi \in \cA$, the bijection $\phi$ is a homeomorphism with respect to the subspace topology on $p^{-1}(U_\phi)$ inherited from $\cT$. It is clear from the definition of $\cT$ that $\phi$ is an open map. Suppose $U \subset U_\phi \times F_\phi$ is open. We want to show that $\phi^{-1}(U)$ is open. Let $\chi \in \cA$ and observe that
\[
\chi(p^{-1}(U_\chi) \cap \phi^{-1}(U)) = (\chi \circ \phi^{-1})[U \cap ((U_\phi \cap U_\chi) \times F_\phi)].
\]
This is open since $U \cap ((U_\phi \cap U_\chi) \times F_\phi)$ is open in $(U_\phi \cap U_\chi) \times F$ and $\chi \circ \phi^{-1}$ is a homeomorphism. It follows that $\phi^{-1}(U)$ is open in $E$, hence open in $p^{-1}(U_\phi)$. Thus, $\phi$ is continuous, and therefore a homeomorphism since it is also open. 

Suppose $\cT_1$ and $\cT_2$ are any two topologies on $E$ such that $U_\phi$ is open and $\phi$ is a homeomorphism for all $\phi \in \cA$. Then for every $\phi \in \cA$, the subspace topologies on $p^{-1}(U_\phi)$ induced by $\cT_1$ and $\cT_2$ coincide because $\phi$ is a homeomorphism with respect to both topologies. Furthermore, $p^{-1}(U_\phi) \in \cT_1 \cap \cT_2$ since $p$ is continuous with respect to both topologies. Thus, for any $U \in \cT_1$, we know $U \cap p^{-1}(U_\phi)$ is open in the subspace topology on $p^{-1}(U_\phi)$ induced by $\cT_1$, hence in the subspace topology induced by $\cT_2$. Since $p^{-1}(U_\phi)$ is open in $\cT_2$, we know $U \cap p^{-1}(U_\phi) \in \cT_2$. Since $U = \bigcup_{\phi \in \cA} U \cap p^{-1}(U_\phi)$, we see that $U \in \cT_2$. By the same argument we obtain $\cT_2 \subset \cT_1$, so the two topologies are equal. This proves uniqueness.
\end{proof}

There are several different equivalent ways of saying what a principal fiber bundle is. Below we try to give the simplest definition. First we establish some notation.

\begin{notation}
Let $E$ and $X$ be topological spaces and let $p:E \rightarrow X$ be a continuous function. Given $x \in X$, denote $E_x \defeq p^{-1}(\qty{x})$. If $\phi:p^{-1}(U) \rightarrow U \times F$ is a local trivialization of $p$, then observe that for every $x \in U$ the composition 
\[
E_x \hookrightarrow p^{-1}(U) \xrightarrow{\phi} U \times F \xrightarrow{\proj_F} F
\]
is a homeomorphism, which we denote $\varphi_x:E_x \rightarrow F$. 
\end{notation}

\begin{definition}\label{def:principal_G_bundle}
Let $G$ be a topological group. A \textdef{principal $G$-bundle} consists of topological spaces $E$ and $X$ and a continuous function $p:E \rightarrow X$ such that:
\begin{enumerate}
  \item for every $x \in X$, the fiber $E_x$ is equipped with a continuous right group action $E_x \times G \rightarrow E_x$,
  \item\label{ite:fiberwise_equivariant} for every $x_0 \in X$, there exists a local trivialization $\phi:p^{-1}(U) \rightarrow U \times G$ of $p$ such that $x_0 \in U$ and for every $x \in U$ the homeomorphism $\phi_x:E_x \rightarrow G$ is equivariant, where $G$ acts on itself by right multiplication. 
\end{enumerate}
In other words, \ref{ite:fiberwise_equivariant} says that $G \rightarrow E \xrightarrow{p} X$ is a fiber bundle and there exists a trivializing atlas  consisting of fiberwise equivariant local trivializations. If $G \rightarrow E \xrightarrow{p} X$ is a principal $G$-bundle, then the actions of $G$ on the fibers $E_x$ combine into a continuous right action of $G$ on the total space $E$. Continuity of this action follows from \ref{ite:fiberwise_equivariant}. 
\end{definition}

One way we will frequently obtain principal $G$-bundles is by constructing sections of transitive group actions. Again our presentation is a slight variation on a well-known result \cite[Sec.~7.4]{Steenrod}. Below, by a \textdef{continuous local section} of a continuous function $p:E \rightarrow X$ we mean a continuous function $s:U \rightarrow E$ defined on an open subset $U \subset X$ such that $p \circ s$ is the inclusion $U \hookrightarrow X$.

\begin{theorem}
Let $G$ be a topological group and let $G$ act continuously and transitively on a topological space $X$. Fix $x_0 \in X$ and let $p:G \rightarrow X$ be the map $p(g) = gx_0$. Given $x \in X$, let $G_x = p^{-1}(x)$ and observe that the stabilizer $G_{x_0}$ acts continuously on $G_{x}$ by right multiplication. If there exists a continuous local section $s:U \rightarrow G$ of $p$ for some neighborhood $x_0 \in U \subset X$, then $G_{x_0} \rightarrow G \xrightarrow{p} X$ is a principal $G_{x_0}$-bundle.
\end{theorem}

\begin{proof}
Let $x \in X$ and choose $g_0 \in G_{x}$. Then $g_0U$ is an open neighborhood of $x$. Define 
\[
\phi: p^{-1}(g_0U) \rightarrow g_0U \times G_{x_0}, \quad \phi(g) = (p(g), s(g_0^{-1}p(g))^{-1}g_0^{-1}g)
\]
Observe that
\[
s(g_0^{-1}p(g))x_0 = p(s(g_0^{-1}p(g))) = g_0^{-1}p(g) = g_0^{-1}gx_0,
\]
so that $s(g_0^{-1}p(g))^{-1}g_0^{-1}g$ is indeed in $G_{x_0}$. We note that $\phi$ is manifestly continuous and fiberwise $G_{x_0}$-equivariant.

Finally, define
\[
\chi :g_0U \times G_{x_0} \rightarrow p^{-1}(g_0U), \quad \chi(c, h) = g_0s(g_0^{-1}c)h.
\]
Observe that 
\[
g_0s(g_0^{-1}c)hx_0 = g_0s(g_0^{-1}c)x_0 = g_0g_0^{-1}c = c,
\]
which proves that indeed $\chi(c,h) \in p^{-1}(g_0U)$. Furthermore, $\chi$ is manifestly continuous. Now we note that
\begin{align*}
\chi(\phi(g)) &= \chi(p(g), s(g_0^{-1}p(g))^{-1}g_0^{-1}g) \\
&= g_0 s(g_0^{-1}p(g))s(g_0^{-1}p(g))^{-1}g_0^{-1}g = g
\end{align*}
and
\begin{align*}
\phi(\chi(y,h)) &= \phi(g_0s(g_0^{-1}y)h)\\
&= (p(g_0s(g_0^{-1}y)h), s(g_0^{-1}p(g_0s(g_0^{-1}y)h))^{-1}g_0^{-1} g_0s(g_0^{-1}y)h)\\
&= (c, s(g_0^{-1}y)^{-1} s(g_0^{-1}y)h  ) = (y, h).
\end{align*}
Thus, $\chi$ is a two-sided continuous inverse for $\phi$, so $\phi$ is a homeomorphism. Thus, $\phi$ is a fiberwise equivariant local trivialization whose base contains the arbitrary point $x \in X$, so Definition \ref{def:principal_G_bundle} is fulfilled.
\end{proof}

\begin{definition}
Let $G \rightarrow E_1 \xrightarrow{p_1} X$ and $G \rightarrow E_2 \xrightarrow{p_2} X$ be principal $G$-bundles over the same base space $X$. A \textdef{$G$-bundle morphism} between them is a continuous equivariant function $f:E_1 \rightarrow E_2$ such that the diagram below commutes.
\[
\begin{tikzcd}
E_1 \arrow[rr,"f"]\arrow[dr,"p_1"']&& E_2\arrow[dl,"p_2"]\\
&X&
\end{tikzcd}
\]
For fixed $G$ and $X$, principal $G$-bundles over $X$ together with $G$-bundle morphisms form a category. One can show that every $G$-bundle morphism is a homeomorphism, and therefore an isomorphism in this category. As with local trivializations, a $G$-bundle morphism $f$ restricts to a homeomorphism $f_x: E_{1,x} \rightarrow E_{2,x}$ for all $x \in X$.
\end{definition}

\subsection{Non-Abelian \v{C}ech Cohomology}
\label{subsec:non_abelian_cech_cohomology}

Non-abelian \v{C}ech cohomology arises naturally in the study of principal fiber bundles in the following way. Let $X$ be a topological space and let $G$ be a topological group. Suppose $G \rightarrow E \xrightarrow{p} X$ is a principal $G$-bundle and let $\cA = (\phi_i)_{i \in I}$ be a trivializing atlas of $p$ indexed by a set $I$, consisting of  equivariant local trivializations $\phi_i:p^{-1}(U_i) \rightarrow U_i \times G$. Suppose $i,j \in I$ and $U_i \cap U_j \neq \varnothing$. Then for $x \in U_i \cap U_j$ we may consider the composition of homeomorphisms
\[
\phi_{i,x}^{\phantom{-1}} \circ \phi_{j,x}^{-1} :G \rightarrow  E_x \rightarrow G
\]
Since $\phi_{i,x}^{\phantom{-1}}$ and $\phi_{j,x}^{-1}$ are equivariant, we see that for any $g \in G$,
\[
(\phi_{i,x}^{\phantom{-1}} \circ \phi_{j,x}^{-1})(g) = \phi^{\phantom{-1}}_{i,x}(\phi^{-1}_{j,x}(1)g) = \phi^{\phantom{-1}}_{i,x}(\phi^{-1}_{j,x}(1))g
\]
where $1 \in G$ is the identity. Thus, $\phi_{i,x}^{\phantom{-1}} \circ \phi_{j,x}^{-1}$ is completely determined by its value at the identity.

Moreover, the \textdef{transition function}
\[
g_{ij} : U_i \cap U_j \rightarrow G, \quad g_{ij}(x) =  \phi^{\phantom{-1}}_{i,x}(\phi^{-1}_{j,x}(1))
\]
is continuous since we can write it as
\[
\phi^{\phantom{-1}}_{i,x}(\phi^{-1}_{j,x}(1)) = (\proj_G \circ \phi_i^{\phantom{1}} \circ \phi_{j}^{-1})(x, 1)
\]
and the right hand side is manifestly continuous in $x$. Finally, we observe that for all $i,j,k \in I$ such that $U_i \cap U_j \cap U_k \neq \varnothing$ and for every $x \in U_i \cap U_j \cap U_k$, equivariance of the local trivializations implies
\begin{align*}
g_{ij}(x)g_{jk}(x) &= \phi^{\phantom{-1}}_{i,x}(\phi^{-1}_{j,x}(1))\phi^{\phantom{-1}}_{j,x}(\phi^{-1}_{k,x}(1)) \\
&= \phi^{\phantom{-1}}_{i,x}(\phi^{-1}_{j,x}(  \phi^{\phantom{-1}}_{j,x}(\phi^{-1}_{k,x}(1)) )) \\
&= \phi^{\phantom{-1}}_{i,x}(\phi^{-1}_{k,x}(1)) = g_{ik}(x).
\end{align*}
This motivates the definition of a \v{C}ech 1-cocycle.

\begin{definition}
Let $G$ be a topological group, let $X$ be a topological space, and let $\cU = \qty(U_i)_{i \in I}$ be an indexed open cover of $X$. Given $i,j,k \in I$, denote $U_{ij} = U_{i} \cap U_j$ and $U_{ijk} = U_i \cap U_j \cap U_k$. 
\begin{itemize}
  \item A \textdef{$\bm{0}$-cochain} $\lambda$ of $\cU$ is a collection of continuous maps $\lambda_i :U_i \rightarrow G$ defined for each $i \in I$. The set of $0$-cochains may be denoted $\check{C}^0(\cU; G)$. Note that $\check{C}^0(\cU; G)$ has a group structure given by pointwise multiplication and inversion.
  \item A \textdef{$\bm{1}$-cochain} $g$ of $\cU$ is a collection of continuous maps $g_{ij}: U_{ij} \rightarrow G$ defined for each pair of indices $i,j \in I$. The set of $1$-cochains may be denoted $\check{C}^1(\cU; G)$. Again, note that $\check{C}^1(\cU; G)$ has a group structure given by pointwise multiplication and inversion.
  \item A \textdef{$\bm{0}$-cocycle} is a $0$-chain $\lambda$ such that
  \[
  \lambda_i(x) = \lambda_j(x)
  \]
  for all $x \in U_{ij}$. The set of $0$-cocycles may be denoted $\check{Z}^0(\cU; G)$. The $0$-cocycles are also by definition the cohomology classes in degree zero, so we write $\check{Z}^0(\cU; G) = \check{H}^0(\cU; G)$ as well. Note that $\check{H}^0(\cU; G)$ is a subgroup of $\check{C}^0(\cU; G)$.
  \item A \textdef{$\bm{1}$-cocycle} is a 1-cochain $g$ such that
  \[
  g_{ij}(x) g_{jk}(x) = g_{ik}(x)
  \]
  for all $i,j,k \in I$ and $x \in U_{ijk}$. The set of $1$-cocycles may be denoted $\check{Z}^1(\cU; G)$.
  \item A \textdef{$\bm{1}$-coboundary} is a 1-cochain $g$ of the form
  \[
  g_{ij}(x) = \lambda_i(x)\lambda_j(x)^{-1}
  \]
  for some $0$-cochain $\lambda$. The set of $1$-coboundaries may be denoted $\check{B}^1(\cU; G)$.
\end{itemize}
\end{definition}

\begin{remark}
\v{C}ech cohomology can be defined more generally to take values in a sheaf of non-abelian groups. This reduces to our definition when using the sheaf of continuous functions from open sets of $X$ into $G$. We will not need the generality of working with arbitrary sheaves, so we stick to the definition above.
\end{remark}

\begin{remark}
Given a $0$-cocycle $(\lambda_i)_{i \in I}$, there exists a unique continuous map $\lambda:X \rightarrow G$ such that $\lambda|_{U_i} = \lambda_i$ for all $i \in I$. Conversely, given a continuous map $\lambda:X \rightarrow G$, one clearly gets a 0-cocycle by defining $\lambda_i = \lambda|_{U_i}$. This correspondence gives a group isomorphism of $\check{H}^0(\cU; G)$ with the group of continuous functions $X \rightarrow G$. We therefore define $\check{H}^0(X;G)$ to be the group of continuous functions $X \rightarrow G$.
\end{remark}

For ease of notation, we will typically drop the argument $x$ and any notation relating to the restricting of functions, so for example the cocycle condition can be written $g_{ij} g_{jk} = g_{ik}$, where this will be understood as an equality of functions defined on $U_{ijk}$. 

\begin{remark}
If $g$ is a 1-cocycle, then $g_{ii} = 1$ and $g_{ji} = g_{ij}^{-1}$ for all $i,j \in I$. Furthermore, if $\lambda$ is a 0-cochain, then the collection of maps 
\begin{equation}\label{eq:Cech_cohomologous_deg_1}
h_{ij} = \lambda_i g_{ij}\lambda_j^{-1}
\end{equation}
form a 1-cocycle, as can easily be checked. Finally, we note that any 1-coboundary is a 1-cocycle.
\end{remark}

\begin{comment}
\begin{proposition}\label{prop:cocycle_index_switch}
If $\cU = \qty{U_i}_{i \in I}$ is an open cover of $X$ and $g \in Z^1(\cU; G)$, then
\begin{enumerate}
  \item $g_{ii} = 1$ for all $i \in I$,
  \item $g_{ji} = g_{ij}^{-1}$ for all $i,j \in I$.
\end{enumerate}
\end{proposition}

\begin{proof}
(a) By the cocycle condition, 
\[
g_{\alpha \alpha}g_{\alpha \alpha} = g_{\alpha \alpha},
\]
hence $g_{\alpha \alpha} = 1$. 

(b) By the cocycle condition and part (a),
\[
g_{\alpha \beta}g_{\beta \alpha} = g_{\alpha \alpha} = 1,
\]
hence $g_{\alpha \beta} = g_{\beta \alpha}^{-1}$.
\end{proof}
\end{comment}

\begin{definition}\label{def:equiv_relation_Z1(U,G)}
Let $\cU = \qty(U_i)_{i \in I}$ be an indexed open cover of $X$. Given $g, h \in \check{Z}^1(\cU; G)$, we write $g \sim h$ if there exists a $0$-cochain $\lambda$ such that \eqref{eq:Cech_cohomologous_deg_1} holds
for all $i,j \in I$. It is easy to check that ${\sim}$ is an equivalence relation. Thus, we define the first \v{Cech} cohomology group of $\cU$ as the set 
\[
\check{H}^1(\cU; G) = \check{Z}^1(\cU; G)/{\sim}.
\]
The equivalence class of $g \in \check{Z}^1(\cU; G)$ will be denoted $\class{g}$. 
\end{definition}

Working with a non-abelian group $G$ causes issues for defining cohomology in degrees higher than one. In general, there is not even a sensible group structure we can put on $\check{Z}^1(\cU; G)$ or $\check{H}^1(\cU; G)$!  However, $\check{H}^1(\cU; G)$ does have a sensible ``zero'' element. Namely, $\check{B}^1(\cU;G)$ is an equivalence class under the equivalence relation defined in Definition \ref{def:equiv_relation_Z1(U,G)}. We call $\check{B}^1(\cU;G)$ the \textdef{trivial class}. Thus, $\check{H}^1(\cU;G)$ is a pointed set.

We would like a notion of cohomology that depends only on the topological space $X$ and the topological group $G$, not on the choice of open cover. The way to achieve this is to take successively refined open covers. Recall that if $\cU = (U_i)_{i \in I}$ and $\cV = (V_j)_{j \in J}$ are indexed open covers of $X$, then $\cV$ is a \textdef{refinement} of $\cU$ if for every $j \in J$ there exists $i \in I$ such that $V_j \subset U_i$. The following theorem, in particular part (d), is somewhat difficult to find in the literature. A version for abelian groups can be found in \cite[Lem~12.3 \& Lem.~12.4]{Forster} and the proof is fortunately more or less the same in the nonabelian case. We have included the full proof.

\begin{theorem}\label{thm:i_VU_injection}
Let $\cU = \qty(U_i)_{i \in I}$ and $\cV = \qty(V_{j})_{j \in J}$ be indexed open covers of $X$ and let $\cV$ be a refinement of $\cU$. Choose a function $\refine:J \rightarrow I$ such that $V_j \subset U_{\refine(j)}$ for all $j \in J$. Given $g \in \check{C}^1(\cU; G)$,  define $g^\refine \in \check{C}^1(\cV; G)$ by
\[
g_{ij}^\refine = g_{\refine(i)\refine(j)}
\]
for all $i,j \in J$, where $g_{\refine(i)\refine(j)}$ is implicitly restricted to $V_{ij} \subset U_{ij}$. The following hold.
\begin{enumerate}
  \item If $g \in \check{Z}^1(\cU;G)$, then $g^\refine \in \check{Z}^1(\cV; G)$.
  \item If $g \in \check{Z}^1(\cU;G)$, then the cohomology class of $g^\refine$ is independent of the choice of function $\refine$.
  \item If $g, h \in \check{Z}^1(\cU; G)$ and $g \sim h$, then $g^\refine \sim h^\refine$.
  \item\label{ite:i_UV_injective} The induced map 
  \[
  \iota_{\cV \cU}:\check{H}^1(\cU; G) \rightarrow \check{H}^1(\cV; G), \quad \iota_{\cV\cU}\class{g} = \class{g^\refine}
  \]
  is injective.
  \item The induced map $\iota_{\cV\cU}$ sends the trivial class to the trivial class.
\end{enumerate}
\end{theorem}

\begin{proof}
(a) Given $i,j,k \in J$,
\begin{align*}
g_{ij}^\refine g_{jk}^\refine =g_{\refine(i)\refine(j)}g_{\refine(j)\refine(k)} = g_{\refine(i)\refine(k)} = g_{ik}^\refine
\end{align*}
Thus, the cocycle condition for $g^\refine$ follows from the cocycle condition for $g$, so $g^\refine \in \check{Z}^1(\cV; G)$.

(b) Choose another function $\refine':J \rightarrow I$ such that $V_j \subset U_{\refine'(j)}$ for all $j \in J$. We may define a 0-cochain $\lambda \in \check{C}^0(\cV; G)$ by
\[
\lambda_j = g_{\refine'(j)\refine(j)}
\]
for all $j \in J$.
Moreover, for any $i,j \in J$, the cocycle condition for $g$ implies
\[
g^{\refine'}_{ij} = g_{\refine'(i)\refine'(j)} = g_{\refine'(i)\refine(i)}g_{\refine(i)\refine(j)}g_{\refine(j)\refine'(j)} = \lambda_i g^\refine_{i \beta} \lambda_j^{-1}
\]
so $g^{\refine'} \sim g^{\refine}$, as desired.

(c) Let $\lambda \in \check{C}^0(\cU; G)$ such that $h_{ij} = \lambda_i g_{ij}\lambda_j^{-1}$ for all $i,j \in I$. Define $\lambda^\refine \in \check{C}^0(\cV; G)$ by $\lambda_j^\refine = \lambda_{\refine(j)}$. Then observe that
\[
h_{ij}^\refine = h_{\refine(i)\refine(j)} = \lambda_{\refine(i)}g_{\refine(i)\refine(j)}\lambda_{\refine(j)}^{-1} = \lambda_{i}^\refine g_{ij}^\refine (\lambda_j^\refine)^{-1},
\]
which proves that $h^\refine \sim g^\refine$.

(d) Suppose $g, h \in \check{Z}^1(\cU; G)$ and $g^\refine \sim h^\refine$. Let $\lambda \in \check{C}^0(\cV; G)$ such that $h_{ij}^\refine = \lambda_i g_{ij}^\refine \lambda_j^{-1}$ for all $i,j \in J$. Fix $k \in I$. For each pair of indices $i,j \in J$ such that $U_k \cap V_i \cap V_j \neq \varnothing$ we have
\begin{align*}
h_{\refine(i)k}h_{k \refine(j)} = h_{\refine(i)\refine(j)}  =  \lambda_i g^\refine_{ij} \lambda_j^{-1} = \lambda_i g_{\refine(i)k}g_{k \refine(j)}\lambda_j^{-1}
\end{align*}
on $U_k \cap V_i \cap V_j$. Rearranging yields
\begin{equation}\label{eq:construct_0-chain_injective}
h_{k \refine(j)}\lambda_j g_{\refine(j) k} = h_{k \refine(i)}\lambda_i g_{\refine(i) k} 
\end{equation}
We may therefore define a function $\mu_k : U_k \rightarrow G$ by
\[
\mu_k(x) = h_{k \refine(j)}(x)\lambda_j(x)g_{\refine(j) k}(x)
\]
where $j \in J$ is chosen so that $x \in V_j$. By \eqref{eq:construct_0-chain_injective}, the value $\mu_k(x)$ is independent of the choice of $j \in J$ satisfying $x \in V_j$. Thus, $\mu_k |_{U_k \cap V_j}$ is continuous for each $j \in J$ and continuity of $\mu_k$ follows since the sets $U_k \cap V_j$ cover $U_k$. Thus, we have a 0-cochain $\mu \in \check{C}^0(\cU; G)$. Finally, given $k, l \in I$ and $x \in U_{kl}$, we choose $j \in J$ such that $x \in V_j$, whence
\begin{align*}
h_{kl}(x) &= h_{k \refine(j)}(x)h_{\refine(j) l}(x) = \qty(\mu_k(x)g_{\refine(j)k}(x)^{-1}\lambda_i(x)^{-1}) \qty(\mu_l(x)g_{\refine(j)l}(x)^{-1}\lambda_j(x)^{-1})^{-1}\\
&= \mu_k(x)g_{k \refine(j)}(x)g_{\refine(j)l}(x)\mu_l(x)^{-1} = \mu_k(x) g_{k l}(x) \mu_l(x)^{-1},
\end{align*}
Therefore, $g \sim h$, as desired.

(e) Let $\lambda \in \check{C}^0(\cU; G)$ and let $g \in \check{B}^1(\cU; G)$ be the associated 1-coboundary $g_{ij} = \lambda_i \lambda_j^{-1}$. Define $\lambda^\refine \in \check{C}^0(\cV; G)$ by $\lambda_j^\refine = \lambda_{\refine(j)}$. Then observe that
\[
g^\refine_{ij} = g_{\refine(i)\refine(j)} = \lambda_{\refine(i)}\lambda_{\refine(j)}^{-1} = \lambda^\refine_i (\lambda^\refine_j)^{-1},
\]
so $g^\refine$ is a 1-coboundary.
\end{proof}

\begin{definition}\label{def:H^1(X;G)_nonabelian}
If $\cU = (U_i)_{i \in I}$, $\cV = (V_j)_{j \in J}$, and $\cW = (W_k)_{k \in K}$ are indexed open covers of $X$ such that $\cW$ is a refinement of $\cV$ and $\cV$ is a refinement of $\cU$, then it is straightforward to check that
\[
\iota_{\cW \cU} = \iota_{\cW \cV} \circ \iota_{\cV \cU}.
\]
It is even more straightforward to check that $\iota_{\cU\cU}$ is the identity on $\check{H}^1(\cU;G)$. It therefore appears that the sets $\check{H}^1(\cU;G)$ and functions $\iota_{\cU \cV}$ form a directed system in the category $\Set_*$ of pointed sets and basepoint-preserving functions and we would like to define $\check{H}^1(X;G)$ to be the directed colimit. But we encounter a logical subtlety at this point, namely that the collection of all indexed open covers of $X$ does not form a set since we have allowed the indexing set $I$ to be arbitrary. We can in general only take directed colimits over directed systems indexed by a set.

To deal with this, we observe that given an indexed open cover $\cU = (U_i)_{i \in I}$ we can index the underlying set $\qty{U_i:i \in I}$ by itself, with indexing function equal to the identity function,  obtaining a new indexed open cover $\cU'$. Since $\cU$ and $\cU'$ are refinements of each other, the map $\iota_{\cU'\cU}$ is a bijection with inverse $\iota_{\cU\cU'}$. The collection of all open covers indexed by themselves does form a set, and we can take the directed colimit
\[
\check{H}^1(X;G) = \colim_{\cU'} \check{H}^1(\cU';G)
\]
over all open covers $\cU'$ of $X$ indexed by themselves.

Having defined things properly we can now pretty much forget about this logical subtlety since $\check{H}^1(X;G)$ behaves like we would expect a directed colimit over all indexed open covers to behave. For every indexed open cover $\cU = (U_i)_{i \in I}$, there exists a canonical injection $\iota_\cU:\check{H}^1(\cU;G) \rightarrow \check{H}^1(X; G)$ defined by $\iota_{\cU} = \iota_{\cU'} \circ \iota_{\cU'\cU}$, where $\cU'$ is as above and $\iota_{\cU'}$ is the canonical embedding into the directed colimit. If $\cU$ and $\cV$ are indexed open covers and we have classes $\class{g} \in \check{H}^1(\cU;G)$ and $\class{h} \in \check{H}^1(\cV;G)$ satisfying $\iota_{\cU}(\class{g}) = \iota_\cV(\class{g})$, then there exists a refinement $\cW$ of $\cU$ and $\cV$ such that $\iota_{\cW\cU}\class{g} = \iota_{\cW \cV}\class{h}$. If $\cV$ is a refinement of $\cU$, then $\iota_{\cV} \circ \iota_{\cV \cU} = \iota_\cU$. 
\end{definition}

Let us return to the topic of principal fiber bundles. At the start of this section we showed how a principal $G$-bundle $G \rightarrow E \xrightarrow{p} B$ and a trivializing atlas $\cA = (\phi_i)_{i \in I}$ with bases $\cU = \qty(U_i)_{i \in I}$ give rise to a \v{C}ech 1-cocycle $g \in \check{Z}^1(\cU;G)$. This is in fact the beginning of a bijective correspondence between isomorphism classes of principal $G$-bundles and $\check{H}^1(X;G)$. It is again difficult to find a detailed proof of the result below, although it is easy to find the statement, see e.g., \cite{Brylinski,Hirzebruch,Lawson}. This is probably because the proof is somewhat long but largely straightforward. Let us nonetheless provide a detailed proof for the convenience of the reader.

\begin{theorem}\label{thm:PrinG(X)=H^1(X,G)}
Let $G \rightarrow E \xrightarrow{p} X$ be a principal $G$-bundle and let $\cA = (\phi_i)_{i \in I}$ be a trivializing atlas of equivariant local trivializations with bases $\cU = (U_i)_{i \in I}$. For each $i,j \in I$ and $x \in U_{ij}$, define
\[
g_{ij}(x) = (\phi_{i,x}^{\phantom{-1}} \circ \phi_{j,x}^{-1})(1)
\]
Then $g \in \check{Z}^1(\cU;G)$. Furthermore, the following hold.
\begin{enumerate}
  \item\label{ite:refine_G-bundle_cycle} If $\cV = (V_j)_{j \in J}$ is a refinement of $\cU$ and $\refine:J \rightarrow I$ is a function such that $V_j \subset U_{\refine(j)}$ for all $j \in J$, then the trivializing atlas $\cB = (\chi_j)_{j \in J}$, where $\chi_j:p^{-1}(V_j) \rightarrow V_j \times G$ is defined as the restriction of $\phi_{\refine(j)}$, gives rise to the cocycle $g^r$.
  \item\label{ite:atlas_independence_cycle} The class $\iota_\cU\class{g} \in \check{H}^1(X;G)$ is independent of the choice of trivializing atlas $\cA$.
  \item\label{ite:iso_class_cycle_independence} Isomorphic principal $G$-bundles give rise to the same class in $\check{H}^1(X;G)$.
\end{enumerate}
Thus, we have a map from isomorphism classes of principal $G$-bundles to $\check{H}^1(X;G)$. This map is a bijection.
\end{theorem}

\begin{proof}
\ref{ite:refine_G-bundle_cycle} We observe that $\chi_{j,x} = \phi_{\refine(j),x}$ for all $j \in J$ and $x \in V_j$. Thus, 
\[
(\chi_{i,x}^{\phantom{-1}} \circ \chi_{j,x}^{-1})(1) = (\phi_{\refine(i),x}^{\phantom{-1}} \circ \phi_{\refine(j),x}^{-1})(1) = g_{ij}^r(x)
\]
for all $x \in V_{ij}$, as desired.

\ref{ite:atlas_independence_cycle} Suppose $\cB = (\chi_j)_{j \in J}$ is another trivializing atlas of equivariant local trivializations with bases $\cV = (V_j)_{j \in J}$. It follows from \ref{ite:refine_G-bundle_cycle} that the class $\iota_\cU\class{g}$ is unchanged by taking a refinement of the set of bases and restricting the local trivializations. Therefore, by taking a mutual refinement $\cW = (W_k)_{k \in K}$ of $\cU$ and $\cV$ we may assume that our trivializing atlases have a common indexing set and base, i.e., $\cA = (\phi_k)_{k \in K}$ and $\cB = (\chi_k)_{k \in K}$ and $W_k$ is the base of both $\phi_k$ and $\chi_k$. 

Now, for each $k \in K$ we define 
\[
\lambda_k(x) = (\phi_{k,x}^{\phantom{-1}} \circ \chi_{k,x}^{-1})(1)
\]
for all $x \in W_k$. Equivariance of the local trivializations implies $\lambda_k(x)^{-1} = (\chi_{k,x}^{\phantom{-1}} \circ \phi_{k,x}^{-1})(1)$. Furthermore, for $i,j \in K$ equivariance of the local trivializations implies
\begin{align*}
\lambda_{i}(x)\cdot (\chi_{i,x}^{\phantom{-1}} \circ \chi_{j,x}^{-1})(1) \cdot \lambda_{j}(x)^{-1} = (\phi_{i,x}^{\phantom{-1}} \circ \phi_{j,x}^{-1})(1)
\end{align*}
This proves that $\cA$ and $\cB$ give rise to cohomologous cocycles.

\ref{ite:iso_class_cycle_independence} Let $G \rightarrow E' \xrightarrow{p'} X$ be another principal $G$-bundle and let $f:E \rightarrow E'$ be an isomorphism. Let $\cA = (\phi_i)_{i \in I}$ be a trivializing atlas of $p'$ consisting of equivariant local trivializations with bases $\cU = (U_i)_{i \in I}$. We observe that $f$ restricts to a homeomorphism $p^{-1}(U_i) \rightarrow p'^{-1}(U_i)$ for all $i \in I$ and $\phi_i \circ f_i :p^{-1}(U_i) \rightarrow U_i \times G$ is a local trivialization for $p$. Thus, $(\phi_i \circ f_i)_{i \in I}$ is a trivializing atlas for $p$ and for all $i,j \in I$ we have
\[
\qty((\phi_{i} \circ f)_x \circ (\phi_{j} \circ f)_x^{-1})(1) = (\phi_{i,x}^{\phantom{-1}} \circ f_{x}^{\phantom{1}} \circ f_{x}^{-1} \circ \phi_{j,x}^{-1})(1) = (\phi_{i,x}^{\phantom{-1}} \circ \phi_{j,x}^{-1})(1).
\] 
Therefore $p$ and $p'$ define the same class in $\check{H}^1(X;G)$.

It remains to show that the correspondence between isomorphism classes of principal $G$-bundles and $\check{H}^1(X;G)$ is a bijection. Suppose and $G \rightarrow E \xrightarrow{p} X$ and $G \rightarrow E' \xrightarrow{p'} X$ define the same class in $\check{H}^1(X;G)$. By refining as in part \ref{ite:refine_G-bundle_cycle} if necessary, we can find trivializing atlases $\cA = (\phi_i)_{i \in I}$ for $p$ and $\cB = (\chi_i)_{i \in I}$ for $p'$ with common bases $\cU = (U_i)_{i \in I}$ such that the cocycle $g \in \check{Z}^1(\cU;G)$ defined by $\cA$ is cohomologous to the cocycle $h \in \check{Z}^1(\cU;G)$ defined by $\cB$. Thus, there exists $\lambda \in \check{C}^0(\cU;G)$ such that $h_{ij} = \lambda_i g_{ij} \lambda_j^{-1}$ for all $i,j \in I$. 

Given $i \in I$, define $\hat f_i :U_i \times G \rightarrow U_i \times G$ as 
\[
 \hat f_i(x,g) = (x, \lambda_i(x)g).
\]
This is clearly continuous. We then define a function $f_i : p^{-1}(U_i) \rightarrow E'$ as
\begin{equation}\label{eq:f_i_def}
 f_i(a) = \chi_i^{-1}(\hat f(\phi_i^{\phantom{1}}(a))) = \chi_{i,x}^{-1}\qty(1)\lambda_i^{\phantom{1}}(x)\phi_{i,x}^{\phantom{-1}}(a)
\end{equation}
where $x \defeq p(a)$ and we have used equivariance of $\chi_{i,x}^{-1}$. We show that $f_i|_{p^{-1}(U_{ij})} = f_j|_{p^{-1}(U_{ij})}$. If $a \in p^{-1}(U_{ij})$ and $x \defeq p(a)$, then
\begin{align*}
f_i(a) &= \chi_{j,x}^{-1}(\chi_{j,x}^{\phantom{-1}}(\chi_{i,x}^{-1}(1))) \lambda_i^{\phantom{1}}(x) \phi_{i,x}^{\phantom{-1}}(a)\\
&= \chi_{j,x}^{-1}(h_{ji}^{\phantom{1}}(x))\lambda_i^{\phantom{1}}(x)\phi_{i,x}^{\phantom{-1}}(a)\\
&= \chi_{j,x}^{-1}(1)\lambda_j(x)g_{ji}(x)\phi_{i,x}^{\phantom{-1}}(a)\\
&= \chi_{j,x}^{-1}(1)\lambda_j(x)\phi_{j,x}(a) = f_j(a)
\end{align*}
as desired. Thus, there exists a continuous function $f:E \rightarrow E'$ such that $f|_{p^{-1}(U_i)} = f_i$. From \eqref{eq:f_i_def} we see that $f_i$ commutes with the projections $p$ and $p'$ and that $f_i$ is equivariant, so the same is true for $f$. Thus, $f$ is a morphism, hence and isomorphism, of principal $G$-bundles.

Finally, we show that every class in $\check{H}^1(X;G)$ corresponds to a principal $G$-bundle. Fix a class in $\check{H}^1(X;G)$ and represent it by $(g_{ij})_{i,j \in I} \in \check{Z}^1(\cU;G)$ for some indexed open cover $\cU = (U_i)_{i \in I}$. Define
\[
\tilde E = \bigsqcup_{i \in I} U_i \times G
\]
Given $(i, (x, g)), (j, (y, h)) \in \tilde E$, set $(i, (x, g)) \sim (j, (y, h))$ if $x = y$ and $g = g_{ij}(x)h$. Reflexivity follows from $g_{ii} = 1$, symmetry follows from $g_{ij} = g_{ji}^{-1}$, and transitivity follows from the cocycle condition, so ${\sim}$ is an equivalence relation. Let $E = E/{\sim}$. Evidently the projection $\tilde E \rightarrow B$, $(x, g) \mapsto x$ factors through a map $p:E \rightarrow B$, $p[(i,(x,g))] = x$, where $[(i,(x,g))]$ is the equivalence class of $(i,(x,g)) \in \tilde E$.

Given $i \in I$, we claim that the map $\phi_i:p^{-1}(U_i) \rightarrow U_i \times G$
\[
\phi_i[(j,(x,g))] = (x, g_{ij}(x)g)
\]
is well-defined. If $(j, (x, g)) \sim (k, (x, h))$, then 
\[
g_{ij}(x)g = g_{ij}(x)g_{jk}(x)h = g_{ik}(x)h,
\]
which proves that $\phi_i$ is well-defined. We show that $\phi$ is bijective. If $\phi_i[(j,(x,g))] = \phi_i[(k, (x, h))]$, then $g_{ij}(x)g = g_{ik}(x)h$, hence 
\[
g = g_{ji}(x)g_{ik}(x)h = g_{jk}(x)h,
\]
so $(j,(x,g)) \sim (k, (x, h))$. Therefore $\phi_i$ is injective. Given $(x, g) \in U_i \times G$, we have $\phi_i[(i, (x, g))] = (x, g_{ii}(x)g) = (x, g)$, so $\phi_i$ is surjective. 

It is clear from the definition that $\phi_i$ commutes with the projections onto $U_i$. Given $i,j \in I$ and $(x, g) \in U_{ij} \times G$, we observe that
\[
(\phi_i^{\phantom{1}} \circ \phi_j^{-1})(x, g) = \phi_i[(j, (x, g))] = (x, g_{ij}(x)g)
\]
This is manifestly continuous. Thus, Theorem \ref{thm:topologize_total_space} yields a unique topology on $E$ such that $p$ is continuous, each $\phi_i$ is a homeomorphism, and $(E, X, p)$ is a fiber bundle. We observe that for any $i,j \in I$, $x \in U_{ij}$, and $g \in G$ we have
\begin{equation}\label{eq:cocycle_achieved}
(\phi_{i,x}^{\phantom{-1}} \circ \phi_{j,x}^{-1})(g) = g_{ij}(x)g.
\end{equation}

To make this a principal $G$-bundle we need to define a right action on each fiber $E_x$ and show that each $\phi_i$ is fiberwise equivariant. Fix $x \in X$ and choose $i \in I$ such that $x \in U_i$. Given $a \in E_x$ and $g \in G$ we define
\begin{equation}\label{eq:action_def}
ag = \phi_{i,x}^{-1}(\phi_{i,x}^{\phantom{-1}}(a)g)
\end{equation}
This is jointly  continuous in $a$ and $g$ since $\phi_{i,x}$ is a homeomorphism. We observe that for any $a \in E_x$ and $g, h \in G$ we have
\[
(ag)h = \phi_{i,x}^{-1}(\phi_{i,x}^{\phantom{-1}}(ag)h) = \phi_{i,x}^{-1}(\phi^{\phantom{-1}}_{i,x}(\phi_{i,x}^{-1}(\phi_{i,x}^{\phantom{-1}}(a)g))h) = \phi_{i,x}^{-1}(\phi_{i,x}^{\phantom{-1}}(a)gh) = a(gh),
\]
so this is a group action. If $j \in J$ such that $x \in U_j$, then
\begin{align*}
\phi_{i,x}^{-1}(\phi_{i,x}^{\phantom{-1}}(a)g) &= \phi_{j,x}^{-1}(\phi_{j,x}^{\phantom{-1}}(\phi_{i,x}^{-1}(\phi_{i,x}^{\phantom{-1}}(a)g)))\\
&= \phi_{j,x}^{-1}(g_{ji}^{\phantom{1}}(x)\phi_{i,x}^{\phantom{-1}}(a)g) \\
&= \phi_{j,x}^{-1}(g_{ji}^{\phantom{1}}(x)\phi_{i,x}^{\phantom{-1}}(\phi_{j,x}^{-1}(\phi_{j,x}^{\phantom{-1}}(a)))g)\\
&= \phi_{j,x}^{-1}(g_{ji}^{\phantom{1}}(x)g_{ij}^{\phantom{1}}(x)\phi_{j,x}^{\phantom{-1}}(a)g)\\
&= \phi_{j,x}^{-1}(\phi_{j,x}^{\phantom{-1}}(a)g).
\end{align*}
Thus, the definition of the action is independent of the choice of $i \in I$ such that $x \in U_i$. Therefore for any $j \in J$, $x \in U_j$, $a \in E_x$, and $g \in G$ we have $\phi_{j,x}(ag) = \phi_{j,x}(a)g$, so each $\phi_j$ is fiberwise equivariant. Thus, $G \rightarrow E \xrightarrow{p} X$ is a principal $G$-bundle with trivializing atlas $\cA = (\phi_i)_{i \in I}$, and $\cA$ induces the desired cocycle $(g_{ij})_{i,j \in I} \in \check{Z}^1(\cU;G)$. 
\end{proof}

A continuous function $f:X \rightarrow Y$ between topological spaces induces a continuous function $f^*:\check{H}^1(Y;G) \rightarrow \check{H}^1(X;G)$ in the following way. Given a 1-cocycle $g \in \check{Z}^1(\cU;G)$ with respect to an open cover $\cU = (U_i)_{i \in I}$ of $Y$, we note that $f^{-1}(\cU) = (f^{-1}(U_i))_{i \in I}$ is an open cover of $X$ and we define $f^*g \in \check{C}^1(f^{-1}(\cU);G)$ by $(f^*g)_{ij} = g_{ij} \circ f$. It is straightforward to show that $f^*g$ is a cocycle and $f^*$ descends to a map on cohomology $f^*:\check{H}^1(X;G) \rightarrow \check{H}^1(Y;G)$ satisfying $f^*\iota_\cU \class{g} = \iota_{f^{-1}(\cU)}\class{f^*g}$ for all open covers $\cU$ of $Y$ and $g \in \check{Z}^1(\cU;G)$. The theorem below gives conditions under which homotopic maps induce the same map on cohomology. It follows from \cite[Thm.~11.5]{Steenrod} and the correspondence Theorem \ref{thm:PrinG(X)=H^1(X,G)} (see also \cite[Appendix A]{Lawson}).

\begin{theorem}\label{thm:homotopy_invariance_Cech_1}
Let $X$ be locally compact, normal, and Lindel\"of. If $f_1:X \rightarrow Y$ and $f_2:X \rightarrow Y$ are homotopic, then the induced maps on cohomology $f_1^*, f_2^*:\check{H}^1(Y;G) \rightarrow \check{H}^1(X;G)$ are equal.
\end{theorem}

\subsection{Exact Sequences}
\label{subsec:exact_sequences}

When $G$ is an abelian topological group, we can define \v{C}ech cohomology in all degrees. 

\begin{definition}
Let $X$ be a topological space and let $G$ be an abelian topological group. Let $\cU = (U_i)_{i \in I}$ be an indexed open cover of $X$. Given $n \in \bbN \cup \qty{0}$, an \textdef{$n$-cochain} is a collection of continuous maps $g_{i_0 \cdots i_n}:U_{i_0 \cdots i_n} \rightarrow G$ defined for all collections of $n+1$ indices $i_0,\ldots, i_n \in I$, where $U_{i_0 \cdots i_n} = U_{i_0} \cap \cdots \cap U_{i_n}$. The set of $n$-cochains is denoted $\check{C}^n(\cU;G)$ and forms an abelian group with respect to pointwise addition of functions.

For each $n \in \bbN \cup \qty{0}$, we define a \textdef{differential} $d:\check{C}^n(\cU;G) \rightarrow \check{C}^{n+1}(\cU;G)$ by
\[
dg_{i_0\cdots i_{n+1}} = \sum_{k=0}^{n+1} (-1)^k g_{i_0 \cdots \widehat{i_k} \cdots i_{n+1}}
\]
where the $\widehat{i_k}$ indicates that $i_k$ is deleted from the list of indices.  Each of the addends is implicitly restricted to $U_{i_0 \cdots i_{n+1}}$. It is clear that $d$ is a homomorphism and one can show that $d^2 = 0$, so the groups $\check{C}^n(\cU;G)$ form a cochain complex. We define $\check{H}^n(\cU;G)$ to be the cohomology of this cochain complex.

If $\cV$ is a refinement of $\cU$, one can define a natural homomorphism $\iota_{\cV \cU}:\check{H}^n(\cU;G) \rightarrow \check{H}^n(\cV;G)$ and then define $\check{H}^n(X;G)$ as a directed colimit with canonical maps $\iota_\cU:\check{H}^n(\cU;G) \rightarrow \check{H}^n(X;G)$, as in Definition \ref{def:H^1(X;G)_nonabelian}.
\end{definition}

\begin{remark}
One can define \v{C}ech cohomology for pairs $(X, A)$, where $X$ is a topological space and $A$ is a subspace of $X$, and show that \v{C}ech cohomology satisfies all the Eilenberg--Steenrod axioms \cite{EilenbergSteenrod}. 
\end{remark}

If $G$ and $H$ are topological groups and $\eta:G \rightarrow H$ is a continuous group homomorphism, then we have an induced homomorphism $\eta_*:\check{C}^n(\cU;G) \rightarrow \check{C}^n(\cU;H)$, $\eta_*g_{i_0 \cdots i_n} = \eta \circ g_{i_0 \cdots i_n}$. This commutes with the differential and therefore descends to a map on cohomology $\eta_* :\check{H}^n(\cU;G) \rightarrow \check{H}^n(\cU;H)$. The map on cohomology commutes with the refinement maps $\iota_{\cV \cU}$ and therefore $\eta_*$ descends to a map $\eta_*:\check{H}^n(X;G) \rightarrow \check{H}^n(X;H)$. The following theorem is a more specialized version of \cite[Thm.~2.10.1]{Hirzebruch}, tailored to our purposes.

\begin{theorem}\label{thm:Cech_long_exact_sequence_abelian}
Let $X$ be paracompact Hausdorff and let
\[
0 \longrightarrow F \xrightarrow{\,\,\,\zeta\,\,\,} G \xrightarrow{\,\,\,\eta\,\,\,} H \longrightarrow 0
\]
be a short exact sequence of topological abelian groups and continuous homomorphisms such that $F \longrightarrow G$ is a topological embedding and $G \longrightarrow H$ is a fiber bundle with typical fiber $F$. There exists a long exact sequence
\begin{align*}
\cdots\, \longrightarrow \check{H}^n(X;F) \xrightarrow{\,\,\zeta_*\,\,} \check{H}^n(X;G) \xrightarrow{\,\,\eta_*\,\,} \check{H}^n(X;H) \xrightarrow{\,\,\delta\,\,} \check{H}^{n+1}(X;F) \longrightarrow \, \cdots 
\end{align*}
\end{theorem}

The connecting homomorphisms $\delta:\check{H}^n(X;H) \rightarrow \check{H}^{n+1}(X;F)$ can be described as follows. Given $\xi \in \check{H}^n(X;H)$, write $\xi = \iota_\cU\class{h}$ where $\cU= (U_i)_{i \in I}$ is an indexed open cover of $X$ and $h$ is an $n$-cocycle such that $h_{i_0\cdots i_n}$ has image contained in the base of a local trivialization for each collection of indices $i_0,\ldots, i_n \in I$ (this is possible since $X$ is paracompact Hausdorff). Lift $h$ to an $n$-cochain $g \in \check{C}^n(\cU;G)$ such that $\eta \circ g = h$. Observe that $\eta \circ dg = dh = 0$ and therefore there exists an $f \in \check{C}^{n+1}(\cU;F)$ such that $\zeta \circ f = dg$. We observe that $\zeta \circ df = d^2 g = 0$ so that $f$ is a cocycle, and we define $\delta \xi = \iota_\cU\class{f}$.

When $F$, $G$, and $H$ are nonabelian, one can obtain a weaker form of the exact sequence above. The sequence stops after $n = 1$ since cohomology is not defined for $n \geq 2$. Furthermore, since $\check{H}^1(X;G)$ is only a pointed set in general, we have an exact sequence of pointed sets and basepoint-preserving maps. Precisely, we say a sequence of pointed sets and basepoint-preserving maps
\[
(A, a) \xrightarrow{\,\,f\,\,} (B,b)  \xrightarrow{\,\,g\,\,} (C, c)
\]
is exact if $f(A) = g^{-1}(\qty{c})$. The following theorem is then a specialization of \cite[Thm.~4.1.3, Thm.~4.1.4, \& Thm.~4.18]{Brylinski} tailored for our purposes.

\begin{theorem}\label{thm:Cech_long_exact_sequence}
Let $X$ be paracompact Hausdorff and let
\[
0 \longrightarrow F \xrightarrow{\,\,\zeta\,\,} G \xrightarrow{\,\,\eta\,\,} H \longrightarrow 0
\]
be a short exact sequence of (not-necessarily-abelian) topological groups and continuous homomorphisms such that $F \longrightarrow G$ is a topological embedding and $G \longrightarrow H$ is a fiber bundle with typical fiber $F$. There exists an exact sequence
\begin{align*}
0 \longrightarrow \check{H}^0(X;F) &\xrightarrow{\,\,\zeta_*\,\,} \check{H}^n(X;G) \xrightarrow{\,\,\eta_*\,\,} \check{H}^0(X;H) \\
\xrightarrow{\,\,\delta_0\,\,} \check{H}^1(X;F) &\xrightarrow{\,\,\zeta_*\,\,} \check{H}^1(X;G) \xrightarrow{\,\,\eta_*\,\,} \check{H}^1(X;H)
\end{align*}
of pointed sets and basepoint-preserving maps. If $\zeta(F)$ is in the center of $G$, then the sequence continue one step further.
\begin{align*}
\phantom{\check{H}^2(X;F) \longrightarrow} 0 \longrightarrow \check{H}^0(X;F) &\xrightarrow{\,\,\zeta_*\,\,} \check{H}^n(X;G) \xrightarrow{\,\,\eta_*\,\,} \check{H}^0(X;H) \\
\xrightarrow{\,\,\delta_0\,\,} \check{H}^1(X;F) &\xrightarrow{\,\,\zeta_*\,\,} \check{H}^1(X;G) \xrightarrow{\,\,\eta_*\,\,} \check{H}^1(X;H) \xrightarrow{\,\,\delta_1\,\,} \check{H}^2(X;F)
\end{align*}
\end{theorem}

The connecting homomorphisms $\delta_0$ and $\delta_1$ are defined similarly to the abelian case. In particular, $\delta_1$ is defined as follows. Given $\xi \in \check{H}^1(X;H)$, we choose an open cover $\cU$ and a 1-cocycle $h$ such that $\xi = \iota_\cU \class{h}$ and the image of $h_{ij}$ is contained in the base of a local trivialization for all $i,j \in I$. We can then find a lift $g \in \check{C}^1(\cU;G)$ such that $\eta \circ g_{ij} = h_{ij}$ for all $i,j \in I$. We then define $f_{ijk} = g_{ik}^{-1}g_{ij}g_{jk}$ for all $i,j,k \in I$. One can show that $f$ is a 2-cocycle, and $\delta_1\xi = \iota_\cU\class{f}$.

Two important examples of the sequence 
\[
0 \longrightarrow F \xrightarrow{\,\,\,\zeta\,\,\,} G \xrightarrow{\,\,\,\eta\,\,\,} H \longrightarrow 0
\]
are provided by 
\begin{equation}\label{eq:Z->R->U(1)}
0 \longrightarrow \bbZ \xrightarrow{\,\,\,\zeta\,\,\,} \bbR \xrightarrow{\,\,\,\eta\,\,\,} \Unitary(1) \longrightarrow 0, \qquad \zeta(x) = x, \qquad \eta(x) = e^{2\pi ix}
\end{equation}
and
\begin{equation}\label{eq:U(1)->U(H)->PU(H)}
0 \longrightarrow \Unitary(1) \xrightarrow{\,\,\,\zeta\,\,\,} \Unitary(\hilbH) \xrightarrow{\,\,\,\eta\,\,\,} \PU(\hilbH) \longrightarrow 0, \qquad \zeta(\lambda) = \lambda \1, \qquad \eta(U) = \bbU.
\end{equation}
In \eqref{eq:U(1)->U(H)->PU(H)}, $\hilbH$ is a complex Hilbert space, $\Unitary(\hilbH)$ is its unitary group, $\1$ is the identity operator, $\PU(\hilbH) = \Unitary(\hilbH)/\Unitary(1)$ is the projective unitary group, and $\bbU$ is the coset represented by $U$. We will review the projective unitary group and the fiber bundle $\Unitary(1) \rightarrow \Unitary(\hilbH) \rightarrow \PU(\hilbH)$ in detail in Section \ref{subsec:PU(H)}. For now, we note that $\Unitary(\hilbH)$ is a topological group in the strong operator topology \cite{SchottenloherUnitaryStrongTopology}; we endow $\Unitary(\hilbH)$ with this topology and we give $\PU(\hilbH)$ the quotient topology induced by the projection $\Unitary(\hilbH) \rightarrow \PU(\hilbH)$. Using \eqref{eq:Z->R->U(1)} and \eqref{eq:U(1)->U(H)->PU(H)} in Theorems \ref{thm:Cech_long_exact_sequence_abelian} and \ref{thm:Cech_long_exact_sequence}, respectively, we obtain maps
\begin{equation}\label{eq:DD_sequence}
\check{H}^1(X;\PU(\hilbH)) \longrightarrow \check{H}^2(X;\Unitary(1)) \longrightarrow \check{H}^3(X;\bbZ).
\end{equation}

Using a partition of unity, one can show that $\check{H}^n(X;\bbR) = 0$ for all $n \geq 1$ when $X$ is paracompact Hausdorff \cite[Example~A.5]{Lawson}. Thus, Theorem \ref{thm:Cech_long_exact_sequence_abelian} implies that the map $\check{H}^2(X;\Unitary(1)) \rightarrow \check{H}^3(X;\bbZ)$ is an isomorphism. Following Dixmier--Douady \cite{DixmierDouady}, Brylinski shows that for a contractible group $G$, the map $\check{H}^1(X;G) \rightarrow \check{H}^2(X;H)$ in Theorem \ref{thm:Cech_long_exact_sequence} is a bijection \cite[Thms.~4.1.6, 4.1.7, 4.1.8]{Brylinski}. 

It is well-known that $\Unitary(\hilbH)$ is contractible in the norm topology for a separable infinite-dimensional Hilbert space; this is known as Kuiper's theorem \cite{Kuiper}. This can also be shown to hold for the compact-open topology on $\Unitary(\hilbH)$ \cite{AtiyahSegalTwistedKTheory}. Schottenloher has shown that the compact-open topology coincides with the strong operator topology on $\Unitary(\hilbH)$ and therefore $\Unitary(\hilbH)$ is contractible with the strong operator topology for a separable infinite-dimensional Hilbert space \cite{SchottenloherUnitaryStrongTopology}. We conclude that both maps in \eqref{eq:DD_sequence} are bijections.

\section{The Topology of Projective Hilbert Space}
\label{sec:PH}

In the study of parametrized phases, it is often important to work with projective Hilbert space, rather than Hilbert space, as the space of quantum states. We conduct a review of projective Hilbert space and its topology here. 

\subsection{Metrics}
\label{subsec:metrics}

There are many ways of defining the topology on projective Hilbert space and we will show that they are all equivalent. This section originated in \cite{PflaumGeometryCQF}; the version that appears below is taken from \cite{Spiegel} with minor edits.

Throughout this section, let $\hilbH$ be a nonzero complex Hilbert space with unit sphere denoted by $\bbS \hilbH = \qty{\Psi \in \hilbH: \norm{\Psi} = 1}$. 

\begin{definition}
The \textdef{projective Hilbert space of $\hilbH$} is the set
\[
\bbP \hilbH = \qty{\bbC \Psi: \Psi \in \hilbH \setminus \qty{0}},
\]
where $\bbC \Psi = \qty{\lambda \Psi : \lambda \in \bbC}$ is the one-dimensional subspace of $\hilbH$ generated by $\Psi$.  The projective Hilbert space may also be denoted $\bbP(\hilbH)$. The elements of $\bbP\hilbH$ are called \textdef{rays}. When we don't need a representative $\Psi$ of a ray, we will label rays with lowercase script letters $\scrj$, $\scrk$, $\scrl$, \ldots 

There is a canonical projection
\[
p:\hilbH \setminus \qty{0} \rightarrow \bbP \hilbH, \quad p(\Psi) = \bbC \Psi
\]
We endow $\bbP \hilbH$ with the quotient topology obtained from this projection, where $\hilbH \setminus \qty{0}$ is given the subspace topology inherited from $\hilbH$.
\end{definition}

\begin{proposition}
The restriction $p|_{\bbS \hilbH}:\bbS \hilbH \rightarrow \bbP \hilbH$ is surjective and the quotient topology on $\bbP \hilbH$ obtained from $p|_{\bbS \hilbH}$ coincides with the quotient topology obtained from $p$. Both $p$ and $p|_{\bbS \hilbH}$ are open maps.
\end{proposition}

\begin{proof}
It is obvious that $p|_{\bbS \hilbH}$ is surjective. Let $\cT$ and $\cT_{\bbS \hilbH}$ be the quotient topologies obtained from $p$ and $p|_{\bbS \hilbH}$, respectively. Consider the identity map $\id:(\bbP \hilbH, \cT) \rightarrow (\bbP \hilbH, \cT_{\bbS \hilbH})$, as well as the map 
\[
f:\hilbH \setminus \qty{0} \rightarrow \bbS \hilbH, \quad f(\Psi) = \frac{\Psi}{\norm{\Psi}},
\]
which is manifestly continuous. Observe that $\id \circ p = p|_{\bbS \hilbH} \circ f$, which is continuous as a map $\hilbH \setminus \qty{0} \rightarrow (\bbP \hilbH, \cT_{\bbS \hilbH})$. Thus, $\id$ is continuous by the universal property of the quotient topology $\cT$. Furthermore, $\id^{-1} \circ p|_{\bbS \hilbH} = p \circ \iota$, where $\iota:\bbS \hilbH \rightarrow \hilbH \setminus \qty{0}$ is the inclusion. Thus, $\id^{-1} \circ p|_{\bbS \hilbH}$ is continuous as a map $\bbS \hilbH \rightarrow (\bbP \hilbH, \cT)$, so $\id^{-1}$ is continuous by the universal property of the quotient topology $\cT_{\bbS \hilbH}$. Since $\id$ is a homeomorphism, we conclude that $\cT = \cT_{\bbS \hilbH}$.

Now we show that $p$ and $p|_{\bbS \hilbH}$ are open. Let $U \subset \hilbH \setminus \qty{0}$ be open. It is easily checked that
\[
p^{-1}(p(U)) = \bigcup_{\lambda \in \bbC^\times} \lambda U,
\]
which is open since $\lambda U$ is open for every $\lambda \in \bbC^\times$. Here $\bbC^\times = \bbC \setminus \qty{0}$. Thus, $p(U)$ is open by definition of $\cT$, so $p$ is an open map. Restricting to $\bbS \hilbH$, we note that an arbitrary open subset of $\bbS \hilbH$ can be written as $U \cap \bbS \hilbH$ where $U \subset \hilbH \setminus \qty{0}$ is open, and
\[
p|_{\bbS \hilbH}^{-1}\qty(p|_{\bbS \hilbH}^{\phantom{-1}}(U \cap \bbS \hilbH)) = \bbS \hilbH \cap \bigcup_{\lambda \in \Unitary(1)} \lambda U.
\]
This is open in $\bbS \hilbH$, so we conclude that $p|_{\bbS \hilbH}$ is open by definition of $\cT_{\bbS \hilbH}$.
\end{proof}

While $\bbP \hilbH$ is not a vector space, it still inherits a lot of structure from $\hilbH$.

\begin{definition}
Given two rays $\bbC \Psi, \bbC \Omega \in \bbP \hilbH$ represented by $\Psi, \Omega \in \hilbH \setminus \qty{0}$, we define their \textdef{ray product} or \textdef{transition amplitude} as
\[
\ray{\bbC\Psi, \bbC \Omega}  = \frac{\abs{\ev{\Psi,  \Omega}}}{\norm{\Psi}\norm{\Omega}}.
\]
Note that by the Cauchy-Schwarz inequality we have $\ray{\bbC \Psi, \bbC \Omega} \leq 1$ with equality if and only if $\bbC \Psi = \bbC \Omega$. The ray product is clearly independent of the choice of representatives $\Psi$ and $\Omega$ for the rays, and therefore gives a well-defined binary operation $\bbP \hilbH \times \bbP \hilbH \rightarrow [0,1]$. The square of the ray product is called the \textdef{transition probability} between $\bbC \Psi$ and $\bbC \Omega$.
\end{definition}

We now define three metrics on $\bbP \hilbH$ which at first glance appear quite different. They are all equivalent, however, and we will relate them to each other and to the ray product.

\begin{definition}\label{def:PH_metrics}
The \textdef{chord metric} or \textdef{chordial metric} is defined for two rays $\bbC \Psi$, $\bbC \Omega \in \bbP \hilbH$ with normalized representatives $\Psi, \Omega \in \bbS \hilbH$ by
\[
d_\chd(\bbC \Psi, \bbC \Omega) = \inf_{\lambda \in \Unitary(1)} \norm{\Psi - \lambda \Omega}.
\]
One can again easily check that this is independent of the representatives $\Psi$ and $\Omega$, and is therefore well-defined. We note that the infimum is achieved for some $\lambda \in \Unitary(1)$ since $\Unitary(1)$ is compact and $\lambda \mapsto \norm{\Psi - \lambda \Omega}$ is continuous. Next, the \textdef{Fubini-Study metric} is defined by
\[
d_\FS(\bbC \Psi, \bbC \Omega) = \cos^{-1} \ray{\bbC \Psi, \bbC \Omega}.
\]
To describe our last metric, we introduce a map $P:\bbP \hilbH \rightarrow \mathcal{ROP}(\hilbH)$, where $\mathcal{ROP}(\hilbH)$ is the space of rank-one projections on $\hilbH$, topologized as a subspace of the space $\cB(\hilbH)$ of bounded linear operators with the operator norm topology. We define $P$ as the map that associates to each ray $\bbC \Psi$ the orthogonal projection $P(\bbC \Psi)$ onto $\bbC \Psi$. In particular, when $\Psi \in \bbS \hilbH$, we have $P(\bbC \Psi) = \ketbra{\Psi}$. Then we define the \textdef{gap metric} as
\[
d_\gap(\bbC \Psi, \bbC \Omega) = \norm{P(\bbC\Psi) - P(\bbC \Omega)}.
\]
\end{definition}

\begin{theorem}\label{thm:metric_equivalences}
Each of the metrics introduced in Definition \ref{def:PH_metrics} is indeed a metric. Moreover, the following hold.
\begin{enumerate}
	\item\label{ite:chd} The chord metric is complete and induces the quotient topology on $\bbP \hilbH$. Furthermore,
	\begin{equation}\label{eq:relation-transition-metric-amplitudes}
	d_\chd(\bbC \Psi, \bbC \Omega)^2 = 2 - 2\ray{\bbC \Psi, \bbC \Omega}
	\end{equation}
	for all $\Psi, \Omega \in \hilbH \setminus \qty{0}$.
	\item\label{ite:FS} The Fubini-Study metric is equivalent to the chord metric. More precisely,
	\begin{equation}\label{eq:equivalence-metric-projective-space-fubini-study-distance}
	d_\chd(\bbC \Psi, \bbC \Omega) \leq d_\FS(\bbC \Psi, \bbC \Omega)\leq \frac{\pi\sqrt{2}}{4}d_\chd(\bbC \Psi, \bbC \Omega),
	\end{equation}
	for all $\Psi, \Omega \in \hilbH \setminus \qty{0}$.
	\item\label{ite:gap} The gap metric is equivalent to the chord metric. More precisely,
	\begin{equation}\label{eq:equivalence-metric-projective-space-gap-metric}
	\frac{1}{\sqrt{2}}d_\chd(\bbC \Psi, \bbC \Omega) \leq d_\gap(\bbC \Psi, \bbC \Omega) = \sqrt{1 - \ray{\bbC \Psi, \bbC \Omega}^2} \leq d_\chd(\bbC \Psi, \bbC \Omega).
	\end{equation}
	for all $\Psi, \Omega \in \hilbH \setminus \qty{0}$. In particular, the map $P$ introduced above is a homeomorphism.
\end{enumerate}
\end{theorem}

\begin{proof}
\ref{ite:chd} We show $d_\chd$ is a metric. From the definition we see that $d_\chd$ is non-negative and symmetric. To show positive definiteness assume that $d_\chd(\bbC\Psi,\bbC\Omega) = 0$ with $\Psi,\Omega\in \bbS\hilbH$. By compactness of $\Unitary(1)$, there exists $\lambda \in \Unitary(1)$ such that $\norm{\Psi -\lambda \Omega}=0$. Hence $\bbC\Psi = \bbC\Omega$. To verify the triangle inequality, let $\Phi, \Psi, \Omega \in \bbS \hilbH$. There exists $\mu \in \Unitary(1)$ such that $d_\chd(\bbC \Phi, \bbC \Psi) = \norm{\Phi - \mu \Psi}$. Then
\begin{align*}
d_\chd(\bbC \Phi, \bbC \Omega) &= \inf_{\lambda \in \Unitary(1)} \norm{\Phi - \lambda \Omega}\\
&\leq \inf_{\lambda \in \Unitary(1)} \qty(\norm{\Phi - \mu \Psi} + \norm{\mu \Psi - \lambda\Omega})\\
&= d_\chd(\bbC \Phi, \bbC \Psi) + \inf_{\lambda \in \Unitary(1)} \norm{\mu \Psi - \lambda \Omega}\\
&= d_\chd(\bbC \Phi, \bbC \Psi) + d_\chd(\bbC \Psi, \bbC \Omega).
\end{align*}
Thus, $d_\chd$ is a metric.

Next we prove formula \eqref{eq:relation-transition-metric-amplitudes}. Given $\Psi,\Omega\in  \bbS \hilbH$, observe that for all $\lambda \in \Unitary(1)$ the estimate
\[
\norm{\Psi - \lambda \Omega}^2 = 2 - 2 \Re \qty(\lambda\ev{\Psi,  \Omega}) \geq  2  -  2\abs{\ev{\Psi, \Omega}}
\]
holds true. Putting
  \[
    \lambda_0 =
    \begin{cases}
      \frac{\ev{\Omega, \Psi}}{\abs{\ev{\Psi, \Omega}}} &:
      \langle {\Psi}, {\Omega} \rangle \neq 0 \ , \\
      1 &: \langle {\Psi}, {\Omega} \rangle = 0 \  ,
    \end{cases}
  \]
  then minimizes the functional $\| {\Psi} -\lambda {\Omega} \|$, hence 
  \[
    d_{\textup{chd}}(\bbC \Psi, \bbC \Omega)^2 = \|{\Psi} -\lambda_0 {\Omega}\|^2 =
    2  -  2|\langle {\Psi}, {\Omega} \rangle | =
    2 -  2\ltrans \bbC \Psi, \bbC \Omega \rtrans . 
  \]

  % With formula \eqref{eq:relation-transition-metric-amplitudes} we can show that $d_{\textup{chd}}$ induces the quotient topology on $\bbP \hilbH$.
  Given $\Psi \in \bbS \hilbH$, the function $d_{\textup{chd}}(\bbC \Psi, -):\bbP \hilbH \rightarrow \qty[0,\sqrt{2}]$ is continuous because by \eqref{eq:relation-transition-metric-amplitudes} its
  composition with $p|_{\bbS \hilbH}:\bbS \hilbH \rightarrow \bbP \hilbH$ is continuous, and this implies continuity by the characteristic property of the quotient topology. It follows that the metric topology is coarser than the quotient topology. Conversely, it follows from the definition of $d_{\textup{chd}}$ that for any $\Psi \in \bbS \hilbH$ and $\varepsilon > 0$, we have 
  \[
  \bbB_{\varepsilon,{\textup{chd}}}(\bbC \Psi) = p|_{\bbS \hilbH}\qty(\bbB_{\varepsilon}(\Psi) \cap \bbS \hilbH),
  \]
  where $\bbB_{\varepsilon,{\textup{chd}}}(\bbC \Psi)$ is the ball of radius $\varepsilon$ centered on $\bbC \Psi$ with respect to
  the chord metric and similarly $\bbB_\varepsilon(\Psi)$ is the ball of radius $\varepsilon$ centered on $\Psi$.
  % Thus, for any open set $U \subset \bbP \hilbH$ and $\bbC \Psi \in U$ with $\Psi \in \bbS \hilbH$, we can find $\varepsilon > 0$ such that $\bbB_\varepsilon(\Psi) \cap \bbS \hilbH \subset p^{-1}(U)$, hence $\bbC \Psi \in  \bbB_{\varepsilon,{\textup{chd}}}(\bbC \Psi) \subset U$.
  It follows that the quotient topology is coarser than the metric topology, so the two topologies are equal.

  To verify completeness observe first that for a given element $\Psi \in  \bbS \hilbH$ and ray
  $\scrl \in \bbP \hilbH$ with $\ltrans \bbC \Psi, \scrl \rtrans \neq 0$ there exists a unique representative
  $\Omega_{\Psi,\scrl} \in  \scrl \cap \bbS \hilbH$ % which we call distinguished
  such that $\langle \Psi, \Omega_{\Psi,\scrl} \rangle = \ltrans \bbC \Psi, \scrl \rtrans$.
  Now assume that $(\scrl_n)_{n\in \bbN}$   is a Cauchy sequence in $\bbP \hilbH$.
  Then there exists a strictly increasing sequence of natural numbers $(n_k)_{k\in \bbN}$ such that 
    \[
       d_{\textup{chd}} (\scrl_m ,  \scrl_n) < \frac{1}{2^{k}} \quad \text{for all } m,n \geq n_k  .
    \]
  Pick a representative $\Psi_1 \in  \scrl_{n_1} \cap \bbS \hilbH$ and
  % let $\Omega_1\in \bbS \hilbH$ be the distinguished representative $\Omega_{\Omega_0,\scrl_{n_1}}$ of $ \scrl_{n_1}$.
  define the sequence $(\Psi_k)_{k\in \bbN}$ of unit vectors recursively by
  \[
    \Psi_{k+1} =  \Psi_{\Psi_k,\scrl_{n_{k+1}}}  .
  \]
  Then, for all $k\in \bbN$,
  \begin{equation*}
    \begin{split}
      \| \Psi_{k+1}- \Psi_k \| \, & =
      \sqrt{2-2\Re \langle \Psi_{k+1},\Psi_k \rangle} =
      \sqrt{2- 2\ltrans \scrl_{n_{k+1}}, \scrl_{n_k} \rtrans } \\
      & = d_{\textup{chd}}( \scrl_{n_{k+1}}, \scrl_{n_k} ) < \frac{1}{2^{k}}  . 
    \end{split}
  \end{equation*}
  So $(\Psi_k)_{k\in \bbN}$ is a Cauchy sequence in $\bbS\hilbH$, hence converges to a vector $\Psi \in \bbS\hilbH$.
  Let $\scrl =\bbC\Psi$. Since $d_{\textup{chd}}(\scrl_{n_k},\scrl)\leq \| \Psi_k - \Psi \|$, the subsequence
  $(\scrl_{n_k})_{k\in \bbN}$  converges to $\scrl$, hence $(\scrl_n)_{n\in \bbN}$ does so too. Therefore
  $(\bbP\hilbH,d_{\textup{chd}})$ is  a complete metric space.
%  
%(\ref{ite:FS})
%  To prove that the gap metric is equivalent to the Fubini--Study distance
%  consider the functions 
%   \[
%     f : [ 0,\sqrt{2}] \to \R , \: s \mapsto \arccos\left( 1 -\frac{s^2}{2} \right) \quad\text{and}\quad 
%     g: [ 0,\frac{\pi}{2} ] \to \R , \:  t \mapsto \sqrt{2( 1 -\cos t) } \ .
%   \] 
%   Both functions are continuous and differentiable on the interior of
%   their domains. Observe that $f(0)=g(0)=0$ and compute
%   \[
%     f'(s) = \frac{s}{\sqrt{1-\left(1-\frac{s^2}{2}\right)^2}} = \frac{s}{\sqrt{s^2-\frac{s^4}{4}}} = \frac{2}{\sqrt{4-s^2}}
%     \leq \sqrt{2} \quad\text{for } s\in ( 0,\sqrt{2}) 
%   \]
%   and 
%   \[
%     g'(t) = \frac{\sqrt{2}}{2} \, \frac{\sin t}{\sqrt{1-\cos t}} = \frac{\sqrt{2}}{2} \, \sqrt{1+\cos t} \leq 1
%     \quad\text{for } t\in ( 0,\frac{\pi}{2}) \ .
%   \]
%   By definition of $d_\textup{FS}$ and \eqref{eq:relation-transition-metric-amplitudes},
%   the mean-value theorem then entails for all $\scrk, \scrl \in \bbP\hilbH$ 
%   \[
%      d_{\textup{gap}}( \scrk, \scrl) = g \left( d_\textup{FS} ( \scrk, \scrl) \right) \leq  d_\textup{FS} ( \scrk, \scrl)  =% 
%      f \left( d_{\textup{gap}} ( \scrk, \scrl) \right) \leq \sqrt{2} \, d_{\textup{gap}}( \scrk, \scrl) \ .
%   \] 
%   Hence the estimate \eqref{eq:equivalence-metric-projective-space-fubini-study-distance}
%  is proved and the metrics $d_{\textup{gap}}$ and $d_\textup{FS}$ are equivalent. 
%
  
  \ref{ite:FS} From the definition we see that $d_\FS$ is non-negative, symmetric, and positive-definite. We prove the triangle inequality. Let $\bbC \Phi, \bbC \Psi, \bbC \Omega \in \bbP\hilbH$ with representatives $\Phi, \Psi, \Omega \in \bbS \hilbH$. If any two of $\Phi$, $\Psi$, and $\Omega$ are linearly dependent, then the triangle inequality is trivial, so suppose otherwise. By multiplying $\Phi$ and $\Omega$ by a phase if necessary, we can achieve $\ev{\Phi, \Psi} > 0$ and $\ev{\Psi, \Omega} > 0$. Let $\theta = \cos^{-1} \ev{\Phi, \Psi}$ and let $\phi = \cos^{-1} \ev{\Psi, \Omega}$, so $\theta = d_\FS(\bbC \Phi, \bbC \Psi)$ and $\phi = d_\FS(\bbC \Psi, \bbC \Omega)$. 

  Define
  \begin{equation}\label{eq:orthogonalize_FS}
  \Phi' = \frac{\Phi - \ev{\Psi, \Phi}\Psi}{\norm{\Phi - \ev{\Psi, \Phi}\Psi}} \qqtext{and} \Omega' = \frac{\Omega - \ev{\Psi, \Omega}\Psi}{\norm{\Omega - \ev{\Psi, \Omega}\Psi}}.
  \end{equation}
  Note that $\norm{\Phi - \ev{\Psi, \Phi}\Psi} = \sqrt{1 - \ev{\Psi, \Phi}^2} = \sin \theta$ and $\norm{\Omega - \ev{\Psi, \Omega}\Psi} = \sqrt{1 - \ev{\Psi, \Omega}^2} = \sin \phi$. Thus, rearranging \eqref{eq:orthogonalize_FS} yields
  \[
  \Phi = \sin \theta \: \Phi' + \cos \theta\: \Psi \qqtext{and} \Omega = \sin \phi\: \Omega' + \cos \phi \:\Psi.
  \]
  The reverse triangle inequality and the Cauchy-Schwarz now yield
  \begin{align*}
  \abs{\ev{\Phi, \Omega}} &= \abs{\cos \theta \cos \phi + \sin \theta \sin \phi \ev{\Phi', \Omega'}}\\
  &\geq \cos \theta \cos \phi - \sin \theta \sin \phi \abs{\ev{\Phi', \Omega'}}\\
  &\geq \cos \theta \cos \phi - \sin \theta \sin \phi = \cos (\theta + \phi).
  \end{align*}
  Since $\cos^{-1}$ is monotonically decreasing, we have
  \[
  d_\FS(\bbC \Phi, \bbC \Omega) = \cos^{-1}\abs{\ev{\Phi, \Omega}} \leq \theta + \phi = d_\FS(\bbC \Phi, \bbC \Psi) + d_\FS(\bbC \Psi, \bbC \Omega).
  \]

  Now we prove \eqref{eq:equivalence-metric-projective-space-fubini-study-distance}. Consider the function 
  \[
    f:[0,1] \rightarrow \bbR, \quad f(x)  =
    \begin{cases}
       \frac{\cos^{-1} x}{\sqrt{2 - 2x}} &:  x \in [0, 1) \\  
        1 &: x = 1 
    \end{cases}  .
  \]
  This function is continuous on $[0,1]$ and differentiable on $(0,1)$, with
  \[
    f'(x) = \frac{1}{\sqrt{2}} \, (1 - x)^{- 3/2} \left( \frac{\cos^{-1} x}{2} - \sqrt{\frac{1 - x}{1 + x}} \right) \ .
    % \: \text{ for } x \in (0, 1)\ .
  \]
  Given $x \in (0, 1)$, put $\theta = \cos^{-1} x$.
  %there exists $\theta \in (0, \pi/2)$ such that $x = \cos \theta$.
  Using trigonometric power reducing identities yields
  \[
  f'(x) = \frac{1}{\sqrt{2}} \, (1 - x)^{- 3/2} \left( \frac{\theta}{2} - \tan\qty(\frac{\theta}{2}) \right) < 0 \ .
  \]
  Therefore $f$ is monotonically decreasing, so $1 = f(1) \leq f(x) \leq f(0) = \frac{\pi\sqrt{2}}{4}$ for all $x \in [0,1]$.
  Since
  \[
    f(\ltrans \bbC \Psi, \bbC \Omega \rtrans) =
    \frac{d_\tn{FS}( \bbC \Psi, \bbC \Omega )}{d_\tn{chd}( \bbC \Psi, \bbC \Omega )}
  \]
  provided $\ltrans \bbC \Psi, \bbC \Omega \rtrans < 1$, this proves
  \eqref{eq:equivalence-metric-projective-space-fubini-study-distance}.

  \ref{ite:gap}
   That $d_\gap$ is a metric is immediate from the definition and the properties of the norm on $\cB(\hilbH)$. We prove that $d_\gap(\bbC \Psi, \bbC \Omega) = \sqrt{1 - \ray{\bbC\Psi, \bbC \Omega}^2}$ for all rays $\bbC \Psi$ and $\bbC \Omega$. We may choose the representatives $\Psi$ and $\Omega$ to be unit vectors satisfying $\ev{\Psi, \Omega} \geq 0$. If $\Psi$ and $\Omega$ are linearly dependent, then the formula is trivial, so suppose otherwise.  We define 
   \[
   \Omega' = \frac{\Omega - \ev{\Psi, \Omega}\Psi}{\norm{\Omega - \ev{\Psi, \Omega}\Psi}} = \frac{\Omega - \ev{\Psi, \Omega}\Psi}{\sqrt{1 - \ev{\Psi, \Omega}^2}},
   \]
   For convenience we set $\theta = \cos^{-1} \ev{\Psi, \Omega} \in (0, \pi/2]$, whence the above line rearranges to
   \[
   \Omega = \cos \theta \: \Psi + \sin \theta \: \Omega'
   \]
   Now observe that
   \begin{equation}\label{eq:projection_difference_gap}
   \begin{split}
   P(\bbC \Psi) - P(\bbC \Omega) &= \ketbra{\Psi} - \ketbra{\Omega} \\
   &= \sin^2 \theta \ketbra{\Psi} - \sin^2 \theta \ketbra{\Omega'} -\sin \theta \cos \theta \qty(\ketbra{\Psi}{\Omega'} + \ketbra{\Omega'}{\Psi}).
   \end{split}
   \end{equation}
   We must compute the norm of this operator.

   Let $\Phi \in \bbS \hilbH$ be arbitrary. We have a decomposition $\Phi = \Phi^\parallel + \Phi^\perp$, where $\Phi^\parallel \in \vecspan\qty{\Psi, \Omega}$ and $\Phi^\perp \in \vecspan\qty{\Psi, \Omega}^\perp$. We expand $\Phi^\parallel$ as a linear combination
   \[
   \Phi^\parallel = \alpha \Psi + \beta \Omega'
   \]
   and note that $\norm{\Phi^\parallel}^2 = \abs{\alpha}^2 + \abs{\beta}^2 \leq 1$ since $\norm{\Phi} = 1$. Then plugging $\Phi$ into \eqref{eq:projection_difference_gap} and taking the norm yields
   \begin{align*}
   \norm{P(\bbC \Psi)\Phi - P(\bbC \Omega)\Phi}^2 &= \norm{\qty(\alpha \sin^2 \theta - \beta \sin \theta \cos \theta)\Psi - \qty(\beta \sin^2 \theta + \alpha \sin \theta \cos \theta)\Omega'}^2\\
   &= \abs{\alpha \sin^2 \theta - \beta \sin \theta \cos \theta}^2 + \abs{\beta \sin^2 \theta + \alpha \sin \theta \cos \theta}^2\\
   %&= \abs{\alpha}^2 \sin^4 \theta - \alpha \beta^* \sin^3 \theta \cos \theta - \alpha^* \beta \sin^3 \theta \cos \theta + \abs{\beta}^2 \sin^2 \theta \cos^2 \theta\\
   %&\quad + \abs{\beta}^2 \sin^4 \theta + \alpha \beta^* \sin^3 \theta \cos \theta + \alpha^* \beta \sin^3 \theta \cos \theta + \abs{\alpha}^2 \sin^2 \theta \cos^2 \theta\\
   &= \norm*{\Phi^\parallel}^2\qty( \sin^4 \theta + \sin^2 \theta \cos^2 \theta)  = \norm*{\Phi^\parallel}^2\sin^2 \theta,
   \end{align*}
   where cross terms have canceled after expanding the middle line. It follows that
   \[
   \norm{P(\bbC \Psi) - P(\bbC \Omega)} = \sin \theta = \sqrt{1 - \ev{\Psi, \Omega}^2},
   \]
   as desired. 

   The inequalities in \eqref{eq:equivalence-metric-projective-space-gap-metric} follow from \eqref{eq:relation-transition-metric-amplitudes} and the inequalities
   \[
   1 - x \leq 1 - x^2 \leq 2 - 2x \qquad \tn{for all $x \in [0,1]$.}
   \]
   This completes the proof.
\end{proof}

\begin{corollary}\label{cor:gap_metric_unit_ball}
Let $\Psi \in \sphere \hilbH$ and let $\ball_1(\bbC \Psi)$ be the ball of radius one around $\bbC \Psi$ with respect to the gap metric. Then
\[
\ball_1(\bbC \Psi) = \qty{\bbC \Omega \in \bbP \hilbH : \ray{\bbC\Psi, \bbC\Omega} > 0}.
\]
\end{corollary}

\begin{proof}
This is immediate from \eqref{eq:equivalence-metric-projective-space-gap-metric}.
\end{proof}

\subsection{Fiber Bundles in Hilbert Space}\label{subsec:fiber_bundle_PH}

We continue to let $\hilbH$ be a nonzero complex Hilbert space. 
We review the construction of a couple important fiber bundles over $\bbP \hilbH$ and $\sphere \hilbH$.  Some of the major results of this thesis will be the generalization of these fiber bundles to the $C^*$-algebraic setting. We will use the abbreviation $F \rightarrow E \rightarrow B$ to denote  a fiber bundle with total space $E$, base space $B$, and fiber $F$. We refer to Steenrod \cite{Steenrod} for an introduction to the theory of fiber bundles.

First we show how $p|_{\sphere \hilbH}:\sphere \hilbH \rightarrow \bbP \hilbH$ is a fiber bundle with model fiber $\Unitary(1)$. The construction of this fiber bundle is adapted from \cite{Spiegel}. Given $\Psi \in \sphere \hilbH$, denote by $\bbC \Psi^\perp$ the orthogonal complement in $\hilbH$ of the one-dimensional subspace $\bbC \Psi$. 

\begin{lemma}
If $\Psi \in \bbS\hilbH$ and $\bbB_1 (\bbC\Psi)$ is the open unit ball around $\bbC \Psi$ with respect to the gap metric, then
\[
p|_{\sphere \hilbH}^{-1}(\ball_1(\bbC \Psi)) = \sphere \hilbH \setminus (\sphere \hilbH \cap \bbC \Psi^\perp).
\]
\end{lemma}

\begin{proof}
This is immediate from Corollary \ref{cor:gap_metric_unit_ball}.
\end{proof}

\begin{lem}\label{lem:PH_trivializations}
  Let $\Psi \in \bbS\hilbH$ and let $\bbB_1 (\bbC\Psi)$ be the open unit ball around $\bbC \Psi$
  with respect to the gap metric. The maps
  $\sigma_\Psi: \bbS \hilbH \setminus (\sphere \hilbH \cap \bbC \Psi^\perp) \rightarrow \bbC \Psi^\perp \times \Unitary(1)$ and
  $\tau_\Psi:\bbB_1 (\bbC\Psi) \rightarrow \bbC \Psi^\perp$ given by
\[
\sigma_\Psi\qty(\Omega) = \qty(\frac{\Omega}{\ev{\Psi, \Omega}} - \Psi, \frac{\ev{\Psi, \Omega}}{\abs{\ev{\Psi, \Omega}}})
\qqtext{and} 
\tau_\Psi (\bbC\Omega) = \frac{\Omega}{\ev{\Psi,\Omega}} - \Psi 
\]
are well-defined homeomorphisms. 
\end{lem}

Note that in the definition of $\tau_\Psi(\bbC \Omega)$,  the representative vector $\Omega$ need not be normalized.

\begin{proof}
From the definition of the domain of $\sigma_\Psi$ we see that $\ev{\Psi, \Omega} \neq 0$ for all $\Omega \in \sphere \hilbH \setminus (\sphere \hilbH \cap \bbC \Psi^\perp)$, so no division by zero occurs in the definition of $\sigma_\Psi(\Omega)$. From Corollary \ref{cor:gap_metric_unit_ball} we see that $\ev{\Psi, \Omega} \neq 0$ for all $\Omega \in \hilbH \setminus \qty{0}$ such that $\bbC \Omega \in \ball_1(\bbC \Psi)$, so no division by zero occurs in the definition of $\tau_\Psi(\bbC \Omega)$. It is also clear that the definition of $\tau_\Psi(\bbC \Omega)$ is independent of the choice of representative $\Omega$.

We show that $\sigma_\Psi$ indeed maps into $\bbC \Psi^\perp \times \Unitary(1)$. That the second component is in $\Unitary(1)$ is clear. 
  If $\Omega \in \bbS \hilbH \setminus (\bbS \hilbH \cap \bbC \Psi^\perp)$, then
  \begin{equation}\label{eq:sigma_tau_well_defined}
    \ev{\Psi, \frac{\Omega}{\ev{\Psi,\Omega}} - \Psi} = 1 - \ev{\Psi,\Psi} = 0
  \end{equation}
  so the first component of $\sigma_\Psi(\Omega)$ is indeed in $\bbC \Psi^\perp$.
Continuity of $\sigma_\Psi$ is manifest. We now define a continuous map
$\sigma^{-1}_\Psi: \bbC\Psi^\perp \times \Unitary(1) \rightarrow \bbS \hilbH \setminus \qty(\bbS \hilbH \cap \bbC\Psi^\perp)$ by
\[
  \sigma^{-1}_\Psi(\Phi, \lambda) = \frac{\lambda (\Phi + \Psi)}{\sqrt{1 + \norm{\Phi}^2}} \ .
\]
% This is also clearly well-defined and continuous.
It is straightforward to check that $\sigma_\Psi^{-1}$ is indeed a well-defined two-sided inverse
for $\sigma_\Psi$, and $\sigma_\Psi^{-1}$ is also manifestly continuous. 

Next we consider $\tau_\Psi$. Given $\bbC \Omega \in \ball_1(\bbC \Psi)$, we may choose the representative $\Omega$ such that $\Omega \in \sphere \hilbH \setminus (\sphere \hilbH \cap \bbC \Psi^\perp)$. Then \eqref{eq:sigma_tau_well_defined} shows that indeed $\tau_\Psi(\bbC \Omega) \in \bbC \Psi^\perp$, so $\tau_\Psi$ is well-defined. Since composition of $\tau_\Psi$ with the canonical projection 
$p: \bbS \hilbH \setminus \qty(\bbS \hilbH \cap \bbC \Psi^\perp)  \to \bbB_1 (\bbC \Psi)$
is continuous, $\tau_\Psi$ is continuous as well since $p:\bbS \hilbH \setminus \qty(\bbS \hilbH \cap \bbC \Psi^\perp)  \to \bbB_1 (\bbC \Psi)$ is a quotient map.
Now define $\tau_\Psi^{-1}:\bbC \Psi^\perp \rightarrow \ball_1(\bbC \Psi)$ as
\begin{equation}
  \label{eq:inverse-chart-projective-hilbert-space}
   \tau_\Psi^{-1}(\Phi) = \bbC(\Phi + \Psi).
\end{equation} 
It is clear that $\Psi + \Phi \neq 0$ for such $\Phi \in \bbC \Psi^\perp$, so
$\bbC(\Psi + \Phi) \in \bbP \hilbH$ and the ray product
$\ltrans \tau^{-1}_\Psi(\Phi), \bbC \Psi \rtrans = \norm{\Phi + \Psi}^{-1}$
is positive. Hence
% , \[ d_{\textup{gap}}(\tau_\Psi^{-1}(\Phi), \bbC \Psi) < 1,\] 
$\tau_\Psi^{-1}(\Phi) \in \bbB_1 (\bbC\Psi)$. Continuity of 
the map $\tau_\Psi^{-1}: C_\Psi \to \bbB_1 (\bbC\Psi)$ is manifest, and it is straightforward to check that it is a two-sided inverse for $\tau_\Psi$.
\end{proof}

\begin{theorem}\label{thm:local_triv_rho}
Let $\Psi \in \sphere \hilbH$ and let $\ball_1(\bbC \Psi)$ be the open unit ball around $\bbC \Psi$ with respect to the gap metric. Then the map
\[
\rho_\Psi : p|_{\sphere \hilbH}^{-1}(\ball_1(\bbC \Psi)) \rightarrow \ball_1(\bbC \Psi) \times \Unitary(1), \quad \rho_\Psi(\Omega) = \qty(\bbC \Omega, \frac{\ev{\Psi, \Omega}}{\abs{\ev{\Psi, \Omega}}})
\]
is a homeomorphism.  The collection of all $\rho_\Psi$ form a trivializing atlas for $p|_{\sphere \hilbH}$, thereby endowing $\sphere \hilbH$ with the structure of a fiber bundle over $\bbP \hilbH$ with model fiber $\Unitary(1)$.
\end{theorem}

\begin{proof}
Observe that $\rho_\Psi = (\tau_\Psi^{-1} \times \id_{\Unitary(1)}) \circ \sigma_\Psi$, where $\sigma_\Psi$ and $\tau_\Psi$ are as in Lemma \ref{lem:PH_trivializations}. Since $\sigma_\Psi$ and $\tau_\Psi$ are homeomorphisms by Lemma \ref{lem:PH_trivializations}, it is immediate that $\rho_\Psi$ is a homeomorphism.

Letting $\pi_1:\ball_1(\bbC \Psi) \times \Unitary(1) \rightarrow \ball_1(\bbC \Psi)$ be the projection onto the first factor, it is clear that the diagram below commutes.
\[
\begin{tikzcd}
\sphere \hilbH \setminus (\sphere \hilbH \cap \bbC \Psi^\perp) \arrow[rr,"\rho_\Psi"] \arrow[dr,"p|_{\sphere \hilbH}"'] && \ball_1(\bbC \Psi) \times \Unitary(1) \arrow[dl,"\pi_1"]\\
&\ball_1(\bbC \Psi) &
\end{tikzcd}
\]
It is also clear that the collection of balls $\ball_1(\bbC \Psi)$ cover $\bbP \hilbH$. This shows that the collection of all $\rho_\Psi$ is a trivializing atlas for $p|_{\sphere \hilbH}$.
\end{proof}

In fact, if we consider $\Unitary(1)$ to act on $\sphere \hilbH$ and on itself by multiplication, then $\rho_\Psi$ is clearly fiberwise equivariant with respect to this action. Thus, $\Unitary(1) \rightarrow \sphere \hilbH \rightarrow \bbP \hilbH$ is in fact a principal $\Unitary(1)$-bundle.

The fiber bundle $p|_{\sphere \hilbH}: \sphere \hilbH \rightarrow \bbP \hilbH$ has the following local sections, which we will use frequently.

\begin{cor}\label{cor:PH_mapsto_SH}
  Let $\Psi \in \sphere \hilbH$ and let $\ball_1(\bbC \Psi)$ be the open unit ball around $\bbC \Psi$ with respect to the gap metric. The function
  $\ball_1 (\bbC\Psi) \rightarrow \sphere \hilbH$ which maps a ray $\bbC\Omega \in\bbB_1 (\bbC\Psi)$ to the unique normalized representative
  $\Omega' \in \bbC \Omega \cap \sphere \hilbH$ such that  $\ev{\Psi, \Omega'} > 0$  is continuous.
\end{cor}

\begin{proof}
  The stated map is 
  \[
  \bbC \Omega \mapsto \rho_\Psi^{-1}(\bbC\Omega, 1) = \frac{\abs{\ev{\Psi,\Omega}}}{\ev{\Psi,\Omega}} \frac{\Omega}{\norm{\Omega}}
  \]
   which is continuous by Theorem \ref{thm:local_triv_rho}. We have used Lemma \ref{lem:PH_trivializations} (and its proof) and the fact that $\rho_\Psi = (\tau_\Psi^{-1} \times \id_{\Unitary(1)}) \circ \sigma_\Psi$ to give the explicit form of $\rho_\Psi^{-1}(\bbC \Omega, 1)$.
\end{proof}

We move on to our next bundle. We shall often encounter the following situation. Suppose $G$ is a topological group, $X$ is a topological space, and $G \times X \rightarrow X$ is a continuous and transitive left group action. Fix $x \in X$ and let $G_x$ be the stabilizer of $x$. Consider the map $p:G \rightarrow X$ defined as $p(g) = gx$. If there exists a neighborhood $O$ of $x$ and a continuous local section $\sigma:O \rightarrow G$ of $p$ (meaning $p \circ \sigma$ is the inclusion $O \hookrightarrow X$), then $p:G \rightarrow X$ has the structure of a principal $G_x$-bundle, with an atlas of local trivializations given by all maps of the form
\[
\rho_g:p^{-1}(gU) \rightarrow gU \times G_x, \quad \rho_g(h) = (hx, \sigma(g^{-1}hx)^{-1}g^{-1}h)
\]
across all $g \in G$ \cite[Sec.~7.3 \& 7.4]{Steenrod}. Construction of a continuous local section $\sigma:O \rightarrow G$ is therefore the only thing we need to do to construct a principal fiber bundle in this situation.

Observe that $\Unitary(\hilbH)$ with the norm topology is a topological group and $\Unitary(\hilbH) \times \sphere \hilbH \rightarrow \sphere \hilbH$, $(U, \Psi) \mapsto U\Psi$ is a continuous and transitive left group action. Fix $\Psi \in \sphere \hilbH$ and let $\Unitary(\hilbH)_\Psi$ be the stabilizer of $\Psi$. We construct the necessary section to show that $\Unitary(\hilbH)_\Psi \rightarrow \Unitary(\hilbH) \rightarrow \sphere \hilbH$ is a principal $\Unitary(\hilbH)_\Psi$-bundle. The following lemmas gives us the desired section and more.

\begin{lemma}\label{lem:unitary_rotator}
Let $\Psi, \Omega \in \sphere \hilbH$ and set $\hilbH_{\Psi, \Omega} = \vecspan\qty{\Psi,\Omega}$. The function $U_{\Psi,\Omega}:\hilbH_{\Psi,\Omega} \rightarrow \hilbH_{\Psi,\Omega}$ defined as
\[
U_{\Psi,\Omega}\Phi = \ev{\Omega, \Psi}\Phi - \ev{\Omega,\Phi}\Psi + \ev{\Psi, \Phi}\Omega
\]
is unitary and satisfies $U_{\Psi,\Omega}\Psi = \Omega$ and $U_{\Psi,\Omega}\Omega = 2\Re \ev{\Psi,\Omega} \Omega - \Psi$.
\end{lemma}

\begin{proof}
Clearly, $U_{\Psi,\Omega}$ is linear. It is easily checked that $U_{\Psi,\Omega}$ maps $\Psi$ and $\Omega$ to the claimed vectors. If $\Psi$ and $\Omega$ are linearly dependent, then linearity of $U_{\Psi,\Omega}$ and the fact that $U_{\Psi,\Omega}\Psi = \Omega$ imply that $U_{\Psi,\Omega}$ is multiplication by $\ev{\Psi, \Omega}$, which has modulus one in this case.

Now suppose $\Psi$ and $\Omega$ are linearly independent; we show that $U_{\Psi,\Omega}$ is unitary. Observe that
\begin{align*}
\ev{U_{\Psi, \Omega}\Psi, U_{\Psi,\Omega}\Psi} &= \ev{\Omega, \Omega} = 1 = \ev{\Psi,\Psi} \\
\ev{U_{\Psi, \Omega}\Psi, U_{\Psi,\Omega}\Omega} &= \ev{\Omega, 2\Re \ev{\Psi, \Omega}\Omega - \Psi}  = 2\Re \ev{\Psi,\Omega} - \ev{\Omega, \Psi} = \ev{\Psi,\Omega}\\
\ev{U_{\Psi,\Omega}\Omega, U_{\Psi,\Omega}\Omega} &= \ev{2\Re \ev{\Psi,\Omega}\Omega - \Psi, 2\Re \ev{\Psi,\Omega}\Omega - \Psi} \\
&= \qty(2\Re \ev{\Psi,\Omega})^2 - \qty(2\Re \ev{\Psi, \Omega})\ev{\Omega, \Psi} - \qty(2 \Re \ev{\Psi, \Omega})\ev{\Psi,\Omega} + 1 \\
&= 1 = \ev{\Omega, \Omega}.
\end{align*}
Since $\Psi$ and $\Omega$ span $\hilbH_{\Psi,\Omega}$, it now follows easily that $U_{\Psi,\Omega}$ preserves inner products, so $U_{\Psi,\Omega}$ is unitary.
\end{proof}

\begin{lemma}\label{lem:unitary_rotator_continuous}
Given $\Psi, \Omega \in \sphere \hilbH$ such that $\ev{\Psi,\Omega} \neq 0$, extend $U_{\Psi,\Omega}$ to a unitary on $\hilbH$ by defining it to act as multiplication by $\ev{\Omega, \Psi}/\abs{\ev{\Omega,\Psi}}$ on $\hilbH_{\Psi,\Omega}^\perp$. Then the map
\[
\qty{(\Psi,\Omega) \in \sphere\hilbH \times \sphere \hilbH: \ev{\Psi,\Omega} \neq 0} \rightarrow \Unitary(\hilbH), \quad (\Psi,\Omega) \mapsto U_{\Psi,\Omega}
\]
is norm-continuous.
\end{lemma}

\begin{proof}
Given $(\Psi, \Omega) \in \sphere \hilbH \times \sphere \hilbH$ such that $\ev{\Psi, \Omega} \neq 0$, define
\[
\lambda(\Psi, \Omega) = \frac{\ev{\Omega, \Psi}}{\abs{\ev{\Omega, \Psi}}} \qqtext{and} \mu(\Psi,\Omega) = \frac{1}{1 + \abs{\ev{\Omega,\Psi}}}.
\]
It is clear that $\lambda$ and $\mu$ are continuous on $\qty{(\Psi,\Omega) \in \sphere\hilbH \times \sphere \hilbH: \ev{\Psi,\Omega} \neq 0}$. It is straightforward to check that
\begin{equation}\label{eq:unitary_rotator_explicit_form}
U_{\Psi,\Omega} = \lambda \1  - \mu \ketbra{\Psi}{\Omega} - \lambda \mu\ketbra{\Psi} - \lambda \mu \ketbra{\Omega} + \qty(1 + \lambda \mu \ev{\Omega,\Psi} )\ketbra{\Omega}{\Psi}
\end{equation}
where we have abbreviated $\lambda = \lambda(\Psi,\Omega)$ and $\mu = \mu(\Psi,\Omega)$. For example, this can be verified by checking that the operator on the right hand side agrees with $U_{\Psi,\Omega}$ on $\Psi$ and $\Omega$, and that it acts as multiplication by $\lambda(\Psi,\Omega) = \ev{\Omega, \Psi}/\abs{\ev{\Omega,\Psi}}$ on the orthogonal complement of $\hilbH_{\Psi,\Omega}^\perp$. Linearity then implies \eqref{eq:unitary_rotator_explicit_form}. In the form of the right hand side of \eqref{eq:unitary_rotator_explicit_form}, $U_{\Psi,\Omega}$ is manifestly norm-continuous in $\Psi$ and $\Omega$.
\end{proof}

\begin{theorem}\label{thm:unitary_section}
Let $\Psi \in \sphere \hilbH$. There exists a continuous map
\[
 \sphere \hilbH \setminus (\sphere \hilbH \cap \bbC \Psi^\perp) \rightarrow \Unitary(\hilbH), \quad \Omega \mapsto U_\Omega
\]
such that 
\[
U_\Omega \Psi = \Omega
\]
for all $\Omega \in \sphere \hilbH \setminus (\sphere \hilbH \cap \bbC \Psi^\perp)$. The inverse $\Omega \mapsto U_\Omega^{-1}$ is continuous and satisfies $U_\Omega^{-1}\Omega = \Psi$ for all $\Omega \in \sphere \hilbH \setminus (\sphere \hilbH \cap \bbC \Psi^\perp)$. The map $\Omega \mapsto U_\Omega$ is a local section of the map $\Unitary(\hilbH) \rightarrow \sphere \hilbH$, $V \mapsto V\Psi$, so $\Unitary(\hilbH) \rightarrow \sphere \hilbH$ is a principal $\Unitary(\hilbH)_\Psi$-bundle. 
\end{theorem}

\begin{proof}
The composition $\Omega \mapsto (\Psi,\Omega) \mapsto U_{\Psi,\Omega}$ has the desired properties by Lemmas \ref{lem:unitary_rotator} and \ref{lem:unitary_rotator_continuous}.
\end{proof}

\subsection{The Projective Unitary Group}
\label{subsec:PU(H)}

The topology of the projective unitary group was first studied in depth by D.\ J.\ Simms  \cite{Simms_1970}. We review some important points from \cite{Simms_1970} here. We will provide full proofs using a slightly different method from \cite{Simms_1970}. Throughout this section, let $\hilbH$ be a nonzero complex Hilbert space.

\begin{definition}
Observe that the unitary group $\Unitary(\hilbH)$ has a copy of $\Unitary(1)$ inside it in the form of the subgroup $\qty{\lambda \1 : \lambda \in \Unitary(1)}$. This is clearly a normal subgroup of $\Unitary(\hilbH)$ and the quotient group $\Unitary(\hilbH)/\Unitary(1)$ is called the \textdef{projective unitary group of $\hilbH$}, denote $\PU(\hilbH)$. Given $U \in \Unitary(\hilbH)$, we denote by $\bbU$ its image in $\PU(\hilbH)$ under the canonical projection $\Unitary(\hilbH) \rightarrow \PU(\hilbH)$. We may also use the blackboard font to denote elements $\bbU, \bbV, \bbW \in \PU(\hilbH)$ without necessarily providing preferred representatives.
\end{definition}

Given $\bbU \in \PU(\hilbH)$ and $\scrl \in \bbP\hilbH$, define
\[
\bbU \scrl = \bbC U\Psi
\]
where $U \in \Unitary(\hilbH)$ is any representative of $\bbU$ and $\Psi \in \scrl \setminus \qty{0}$ is any representative of $\scrl$; the ray $\bbU\scrl$ is  independent of the choice of representatives. For $\Psi \in \hilbH \setminus \qty{0}$, this can be rewritten as
\[
\bbU \bbC \Psi = \bbC U \Psi
\]
for any representative $U$ of $\bbU$. It is clear that this defines a group action $\bbP\Unitary(\hilbH) \times \bbP \hilbH \rightarrow \bbP \hilbH$.

Given $\scrl \in \bbP\hilbH$, define $\hat{\scrl}:\PU(\hilbH) \rightarrow \bbP\hilbH$ as $\hat{\scrl}(\bbU) = \bbU\scrl$. Observe that for any nonzero vector $\Psi \in \hilbH \setminus \qty{0}$, the map $\widehat \Psi: \Unitary(\hilbH) \rightarrow \hilbH$, $\widehat \Psi(U) = U\Psi$ commutes with $\widehat{\bbC \Psi}$ and the projections according to the commutative diagram below.
\begin{equation}\label{cd:projective_unitary}
\begin{tikzcd}
\Unitary(\hilbH) \rar["\widehat \Psi"]\dar& \hilbH \dar\\
\PU(\hilbH) \rar["\widehat{\bbC \Psi}"] & \bbP \hilbH
\end{tikzcd}
\end{equation}
In other words, $\widehat{\bbC \Psi}\qty(\bbU) =\bbU\bbC\Psi =  \bbC U\Psi$ for any $\bbU \in \PU(\hilbH)$ with representative $U \in \Unitary(\hilbH)$.

We would like to define a sensible ``strong'' topology on $\PU(\hilbH)$. A priori, there appear to be two ways to do this. The first is to give $\PU(\hilbH)$ the quotient topology induced by the projection $\Unitary(\hilbH)_\tn{s} \rightarrow \PU(\hilbH)$ (where $\Unitary(\hilbH)_\tn{s}$ is the unitary group with the strong operator topology). Since $\Unitary(\hilbH)_\tn{s}$ is a topological group \cite{SchottenloherUnitaryStrongTopology}, $\PU(\hilbH)$ is a topological group with this quotient topology. Alternatively, we could give $\PU(\hilbH)$ the initial topology induced by the maps $\hat{\scrl}$ across all $\scrl \in \bbP\hilbH$. The following theorem, first proven by D. J. Simms \cite{Simms_1970}, shows that these topologies are the same.  We provide a different proof below, but first we introduce a bit of notation. Given $\varepsilon > 0$, unit vectors $\Psi_1,\ldots, \Psi_n \in \sphere\hilbH$, and $U \in \Unitary(\hilbH)$ we denote
\begin{align*}
B_{\varepsilon, \Psi_1,\ldots, \Psi_n}(U) = \qty{V \in \Unitary(\hilbH): \norm{V\Psi_i - U\Psi_i} < \varepsilon \tn { for all $i = 1,\ldots, n$}}.
\end{align*}
Given $\varepsilon > 0$, rays $\scrl_1,\ldots, \scrl_n \in \bbP\hilbH$, and $\bbU \in \PU(\hilbH)$, we denote
\begin{align*}
B_{\varepsilon, \scrl_1,\ldots,  \scrl_n}(\bbU) = \qty{\bbV \in \PU(\hilbH): d_\tn{chd}(\bbU\scrl_i, \bbV\scrl_i) < \varepsilon \tn{ for all $i = 1,\ldots, n$}},
\end{align*}
where $d_\tn{chd}$ is the chord metric on $\bbP\hilbH$. Recall that the chord metric induces the quotient topology on $\bbP \hilbH$ (Theorem \ref{thm:metric_equivalences}), so $B_{\varepsilon,\scrl_1,\ldots, \scrl_n}(\bbU)$ is open in the inital topology on $\PU(\hilbH)$.

\begin{theorem}[{\cite[Thm.~1]{Simms_1970}}]\label{thm:topologies_PU(H)}
Let $\hilbH$ be a nonzero complex Hilbert space. The quotient topology on $\PU(\hilbH)$ obtained from the strong operator topology on $\Unitary(\hilbH)$ coincides with the initial topology on $\PU(\hilbH)$  induced by the functions $\hat{\scrl}$ across all $\scrl \in \bbP\hilbH$.
\end{theorem}

We call this the \textdef{strong operator topology} on $\PU(\hilbH)_\tn{s}$ and denote it by $\PU(\hilbH)_\tn{s}$. 

\begin{proof}
If $\PU(\hilbH)$ is given the quotient topology from $\Unitary(\hilbH)_\tn{s}$, then each map $\hat{\scrl}$ is continuous by the characteristic mapping property of the quotient topology and the fact that the upper path $\Unitary(\hilbH)_\tn{s} \rightarrow \hilbH \rightarrow \bbP \hilbH$ in the commutative diagram \eqref{cd:projective_unitary} is continuous. Therefore the quotient topology is finer than the initial topology. 

For the opposite inclusion, let $O \subset \PU(\hilbH)$ be nonempty and open in the quotient topology. Let $\bbU \in O$, and let $U \in \Unitary(\hilbH)$ such that $q(U) = \bbU$, where $q:\Unitary(\hilbH)_\tn{s} \rightarrow \PU(\hilbH)$ is the quotient map.  Then $q^{-1}(O)$ is open in $\Unitary(\hilbH)_\tn{s}$ and $U \in q^{-1}(O)$, so there exists $\varepsilon > 0$ and unit vectors $\Psi_1,\ldots, \Psi_n \in \sphere\hilbH$ such that
\[
B_{\varepsilon,\Psi_1,\ldots, \Psi_n}(U) \subset q^{-1}(O).
\]
Let $\Omega_1,\ldots, \Omega_m \in \hilbH$ be an orthonormal basis for $\vecspan\qty{\Psi_1,\ldots, \Psi_n}$. Expand each $\Psi_i$ in this basis as $\Psi_i = \sum_{j=1}^m \alpha_{ij}\Omega_j$ where $\alpha_{ij} \in \bbC$. Let $M = \max_{i,j} \abs{\alpha_{ij}}$ and let $\delta = \varepsilon/mM$. Then one easily checks that 
\[
B_{\delta, \Omega_1,\ldots, \Omega_m}(U) \subset B_{\varepsilon,\Psi_1,\ldots, \Psi_n}(U).
\]

We must find a set $O' \subset \PU(\cH)$ which is open in the initial topology and satisfies $\bbU \in  O' \subset O$. Suppose $m > 1$. For $j \in \qty{2,\ldots, m}$, we define
\[
\Phi_j = \frac{\Omega_1 + \Omega_j}{\sqrt{2}}.
\]
Next, we define
\[
\gamma = \frac{\delta}{8 + 4 \sqrt{2}} < \frac{\delta}{2}.
\]
Finally, we define
\[
 O' = B_{\gamma, \bbC \Omega_1,\ldots, \bbC \Omega_m}(\bbU) \cap B_{\gamma, \bbC \Phi_2,\ldots, \bbC\Phi_m}(\bbU)
\]
If $m = 1$, then we define $ O' = B_{\gamma, \bbC \Omega_1}(\bbU)$. In either case, by definition we have $\bbU \in  O'$ and $ O'$ is open in the initial topology. We must show that $ O' \subset O$.

Let $\bbV \in O'$ and let $V \in \Unitary(\hilbH)$ such that $q(V) = \bbV$. By definition of the chord metric, for each $j = 1,\ldots, m$ there exists $\lambda_j \in \Unitary(1)$ such that
\[
\norm{V \Omega_j - \lambda_j U\Omega_j} < \gamma.
\]
Let us redefine $V$ to be $\lambda_1^{-1}V$, so that we may take $\lambda_1 = 1$. If $m = 1$, then $V \in B_{\delta, \Omega_1}(U) \subset q^{-1}(O)$, so $\bbV \in O$, which proves that $ O' \subset O$. 

Now suppose $m > 1$. For $j = 2,\ldots, m$, there also exists $\mu_j \in \Unitary(1)$ such that
\[
\norm{V\Phi_j - \mu_j U\Phi_j} < \gamma.
\]
Substituting in our definition of $\Phi_j$ and using the reverse triangle inequality yields
\begin{align*}
\sqrt{2}\gamma &> \norm{V(\Omega_1 + \Omega_j) - \mu_j U(\Omega_1 + \Omega_j)}\\
&\geq \norm{(1 - \mu_j)U\Omega_1 + (\lambda_j - \mu_j)U\Omega_j} - \norm{V\Omega_1 - U\Omega_1} - \norm{V\Omega_j - \lambda_j U\Omega_j}\\
&\geq \sqrt{\abs{1 - \mu_j}^2 + \abs{\lambda_j - \mu_j}^2} - 2\gamma.
\end{align*}
Rearranging this, we see that
\[
\max\qty{\abs{1 - \mu_j}, \abs{\lambda_j - \mu_j}} \leq (2 + \sqrt{2})\gamma.
\]
Then by the triangle inequality
\[
\max_j \abs{\lambda_j - 1} \leq \qty(4 + 2 \sqrt{2})\gamma = \frac{\delta}{2}.
\]

We now show that $V \in B_{\delta,\Omega_1,\ldots, \Omega_m}(U) \subset q^{-1}(O)$, so $\bbV \in O$. Indeed, for any $j = 1,\ldots, m$, the triangle inequality yields
\begin{align*}
\norm{V\Omega_j - U\Omega_j} \leq \norm{V\Omega_j - \lambda_j U\Omega_j} + \abs{\lambda_j - 1}  < \gamma + \frac{\delta}{2} < \delta.
\end{align*}
This proves $V \in B_{\delta, \Omega_1,\ldots, \Omega_m}(U) \subset q^{-1}(O)$, hence $\bbV \in O$. Thus, $ O' \subset O$, so the proof is complete.
\end{proof}

Now that we have the topology of $\PU(\hilbH)_\tn{s}$ sorted out, we can consider the following corollary.

\begin{corollary}[{\cite[Lem.~1]{Simms_1970}}]\label{cor:U(1)->U(H)->PU(H)}
Given $\Psi \in \hilbH \setminus \qty{0}$, define
\[
O_\Psi = \qty{U \in \Unitary(\hilbH): \ev{\Psi, U\Psi} > 0}
\]
and let $q:\Unitary(\hilbH)_\tn{s} \rightarrow \PU(\hilbH)_\tn{s}$ be the projection. The set $q(O_\Psi)$ is a neighborhood of $\1 \in \PU(\hilbH)_\tn{s}$ and the map
\begin{equation}\label{eq:PU(H)->U(H)_section}
q(O_\Psi) \rightarrow \Unitary(\hilbH)_\tn{s}, \quad \bbU \mapsto \frac{\abs{\ev{\Psi, U\Psi}}}{\ev{\Psi,U\Psi}}U
\end{equation}
is a continuous local section of $q$, where $U \in \Unitary(\hilbH)$ is any representative of $\bbU$. Thus, the projection $q:\Unitary(\hilbH)_\tn{s} \rightarrow \PU(\hilbH)_\tn{s}$ has the structure of a prinicpal $\Unitary(1)$-bundle.
\end{corollary}

\begin{proof}
Since $\PU(\hilbH)_\tn{s}$ has the quotient topology is a quotient of the topological group $\Unitary(\hilbH)_\tn{s}$, we know $q$ is an open map. Thus, $q(O_\Psi)$ is open since $O_\Psi$ is open in $\Unitary(\hilbH)_\tn{s}$. The restriction $q:O_\Psi \rightarrow q(O_\Psi)$ is therefore also an open map, hence a quotient map. The composition of $q:O_\Psi \rightarrow q(O_\Psi)$ with \eqref{eq:PU(H)->U(H)_section} is manifestly continuous by definition of $\Unitary(\hilbH)_\tn{s}$, so \eqref{eq:PU(H)->U(H)_section} is continuous. Since $q$ has a local section around the identity, it is a principal fiber bundle with structure group equal to the divisor $\Unitary(1)$.
\end{proof}

\begin{remark}
We can also topologize $\PU(\hilbH)$ as a quotient of $\Unitary(\hilbH)_\tn{n}$. We denote this by $\PU(\hilbH)_\tn{n}$ and note that $\PU(\hilbH)_\tn{n}$ is also a topological group. We shall have less use for this topology, however. We do note that this topology is finer than the topology on $\PU(\hilbH)_\tn{s}$. Indeed, if $O$ is open in $\PU(\hilbH)_\tn{s}$, then $q^{-1}(O)$ is open in $\Unitary(\hilbH)_\tn{s}$, hence in $\Unitary(\hilbH)_\tn{n}$, so $O = q(q^{-1}(O))$ is open in $\PU(\hilbH)_\tn{n}$ since $q:\Unitary(\hilbH)_\tn{n} \rightarrow \PU(\hilbH)_\tn{n}$ is an open map.
\end{remark}

We can now also easily consider continuity of the action of $\PU(\hilbH)$ on $\bbP\hilbH$.

\begin{prop}\label{prop:norm_unitary_PH_action}
Let $\cB(\hilbH)_{1,\tn{s}}$ denote the closed unit ball of $\cB(\hilbH)$ with the strong operator topology. The maps
\begin{alignat*}{2}
\cB(\hilbH)_{1,\tn{s}} \times \hilbH &\rightarrow \hilbH, &\quad (A, \Psi) &\mapsto A\Psi \\
\cB(\hilbH)_{1,\tn{s}} \times \bbP \hilbH &\rightarrow \bbP \hilbH, &\quad (A, \bbC\Psi) &\mapsto \bbC A\Psi\\
\PU(\hilbH)_\tn{s} \times \bbP \hilbH &\rightarrow \bbP \hilbH, &\quad (\bbU, \scrl) &\mapsto \bbU\scrl
\end{alignat*}
are all continuous. In particular, $\PU(\hilbH)_\tn{s}$, $\Unitary(\hilbH)_\tn{s}$, $\PU(\hilbH)_\tn{n}$, and $\Unitary(\hilbH)_\tn{n}$ all act continuously on $\bbP\hilbH$.
\end{prop}

\begin{proof}
Fix $A \in \cB(\hilbH)_{1,\tn{s}}$ and $\Psi \in \hilbH$. For any other $B \in \cB(\hilbH)_{1,\tn{s}}$ and $\Omega \in \hilbH$, we have
\[
\norm{B\Omega - A\Psi} \leq \norm{B \Omega - B\Psi} + \norm{B\Psi - A\Psi} = \norm{\Omega - \Psi} + \norm{B\Psi - A\Psi}.
\]
We see that by choosing $\Omega$ close to $\Psi$ and $B$ such that $B\Psi$ is close to $A\Psi$, we can make $B\Omega$ arbitrarily close to $A\Psi$. This proves continuity of $\cB(\hilbH)_{1,\tn{s}} \times \hilbH \rightarrow \hilbH$. Continuity of $\cB(\hilbH)_{1,\tn{s}} \times \bbP\hilbH \rightarrow \bbP\hilbH$ follows by choosing a continuous local section $\bbP \hilbH \rightarrow \sphere \hilbH$ and appealing to continuity of $\cB(\hilbH)_{1,\tn{s}} \times \hilbH \rightarrow \hilbH$.  Continuity of $\PU(\hilbH)_\tn{s} \times \bbP\hilbH \rightarrow \bbP \hilbH$ follows by choosing continuous local sections $\PU(\hilbH)_\tn{s} \rightarrow \Unitary(\hilbH)_\tn{s}$ and $\bbP\hilbH \rightarrow \sphere \hilbH$ of the canonical projections and appealing to continuity of the action of $\Unitary(\hilbH)_\tn{s}$ on $\sphere \hilbH$. We can choose local sections since $\Unitary(1) \rightarrow \Unitary(\hilbH)_\tn{s} \rightarrow \PU(\hilbH)_\tn{s}$ and $\Unitary(1) \rightarrow \sphere \hilbH \rightarrow \bbP\hilbH$ are fiber bundles. That $\Unitary(\hilbH)_\tn{s}$, $\PU(\hilbH)_\tn{n}$, and $\Unitary(\hilbH)_\tn{n}$ all act continuously on $\bbP\hilbH$ follows from continuity of all the canonical homomorphisms of these spaces in $\PU(\hilbH)_\tn{s}$.
\end{proof}

\newpage
\section{A Crash Course in \texorpdfstring{$C^*$}{C*}-Algebras}
\label{sec:crash_course}

We provide a concise introduction to $C^*$-algebras, with most proofs left to the literature. This abstract theory will be necessary to provide a mathematically rigorous treatment of systems on infinite lattices. We refer the reader to \cite{Murphy} for an excellent introduction to the subject and to \cite{KadisonRingroseI,KadisonRingroseII,PedersenCAlgAutomorphisms,DixCA} for more advanced treatments.

\subsection{Basics}

We begin by building up to the definition.

\begin{definition}
A \textdef{complex associative algebra} is a complex vector space $\fA$ with a bilinear and associative operation 
\[
\fA \times \fA \rightarrow \fA, \quad (A, B) \mapsto AB
\]
called \textdef{multiplication}. We note that multiplication is not necessarily commutative. A complex associative algebra is \textdef{unital} if it has a nonzero two-sided multiplicative identity $\1 \in \fA$. In particular, we do not consider $\fA = \qty{0}$ to be unital. If $\fA$ is unital, then we denote by $\fA^\times$ the set of (two-sided) invertible elements of $\fA$.

A \textdef{$\bm{*}$-algebra} is a complex associative algebra $\fA$ equipped with an additional operation
\[
\fA \rightarrow \fA, \quad A \mapsto A^*,
\]
called \textdef{involution} or simply \textdef{the $\bm{*}$-operation}, satisfying the following axioms:
\begin{alignat*}{3}
A^{**} &= A &&\quad \tn{for all $A \in \fA$,}\\ 
(AB)^* &= B^*A^* &&\quad \tn{for all $A, B \in \fA$,}\\
(\alpha A + \beta B)^* &= \alpha^*A^* + \beta^* B^* &&\quad \tn{for all $A, B \in \fA$ and $\alpha, \beta \in \bbC$.}
\end{alignat*}
Given $A \in \fA$, the element $A^*$ is called the \textdef{adjoint} of $A$. An element $A \in \fA$ is called \textdef{self-adjoint} if $A^* = A$. The set of self-adjoint elements of $\fA$ is denoted $\fA_\sa$. An element $A \in \fA$ is said to be \textdef{normal} if it commutes with its adjoint. If $\fA$ is a unital $*$-algebra, then an element $A \in \fA$ is called \textdef{unitary} if $A^*A = AA^* = \1$. The set of unitary elements of $\fA$ is denoted $\Unitary(\fA)$. Both self-adjoint elements and unitary elements are normal.

A \textdef{complex Banach space} is a complex vector space $\fA$ equipped with a norm $\norm{\cdot}:\fA \rightarrow \bbR$ such that the metric $d(A, B) = \norm{A - B}$ induced by the norm is complete, meaning every Cauchy sequence converges. Note that the metric induces a topology on $\fA$. A \textdef{complex Banach algebra} is a complex vector space that is both a complex associative algebra and a complex Banach space, such that the norm and multiplication operation satisfy \textdef{submultiplicativity}:
\begin{equation}\label{eq:C*-submultiplicativity}
\norm{AB} \leq \norm{A}\norm{B} \quad \tn{for all $A, B \in \fA$}.
\end{equation}
Finally, a \textdef{$\bm{C^*}$-algebra} is a complex vector space $\fA$ that is both a complex Banach algebra and a $*$-algebra, such that the norm and $*$-operation satisfy the \textdef{$\bm{C^*}$-identity} or \textdef{$\bm{C^*}$-property}:
\begin{equation}\label{eq:C*-property}
\norm{A^*A} = \norm{A}^2 \quad \tn{for all $A \in \fA$}.
\end{equation}
\end{definition}

$C^*$-algebras are truly majestic creatures. They are an abundant and extremely powerful structure. We will get a glimpse of their beauty below. A common example is the set $\cB(\hilbH)$ of bounded linear operators on a Hilbert space $\hilbH$, where the multiplication operation is composition of operators, the involution is the adjoint operation, and the norm is the operator norm:
\[
\norm{A} = \sup_{\Psi \in \bbS\hilbH} \norm{A\Psi}.
\]
A commutative example is furnished by the set 
\[
C(X) = \qty{f:X \rightarrow \bbC \mid \tn{$f$ is continuous}}
\]
of continuous complex-valued functions on a compact Hausdorff space $X$, where the algebraic operations are defined pointwise and the norm is given by
\[
\norm{f} = \sup_{x \in X} \abs{f(x)}.
\]

The next thing to do is define notions of subalgebra and homomorphism between $C^*$-algebras.

\begin{definition}
Let $\fA$ be a $C^*$-algebra. A \textdef{$C^*$-subalgebra} of $\fA$ is a topologically closed subset $\fB \subset \fA$ that is also closed under all algebraic operations (addition, scalar multiplication, multiplication, and involution). In particular, $\fB$ is itself a $C^*$-algebra when the operations on $\fA$ are restricted to $\fB$. If $\fA$ is a unital $C^*$-algebra, then we say $\fB$ is a \textdef{unital $C^*$-subalgebra} of $\fA$ if $\fB$ is a $C^*$-subalgebra of $\fA$ and $\1 \in \fB$, where $\1$ is the unit of $\fA$.

Observe that the intersection of an arbitrary collection of $C^*$-subalgebras of $\fA$ is a $C^*$-subalgebra of $\fA$. Given a subset $S \subset \fA$, the intersection of all $C^*$-subalgebras of $\fA$ containing $S$ is called the \textdef{$C^*$-subalgebra generated by $S$.}
\end{definition}

\begin{definition}
Let $\fA$ and $\fB$ be $*$-algebras. A linear map $\pi:\fA \rightarrow \fB$ is a \textdef{$*$-homomorphism} if, along with addition and scalar multiplication, it preserves multiplication and the $*$-operation:
\begin{alignat*}{3}
\pi(AB) &= \pi(A)\pi(B)  &&\quad \tn{for all $A, B \in \fA$,}\\
\pi(A^*) &= \pi(A)^* &&\quad \tn{for all $A \in \fA$}.
\end{alignat*}
If $\fA$ and $\fB$ are unital, then we say a $*$-homomorphism $\pi:\fA \rightarrow \fB$ is \textdef{unital} if $\pi(\1) = \1$. A bijective $*$-homomorphism $\pi$ is a called a \textdef{$*$-isomorphism}. It is easily checked that the inverse of a $*$-isomorphism is a $*$-isomorphism, so the name is appropriate. A $*$-isomorphism $\pi:\fA \rightarrow \fA$ is called an \textdef{automorphism} of $\fA$.
\end{definition}

One of the most astonishing things about the theory of $C^*$-algebras is that one can often derive topological conclusions from algebraic hypotheses. The following theorems are examples of this phenomenon.

\begin{theorem}\label{thm:*-hom_contractive}
Let $\fA$ and $\fB$ be $C^*$-algebras. If $\pi:\fA \rightarrow \fB$ is a $*$-homomorphism, then $\pi$ is continuous. Moreover,
\begin{equation}\label{eq:*-hom_contractive}
\norm{\pi(A)} \leq \norm{A} \quad \tn{for all $A \in \fA$,}
\end{equation}
and $\pi$ is injective if and only if equality holds in \eqref{eq:*-hom_contractive} for all $A \in \fA$.
\end{theorem}

\begin{theorem}
Let $\fA$ and $\fB$ be $C^*$-algebras. If $\pi:\fA \rightarrow \fB$ is a $*$-homomorphism, then $\pi(\fA)$ is closed in $\fB$, hence $\pi(\fA)$ is a $C^*$-subalgebra of $\fB$.
\end{theorem}

\begin{theorem}[{cf.~\cite[Cor.~10.1.8]{KadisonRingroseII}}]\label{thm:contractions_lift_to_contractions}
Let $\fA$ and $\fB$ be $C^*$-algebras. If $\pi:\fA \rightarrow \fB$ is a $*$-homomorphism and $B \in \pi(\fA)$, then there exists $A \in \fA$ such that $\pi(A) = B$ and $\norm{A} \leq \norm{B}$. If $B$ is self-adjoint, then we can choose $A$ to be self-adjoint. 
\end{theorem}

\begin{remark}
The statement about self-adjointness in Theorem \ref{thm:contractions_lift_to_contractions} follows immediately from the general case. Indeed, if $B$ is self-adjoint and $\pi(A) = B$ with $\norm{A} \leq \norm{B}$, then we can define $C = (A + A^*)/2$ and observe that $C$ is self-adjoint, $\pi(C) = B$, and $\norm{C} \leq \norm{B}$.

The statement of Theorem \ref{thm:contractions_lift_to_contractions} provided by \cite[Cor.~10.1.8]{KadisonRingroseII} is that $\pi$ maps the closed unit ball of $\fA$ onto the closed unit ball of $\pi(\fA)$, provided $\fA$, $\fB$, and $\pi$ are unital (In \cite{KadisonRingroseII}, all $C^*$-algebras and $*$-homomorphisms are defined to be unital). By scaling $B$ we get the result in the form above for unital $C^*$-algebras. The non-unital case of Theorem \ref{thm:contractions_lift_to_contractions} is immediate from the unital case by extending $\pi$ to a unital $*$-homomorphism $\tilde \pi:\tilde \fA \rightarrow \tilde \fB$, where $\tilde \fA$ and $\tilde \fB$ are the unitizations of $\fA$ and $\fB$. Theorem \ref{thm:contractions_lift_to_contractions} is also stated in the non-unital case and proved in \cite{CourtneryGillaspy}, although without the statement about self-adjointness.
\end{remark}

$C^*$-algebras have a rich spectral theory. We touch on this now.

\begin{definition}
Let $\fA$ be a unital complex Banach algebra. The \textdef{spectrum} of an element $A \in \fA$ is the set
\[
\sigma(A) = \qty{\lambda \in \bbC: A - \lambda \1 \tn{ is not invertible}}.
\]
An element of $\sigma(A)$ is called a \textdef{spectral value} of $A$. The \textdef{spectral radius} of $A$ is the number
\[
r(A) = \sup_{\lambda \in \sigma(A)} \abs{\lambda}.
\]
\end{definition}

If $A \in \fA$ had an empty spectrum, then its spectral radius would be $r(A) = \sup \varnothing = -\infty$. That would be a little icky. Thankfully this is never the case!

\begin{theorem}
Let $\fA$ be a unital complex Banach algebra. Given $A \in \fA$, the spectrum $\sigma(A)$ is nonempty and compact in $\bbC$.
\end{theorem}

\begin{theorem}
Let $\fA$ be a unital $C^*$-algebra. If $A \in \fA$ is normal, then
\[
r(A) = \norm{A}.
\]
An element $A \in \fA$ is self-adjoint if and only if $A$ is normal and $\sigma(A) \subset \bbR$. An element $A \in \fA$ is unitary if and only if $A$ is normal and $\sigma(A) \subset \Unitary(1)$. 
\end{theorem}

There is one more important type of normal element we would like to discuss. These are the positive elements, which are a subset of the self-adjoint elements.

\begin{thmdef}
Let $\fA$ be a unital $C^*$-algebra and let $A \in \fA$. The following are equivalent.
\begin{enumerate}
	\item There exists $B \in \fA$ such that $A = B^*B$.
	\item The element $A$ is normal and $\sigma(A) \subset [0,\infty)$.
\end{enumerate}
Such an element $A$ is said to be \textdef{positive}, and the set of positive elements of $\fA$ is denoted $\fA_+$.
\end{thmdef}

\begin{theorem}[Spectral Permanence]
Let $\fA$ be a unital $C^*$-algebra and let $\fB$ be a unital $C^*$-subalgebra of $\fA$. If $A \in \fB$, then the spectrum of $A$ when regarded as an element of $\fB$ is the same as the spectrum of $A$ when regarded as an element of $\fA$.
\end{theorem}

\begin{theorem}
Let $\fA$ and $\fB$ be unital $C^*$-algebras. If $\pi:\fA \rightarrow \fB$ is a unital $*$-homomorphism, then
\[
\sigma(\pi(A)) \subset \sigma(A)
\]
for all $A \in \fA$, with equality if $\pi$ is injective.
\end{theorem}

\begin{theorem}[Continuous Functional Calculus]
Let $\fA$ be a unital $C^*$-algebra, let $A \in \fA$ be normal, and let $\fB$ be the $C^*$-subalgebra of $\fA$ generated by $A$ and $\1$. There exists a unique unital $*$-isomorphism 
\[
C(\sigma(A)) \rightarrow \fB, \quad f \mapsto f(A)
\]
such that $\iota(A) = A$, where $\iota:\sigma(A) \rightarrow \bbC$ is the inclusion $\iota(z) = z$.
\end{theorem}

\subsection{States and Representations}

\begin{definition}
Let $\fA$ be a $C^*$-algebra. A linear functional $\omega:\fA \rightarrow \bbC$ is \textdef{positive} if
\[
\omega(A^*A) \geq 0 \quad \tn{for all $A \in \fA$.}
\]
\end{definition}

The following is another example of algebraic properties yielding topological conclusions. 

\begin{theorem}\label{thm:pos_lin_funct_norm=omega(1)}
Let $\fA$ be a $C^*$-algebra. If $\omega:\fA \rightarrow \bbC$ is a positive linear functional, then $\omega$ is continuous. If $\fA$ is unital and $\omega$ is a positive linear functional on $\fA$, then
\[
\norm{\omega} = \omega(\1).
\]
\end{theorem}

Thus, any positive linear functional on $\fA$ is an element of the dual space $\fA^*$. The norm above is the operator norm on $\fA^*$, namely
\[
\norm{\omega} = \sup_{\norm{A} \leq 1} \abs{\omega(A)}.
\]

\begin{definition}
Let $\fA$ be a $C^*$-algebra. A \textdef{state} on $\fA$ is a positive linear functional $\omega:\fA \rightarrow \bbC$ with $\norm{\omega} = 1$. The set of states of $\fA$ is denoted $\state(\fA)$; it is a subset of the dual space $\fA^*$. A state $\omega$ is \textdef{pure} if it is an extreme point of $\state(\fA)$, meaning it cannot be written as a nontrivial convex combination of two distinct states of $\state(\fA)$.
\end{definition}

The dual space $\fA^*$ has two important topologies which are inherited by $\state(\fA)$. The first is the metric topology induced by the norm on $\fA^*$, and the second is the weak* topology, defined as the coarsest topology such that the evaluation maps
\[
\hat A :\fA^* \rightarrow \bbC, \quad \hat A(\omega) = \omega(A)
\]
are continuous for all $A \in \fA$. A key property of this topology is that a function $f:X \rightarrow \fA^*$ from a topological space $X$ into $\fA^*$ is continuous if and only if $\hat A \circ f$ is continuous for all $A \in \fA$. We will use the prefixes ``norm-'' or ``weak*-'' to delineate properties pertaining to the two topologies.

\begin{theorem}\label{thm:states_nonempty_convex_compact}
Let $\fA$ be a nonzero $C^*$-algebra. The set of states $\state(\fA)$ is nonempty and convex in $\fA^*$. The set of pure states $\pstate(\fA)$ is nonempty. If $\fA$ is unital, then $\state(\fA)$ is weak*-compact and $\state(\fA)$ is the weak*-closed convex hull of $\pstate(\fA)$.
\end{theorem}

When $\fA$ is unital, one can use Theorem \ref{thm:pos_lin_funct_norm=omega(1)} to show that $\state(\fA)$ is weak*-closed, hence weak*-compactness follows from the Banach-Alaoglu theorem. The last clause of Theorem \ref{thm:states_nonempty_convex_compact} then follows from the Krein-Milman theorem.

In the context of quantum mechanics, \textbf{physical states are represented as states on the $C^*$-algebra of observables}. If $\fA$ is the $C^*$-algebra of observables and $\omega$ represents a physical state on $\fA$, then for any observable $A \in \fA$, the value $\omega(A)$ is interpreted as the expectation value of $A$ in the state $\omega$. To connect back to the usual formulation of quantum mechanics, we note that if $\fA = \cB(\hilbH)$ for some Hilbert space $\hilbH$ and if $\Psi \in \bbS \hilbH$, then the functional
\[
\omega:\cB(\hilbH) \rightarrow \bbC, \quad \omega(A) = \ev{\Psi, A\Psi}
\]
is a state on $\cB(\hilbH)$, as can easily be verified. This state is, in fact, pure (this will follow from the results below on representation theory). More generally, given a positive trace-class operator $\varrho \in \cB(\hilbH)$ with $\tr(\varrho) = 1$, the functional
\[
\omega:\cB(\hilbH) \rightarrow \bbC, \quad \omega(A) = \tr(\varrho A)
\]
is a state on $\cB(\hilbH)$. A positive trace-class operator $\varrho \in \cB(\hilbH)$ with $\tr(\varrho) = 1$ is called a \textdef{density operator}.

So far we have reformulated some basic axioms of quantum mechanics in terms of $C^*$-algebras and states on $C^*$-algebras. Hilbert spaces need not be present in this foundation, but may be reached through representations of the $C^*$-algebra. We define these now.

\begin{definition}
A \textdef{representation} of a $*$-algebra $\fA$ is a $*$-homomorphism $\pi:\fA \rightarrow \cB(\hilbH)$ for some Hilbert space $\hilbH$. We will sometimes denote a representation as $(\hilbH, \pi)$ instead of $\pi:\fA \rightarrow \cB(\hilbH)$. A representation $(\hilbH,\pi)$ is described as:
\begin{enumerate}
	\item \textdef{nonzero} if $\pi(\fA)$ contains a nonzero operator,
	\item \textdef{faithful} if $\pi$ is injective,
	\item \textdef{nondegenerate} if for every nonzero $\Psi \in \hilbH$ there exists $A \in \fA$ such that $\pi(A)\Psi \neq 0$,
	\item \textdef{cyclic} if there exists $\Psi \in \hilbH$ such that $\pi(\fA)\Psi = \qty{\pi(A)\Psi: A \in \fA}$ is dense in $\hilbH$. In this case the vector $\Psi$ is said to be a \textdef{cyclic vector for $\bm{\pi}$}. We will sometimes denote a cyclic representation as $(\hilbH, \pi, \Psi)$.
\end{enumerate}
Finally, a subspace $\hilbK \subset \hilbH$ is \textdef{invariant under $\pi(\fA)$} if $\pi(\fA)\hilbK \subset \hilbK$, where 
\[
\pi(\fA)\hilbK = \qty{\pi(A)\Psi: A \in \fA \tn{ and } \Psi \in \hilbK}.
\]
Then we say $\pi$ is:
\begin{enumerate}[resume]
	\item \textdef{irreducible} if the only closed invariant subspaces are $\qty{0}$ and $\hilbH$.
\end{enumerate}
\end{definition}

\begin{theorem}
Let $\pi:\fA \rightarrow \cB(\hilbH)$ be a representation of a $C^*$-algebra $\fA$.
\begin{enumerate}
	\item If $\fA$ is unital, then $\pi$ is nondegenerate if and only if $\pi(\1) = \1$.
	\item If $\pi$ is cyclic, then it is nondegenerate.
	\item If $\pi$ is nonzero and irreducible, then every nonzero $\Psi \in \hilbH$ is a cyclic vector for $\pi$.
\end{enumerate}
\end{theorem}

\begin{theorem}[Schur's Lemma]
Let $\fA$ be a $*$-algebra. A representation $\pi:\fA \rightarrow \cB(\hilbH)$ is irreducible if and only if
\[
\pi(\fA)' = \bbC \1,
\]
where 
\[
\pi(\fA)' = \qty{B \in \cB(\hilbH): [\pi(A),B] = 0 \tn{ for all $A \in \fA$}}
\]
is the \textdef{commutant} of $\pi(\fA)$.
\end{theorem}

\begin{theorem}[Gelfand--Naimark--Segal]
Let $\fA$ be a nonzero $C^*$-algebra and let $\omega \in \state(\fA)$. 
\begin{enumerate}
\item The set
\[
\fN_\omega = \qty{A \in \fA: \omega(A^*A) = 0}
\]
is a closed left ideal in $\fA$ called the \textdef{Gelfand ideal} of $\omega$. 
\item The equation
\[
\ev{A + \fN_\omega, B+ \fN_\omega} = \omega(A^*B)
\]
defines an inner product on $\fA/\fN_\omega$. Let $\hilbH_\omega$ be the completion of $\fA/\fN_\omega$ with respect to this inner product.
\item There exists a unique representation $\pi_\omega:\fA \rightarrow \cB(\hilbH_\omega)$ such that
\[
\pi_\omega(A)(B + \fN_\omega) = AB + \fN_\omega
\]
for all $A, B \in \fA$.
\item There exists a unique cyclic unit vector $\Omega_\omega \in \hilbH_\omega$ such that
\[
\pi_\omega(A)\Omega_\omega = A + \fN_\omega \qqtext{and} \omega(A) = \ev{\Omega_\omega, \pi_\omega(A)\Omega_\omega}
\]
for all $A \in \fA$. The representation $(\hilbH_\omega, \pi_\omega, \Omega_\omega)$ is called the \textdef{Gelfand-Naimark-Segal or GNS representation of $\omega$.}
\item If $(\hilbK, \rho)$ is another representation of $\fA$ with cyclic unit vector $\Psi \in \hilbK$ such that $\omega(A) = \ev{\Psi, \rho(A)\Psi}$ for all $A \in \fA$, then there exists a unique unitary $U:\hilbH \rightarrow \hilbK$ such that
\[
U\pi_\omega(A) = \rho(A)U \qqtext{and} U\Omega_\omega = \Psi.
\]
for all $A \in \fA$.
\item The GNS representation of $\omega$ is irreducible if and only if $\omega$ is pure.
\end{enumerate}
\end{theorem}

\begin{comment}
\begin{thmdef}
Let $\fA$ be a $C^*$-algebra and let $\omega \in \sS(\fA)$. The following are equivalent.
\begin{enumerate}
	\item The state $\omega$ is an extreme point of $\sS(\fA)$, meaning it cannot be written as a nontrivial convex combination of two distinct states of $\sS(\fA)$.
	\item For every positive linear functional $\psi:\fA \rightarrow \bbC$ satisfying $\psi(A^*A) \leq \omega(A^*A)$ for all $A \in \fA$, there exists $t \in [0,1]$ such that $\psi = t \omega$.
\end{enumerate}
Such a state is said to be \textdef{pure}. The set of pure states of $\fA$ is denoted $\sP(\fA)$.
\end{thmdef}
\end{comment}

\subsection{The Kadison Transitivity Theorem}
\label{sec:Kadison_transitivity}

We now provide a key theorem, called the Kadison transitivity theorem, that we will use time and time again. Indeed, the Kadison transitivity theorem is the basis of one of the major results of this thesis. 

\begin{theorem}[Kadison transitivity]\label{thm:Kadison_transitivity_theorem}
Let $\fA$ be a nonzero $C^*$-algebra with a nonzero irreducible representation $(\hilbH, \pi)$. Given linearly independent vectors $x_1,\ldots, x_n \in \hilbH$ and arbitrary vectors $y_1,\ldots, y_n \in \hilbH$, there exists $A \in \fA$ such that 
\[
\pi(A)x_i = y_i
\]
for all $i = 1,\ldots, n$. If there exists $T \in \cB(\hilbH)$ such that $Tx_i = y_i$ for all $i$, then we may choose $A$ to satisfy $\norm{A} \leq \norm{T}$. If there exists a self-adjoint (resp.\ positive) $T \in \cB(\hilbH)$ such that $Tx_i = y_i$ for all $i$, then we may choose $A$ to be self-adjoint (resp.\ positive) and satisfy $\norm{A} \leq \norm{T}$. If $\fA$ is unital and there exists a unitary $V \in \Unitary(\hilbH)$ such that $Vx_i = y_i$ for all $i$, then there exists a self-adjoint $B \in \fA$ such that $\pi(e^{iB})x_i = y_i$ for all $i$; in particular, $A$ may be chosen to be unitary.
\end{theorem}

Unfortunately, I am not aware of anywhere in the literature where the theorem is given exactly as above. The result first appeared in \cite{Kadison1957IRREDUCIBLEOA}, without the norm bounds on $A$ and without the unitary version of the theorem. The unitary version was published in \cite{GlimmKadisonUOCA} and the norm bounds on $A$ were stated in \cite{Kadison:1967zta}. The theorem has been included in many textbooks since then, sometimes with slight variations or generalizations. For example Kadison always takes $C^*$-algebras to be unital, but \cite[Thm.~5.2.2]{Murphy} does not require $\fA$ to be unital (except for in the unitary version, which does not make sense for non-unital $\fA$). However, \cite[Thm.~5.2.2]{Murphy} does not provide the norm bounds. A version for non-unital $C^*$-algebras with slightly weaker norm bounds is provided in \cite[Thm.~4.18]{takesaki2001theory}, wherein norm bounds of the form $\norm{\1 - e^{iB}} \leq \norm{\1 - V} + \varepsilon$ are also provided in the unitary case. The theorem is stated for non-unital $C^*$-algebras with the desired norm bounds in \cite{BratteliRobinsonOAQSMI}, but no proof is provided. Every statement of the theorem I am aware of takes $\fA$ to be a $C^*$-subalgebra of $\cB(\hilbH)$ acting irreducibly on $\hilbH$ (meaning the inclusion $\fA \hookrightarrow \cB(\hilbH)$ is an irreducible representation), rather than letting $\fA$ be an abstract $C^*$-algebra with a nonzero irreducible representation, in which case $\pi(\fA)$ is a $C^*$-subalgebra of $\cB(\hilbH)$ acting irreducibly on $\hilbH$. The variation provided in Theorem \ref{thm:Kadison_transitivity_theorem} can be obtained almost immediately from \cite[Thm.~2.7.5]{PedersenCAlgAutomorphisms} and Theorem \ref{thm:contractions_lift_to_contractions}, but \cite[Thm.~2.7.5]{PedersenCAlgAutomorphisms} is phrased abstractly enough that this might not be obvious to the untrained eye. 

Therefore, for the sake of completeness I will prove the desired variation Theorem \ref{thm:Kadison_transitivity_theorem}, starting from the form given in \cite[Thm.~5.2.2]{Murphy}, since this is perhaps the most accessible place to find the theorem. We have reproduced \cite[Thm.~5.2.2]{Murphy} below verbatim (modulo changes in notation) for the convenience of the reader.

\begin{theorem}[{\cite[Thm.~5.2.2]{Murphy}}]
Let $\fB$ be a nonzero $C^*$-algebra acting irreducibly on a Hilbert space $\hilbH$, and suppose that $x_1,\ldots, x_n$ and $y_1,\ldots, y_n$ are elements of $\hilbH$ and that $x_1,\ldots, x_n$ are linearly independent. Then there exists an operator $B \in \fB$ such that $Bx_i = y_i$ for all $i = 1,\ldots, n$. If there is a self-adjoint operator $T$ on $\hilbH$ such that $Tx_i = y_i$ for $i = 1,\ldots, n$, then we may choose $B$ to be self-adjoint also. If $\fB$ contains $\1$ and there is a unitary $V$ on $\hilbH$ such that $Vx_i = y_i$ for all $i = 1,\ldots, n$, then we may choose $B$ to be unitary also---we may even suppose that $B = e^{iC}$ for some element $C \in \fB_\tn{sa}$.
\end{theorem}

From this, we now give the proof of Theorem \ref{thm:Kadison_transitivity_theorem}, following the essential idea of \cite[Thm.~2.7.5]{PedersenCAlgAutomorphisms}. We set $\fB = \pi(\fA)$.

\begin{proof}
First suppose $T$ is self-adjoint (resp.\ positive). Let $P \in \cB(\hilbH)$ be the orthogonal projection onto the subspace $\vecspan\qty{x_1,\ldots, x_n, y_1,\ldots, y_n}$ and define $S = PTP$, which is self-adjoint (resp.\ positive), satisfies $\norm{S} \leq \norm{T}$, and has $Sx_i = Tx_i = y_i$ for all $i \leq n$. Extend $x_1,\ldots, x_n$ to a basis $x_1,\ldots, x_m$ of the above span. By \cite[Thm.~5.2.2]{Murphy}, there exists $B \in \pi(\frA)_\tn{sa}$ such that $Bx_i = PTPx_i = PTx_i$ for all $i = 1,\ldots, m$, hence $BP = S$. There exists $A_0 \in \fA$ such that $\pi(A_0) = B$, and defining $A = (A_0 + A_0^*)/2$, we have $\pi(A) = B$ and $A \in \fA_\tn{sa}$. The equation $BP = S$ becomes $\pi(A)P = S$.  Assuming $\pi(A)^kP = S^k$ for some $k \in \bbN$, we have
\[
\pi(A)^{k+1}P = \pi(A)^k S = \pi(A)^k PS = S^{k+1},
\]
so $\pi(A)^{k}P = S^k$ for all $k \in \bbN$ by induction.

If $T$ is only self-adjoint, define a continuous function $f:\bbR \rightarrow \bbR$ by
\[
f(t) = \begin{cases} \norm{S} &: t \geq \norm{S} \\ t &: \abs{t} \leq \norm{S} \\ -\norm{S} &: t \leq -\norm{S}\end{cases}.
\]
If $T$ is positive, instead define $f:\bbR \rightarrow \bbR$ by
\[
f(t) = \begin{cases} \norm{S} &: t \geq \norm{S} \\ t &: 0 \leq t \leq \norm{S} \\ 0 &: t \leq 0 \end{cases}. 
\]
Note that $f(A)$ is well-defined by continuous functional calculus and lies in $\frA$ regardless of whether $\frA$ is unital or not since $f(0) = 0$. Furthermore, observe that $f(A)$ is self-adjoint (resp.\ positive), $\norm{f(A)} \leq \norm{S} \leq \norm{T}$, and $\pi(f(A)) = f(\pi(A))$. We also have $f(S) = S$ since $f$ restricts to the identity on $\sigma(S)$. 

We show that $\pi(f(A))x_i = Sx_i = y_i$ for all $i \leq n$. Given $i \leq n$ and $\varepsilon > 0$, we choose a real polynomial $g$ such that $\abs{f(t) - g(t)} < \varepsilon/2\norm{x_i}$ whenever $\abs{t} \leq \max(\norm{A}, \norm{S})$. Note that $g(\pi(A))x_i = g(S)x_i$ since $\pi(A)^kP = S^k$ for all $k \in \bbN$. Then
\begin{align*}
\norm{\pi(f(A))x_i - Sx_i} &\leq \norm{f(\pi(A))x_i - g(\pi(A))x_i} + \norm{g(S)x_i - Sx_i}\\
&\leq \norm{(f-g)(\pi(A))}\norm{x_i} + \norm{(g - f)(S)}\norm{x_i} < \varepsilon.
\end{align*}
Since $\varepsilon > 0$ was arbitrary, this implies $\pi(f(A))x_i = Sx_i = y_i$, as desired. This proves the self-adjoint and positive versions of the theorem.

For the general case, we again consider $S = PTP$, where $P$ is defined as before, but $T$ is not necessarily self-adjoint or positive. We still have $\norm{S} \leq \norm{T}$ and $Sx_i = Tx_i$ for all $i \leq n$. The map $\abs{S}(\hilbH) \rightarrow S(\hilbH)$, $\abs{S}x \mapsto Sx$ is a well-defined bijective isometry, and may therefore be extended to a unitary $V: \hilbH \rightarrow \hilbH$ (this is only possible because $\dim \abs{S}(\hilbH) = \dim S(\hilbH) < \infty$). Note that $S = V\abs{S}$. By the self-adjoint case above, there exists $A \in \frA_\tn{sa}$ such that $\norm{A} \leq \norm{\abs{S}} = \norm{S} \leq \norm{T}$ and $\pi(A)x_i = \abs{S}x_i$ for all $i\leq n$. 

We now want to mimic the action of $V$ on the vectors $\abs{S}x_i$ with an element of $\tilde\fA$, where $\tilde \fA$ is the unitization of $\fA$. Let $\tilde \pi$ be the unique extension of $\pi$ to a unital $*$-homomorphism $\tilde \pi:\tilde \fA \rightarrow \cB(\hilbH)$. By \cite[Thm.~5.2.2]{Murphy} applied $\tilde \pi(\tilde \fA)$, there exists $\tilde C \in \tilde \pi(\tilde \fA)_\tn{sa}$ such that $e^{i\tilde C}\abs{S}x_i = V\abs{S}x_i = Sx_i$ for all $i \leq n$ (the vectors $\abs{S}x_i$ may not be linearly independent, but one can replace them by a basis for their span and still achieve the claimed result). There exists $C \in \tilde \fA_\tn{sa}$ such that $\tilde \pi(C) = \tilde C$. Defining $U = e^{iC} \in \Unitary(\tilde \fA)$, we have $\pi(U)\abs{S}x_i = e^{i\pi(C)}\abs{S}x_i = Sx_i$.

Now observe that Then $UA \in \frA$, $\norm{UA} \leq \norm{T}$, and $\pi(UA)x_i = Sx_i = Tx_i$ for all $i$. This proves the general case of the theorem.

Finally, we need the unitary case of Theorem \ref{thm:Kadison_transitivity_theorem}. By \cite[Thm.~5.2.2]{Murphy}, there exists a self-adjoint $C \in \pi(\fA)_\tn{sa}$ such that $e^{iC}x_i = y_i$ for all $i$. There exists a self-adjoint $B \in \fA_\tn{sa}$ such that $\pi(B) = C$, and since $\pi(e^{iB}) = e^{i\pi(B)}$, we have $\pi(e^{iB})x_i = y_i$ for all $i$.
\end{proof}

\subsection{Topologies on \texorpdfstring{$\Aut(\fA)$}{Aut(A)} and \texorpdfstring{$\Inn(\fA)$}{Inn(A)}}
\label{subsec:Aut(A)_Inn(A)}

We review here some basic facts about the automorphism group and inner automorphism group of a $C^*$-algebra. Throughout this section, let $\fA$ be a $C^*$-algebra. Let $\Aut(\fA)$ denote the set automorphisms of $\fA$, that is, the set of $*$-isomorphisms $\alpha:\fA \rightarrow \fA$. Observe that $\Aut(\fA)$ is a group with respect to composition of automorphisms. There are two topologies on $\Aut(\fA)$ that we will work with, namely the \textdef{norm topology} and the \textdef{strong topology}. The norm topology is the topology on $\Aut(\fA)$ inherited from $\cB(\fA)$, the space of bounded linear operators $\fA \rightarrow \fA$ with the operator norm. The strong topology is the initial topology on $\Aut(\fA)$ induced by the evaluation maps $\widehat{A}:\Aut(\fA) \rightarrow \fA$, $\widehat{A}(\alpha) = \alpha(A)$ across all $A \in \fA$. In other words, it is the coarsest topology such that all evaluation maps are continuous. We let $\Aut(\fA)_\tn{n}$ and $\Aut(\fA)_\tn{s}$ denote the automorphism group endowed with the norm and strong topologies, respectively. Note that the strong topology is coarser than the norm topology.

\begin{proposition}\label{prop:aut(A)xA->A_continuous}
The map
\[
\Aut(\fA)_\tn{s} \times \fA \rightarrow \fA, \quad (\alpha, A) \mapsto \alpha(A)
\]
is continuous.
\end{proposition}

\begin{proof}
Fix $(\alpha, A) \in \Aut(\fA) \times \fA$.  Given $(\beta, B) \in \Aut(\fA) \times \fA$, we have
\begin{equation}\label{eq:aut(A)xA->A_continuous}
\norm{\beta(B) - \alpha(A)} \leq \norm{\beta(B) - \beta(A)} + \norm{\beta(A) - \alpha(A)} \leq \norm{B - A} + \norm{\beta(A) - \alpha(A)}.
\end{equation}
Given $\varepsilon > 0$, we can choose neighborhoods $\alpha \in O_\alpha \subset \Aut(\fA)_\tn{s}$ and $A \in O_A \subset \fA$ such that for all $(\beta, B) \in O_\alpha \times O_A$, we have $\norm{B - A} < \varepsilon/2$ and $\norm{\beta(A) - \alpha(A)} < \varepsilon/2$. Continuity then follows from \eqref{eq:aut(A)xA->A_continuous}.
\end{proof}

\begin{proposition}
The automorphism groups $\Aut(\fA)_\tn{n}$ and $\Aut(\fA)_\tn{s}$ are both topological groups.
\end{proposition}

\begin{proof}
First consider $\Aut(\fA)_\tn{n}$. Recall that $\cB(\fA)$ is a Banach algebra and $\Aut(\fA)$ is a subset of the set of invertible elements of $\cB(\fA)$. Composition is norm-continuous on $\cB(\fA)$ and inversion is norm-continuous on the space of invertible elements. Therefore $\Aut(\fA)_\tn{n}$ is a topological group.

Now consider $\Aut(\fA)_\tn{s}$. Fix $A \in \fA$. Observe that 
$\Aut(\fA)_\tn{s} \times \Aut(\fA)_\tn{s} \rightarrow \fA$, $(\alpha, \beta) \mapsto \alpha(\beta(A))$
is given by a composition 
\[
\Aut(\fA)_\tn{s} \times \Aut(\fA)_\tn{s} \rightarrow \Aut(\fA)_\tn{s} \times \fA \rightarrow \fA, \quad (\alpha, \beta) \mapsto (\alpha, \beta(A)) \mapsto \alpha(\beta(A)).
\]
Thus, $(\alpha, \beta) \mapsto \alpha(\beta(A))$ is continuous by definition of the strong topology and by Proposition \ref{prop:aut(A)xA->A_continuous}, so composition is continuous on $\Aut(\fA)_\tn{s}$ by definition of the strong topology. For inversion, fix $\alpha \in \Aut(\fA)$ and note that for any $\beta \in \Aut(\fA)$ we have
\begin{equation}\label{eq:Aut(A)_inversion_continuous}
\norm{\beta^{-1}(A) - \alpha^{-1}(A)} = \norm{A - \beta(\alpha^{-1}(A))} = \norm{\alpha(\alpha^{-1}(A)) - \beta(\alpha^{-1}(A))}.
\end{equation}
Given $\varepsilon > 0$, there exists a neighborhood $\alpha \in O \subset \Aut(\fA)_\tn{s}$ such that $\beta \in O$ implies that \eqref{eq:Aut(A)_inversion_continuous} is less than $\varepsilon$ by continuity of $\widehat{\alpha^{-1}(A)}$. This proves continuity of $\beta \mapsto \beta^{-1}(A)$ at $\alpha$. Since $\alpha$ was arbitrary, we know $\beta \mapsto \beta^{-1}(A)$ is continuous. Continuity of inversion on $\Aut(\fA)_\tn{s}$ follows by definition of the strong topology. 
\end{proof}

Schottenloher has shown that the unitary group of an (infinite-dimensional) Hilbert space with the strong operator topology is a topological group \cite{SchottenloherUnitaryStrongTopology}. Our proof above that $\Aut(\fA)_\tn{s}$ is a topological group is akin to his proof for the unitary group. We suspect the result is well-known, but we are not aware of a reference.

\begin{proposition}\label{prop:dense_span_generates_strong_top}
Let $\fA_0$ be a subset of $\fA$ such that $\vecspan\fA_0$ is dense in $\fA$. The strong topology on $\Aut(\fA)$ coincides with the initial topology induced by the evaluation maps $\widehat{A}$ across all $A \in \fA_0$.
\end{proposition}

\begin{proof}
Let $\cT_{\fA_0}$ be the initial topology induced by $\fA_0$. By definition of the initial topology, we know the $\cT_{\fA_0}$ is coarser than the strong topology. If $A = \sum_{i=1}^n \lambda_i A_i \in \vecspan \fA_0$ with $A_1,\ldots, A_n \in \fA_0$, then $\widehat{A} = \sum_{i=1}^n \lambda_i \widehat{A}_i$ is continuous with respect to $\cT_{\fA_0}$ since addition and scalar multiplication are continuous on $\fA$. 

We show continuity of $\widehat{A}$ with respect to $\cT_{\fA_0}$ for arbitrary $A \in \fA$. Fix $\alpha \in \Aut(\fA)$ and $\varepsilon > 0$. Choose $B \in \vecspan \fA_0$ such that $\norm{A - B} < \varepsilon/3$. Since $\widehat{B}$ is continuous with respect to $\cT_{\fA_0}$, there exists a neighborhood $\alpha \in O \in \cT_{\fA_0}$ such that $\beta \in O$ implies $\norm{\beta(B) - \alpha(B)} < \varepsilon/3$. Thus, $\beta \in O$ implies
\begin{align*}
\norm{\beta(A) - \alpha(A)} &\leq \norm{\beta(A) - \beta(B)} + \norm{\beta(B) - \alpha(B)} + \norm{\alpha(B) - \alpha(A)} \\
&\leq 2\norm{A - B} + \norm{\beta(B) - \alpha(B)} < \varepsilon.
\end{align*}
This proves continuity of $\widehat{A}$ at $\alpha$. Thus, $\widehat{A}$ is continuous with respect to $\cT_{\fA_0}$ for all $A \in \fA$, so the strong topology is coarser than $\cT_{\fA_0}$.
\end{proof}

Now that we know $\Aut(\fA)_\tn{s}$ is a topological group, it is clear that the map from Proposition \ref{prop:aut(A)xA->A_continuous} is a continuous group action.

We now move towards introducing the inner automorphism group. First, we introduce some convenient notation.

\begin{definition}\label{def:adjoint_action_A}
Given $A \in \fA$, we define the \textdef{adjoint action} of $A$ as
\[
\Ad(A):\fA \rightarrow \fA, \quad \Ad(A)(B) = ABA^*.
\]
The reason for putting the star on the right side is that with this definition we get
\begin{equation}\label{eq:Ad_respect_multiplication}
\Ad(A) \circ \Ad(B) = \Ad(AB)
\end{equation}
for all $A, B \in \fA$, as can easily be checked. We note that $\Ad(A)$ is linear and respects the $*$-operation for all $A \in \fA$. 
\end{definition}

\begin{definition}
Suppose $\fA$ is unital. An automorphism $\alpha \in \Aut(\fA)$ is \textdef{inner} if there exists $U \in \Unitary(\fA)$ such that $\alpha = \Ad(U)$. Note also that $\Ad(U) \in \Aut(\fA)$ for all $U \in \Unitary(\fA)$, with inverse $\Ad(U)^{-1} = \Ad(U^*)$. The set of all inner automorphisms of $\fA$ is denoted $\Inn(\fA)$. By \eqref{eq:Ad_respect_multiplication}, $\Inn(\fA)$ is a subgroup of $\Aut(\fA)$. We may equip $\Inn(\fA)$ with the norm or strong topology, denoted $\Inn(\fA)_\tn{n}$ and $\Inn(\fA)_\tn{s}$, respectively. With either topology $\Inn(\fA)$ is a topological group because subgroups of topological groups are topological groups.
\end{definition}

\begin{remark}
If $\fA$ is unital, then $\Unitary(\fA)$ is a topological group with the subspace topology inherited from the norm on $\fA$. For a Hilbert space $\hilbH$, the unitary group $\Unitary(\hilbH)$ has another natural topology, namely the subspace topology inherited from the strong operator topology on $\cB(\hilbH)$. The unitary group with the strong operator topology is a topological group \cite{SchottenloherUnitaryStrongTopology}, which we will denote by $\Unitary(\hilbH)_\tn{s}$ to distinguish from the unitary group with the norm topology, denoted $\Unitary(\hilbH)_\tn{n}$. We will not define a ``strong topology'' on the unitary group of an abstract $C^*$-algebra, so the symbol $\Unitary(\fA)$ will always denote the unitary group with the norm topology.
\end{remark}

\begin{proposition}
Let $\fA$ be unital. If $U, V \in \Unitary(\fA)$, then
\[
\norm{\Ad(U) - \Ad(V)} \leq 2\norm{U - V}.
\]
Thus, $\Ad:\Unitary(\fA) \rightarrow \Inn(\fA)_\tn{n}$ is a continuous surjective group homomorphism.
\end{proposition}

\begin{proof}
Given $A \in \fA$ such that $\norm{A}\leq 1$, observe that
\begin{align*}
\norm{\Ad(U)(A) - \Ad(V)(A)} &= \norm{UAU^* - VAV^*} \\
&\leq \norm{UAU^* - UAV^*} + \norm{UAV^* - VAV^*}\\
&\leq \norm{U^* - V^*} + \norm{U - V} = 2\norm{U - V}.
\end{align*}
Thus, $\norm{\Ad(U) - \Ad(V)} \leq 2\norm{U - V}$.
\end{proof}

\begin{definition}\label{def:adjoint_action_alpha}
Given $\alpha \in \Aut(\fA)$, we define the \textdef{adjoint action} of $\alpha$ as
\[
\Ad(\alpha):\Aut(\fA) \rightarrow \Aut(\fA), \quad \Ad(\alpha)(\beta) = \alpha \circ \beta \circ \alpha^{-1}.
\]
As before, we note that 
\[
\Ad(\alpha) \circ \Ad(\beta) = \Ad(\alpha \circ \beta)
\]
for all $\alpha, \beta \in \Aut(\fA)$.
\end{definition}

\begin{theorem}\label{thm:Aut(A)_Inn(A)_group_action}
Let $\fA$ be unital. If $\alpha \in \Aut(\fA)$ and $U \in \Unitary(\fA)$, then
\begin{equation}\label{eq:Ad(alpha)_Ad(U)}
\Ad(\alpha)(\Ad(U)) = \Ad(\alpha(U)).
\end{equation}
Moreover, the map
\begin{equation}\label{eq:Aut(A)_Inn(A)_group_action}
\Aut(\fA)_\tn{s} \times \Inn(\fA)_\tn{n} \rightarrow \Inn(\fA)_\tn{n}, \quad (\alpha, \eta) \mapsto \Ad(\alpha)(\eta)
\end{equation}
is a continuous group action of $\Aut(\fA)_\tn{s}$ on $\Inn(\fA)_\tn{n}$.
\end{theorem}

\begin{proof}
Given $A \in \fA$, observe that
\begin{align*}
[\Ad(\alpha)(\Ad(U))](A) &= [\alpha \circ \Ad(U) \circ \alpha^{-1}](A) = \alpha(U\alpha^{-1}(A)U^*) \\
&= \alpha(U)A \alpha(U)^* = \qty[\Ad(\alpha(U))](A).
\end{align*}
This proves \eqref{eq:Ad(alpha)_Ad(U)}.

Well-definedness of \eqref{eq:Aut(A)_Inn(A)_group_action} follows from \eqref{eq:Ad(alpha)_Ad(U)}. It is then clear that \eqref{eq:Aut(A)_Inn(A)_group_action} is a group action. We show continuity of \eqref{eq:Aut(A)_Inn(A)_group_action} at an arbitrary element $(\alpha, \Ad(U)) \in \Aut(\fA) \times \Inn(\fA)$. Given an arbitrary $(\beta, \Ad(V)) \in \Aut(\fA) \times \Inn(\fA)$, we observe that
\begin{align*}
\norm{\Ad(\beta)(\Ad(V)) - \Ad(\alpha)(\Ad(U))} &\leq \norm{\Ad(\beta)(\Ad(V)) - \Ad(\beta)(\Ad(U))} \\
&\qquad + \norm{\Ad(\beta(U)) - \Ad(\alpha(U))}\\
&\leq \norm{\Ad(V) - \Ad(U)} + 2\norm{\beta(U) - \alpha(U)}.
\end{align*}
The above quantity can be made arbitrarily small by choosing $\beta$ and $\Ad(V)$ to be in small enough neighborhoods around $\alpha$ and $\Ad(U)$, respectively. This proves continuity at $(\alpha, \Ad(U))$.
\end{proof}

\subsection{The Quasi-Local Algebra of a Quantum Spin System}\label{subsec:Quasi-local_Algebra}

In this dissertation we will be interested in the $C^*$-algebra of observables of a system of spins arranged on a lattice $\Gamma$. For example, if we have a spin-1/2 particle on each lattice site, then the observable algebra associated to the spin on a single site is $M_2(\bbC)$, and the total $C^*$-algebra of observables would be
\[
\bigotimes_{i \in \Gamma} M_2(\bbC).
\]
When $\Gamma$ is infinite, we have an infinite tensor product, and some care is required in giving a sensible and precise definition. We shall formalize this infinite tensor product as a directed colimit in the category of $C^*$-algebras and $*$-homomorphisms. We explain this coarsely below and provide Appendix \ref{app:directed_colimits} for a precise definition and review of directed colimits. For further exposition of spin systems we refer the reader to Naaijken's excellent introductory book \cite{NaaijkensQSSIL} and for further mathematical details we refer to \cite{KadisonRingroseII,Murphy}.

\begin{definition}\label{def:quasi-local_algebra}
Let $\Gamma$ be a nonempty countable set, which we call the \textdef{lattice}, and let $\nfpset(\Gamma)$ be the set of nonempty finite subsets of $\Gamma$. Note that $\nfpset(\Gamma)$ is a directed set, partially ordered by inclusion. 
%We will use the notation $\Lambda \Subset \Gamma$ as a synonym for $\Lambda \in \wp_0(\Gamma)$.
Elements of $\Gamma$ will be called \textdef{sites}. For each site $i \in \Gamma$, we assume to be given a natural number $n_i \geq 2$ representing the dimension of an \textdef{on-site Hilbert space} at site $i$. In particular, if site $i$ consists of a particle of spin $s$, then we would have $n_i = 2s + 1$.

For every nonempty finite subset $\Lambda \in \nfpset(\Gamma)$, define the finite tensor product
\[
\fA_\Lambda = \bigotimes_{i \in \Lambda} M_{n_i}(\bbC).
\]
For $\Lambda_1, \Lambda_2 \in \nfpset(\Gamma)$ such that $\Lambda_1 \subset \Lambda_2$, we have a natural $*$-homomorphism $\iota_{\Lambda_2\Lambda_1}:\fA_{\Lambda_1} \rightarrow \fA_{\Lambda_2}$  defined by tensoring on identity operators to the sites in $\Lambda_2 \setminus \Lambda_1$. In symbols, 
\[
\iota_{\Lambda_2\Lambda_1}\qty(\bigotimes_{v \in \Lambda_1} A_v) = \bigotimes_{v \in \Lambda_2} B_v \qqtext{where} B_v = \begin{cases} A_v &: v \in \Lambda_1 \\ \1 &: v \notin \Lambda_1  \end{cases}.
\]
It is easy to see that $\iota_{\Lambda_2\Lambda_1}$ is unital and it can be shown that $\iota_{\Lambda_2\Lambda_1}$ is injective. 

The finite tensor products $\fA_\Lambda$ together with the inclusions inclusions $\iota_{\Lambda_2 \Lambda_1}$  form a directed system indexed by $\nfpset(\Gamma)$, and we define the \textdef{quasi-local algebra} $\fA$ of the lattice spin system to be the directed colimit of this system. Intuitively, since the embeddings $\iota_{\Lambda_2\Lambda_1}$ for $\Lambda_1 \subset \Lambda_2$ allow us to think of $\fA_{\Lambda_1}$ as a $C^*$-subalgebra of $\fA_{\Lambda_2}$, we can in a  sense take a union
\begin{equation}\label{eq:local_algebra}
\fA_\tn{loc} = \bigcup_{\Lambda \in \nfpset(\Gamma)} \fA_\Lambda.
\end{equation}
This is called the \textdef{local algebra} and consists of all local operators, i.e., operators with finite support on the lattice. This is a $*$-algebra with a norm that agrees with the existing norm on $\fA_\Lambda$ for every $\Lambda \in \nfpset(\Gamma)$. However, $\fA_\tn{loc}$ is not complete, so to get a $C^*$-algebra we must take a completion of $\fA_\tn{loc}$. Thus, the quasi-local algebra is the completion
\begin{equation}\label{eq:quasi-local_algebra}
\fA = \overline{\bigcup_{\Lambda \in \nfpset(\Gamma)} \fA_\Lambda},
\end{equation}
consisting of all operators that can be uniformly approximated by local operators, i.e., operators with ``tails'' that get arbitrarily close to the identity the further out one goes on the lattice. We have canonical embeddings $\iota_\Lambda:\fA_\Lambda \rightarrow \fA$ for each $\Lambda \in \nfpset(\Gamma)$, although these are often left implicit (we have already left them implicit in \eqref{eq:local_algebra} and \eqref{eq:quasi-local_algebra}). We may also write the quasi-local algebra as 
\[
\fA = \bigotimes_{i \in \Gamma} M_{n_i}(\bbC),
\]
thinking of it as an infinite tensor product.
\end{definition}

The quasi-local algebra is an example of a uniformly hyperfinite $C^*$-algebra, defined below.

\begin{definition}
A \textdef{uniformly hyperfinite (UHF) algebra} is a unital $C^*$-algebra $\fA$ that has a strictly increasing sequence of unital $C^*$-subalgebras 
\[
\1 \in \fA_1 \subsetneq  \fA_2 \subsetneq \cdots \subsetneq \fA_j \subsetneq \cdots
\]
such that each $\fA_j$ is $*$-isomorphic to a matrix algebra $M_{k_j}(\bbC)$ and $\bigcup_{j=1}^\infty \fA_j$ is dense in $\fA$.
\end{definition}

If we take an increasing sequence $(\Lambda_j)_{j \in \bbN}$ in $\nfpset(\Gamma)$ such that $\Gamma = \bigcup_{j=1}^\infty \Lambda_j$ and note that $\fA_{\Lambda_j}$ is $*$-isomorphic to the algebra of $k_j \times k_j$ matrices where $k_j = \prod_{i \in \Lambda_j} n_i$, then we see that the quasi-local algebra $\fA$ is a UHF algebra. In fact, any UHF algebra is $*$-isomorphic to an infinite tensor product as defined in Definition \ref{def:quasi-local_algebra} \cite{KadisonRingroseII}. UHF algebras were invented by Glimm \cite{Glimm_UHF_Algebras} and later studied by Powers, who proved the following important theorem.

\begin{theorem}[{\cite[Cor.~3.8]{Powers_UHF_Representations}}]
Let $\fA$ be a UHF algebra. For any two pure states $\psi, \omega \in \pstate(\fA)$, there exists an automorphism $\alpha \in \Aut(\fA)$ such that
\[
\psi \circ \alpha = \omega.
\]
\end{theorem}

States on directed colimits of $C^*$-algebras were studied by Takeda \cite{TakedaInductiveLimits} even before the advent of UHF algebras.

\begin{theorem}[{\cite[Thm.~2 \& Prop.~5]{TakedaInductiveLimits}}]\label{thm:takeda}
Let $I$ be a directed set, let $(\fA_i)_{i \in I}$ be a collection of unital $C^*$-algebras, and for each pair $i, j \in I$ with $i \leq j$ let $\iota_{ji}:\fA_i \rightarrow \fA_j$ be an injective unital $*$-homomorphism such that the $\fA_i$ and $\iota_{ji}$ form a directed system. Let $\frA$ together with the $*$-homomorphisms $\iota_i:\fA_i \rightarrow \fA$ be the directed colimit. If $\omega \in \state(\frA)$, then $\omega_i \defeq \omega \circ \iota_i \in \state(\frA_i)$ for all $i \in I$ and 
\begin{equation}\label{eq:state_coherence}
\omega_i = \omega_j \circ f_{ji}
\end{equation}
whenever $i \leq j$. Conversely, given states $\omega_i \in \state(\frA_i)$ for each $i$ satisfying \eqref{eq:state_coherence} whenever $i \leq j$, there exists a unique state $\omega \in \state(\frA)$ such that $\omega_i = \omega \circ f_i$ for all $i$. If each $\omega_i$ is pure, then $\omega$ is pure as well.
\end{theorem}

We caution that purity of $\omega$ does not imply purity of the $\omega_i$. If $\fA$ is a quasi-local algebra, then a state $\omega \in \pstate(\fA)$ is called a \textdef{product state} or is said to be \textdef{completely factorized} if $\omega_\Lambda \defeq  \omega|_{\fA_\Lambda}$ is pure for all $\Lambda \in \nfpset(\Gamma)$. We can also use Theorem \ref{thm:takeda} to construct product states on a quasi-local algebra $\fA$. If for each $i \in \Gamma$ we are given a unit vector $\Psi_i \in \bbC^{n_i}$, then for any $\Lambda \in \nfpset(\Gamma)$ we can define the state $\omega_\Lambda$ represented by the tensor product $\bigotimes_{i \in \Lambda} \Psi_i$. These states are pure and satisfy $\omega_{\Lambda_2} = \omega_{\Lambda_1} \circ \iota_{\Lambda_2 \Lambda_1}$ whenever $\Lambda_1 \subset \Lambda_2$, so Theorem \ref{thm:takeda} yields a unique pure state $\omega \in \pstate(\fA)$ such that $\omega|_{\fA_\Lambda} = \omega_\Lambda$ for all $\Lambda \in \nfpset(\Gamma)$. We intuitively think of this state as represented by the infinite tensor product $\bigotimes_{i \in \Gamma} \Psi_i$, although we will not define a Hilbert space where this expression makes literal sense. 

In the absence of a Hilbert space, the notion of a Hamiltonian as an operator on Hilbert space must be reconsidered. For example, an expression like $H = \sum_{i \in \bbZ} \sigma^z_i$ is not a well-defined element of the quasi-local algebra $\fA$ since the support of such an operator is not approximately local. More precisely, the partial sums do not converge. Hamiltonians are therefore replaced by the notion of an interaction, defined below.

\begin{definition}
An \textdef{interaction} on a quasi-local algebra $\fA$ is a function $\Phi:\nfpset(\Gamma) \rightarrow \fA$ such that $\Phi_\Lambda$ is self-adjoint and $\Phi_\Lambda \in \fA_\Lambda$ for all $\Lambda \in \nfpset(\Gamma)$. 
\end{definition}

We typically think of $\Phi_\Lambda$ as the term in the Hamiltonian involving all sites in $\Lambda$. We will only deal with very simple interactions in this dissertation and therefore will not delve into the theory of unbounded derivations and automorphism groups that often accompany the study of interactions. This theory can be found in \cite{BratteliRobinsonOAQSMI,BratteliRobinsonOAQSMII,Nachtergaele_JvN_lectures,NaaijkensQSSIL} and is necessary for the rigorous $C^*$-algebraic formulation of time-evolution operators and ground states, but for our purposes it will always be intuitively clear what the ground state is. We will see nontrivial physics arise from fairly trivial interactions and ground states when these are allowed to vary over a parameter space with nontrivial global topology. 

One sometimes wishes to act with an operator of the form $\bigotimes_{i \in \Gamma} U_i$, where $U_i \in M_{n_i}(\bbC)$ is a unitary on site $i$. This may not be a well-defined element of $\fA$, again because the support of the operator might not be approximately local. However, we can interpret $\bigotimes_{i \in \Gamma}U_i$ as an automorphism of $\fA$ that we might suggestively write as $\Ad\qty(\bigotimes_{i \in \Gamma} U_i)$. Precisely, there exists an automorphism $\alpha = \Ad\qty(\bigotimes_{i \in \Gamma} U_i)$ of $\fA$ such that for any $A \in \fA_\Lambda$
\[
\alpha(A) = \Ad\qty(\bigotimes_{i \in \Gamma} U_i)(A).
\]
This follows from the universal property of directed colimits and the fact that the maps $\Ad\qty(\bigotimes_{i \in \Lambda} U_i)$ commute with the inclusions $\iota_{\Lambda_2\Lambda_1}$.

\subsection{Noninteracting Lattice System}\label{subsec:1d_system}
\label{subsec:noninteracting}

We give a trivial example of a one-dimensional parametrized system that is nonetheless illustrative of the various topologies and techniques encountered so far. 

First, consider the single-particle Hamiltonian parametrized by the 2-sphere $\sphere^2$:
\[
H:\sphere^2 \rightarrow M_2(\bbC), \quad H(\r) = -\r \cdot \bm{\sigma},
\]
where $\bm{\sigma} = (\sigma^x, \sigma^y, \sigma^z)$ is the vector of Pauli matrices. This Hamiltonian was considered by Berry \cite{Berry} and Simon \cite{PhysRevLett.51.2167} in their studies of Berry phase. We will freely identify $M_2(\bbC) \cong \cB(\bbC^2)$ via the standard basis of $\bbC^2$. To analyze the spectrum and ground states of this Hamiltonian we will introduce a useful unitary transformation. 

Given $\theta, \phi \in \bbR$, define
\begin{equation}\label{eq:U(theta,phi)}
U(\theta, \phi) = \mqty(\cos(\theta/2)e^{i\phi/2} & \sin(\theta/2)e^{-i\phi/2} \\ -\sin(\theta/2)e^{i\phi/2} & \cos(\theta/2)e^{-i\phi/2}).
\end{equation}
%For reference, the adjoint is
%\[
%U(\theta, \phi)^* = \mqty(\cos(\theta/2)e^{-i\phi/2} & -\sin(\theta/2)e^{-i\phi/2} \\ \sin(\theta/2)e^{i\phi/2} & \cos(\theta/2)e^{i\phi/2}).
%\]
It is easy to check that $U(\theta, \phi)$ is indeed unitary. Another easy calculation reveals
\begin{align}\label{eq:2x2_unitary_conjugate_H(r)}
U(\theta, \phi)^*\sigma^z U(\theta, \phi) &= %\mqty(\cos(\theta/2)e^{-i\phi/2}&\sin(\theta/2)e^{-i\phi/2} \\ \sin(\theta/2)e^{i\phi/2}&-\cos(\theta/2)e^{i\phi/2})\mqty(\cos(\theta/2)e^{i\phi/2} & \sin(\theta/2)e^{-i\phi/2}  -\sin(\theta/2)e^{i\phi/2} & \cos(\theta/2)e^{-i\phi/2})\\
 \mqty(\cos \theta & e^{-i\phi}\sin \theta \\ e^{i\phi}\sin \theta & -\cos \theta) = \n(\theta, \phi) \cdot \bm{\sigma},
\end{align}
where 
\[
\n(\theta, \phi) = (\cos \phi \sin \theta, \sin \phi \sin \theta, \cos \theta).
\]

Given $\r \in \bbS^2$ we may choose $\theta, \phi \in \bbR$ so that $\r = \n(\theta, \phi)$. Then
\[
H(\r) = -U(\theta, \phi)^* \sigma^z U(\theta, \phi).
\]
It follows that $\sigma(H(\r)) = \qty{-1, 1}$ and that the ground state of $H(\r)$ is given by
\begin{equation}\label{eq:single_particle_gs}
\Omega(\theta, \phi) = U(\theta, \phi)^*\ket{\uparrow} = \mqty(\cos(\theta/2)e^{-i\phi/2}\\ \sin(\theta/2)e^{i\phi/2}).
\end{equation}
where $\ket{\uparrow} = \mqty(1\\0)$ and $\ket{\downarrow} = \mqty(0\\1)$. Let us consider the continuity properties of this ground state.

\begin{theorem}\label{thm:S^2_CP^1_homeomorphism}
The ground state function $\bbS^2 \rightarrow \bbP(\bbC^2)$, $\r \mapsto \bbC \Omega(\theta, \phi)$ is a homeomorphism.
\end{theorem}

\begin{proof}
To show continuity of $\r \mapsto \bbC \Omega(\theta, \phi)$, it suffices to show that the composition with the map $\bbP(\bbC^2) \rightarrow \cB(\bbC^2)$, $\bbC \Psi \mapsto \ketbra{\Psi}{\Psi}$ is continuous (where $\Psi \in \bbS \hilbH$), since the map $\bbC \Psi \mapsto \ketbra{\Psi}{\Psi}$ is an embedding by Theorem \ref{thm:metric_equivalences}. Observe that
\begin{equation}\label{eq:rank_one_projection_bloch_rep}
\ketbra{\Omega(\theta, \phi)} = U(\theta, \phi)^*\ketbra{\uparrow} U(\theta, \phi) = U(\theta, \phi)^* \qty(\frac{\1+\sigma^z}{2})U(\theta, \phi) = \frac{\1 + \r \cdot \bm{\sigma}}{2}.
\end{equation}
This is manifestly continuous in $\r$. Moreover, this map is injective by linear independence of the matrices $\1$, $\sigma^x$, $\sigma^y$, $\sigma^z$, so we know that $\bbS^2 \rightarrow \bbP(\bbC^2)$ is injective.

Since $\bbS^2$ and $\bbP(\bbC^2)$ are compact Hausdorff, all that remains to do is show surjectivity. It suffices to show that the range of the composition $\bbS^2 \rightarrow \bbP(\bbC^2) \rightarrow \cB(\bbC^2)$ contains all rank one projections. Suppose $\varrho \in \cB(\bbC^2)$ is a rank one projection and expand it as $\varrho = \beta \1 + \bm{\alpha} \cdot \bm{\sigma}$, where $\bm{\alpha} = (\alpha_x, \alpha_y, \alpha_z)$. All coefficients are real since $\varrho$ is self-adjoint. Taking the trace of both sides yields $\beta = 1/2$. It follows from \eqref{eq:2x2_unitary_conjugate_H(r)} that $\bm{\alpha} \cdot \bm{\sigma}$ has eigenvalues $\pm \norm{\bm{\alpha}}$. Since $\varrho$ has eigenvalues $0$ and $1$ and $\frac{1}{2}\1 + \bm{\alpha} \cdot \bm{\sigma}$ has eigenvalues $\frac{1}{2} \pm \norm{\bm{\alpha}}$, we conclude that $\norm{\bm{\alpha}} = \frac{1}{2}$. Thus, $\varrho$ is of the form \eqref{eq:rank_one_projection_bloch_rep} for $\r = 2\bm{\alpha}$.
\end{proof}

\begin{theorem}\label{thm:impossible_ground_state_hairy_ball}
There does not exist a continuous function $\Omega:\bbS^2 \rightarrow \sphere(\bbC^2)$ such that $\Omega(\r)$ is a normalized ground state of $H(\r)$ for all $\r \in \bbS^2$. 
\end{theorem}

\begin{proof}
Suppose such a function $\Omega$ existed. Composing with the projection to $\bbP(\bbC^2)$ yields the homeomorphism $\bbC \Omega(\r)$ from Theorem \ref{thm:S^2_CP^1_homeomorphism}. But then on the second homotopy groups we have that $\bbC \Omega(\r)$ induces the zero homomorphism since $\bbS(\bbC^2) \cong \bbS^3$ and $\pi_2(\bbS^3) = 0$. This contradicts that $\bbC \Omega(\r)$  induces an isomorphism of groups $\pi_2(\bbS^2) \cong \pi_2(\bbP(\bbC^2)) \cong \bbZ$.
\end{proof}

Here is a cuter and more elementary proof that does not rely on knowledge of homotopy groups.

\begin{proof}
Suppose such a function $\Omega$ existed. Define $\Psi(\r) = \Omega(-\r)$ and observe that 
\[
H(\r)\Psi(\r) = -H(-\r)\Omega(-\r) = \Omega(-\r) = \Psi(\r),
\]
so $\Psi(\r)$ is an excited state of $H(\r)$. In particular, $\ev{\Psi(\r), \Omega(\r)} = 0$ for all $\r$. 

Consider the function
\[
U(\r) = \ketbra{\Omega(\r)}{\Psi(\r)} + \ketbra{\Psi(\r)}{\Omega(\r)}.
\]
Evidently, $U(\r)$ is unitary and self-adjoint for all $\r$. If we expand $U(\r)$ in the basis $\1, \sigma^x, \sigma^y, \sigma^z$, then the coefficient of $\1$ is zero since $U(\r)$ is traceless. The other coefficients must form a unit vector $\s(\r) \in \bbS^2$ since $U$ is unitary and self-adjoint. Thus,
\[
U(\r) = \s(\r) \cdot \bm{\sigma}.
\]
The function $U(\r)$ is continuous since $\Omega(\r)$ and $\Psi(\r)$ are continuous, hence $\s(\r)$ is also continuous since the linear functionals $\cB(\bbC^2) \rightarrow \bbC$ dual to the basis $\1, \sigma^x, \sigma^y, \sigma^z$ are continuous.

Now recall the identity
\[
\qty{\bv \cdot \bm{\sigma}, \w \cdot \bm{\sigma}} = 2(\bv \cdot \w) \1
\]
valid for all $\bv, \w \in \bbR^3$. Substituting in $\r$ and $\s(\r)$  and applying this anticommutator to $\Omega(\r)$ yields
\begin{align*}
2(\r \cdot \s(\r))\Omega(\r) = \qty{H(\r), U(\r)}\Omega(\r) = H(\r)\Psi(\r) - U(\r)\Omega(\r) = \Psi(\r) - \Psi(\r) = 0.
\end{align*}
Thus $\r \cdot \s(\r) = 0$ for all $\r$. The existence of such a continuous function $\s:\bbS^2 \rightarrow \bbS^2$ contradicts the hairy ball theorem.
\end{proof}

We now copy this single-particle Hamiltonian above to each vertex of the lattice $\bbZ$. This system is described by the quasi-local $C^*$-algebra
\[
\fA = \overline{\bigcup_{\Lambda \in \nfpset(\bbZ)} \fA_\Lambda} \qqtext{where} \fA_\Lambda = \bigotimes_{i \in \Lambda} M_2(\bbC).
\]
The parameter space remains $\sphere^2$ and we consider the Hamiltonian
\[
H_\tn{1d}(\r) = \sum_{i \in \bbZ} \r \cdot \bm{\sigma}_i,
\]
where $\bm{\sigma}_i = (\sigma^x_i, \sigma^y_i, \sigma^z_i)$ is a vector of Pauli matrices acting on site $i$. More precisely, we consider the parametrized interaction
\[
\Phi_\Lambda(\r) = \begin{cases} -\r \cdot \bm{\sigma}_i &: \Lambda = \qty{i}, \\ 0 &: \tn{otherwise} \end{cases}. 
\]
At every point of $\sphere^2$, the ground state $\omega_\r$ is the product state determined by putting the state $\Omega(\theta, \phi)$ on each site, where again $\theta, \phi \in \bbR$ are chosen so that $\r = \n(\theta, \phi)$. In other words, for any $\Lambda \in \nfpset(\bbZ)$ and simple tensor $\bigotimes_{i \in \Lambda} A_i \in \fA_\Lambda$, we have
\begin{equation}\label{eq:trivial_1d_ground_state}
\omega_\r\qty(\bigotimes_{i \in \Lambda} A_i) = \prod_{i \in \Lambda} \ev{\Omega(\theta, \phi), A_i \Omega(\theta, \phi)}.
\end{equation}

We claim that the ground state function $\omega:\sphere^2 \rightarrow \pstate(\fA)$ is weak*-continuous but not norm-continuous. By Theorem \ref{thm:S^2_CP^1_homeomorphism} we know the quantity \eqref{eq:trivial_1d_ground_state} is continuous in $\r$, even though $\Omega(\theta, \phi)$ cannot be defined to be globally continuous in $\r$. Weak*-continuity of $\omega_\r$ now follows in a manner very similar to Proposition \ref{prop:dense_span_generates_strong_top}, since the linear span of all elements of the form $\bigotimes_{i \in \Lambda} A_i$ is dense in $\fA$. 

We show that $\omega$ is not norm-continuous. Let $\r, \s \in \sphere^2$ and let $\theta, \phi \in \bbR$ such that $\s = \n(\theta, \phi)$. Observe that
\begin{align*}
\ev{\Omega(\theta, \phi), H(\r)\Omega(\theta, \phi)} &= -\frac{1}{2}\tr\qty[(\r \cdot \bm{\sigma})\qty(\1 + \s \cdot \bm{\sigma})] \\
&= -\frac{1}{2}\tr[(\r \cdot \bm{\sigma})(\s \cdot \bm{\sigma})]\\
&= -\frac{1}{2}\tr\qty[(\r \cdot \s)\1 + i(\r \times \s)\cdot \bm{\sigma})]\\
&= -\r \cdot \s.
\end{align*}
Define
\[
H_N(\r) = \bigotimes_{i=1}^N H(\r)
\]
and observe that $H_N(\r)$ is unitary, hence $\norm{H_N(\r)} = 1$. Furthermore,
\[
\omega_\s(H_N(\r)) = \prod_{i=1}^N \ev{\Omega(\theta, \phi), H(\r)\Omega(\theta, \phi)} = (-\r \cdot \s)^N.
\]
Then for any $\r, \s \in \sphere^2$, we have
\begin{align*}
\norm{\omega_\r - \omega_\s} &\geq \abs{\omega_\r(H_N(\r)) - \omega_\s(H(\r))} = \abs{(-1)^N - (-\r \cdot \s)^N}  = \abs{1 - (\r \cdot \s)^N}.
\end{align*}
Assuming $\r \neq \s$ and taking a limit as $N \rightarrow \infty$ yields
\[
\norm{\omega_\r - \omega_\s} \geq 1.
\]
This shows that $\omega$ is not norm-continuous at any point of $\sphere^2$. In fact, one can show using Lemma \ref{lem:sector_is_open} and Theorem \ref{thm:superselection_sector_equivalences} of Chapter \ref{chp:pure_state_topology} that whenever $\r, \s \in \sphere^2$ and $\r \neq \s$, one has that $\norm{\omega_\r - \omega_\s} = 2$ and that $\omega_\r$ and $\omega_\s$ are in different path components of $\pstate(\fA)$ with respect to the norm topology. The full proof of this fact is given in \cite{Spiegel}.

For any $\theta, \phi \in \bbR$ and $\Lambda \in \nfpset(\bbZ)$, we may define
\[
U_\Lambda(\theta, \phi) = \bigotimes_{i \in \Lambda} U(\theta, \phi).
\]
This is a unitary in $\fA_\Lambda \subset \fA$ and defines an inner automorphism $\Ad(U_\Lambda(\theta, \phi))$ of $\fA_\Lambda$ and $\fA$. There exists an automorphism $\gamma(\theta, \phi) \in \Aut(\fA_\Lambda)$ that we might suggestively write as $\Ad(\bigotimes_{i \in \bbZ} U(\theta, \phi))$, which acts as
\begin{equation}\label{eq:gamma_theta_phi_def}
\gamma(\theta, \phi)\qty(\bigotimes_{i \in \Lambda} A_i) = \bigotimes_{i \in \Lambda} U(\theta, \phi)A_i U(\theta, \phi)^*
\end{equation}
for every local simple tensor $\bigotimes_{i \in \bbZ} A_i \in \fA_\Lambda$. If $\r_0 = (0,0,1)$, then we observe that
\begin{align*}
\qty(\omega_{\r_0} \circ \gamma(\theta, \phi))\qty(\bigotimes_{i \in \bbZ} A_i) &= \prod_{i \in \Lambda} \ev{U(\theta, \phi)A_iU(\theta, \phi)^*}{\uparrow} \\
&= \prod_{i \in \Lambda} \ev{\Omega(\theta, \phi), A_i \Omega(\theta, \phi)} = \omega_\r\qty(\bigotimes_{i \in \Lambda} A_i)
\end{align*}
Thus, 
\begin{equation}\label{eq:S^2_gamma_transitivity}
\omega_{\r_0} \circ \gamma(\theta, \phi) = \omega_\r.
\end{equation}

By Theorem \ref{thm:impossible_ground_state_hairy_ball} and \eqref{eq:single_particle_gs} we see that we cannot choose $\theta$ and $\phi$ as functions of $\r$ so as to make $U(\theta, \phi)$ continuous on all of $\sphere^2$. Of course, for any $\r \in \sphere^2 \setminus \qty{(0,0,1), (0,0,-1)}$ we can choose polar coordinates $\theta$ and $\phi$ continuously in a neighborhood of that point. On that neighborhood $U_\Lambda(\theta, \phi)$ is continuous, hence $\Ad(U_\Lambda(\theta, \phi))$ is norm-continuous, and \eqref{eq:gamma_theta_phi_def} and Proposition \ref{prop:dense_span_generates_strong_top} imply that $\gamma(\theta, \phi)$ is strongly continuous. However, $\gamma(\theta, \phi)$ is not norm-continuous, otherwise \eqref{eq:S^2_gamma_transitivity} would imply norm-continuity of $\omega_\r$.

%!TEX root = dissertation.tex

\newpage
\chapter{Topological Properties of Pure State Space}
\label{chp:pure_state_topology}

In this chapter we investigate in detail the norm and weak* topologies on $\pstate(\fA)$. The most important part for later use is the $C^*$-algebraic theory of superselection sectors developed in Section \ref{sec:pure-states}. The notion of a folium can loosely be thought of as a more general version of superselection sector for non-pure states, so we take this as our starting point in Section \ref{sec:folia}. Sections \ref{sec:folia} and \ref{sec:pure-states} pertain to the norm topology. In Section \ref{sec:P(A)_weak*} we review some basic properties of the weak* topology. Sections \ref{sec:folia}, \ref{sec:pure-states}, and \ref{sec:P(A)_weak*} are adapted from our paper \cite{Spiegel} and are a review of existing literature, with new proofs in some places and, in other places, full proofs that were omitted by the original authors. Most parts are reproduced verbatim from \cite{Spiegel}.

Section \ref{sec:UHF_simply_connected} proves the main original result of this section: the pure state space of a UHF algebra has trivial fundamental group with respect to the weak* topology. Combined with the results of Section \ref{sec:P(A)_weak*}, we can say that the pure state space of a UHF algebra is simply connected with respect to the weak* topology. Section \ref{sec:UHF_simply_connected} has been submitted for publication \cite{beaudry2023homotopical}.

\section{Folia}
\label{sec:folia}

%We will denote $\sS (\fA)$ endowed with the norm defined or the weak$^*$ uniformity by $\sS (\fA)_{\textup{n}}$ and $\sS (\fA)_{\textup{w}^*}$, respectively. Later we will give a more detailed account of these uniform spaces.

\begin{comment}
Recall that by the GNS construction every state $\omega$ gives rise to a nontrivial cyclic % (hence non-degenerate)
representation $(\hilbH_\omega, \pi_\omega,\Omega_\omega)$, called the \emph{GNS representation} of $\omega$. The Hilbert space $\hilbH_\omega$ is the completion of the quotient $\fA/\fN_\omega$ by the left ideal 
\[\fN_\omega = \{A \in \fA \mid \omega (A^*A)=0\}\] 
with respect to the inner product
\[
\langle A + \fN_\omega ,B+\fN_\omega \rangle = \omega(A^*B)
\]
for all $A,B\in \fA$. Following Haag \cite{HaagLocalQP2}, we call $\fN_\omega$ the Gelfand ideal associated to  $\omega$. The representation $\pi_\omega : \fA \to \fB(\hilbH_\omega)$
is the $*$-homomorphism defined by $\pi_\omega(A)(B +\fN_\omega) = AB +\fN_\omega$.  If $\fA$ is unital and $I$
denotes the identity, the cyclic unit vector $\Omega_\omega $ coincides by  definition with $I+\fN_\omega$.
In the non-unital case, $\Omega_\omega$ is the limit of the net $(E_\lambda + \fN_\omega)_{\lambda \in \Lambda}$ for any approximate identity $(E_\lambda)_{\lambda \in \Lambda}$ in $\fA$ (all such limits exist and coincide). 
\end{comment}

\begin{definition}\label{def:normal_state}
Given a nondegenerate representation $(\hilbH,\pi)$ of a $C^*$-algebra $\fA$ and a density operator $\varrho$ on $\hilbH$, the functional
\[
\omega:\fA \rightarrow \bbC, \quad \omega(A) = \tr(\varrho \pi(A))
\]
is a state on $\fA$. Nondegeneracy of $\pi$ is necessary in order to ensure that $\norm{\omega} = 1$. Such states are called \textdef{$\bm{\pi}$-normal} states, with the set of all $\pi$-normal states denoted $\state_\pi(\fA)$.  In particular, setting $\varrho = \ketbra{\Omega}$ for some $\Omega \in \bbS \hilbH$ yields a state of the form
\[
\omega(A) = \ev{\Omega, \pi(A)\Omega}.
\]
These states are called \textdef{vector states}; they form a subset of $\state_\pi(\fA)$. If $\omega$ is a $\pi$-normal or vector state, we will say that the corresponding density operator $\varrho$ or unit vector $\Omega$ ``represents'' $\omega$.
\end{definition}

\begin{lem}\label{lem:vector_state}
Let $(\hilbH, \pi)$ be a nondegenerate representation of a $C^*$-algebra $\frA$. If $\Psi, \Omega \in \bbS\hilbH$ are unit vectors representing the vector states $\psi, \omega \in \state_\pi(\frA)$, then
\[
\norm{\psi - \omega} \leq 2 \norm{\Psi - \Omega}.
\]
Thus, the map $\bbS\hilbH \rightarrow \state_\pi(\frA)$ which assigns to a unit vector
the corresponding vector state is norm-continuous.
\end{lem}

\begin{proof}
This follows from the triangle inequality and the Cauchy-Schwarz inequality:
\begin{align*}
\norm{\psi - \omega} &= \sup_{\norm{A}\leq 1} \abs{\ev{\Psi, \pi(A)\Psi} - \ev{\Omega, \pi(A)\Omega}}\\
&\leq \sup_{\norm{A}\leq 1} \abs{\ev{\Psi, \pi(A)(\Psi - \Omega)}} + \sup_{\norm{A} \leq 1} \abs{\ev{\Psi - \Omega, \pi(A)\Omega}}\\
&\leq 2 \norm{\Psi - \Omega},
\end{align*}
where, in the last line, we have used the fact
% the facts that $\norm{\Psi} = \norm{\Omega} = 1$ and
that $\norm{\pi} \leq 1$, as provided by Theorem \ref{thm:*-hom_contractive}.
% since $\pi$ is a $*$-homomorphism.
\end{proof}

In the proof above, we used the fact that a representation $\pi$ of a $C^*$-algebra always satisfies $\norm{\pi} \leq 1$. Recall that Theorem \ref{thm:*-hom_contractive} also entails that $\pi$ is an isometry if and only if it is faithful. When this is not the case, the following lemma may be of service. The proof of this lemma is explained as part of the proof of Lemma 4.4 in \cite{RobertsRoepstorffSBCAQT}. 

\begin{lem}\label{lem:rep_almost_isometry}
Let $\fA$ and $\fB$ be $C^*$-algebras and let $\pi:\fA \rightarrow \fB$ be $*$-homomorphism. Given $A \in \fA$ and $\varepsilon > 0$, there exists $B \in \fA$ such that $\pi(A) = \pi(B)$ and
\[
\norm{B} < \norm{\pi(A)} + \varepsilon.
\]
\end{lem} 

It is in fact true that there exists $B \in \fA$ such that $\pi(A) = \pi(B)$ and $\norm{B} = \norm{\pi(A)}$, i.e., we can set $\varepsilon = 0$ \cite[Cor.~10.1.8]{KadisonRingroseII} (see also \cite{CourtneryGillaspy}). However, this is significantly more difficult to prove, and Lemma \ref{lem:rep_almost_isometry} in the form stated above suffices for our purposes.

%\daniel{K-R has a strengthened version of this theorem where $\norm{B} \leq \norm{\pi(A)}$. However, K-R define $C*$-algebras to be unital (very annoying). Their argument stretches back through many propositions, making it very difficult to tell if we can achieve $\norm{B} \leq \norm{\pi(A)}$ in the non-unital case. Therefore I hesitate to cite them for this theorem. I'm happy to cite R-R.}

\begin{proof}
  Since $\ker \pi$ is a closed two-sided ideal in $\fA$, the quotient  $\fA/\ker \pi$ is a $C^*$-algebra, which
  satisfies the commutative diagram
  \begin{displaymath} 
    \begin{tikzcd}[ampersand replacement=\&]
    \fA \arrow[r,"\pi"] \arrow[d,"q"'] \& \pi(\fA)\\
    \fA/\ker \pi \arrow[ur,"\overline \pi"']
    \end{tikzcd}
  \end{displaymath}
  where $q$ is the quotient map and $\overline \pi:\fA/\ker \pi \rightarrow \pi(\fA)$ is a $*$-isomorphism, hence an
  isometry. By definition of the norm on $\fA/\ker \pi$, we then have
  \[
    \norm{\pi(A)} = \norm{\overline \pi(q(A))} = \norm{q(A)} = \inf_{C \in \ker \pi} \norm{A - C} =
    \inf_{B \in \fA}\qty{\norm{B} : \pi(A) = \pi(B)},
  \]
  and the result follows.
  \end{proof}

The next result uses this lemma to show that the space of  normal states with respect to a representation $\pi$
can be identified with the space of normal states of the induced von Neumann algebra even if
$\pi$ is not faithful. Recall that if $\fR \subset \cB(\hilbH)$ is a von Neumann algebra, then a normal state on $\fR$ is a $\pi$-normal state in the sense of Definition \ref{def:normal_state} for the representation $\pi:\fR \rightarrow \cB(\hilbH)$ defined as $\pi(A) = A$. 
%\todo{Maybe need to define what a von Neumann algebra is, state bicommutant and Kaplansky density theorems.}

\begin{prop}[\emph{cf.} {\cite[Lem.~4.4]{RobertsRoepstorffSBCAQT}}]\label{prop:pullback-representation-isometric-isomorphism-normal-state-spaces}
  Let $(\hilbH,\pi)$ be a nondegenerate representation of a $C^*$-algebra $\fA$ and let
  $\fR = \pi(\fA)'' \subset \cB(\hilbH)$ be the induced von Neumann algebra.
  Denote by $\state_*(\fR)$ the space of normal states on $\fR$.
  Then the pullback map
  $\pi^* : \state_* (\fR)_{\textup{n}} \to \state_\pi(\fA)_{\textup{n}}$, $\omega \mapsto \omega \circ \pi$
  is a bijective isometry. In other words,
  \[
  \norm{\omega \circ \pi - \psi \circ \pi} = \norm{\omega - \psi}
  \] 
  for all $\omega, \psi \in \state_*(\fR)$.
\end{prop}

%\daniel{This Proposition is essentially \cite[Lem.~4.4]{RobertsRoepstorffSBCAQT}. The only difference is that they do not require a non-degenerate representation and they do not specify the range, but they prove this is an isometry.}

\begin{proof}
  By definition, $\pi^*$ is surjective, so it suffices to show that $\pi^*$ is an isometry.
  Clearly, for all $f \in \fR^*$
  \[
    \| \pi^* f \| = \sup\limits_{A \in \fA, \: \| A\|\leq 1}  | f (\pi (A)) |
    \leq \sup\limits_{B \in \fR, \: \| B\|\leq 1}  | f (B) | = \| f \| ,  
  \]
  hence $\pi^*$ is contraction. In particular, $\norm{\pi^*\omega - \pi^*\psi} \leq \norm{\omega - \psi}$ for all $\psi, \omega \in \state_*(\fR)$.

  It remains to show that for all $\psi, \omega \in  \state_* (\fR)$
  \begin{equation}
    \label{eq:pistar-norm-increasing}
    \| \psi - \omega \| \leq \| \pi^* \psi - \pi^* \omega \|   .
  \end{equation}
  To prove this recall that the state $\psi$ is induced by a density operator which means that
  there exists a collection of vectors $(x_i)_{i \in I}$ such that
  $\sum_{i\in I} \norm{x_i}^2 = 1$ and
  $\psi (B) = \sum_{i\in I} \ev{x_i, Bx_i}$ for all $B \in \fR$.
  The (possibly infinite) sum over $i \in I$ is defined as the limit of the net of finite partial sums. The vectors $(x_i)_{i \in I}$ may be taken to be a set of eigenvectors of the density operator, with $\norm{x_i}$ equal to the square root of the corresponding eigenvector.
  Likewise, there exists for $\omega$ a family
  $(y_i)_{i \in I} \subset \hilbH$ such that  $\sum_{i\in I} \norm{y_i}^2 = 1$ and
  $\omega (B) = \sum_{i\in I} \ev{y_i, By_i}$ for all $B \in \fR$.
  
  Now let $\varepsilon >0$ and choose $B\in \fR$ with $\norm{B} \leq 1$ such that 
  \begin{equation}
    \label{eq:canonical-metric-estimate-value-unit-ball-element}
     \| \psi - \omega  \| \leq |\psi(B) -\omega(B)| + \varepsilon  .
  \end{equation}
  Next choose a finite subset $J\subset I$ so that
  $\sum_{i \in I\setminus J} \norm{x_i}^2 < \varepsilon $ and
  $\sum_{i\in I\setminus J} \norm{y_i}^2 < \varepsilon $. By the Kaplansky density theorem
  there exists $A^\prime \in \fA$ with $\norm{\pi(A^\prime)} \leq 1$ such that
  \[
    \sum_{j\in J} \abs{\ev{x_j, (B - \pi(A^\prime))x_j}} < \varepsilon
    \quad\text{and}\quad
    \sum_{j\in J} \abs{\ev{y_j, (B - \pi(A^\prime))y_j}} < \varepsilon \ .
  \]
  Hence
  \begin{equation}
    \label{eq:estimate-kaplansky}
    \abs{\psi (B) - \psi (\pi(A^\prime))} < 3 \varepsilon
    \quad\text{and}\quad
    \abs{\omega (B) - \omega (\pi(A^\prime))} < 3 \varepsilon \ .
  \end{equation}
  By Lemma \ref{lem:rep_almost_isometry}, there exists $A \in \fA$ such that
  $ \pi (A) = \pi (A^\prime)$ and
  \[ \| A \| < \| \pi (A^\prime)\| + \varepsilon \leq 1 +\varepsilon \ . \]
  Together with the estimates \eqref{eq:canonical-metric-estimate-value-unit-ball-element}
  and \eqref{eq:estimate-kaplansky} this finally entails
  \begin{equation*}
  \begin{split}  
    \| \psi - \omega \| & \leq |\psi (B) -\omega  (B)| + \varepsilon  <
    | \psi (\pi(A^\prime)) - \omega (\pi(A^\prime))|  + 7  \varepsilon = \\
    & = | \psi (\pi(A)) - \omega (\pi(A))| + 7  \varepsilon <
    (1 +\varepsilon) \| \pi^*\psi - \pi^* \omega \| + 7  \varepsilon \ .
  \end{split}  
  \end{equation*}
  By passing to the limit $\varepsilon \rightarrow 0$, the estimate 
  \eqref{eq:pistar-norm-increasing} follows.
\end{proof}

The $C^*$-algebra $\fA$ acts in a natural way  on the dual $\fA^*$ by
associating to a pair $(B,\omega)\in \fA \times \fA^*$ the continuous linear functional
$B\cdot \omega : \fA \to\bbC$, $A \mapsto  \omega (B^*AB) $.
Note that if $\omega$ is a state and $B$ fulfills $\omega (B^*B)=1$, then $B\cdot \omega$ is again
a state.  This motivates the notion of a folium, introduced by Haag--Kadison--Kastler
in \cite{HaaKadKasNCSACS} as a tool for the classification of states in local
quantum physics; see also \cite[Sec.~8.6]{LanFQT}.

\begin{defn}
  A \textdef{folium} in the state space $\state(\fA)$ of a $C^*$-algebra $\fA$ is a nonempty subset $\folium \subset \state(\fA)$ which is:
  \begin{enumerate}[label=(F\arabic*)]
  \item\label{ite:folium-norm-closed} norm-closed, % in the metric uniformity,
  \item\label{ite:folium-convex} convex, and
  \item\label{ite:folium-invariance} invariant under the action of $\fA$ in the sense that if
    $\omega \in \folium$ and $B \in \fA$ with  $\omega (B^*B)=1$, then 
    $B\cdot\omega$ lies again in $\folium$.
  \end{enumerate}
  Note that $\state(\fA)$ is norm-closed in $\fA^*$, even if $\fA$ is nonunital. Thus, a set $\folium \subset \state(\fA)$ is norm-closed in $\state(\fA)$ if and only if it is norm-closed in $\fA^*$. In particular, $\state(\fA)$ is a folium. Furthermore, it is clear from the definition that the intersection of arbitrarily many folia is a folium, provided the intersection is nonempty.
\end{defn}

An important observation of Haag--Kadison--Kastler in \cite[\S 1]{HaaKadKasNCSACS}, summarized below, is that the $\pi$-normal states of a nondegenerate representation form
a folium. The proof is omitted in \cite{HaaKadKasNCSACS}, so we provide it;
cf.\ also \cite[Lemma 5.6]{StormerPOVSDASC}.

\begin{thm}
  For every nonzero nondegenerate representation $(\hilbH,\pi)$ of a $C^*$-algebra $\fA$, the space of $\pi$-normal states $\state_\pi(\fA)$ is the smallest folium containing the vector states of $\pi$. Furthermore, any folium coincides with the space $\state_\pi(\fA)$ for some nonzero nondegenerate representation $\pi$ of $\fA$.
\end{thm}

%\daniel{The first sentence of this theorem is Lemma 5.6 in \cite{StormerPOVSDASC} in slightly different words.}

\begin{proof}
  We first show that for every nonzero nondegenerate representation $\pi$ the
  space $\state_\pi(\fA)$ is a folium. 
  Denote by  $\fR = \pi(\fA)'' \subset \cB(\hilbH)$ the von Neumann algebra induced by
  the representation $\pi$. According to
  \cite[7.1.13]{KadisonRingroseII}, the space $\state_* (\fR)$ of normal states on $\fR$
  is norm-closed in $\fR^*$.
  %\todo{Learn this fact.} 
  In particular  $\state_* (\fR)$ is complete as a metric space. 
  Since the pullback $\pi^*: \state_* (\fR) \to \state_\pi(\fA)$ is a bijective isometry by
  Proposition \ref{prop:pullback-representation-isometric-isomorphism-normal-state-spaces},
  its image has to be complete as well. Hence $\state_\pi(\fA)$ is closed, proving
  \ref{ite:folium-norm-closed}.
  
  The set of density matrices on $\cB(\hilbH)$ is convex, hence $\state_\pi(\fA)$ is so, too,
  and \ref{ite:folium-convex} holds. Now let $\omega \in \state_\pi(\fA)$ and 
  $\varrho \in \cB (\hilbH)$ a density matrix such that $\omega (A) = \tr (\varrho \pi(A))$
  for all $ A \in \fA$. Let $B \in \fA$ such that $\omega (B^*B)=1$. Then
  the operator $\pi (B) \varrho \pi(B)^*$ is positive, trace-class (since the trace-class operators are an ideal in $\cB(\hilbH)$), and has trace one
  by the equality $\tr\qty(\pi (B) \varrho \pi(B)^*) = \omega (B^*B)=1$. 
  Moreover, 
  \[
    (B \cdot \omega)(A) =\omega (B^*AB) = \tr\qty( \left(\pi (B) \varrho \pi(B)^*\right) \pi(A) )
     \quad \text{for all } A \in \fA  ,
  \]
  which shows that $B\cdot \omega $ is a $\pi$-normal state and  \ref{ite:folium-invariance}
  is fulfilled. Hence  $\state_\pi(\fA)$ is a folium. 

  Now let $\folium$ be a folium containing the vector states of the representation $\pi$. Consider a positive trace class operator $\varrho \in \cB(\hilbH)$ of trace one, and let $\omega \in \state_\pi(\fA)$ be the corresponding $\pi$-normal state. There exists an orthonormal set $(\Omega_i)_{i \in I}$ in $\hilbH$ such that $\varrho = \sum_{i \in I} \lambda_i \ketbra{\Omega_i}$, where $\lambda_i > 0$ and $\sum_{i \in I} \lambda_i = 1$. In particular, for any $A \in \fA$,
 \begin{equation}\label{eq:folium_weak*-converge}
 \omega(A) = \tr(\varrho \pi(A)) = \sum_{i \in I} \lambda_i \ev{\Omega_i, \pi(A)\Omega_i}  .
 \end{equation}
Let $\omega_i$ denote the vector state corresponding to $\Omega_i$. Denoting by $\nfpset(I)$ the set of all nonempty finite subsets of $I$, \eqref{eq:folium_weak*-converge} shows that the $\qty(\sum_{j \in J} \lambda_j \omega_j)_{J \in \nfpset(I)}$ converges to $\omega$ in the weak* topology. 

We show that the net is convergent in the norm topology and therefore must converge to $\omega$ in the norm topology as well.
 % where, again, the sum is meant in the sense of net convergence.
 Given $\varepsilon > 0$, choose a finite subset $J \subset I$ such that for any finite subset
 $K \subset I$ with $J \cap K = \varnothing$ the estimate $\sum_{k \in K} \lambda_k < \varepsilon$ holds true.
 If $\omega_i$ is the vector state corresponding to $\Omega_i$, then
 \[
 \norm{\sum_{k \in K} \lambda_k \omega_k} \leq \sum_{k \in K} \lambda_k \norm{\omega_k} < \varepsilon
 \]
 since $\norm{\omega_k} = 1$. It follows that the net $\qty(\sum_{j \in J} \lambda_j \omega_j)_{J \in \nfpset(I)}$ converges in the norm topology, as claimed.
 Moreover, the net
 \begin{align}\label{eq:net1}
  \left( \frac{\sum_{j \in J} \lambda_j \omega_j}{\sum_{j \in J} \lambda_j}\right)_{J \in \nfpset (I)}
 \end{align}
 converges to $\omega$ since the denominators converge to one. Since $\omega_i \in \folium$ for all $i$ and each element of the net \eqref{eq:net1} is a convex combination of $\omega_i$, we conclude from \ref{ite:folium-convex} and \ref{ite:folium-norm-closed} that $\omega \in \folium$. This proves that $\state_\pi(\fA) \subset \folium$, proving the claim that $\state_\pi(\fA)$ is the smallest folium containing the vector states of $\pi$.
  
 Finally let $\folium$ be a folium and let $\hilbH = \bigoplus_{\omega \in \folium} \hilbH_\omega$ and $\pi = \bigoplus_{\omega \in \folium} \pi_\omega$, where $(\hilbH_\omega, \pi_\omega, \Omega_\omega)$ denotes the GNS representation of $\omega$.  Clearly, every $\omega \in \folium$ is the $\pi$-normal vector state corresponding to $\Omega_\omega \in \hilbH_\omega \subset \hilbH$, so $\folium \subset \state_\pi(\fA)$. It remains to be shown that $\state_\pi(\frA) \subset \folium$. To this end it suffices to prove that every vector state is in $\folium$, since $\state_\pi(\fA)$ is the minimal folium containing the vector states of $\pi$.
 
If $\omega \in \folium$ and $B \in \frA$ with $\omega(B^*B) = 1$, then the $\pi$-normal vector state corresponding to $\pi(B)\Omega_\omega$ is $B \cdot \omega$, which is in $\folium$ by \ref{ite:folium-invariance}. Given a unit vector $\Psi \in \hilbH_\omega$ and $\varepsilon > 0$, cyclicity of $\pi_\omega$ yields $C \in \frA$ such that 
 \begin{equation}\label{eq:vector_approximate}
 \norm{\Psi - \pi(C)\Omega_\omega} < \min(\varepsilon/4, 1).
 \end{equation}
 Then $\norm{\pi(C)\Omega_\omega}^2 = \omega(C^*C) > 0$, so we may define $B = C/\sqrt{\omega(C^*C)}$, for which $\norm{\pi(B)\Omega_\omega}^2 = \omega(B^*B) = 1$. Note that by \eqref{eq:vector_approximate} and the
 reverse triangle inequality
 \begin{equation}\label{eq:normalized_vector_approximate}
 \norm{\pi(B)\Omega_\omega - \pi(C)\Omega_\omega} = \abs{1 - \sqrt{\omega(C^*C)}} < \frac{\varepsilon}{4} \ .
 \end{equation}
 If $\psi \in \state(\frA)$ is the $\pi$-normal vector state corresponding to $\Psi$, then Lemma \ref{lem:vector_state} entails
 \begin{align}\label{eq:state_approximate}
 \norm{\psi - B \cdot \omega} &\leq 2\norm{\Psi - \pi(B)\Omega_\omega}. 
 \end{align}
 Thus, \eqref{eq:vector_approximate}, \eqref{eq:normalized_vector_approximate}, and \eqref{eq:state_approximate} together imply that $\norm{\psi - B \cdot \omega} < \varepsilon$. Since $\folium$
 is norm closed and $B \cdot \omega \in \folium$, one concludes that $\psi \in \folium$.
 
 Next consider a unit vector of the form $\Psi = \sum_{i=1}^n \lambda_i \Psi_{i}$, where $\lambda_i \in \mathbb{C}$ and $\Psi_{i} \in \hilbH_{\omega_i}$ are unit vectors with distinct $\omega_i \in \folium$. If $\psi$ is the $\pi$-normal vector state corresponding to $\Psi$, then
 \[
 \psi(A) = \sum_{i=1}^n \abs{\lambda_i}^2 \ev{\Psi_i, \pi_{\omega_i}(A)\Psi_i}.
 \]
 Since $\norm{\Psi}^2 = \sum_{i=1}^n \abs{\lambda_i}^2 = 1$, we see that $\psi$ is a convex combination of elements of $\folium$, so $\psi \in \folium$ by \ref{ite:folium-convex}. Since any unit vector in $\hilbH$ is the limit of a net of such finite linear combinations, $\folium$ contains all vector states of $\pi$ by  Lemma \ref{lem:vector_state} and \ref{ite:folium-norm-closed}. This proves that $\state_\pi(\fA) \subset \folium$. 
\end{proof}

\section{Superselection Sectors}
\label{sec:pure-states}
%\todo{This section is done.}

The extreme points of $\state(\fA)$ are called pure states, the set of which we denote by $\pstate(\fA)$. It is well-known that $\pstate(\fA)$ is nonempty when $\fA$ is nonzero and that $\state(\fA)$ coincides with the weak*-closed convex hull of $\pstate(\fA)$ when $\fA$ is unital (both being consequences of the Krein--Milman theorem). The GNS representation $(\hilbH_\omega, \pi_\omega)$ of a state $\omega$ is irreducible if and only if $\omega$ is pure. In this case, the quotient $\fA/\fN_\omega$ is already complete, hence we can identify $\hilbH_\omega = \fA/\fN_\omega$; this follows from the Kadison transitivity theorem \cite[Thm.~5.2.4]{Murphy}.

Given a nondegenerate representation $\pi$, we denote the set of pure $\pi$-normal states on $\fA$ by $\pstate_\pi(\fA) = \pstate(\fA) \cap \state_\pi(\fA)$.

\begin{prop}\label{prop:pure_folia_vector_state}
Let $\fA$ be a $C^*$-algebra with a nondegenerate representation $(\hilbH,\pi)$. 
\begin{enumerate}
\item\label{ite:vector-state} If $\omega \in \pstate_\pi(\fA)$, then $\omega$ is a vector state. 
\item\label{ite:irreducible-vector-state-pure} If $\pi$ is irreducible and $\omega$ is a vector state, then $\omega \in \pstate_\pi(\fA)$. 
\item\label{ite:vectors-states-same-state-linearly-dependent} If $\pi$ is irreducible and $\Psi, \Omega \in \hilbH$ are unit vectors defining the same state, then $\Psi$ and $\Omega$ are linearly dependent.
\end{enumerate}
\end{prop}

\begin{proof}
\ref{ite:vector-state} Suppose $\omega \in \pstate_\pi(\fA)$ and let $\varrho \in \cB(\hilbH)$ be a density matrix such that $\omega(A) = \tr(\varrho \pi(A))$. There exists an orthonormal set $\qty{\Omega_i}_{i \in I}$ in $\hilbH$ such that $\varrho = \sum_{i \in I} \lambda_i \ketbra{\Omega_i}$, where $\lambda_i > 0$ and $\sum_{i \in I} \lambda_i = 1$. Fix $j \in I$ and let $\psi$ be the vector state corresponding to $\Omega_j$. Then for all positive $A \in \fA$,
\[
\lambda_j \psi(A) = \lambda_j \ev{\Omega_j, \pi(A)\Omega_j} \leq \sum_{i \in I} \lambda_i \ev{\Omega_i, \pi(A)\Omega_i} = \omega(A).
\]
Since $\omega$ is pure, there exists $\lambda \in [0,1]$ such that $\lambda_j \psi = \lambda \omega$. Since $\norm{\psi} = \norm{\omega} = 1$, we see that $\lambda_j = \lambda$, which implies that $\omega = \psi$, a vector state.

\ref{ite:irreducible-vector-state-pure} This is given by \cite[Thm.~5.1.7]{Murphy}.

% If $\Omega \in \hilbH$ is a unit vector representing $\omega$,  then the GNS representation $\pi_\omega$ of $\omega$ is unitarily equivalent to $\pi$ by uniqueness (up to unitary equivalence) of the GNS representation. It follows that $\pi_\omega$ is irreducible, which implies that $\omega$ is pure.

\ref{ite:vectors-states-same-state-linearly-dependent} If $\Psi$ and $\Omega$ are linearly independent, then the Kadison transitivity theorem yields $B \in \fA$ such that $\pi(B)\Psi = \Psi$ and $\pi(B) \Omega = 0$. This contradicts the assumption that $\ev{\Psi, \pi(A)\Psi} = \ev{\Omega, \pi(A)\Omega}$ for all $A \in \fA$.
\end{proof}

\begin{defn}\label{def:superselection_sector}
  Given a $C^*$-algebra $\fA$, call two pure states $\psi,\omega \in \pstate(\fA)$
  \textdef{equivalent} %, in signs $\psi\sim\omega$,
  if their GNS representations $\pi_\psi$ and $\pi_\omega$ are unitarily equivalent.
  Let $\sim$ denote the corresponding equivalence relation  on $\pstate(\fA)$.
  By a \textdef{superselection sector} of $\fA$ we understand an equivalence class of pure states with respect
  to $\sim$. 

  Given $\omega \in \pstate(\fA)$, we denote its superselection sector by $\pstate_\omega(\fA)$. Conveniently, it follows from Proposition \ref{prop:pure_folia_vector_state} that $\pstate_{\pi_\omega}(\fA) = \pstate_\omega(\fA)$. Indeed, if $\psi \in \pstate_{\pi_\omega}(\fA)$, then $\psi$ is a vector state of $\pi_\omega$, hence $\pi_\psi$ is unitarily equivalent to $\pi_\omega$ by uniqueness of the GNS representation up to unitary equivalence, so $\psi \in \pstate_\omega(\fA)$. Conversely, if $\psi \in \pstate_\omega(\fA)$, then unitary equivalence of $\pi_\psi$ and $\pi_\omega$ implies that $\psi$ is a vector state of $\pi_\omega$, hence $\psi \in \pstate_{\pi_\omega}(\fA)$.
\end{defn}

In the remainder of this section we will give various characterizations of superselection sectors, including results that develop physical intuition behind the concept and show it is deserving of its name.  Most of the results below can be found in
Roberts--Roepstorff \cite{RobertsRoepstorffSBCAQT}, Glimm--Kadison \cite{GlimmKadisonUOCA}, and Pedersen \cite{PedersenCAlgAutomorphisms}. 

%The proposition is important  the interpretation of superselection sectors in physics.
Before giving our next definition, let us note that given two states 
$\omega$ and $\psi$ there always exists a nondegenerate representation $(\hilbH,\pi)$ such that both  $\omega$ and $\psi$ become
vector states with respect to that representation. For example, one can take $(\hilbH,\pi)$ as the
direct sum of the GNS representations of  $\omega$ and $\psi$. 

\begin{defn}
  Two distinct pure states  $\psi$ and $\omega$  of a $C^*$-algebra $\fA$  are said to
  \textdef{fulfill the superposition principle} or to be \textdef{coherently superposable} if there exists
  a nondegenerate representation $(\hilbH,\pi)$  in which $\psi$ and $\omega$ are represented by the unit
  vectors  $\Psi$ and $\Omega$, respectively, and for all $\alpha,\beta \in \bbC$ such that $\alpha \Psi + \beta \Omega \neq 0$, the vector state
  $\varphi$ corresponding to the unit vector
  \[\Phi = \frac{\alpha\Psi + \beta\Omega}{\|\alpha\Psi + \beta\Omega\|}\] %\widehat{\phantom{\omega}}
  is a pure state. If this is the case, one calls each of the states $\varphi$ obtained in this way a
  \emph{coherent superposition} of $\omega$ and $\psi$. 
  %\[
  %  \fA \to \langle \alpha\Omega + \beta\Psi , \pi(A)( \alpha\Omega + \beta\Psi)\rangle
  %\]
\end{defn}

\hide{
  \begin{prop}\label{prop:intertwiners_iff_equivalent}
  Let $\frA$ be a nonzero $C^*$-algebra and let $(\hilbH_1, \pi_1)$ and $(\hilbH_2, \pi_2)$ be nonzero irreducible representations
  of $\frA$. If $T: (\hilbH_1, \pi_1) \rightarrow (\hilbH_2,\pi_2)$ is a nonzero intertwiner, that is,
  if  $T:\hilbH_1\to \hilbH_2$ is a bounded linear map  such that 
  \[
    T\pi_1(A) = \pi_2(A)T \quad \text{for all } A \in \frA \ ,
  \]
  then there exists $\lambda > 0$  such that $\lambda T$ is unitary.
  In particular, $\pi_1$ and $\pi_2$ are unitarily equivalent.
  \end{prop}

  \begin{proof}
  First we take adjoints to obtain
  \[
    \pi_1(A^*)T^* = T^* \pi_2(A^*)
  \]
  for all $A \in \frA$. Replacing $A$ with $A^*$, we see that $\pi_1(A)T^* = T^* \pi_2(A)$ for all $A \in \frA$. Therefore,
  \[
    \pi_1(A)T^*T = T^*\pi_2(A)T = T^*T \pi_1(A)
  \]
  and
  \[
    \pi_2(A)TT^* = T\pi_1(A)T^* = TT^*\pi_2(A)
  \]
  for all $A \in \fA$. Since $\pi_1$ and $\pi_2$ are irreducible, it follows from Schur's lemma that $T^*T = \alpha I$ and
  $TT^* = \beta I$ for some $\alpha, \beta \in \bbC$. 

  Observe that
  \[
    \alpha T \pi_1(A) = T \pi_1(A)T^*T = TT^* \pi_2(A)T = \beta \pi_2(A)T = \beta T\pi_1(A),
  \]
  so either $\alpha = \beta$ or $T\pi_1(A) = 0$ for all $A \in \frA$. Since $\pi_1$ is a nonzero irreducible representation,
  we know $\pi_1(\frA)\hilbH_1$ is dense in $\hilbH_1$, so $T\pi_1(A) =0$ for all $A \in \frA$ implies that $T = 0$, which is a
  contradiction by hypothesis. Therefore $\alpha = \beta$, and since $T \neq 0$ implies $\norm{T^*T} = \norm{T}^2 > 0$,
  we know $\alpha$ is nonzero. Furthermore, for any $x \in \hilbH_1$ we have
  \[
    \norm{Tx}^2 = \ev{T^*Tx, x} = \alpha^* \norm{x}^2,
  \]
  which implies that $\alpha$ is real and positive.

  Let $\lambda = \alpha^{-1/2}$ and set $U = \lambda T$. For all $x, y \in \hilbH_1$ we have
  \[
    \ev{Ux, Uy} = \ev{\alpha^{-1} T^*Tx, y} = \ev{x, y},
  \]
  so $U$ is an isometry. Furthermore, since $\pi_2$ is nonzero and irreducible, every nonzero vector in $\hilbH_2$ is cyclic,
  which implies that $\pi_2(\frA)U \hilbH_1 = U\pi_1(\frA)\hilbH_1$ is dense in $\hilbH_2$. Thus, $U(\hilbH_1)$ is dense in $\hilbH_2$, and
  since  $U$ is an isometry, we know $U(\hilbH_1)$ is closed. We conclude that $U$ is bijective, hence unitary.
  \end{proof}
}

We will soon show that pure states in different superselection sectors are not coherently superposable. First, we need the following lemma, which is adapted from \cite[Sec.~2.2.2]{DixCA}.

\begin{lemma}\label{lem:intertwiners_iff_equivalent}
Let $\frA$ be a $C^*$-algebra and let $(\hilbH_1, \pi_1)$ and $(\hilbH_2, \pi_2)$ be nonzero irreducible representations of $\frA$. If $T:\hilbH_1 \rightarrow \hilbH_2$ is a nonzero bounded linear operator such that 
\[
T\pi_1(A) = \pi_2(A)T
\]
for all $A \in \frA$, then there exists $\gamma > 0$  such that $\gamma T$ is unitary. In particular, $\pi_1$ and $\pi_2$ are unitarily equivalent.
\end{lemma}

\begin{proof}
First we take adjoints to obtain
\[
\pi_1(A^*)T^* = T^* \pi_2(A^*)
\]
for all $A \in \frA$. Replacing $A$ with $A^*$, we see that $\pi_1(A)T^* = T^* \pi_2(A)$ for all $A \in \frA$. Therefore,
\[
\pi_1(A)T^*T = T^*\pi_2(A)T = T^*T \pi_1(A).
\]
and
\[
\pi_2(A)TT^* = T\pi_1(A)T^* = TT^*\pi_2(A)
\]
for all $A \in \frA$. Since $\pi_1$ and $\pi_2$ are irreducible, it follows from Schur's lemma that $T^*T = \alpha I$ and $TT^* = \beta I$ for some $\alpha, \beta \in \bbC$. 

Observe that
\[
\alpha T \pi_1(A) = T \pi_1(A)T^*T = TT^* \pi_2(A)T = \beta \pi_2(A)T = \beta T\pi_1(A),
\]
so either $\alpha = \beta$ or $T\pi_1(A) = 0$ for all $A \in \frA$. Since $\pi_1$ is a nonzero irreducible representation, we know $\pi_1(\frA)\hilbH_1$ is dense in $\hilbH_1$, so $T\pi_1(A) =0$ for all $A \in \frA$ implies that $T = 0$, which is a contradiction. Therefore $\alpha = \beta$, and since $T \neq 0$ implies $T^*T \neq 0$, we know $\alpha$ and $\beta$ are nonzero. Furthermore, for any $x \in \hilbH_1$ we have
\[
\norm{Tx}^2 = \ev{T^*Tx, x} = \alpha^* \norm{x}^2,
\]
which implies that $\alpha$ is real and positive.

Let $U = \alpha^{-1/2}T$. For all $x, x' \in \hilbH_1$ we have
\begin{equation}
\ev{Ux, Ux'} = \ev{\alpha^{-1} T^*Tx, x'} = \ev{x, x'},
\end{equation}
so $U$ is an isometry. Furthermore, since $\pi_2$ is nonzero and irreducible, every nonzero vector in $\hilbH_2$ is cyclic, which implies that $\pi_2(\frA)T \sH_1 = T\pi_1(\frA)\hilbH_1$ is dense in $\hilbH_2$. Thus, $U\hilbH_1$ is dense in $\sH_2$, and since $U$ is an isometry, we know $U\hilbH_1$ is closed. We conclude that $U$ is bijective.
\end{proof}

The following proposition is a rephrasal and clarification of the ``sufficient'' implication of \cite[Thm.~6.1]{ArakiMTQF}. We restrict our attention here to pure states, although \cite[Thm.~6.1]{ArakiMTQF} is stated for general states.

\begin{prop}\label{prop:inequivalent_superposition}
Let $\frA$ be a nonzero $C^*$-algebra and let $\psi$ and $\omega$ be pure states in different superselection sectors. If $(\hilbH, \pi)$ is a nondegenerate representation with unit vectors $\Psi, \Omega \in \hilbH$ representing $\psi$ and $\omega$, respectively, then $\ev{\Psi, \Omega} = 0$. Furthermore, if $\varphi$ is the vector state corresponding to $\Phi = \alpha \Psi + \beta \Omega$ for any nonzero $\alpha, \beta \in \bbC$ with $\abs{\alpha}^2 + \abs{\beta}^2 = 1$, then 
\[
\varphi = \abs{\alpha}^2 \psi + \abs{\beta}^2 \omega.
\]
In particular, $\varphi \notin \pstate(\fA)$, hence $\psi$ and $\omega$ are not coherently superposable.
\end{prop}

\begin{proof}
Define $\hilbH_\Psi = \overline{\pi(\frA)\Psi}$ and note that $\hilbH_\Psi$ is a closed invariant subspace. Let $P_\Psi :\hilbH \rightarrow \hilbH_\Psi$ be the orthogonal projection and $\iota_\Psi :\hilbH_\Psi \rightarrow \hilbH$ the inclusion. Lastly, define $\pi_\Psi : \frA \rightarrow \cB(\hilbH_\Psi)$ by
\[
\pi_\Psi(A) = P_\Psi \pi(A)\iota_\Psi
\] 
for all $A \in \frA$. Since $\hilbH_\Psi$ is an invariant subspace, we see that $\iota_\Psi \pi_\Psi(A) = \pi(A)\iota_\Psi$ for all $A \in \fA$. The orthogonal complement of an invariant subspace is invariant, so it further follows that $\pi_\Psi(A)P_\Psi = P_\Psi \pi(A)$ for all $A \in \fA$. Note also that $P_\Psi \iota_\Psi = id_{\hilbH_{\Psi}}$. Thus, for all $A, B \in \fA$,
\begin{align*}
\pi_\Psi(AB) &= P_\Psi \pi(A) \pi(B)\iota_\Psi P_\Psi \iota_\Psi = P_\Psi \pi(A) \iota_\Psi \pi_\Psi(B) P_\Psi \iota_\Psi \\
&= P_\Psi \pi(A)\iota_\Psi P_\Psi \pi(B) \iota_\Psi = \pi_\Psi(A)\pi_\Psi(B).
\end{align*}
Furthermore, since $\iota_\Psi$ is an isometry, for any $x, y \in \hilbH_\Psi$ we have
\begin{align*}
\ev{\pi_\Psi(A^*)x, y} &= \ev{\iota_\Psi \pi_\Psi(A^*)x, \iota_\Psi y} = \ev{\pi(A^*)\iota_\Psi x, \iota_\Psi y} \\
&= \ev{\iota_\Psi x, \pi(A)\iota_\Psi y} = \ev{\iota_\Psi x, \iota_\Psi \pi_\Psi(A)y} = \ev{x, \pi_\Psi(A)y}.
\end{align*}
Therefore $\pi_\Psi(A^*) = \pi_\Psi(A)^*$, and $(\hilbH_\Psi, \pi_\Psi)$ is a representation of $\frA$.
Identical arguments and the corresponding notation apply with $\Psi$ replaced by $\Omega$.

The nondegeneracy of $\pi$ implies that $\Psi \in \hilbH_\Psi$, so $\Psi$ is a cyclic vector for $\pi_\Psi$. Furthermore,
\[
\psi(A) = \ev{\Psi, \pi(A)\Psi} = \ev{\iota_\Psi \Psi, \pi(A)\iota_\Psi \Psi} = \ev{\iota_\Psi \Psi, \iota_\Psi \pi_\Psi(A)\Psi} = \ev{\Psi, \pi_\Psi(A)\Psi}.
\]
It follows that $\pi_\Psi$ is unitarily equivalent to the GNS representation of $\psi$. Thus, $\pi_\Psi$ is not unitarily equivalent to $\pi_\Omega$. Since $\psi$ and $\omega$ are pure, their GNS representations are irreducible, and so are $\pi_\Psi$ and $\pi_\Omega$.

Finally, observe that
\[
P_\Omega \iota_\Psi \pi_\Psi(A) = P_\Omega \pi(A)\iota_\Psi = \pi_\Omega(A)P_\Omega \iota_\Psi \quad\text{for all } A \in \frA \ .
\] 
Since $\pi_\Psi$ and $\pi_\Omega$ are not unitarily equivalent, Lemma \ref{lem:intertwiners_iff_equivalent} implies that $P_\Omega \iota_\Psi= 0$.
It follows that $\hilbH_\Psi$ and $\hilbH_\Omega$ are mutually orthogonal. In particular, $\ev{\Psi, \Omega} = 0$ and
\begin{align*}
\varphi(A) = \ev{\alpha \Psi + \beta \Omega, \pi(A)\qty(\alpha \Psi + \beta \Omega)}  = \abs{\alpha}^2 \psi(A) + \abs{\beta}^2 \omega(A)
\end{align*}
as desired. Since $\varphi$ is a nontrivial convex combination of pure states, $\varphi \notin \pstate(\fA)$.
\end{proof}

The following lemma is a rephrasal of Proposition 3.13.4 in \cite{PedersenCAlgAutomorphisms}. It was originally stated for unital $C^*$-algebras in \cite{GlimmKadisonUOCA}.

\begin{lem}\label{lem:sector_is_open}
Let $\frA$ be a $C^*$-algebra and let $\psi, \omega \in \pstate(\fA)$. If $\psi$ and $\omega$ are in different superselection sectors, then $\norm{\psi - \omega} = 2$.
\end{lem}

\begin{proof}
Let $\hilbH = \hilbH_\psi \oplus \hilbH_\omega$ and $\pi = \pi_{\psi} \oplus \pi_{\omega}$ be the direct sum of the two GNS representations. Note that $\pi$ is nondegenerate since $\pi_{\psi}$ and $\pi_{\omega}$ are nondegenerate. Let $P_\psi :\hilbH \rightarrow \hilbH_\psi$ and $\iota_\psi :\hilbH_\psi \rightarrow \hilbH$ be the usual projections and inclusions, and define $P_\omega$ and $\iota_\omega$ similarly. Observe that 
\[
\pi_{\psi}(A) P_\psi = P_\psi \pi(A) \qqtext{and} \iota_\psi \pi_{\psi}(A) = \pi(A) \iota_\psi
\]
for all $A \in \frA$, and similarly with $\psi$ replaced by $\omega$.

Define $U \in \cB(\hilbH)$ by $U(x, y) = (x, -y)$ for all $(x,y) \in \hilbH$. Our goal is to show that $U \in \pi(\fA)''$. Suppose $T \in \pi(\frA)'$. For $i, j \in \qty{\psi, \omega}$, we compute
\[
\pi_{i}(A) P_i T \iota_j  = P_i \pi(A) T \iota_j  = P_i T \pi(A) \iota_j = P_i T \iota_j \pi_{j}(A).
\]
For $i = j$, this implies that $P_i T \iota_i \in \pi_{i}(\frA)'$. Since $\pi_{i}$ is irreducible, we know $P_i T\iota_i = \lambda_i I$ for some $\lambda_i \in \bbC$. For $i \neq j$, Lemma \ref{lem:intertwiners_iff_equivalent} and the assumption that $\pi_\psi$ and $\pi_\omega$ are not unitarily equivalent imply $P_i T\iota_j = 0$. Thus, $T(x,y) = (\lambda_\psi x, \lambda_\omega y)$ for all $(x,y) \in \hilbH$, so
\[
UT(x, y) = (\lambda_\psi x, - \lambda_\omega y) = TU(x, y).
\]
This implies that $U \in \pi(\frA)''$, as desired. Note that clearly $U \in \pi(\frA)'$ as well.

Fix $\varepsilon > 0$. By the von Neumann bicommutant theorem, we know the closure of $\pi(\frA)$ in the strong operator topology on $\cB(\hilbH)$ is equal to $\pi(\frA)''$. Since $U \in \pi(\fA)''$ and $\norm{U} = 1$, the Kaplansky density theorem implies that there exists $A \in \frA$ with $\norm{\pi(A)} \leq 1$ and 
\[
\norm{\pi(A)(\Psi,\Omega) - U(\Psi,\Omega)} < \varepsilon,
\]
where $\Psi$ and $\Omega$ are the cyclic unit vectors corresponding to $\psi$ and $\omega$ in the GNS construction. By Lemma \ref{lem:rep_almost_isometry}, there exists $B \in \fA$ such that $\pi(A) = \pi(B)$ and $\norm{B} < 1 + \varepsilon$. Thus,
\begin{align*}
\abs{(\psi - \omega)\qty(B)} &=  \abs{\ev{\Psi, \pi_\psi(B)\Psi} - \ev{\Omega, \pi_\omega(B)\Omega}}\\
&=  \abs{\ev{U(\Psi, \Omega), \pi(B)(\Psi, \Omega)}}\\
&\geq \abs{\ev{U(\Psi, \Omega),U(\Psi, \Omega)}} - \abs{\ev{U(\Psi, \Omega), (\pi(A) - U)(\Psi, \Omega)}}\\
&\geq 2 - \sqrt{2} \varepsilon.
\end{align*}
Thus,
\[
\norm{\psi - \omega} \geq \abs{(\psi - \omega)\qty(\frac{B}{1 + \varepsilon})} \geq \frac{2 - \sqrt{2}\varepsilon}{1 + \varepsilon}.
\]
Since $\varepsilon$ was arbitrary, this implies that $\norm{\psi - \omega} \geq 2$. Since $\norm{\psi - \omega} \leq 2$ by the triangle inequality, the result is proven.
\end{proof}

The contrapositive of Lemma \ref{lem:sector_is_open} immediately yields the following corollary. 

\begin{cor}\label{cor:sector_is_open}
Let $\fA$ be a $C^*$-algebra. The superselection sectors of $\fA$ are open in $\pstate(\fA)_{\textup{n}}$, i.e., in $\pstate(\fA)$ endowed with the norm topology.
\end{cor}

The following theorem is a rephrasal of Proposition 4.6 in \cite{RobertsRoepstorffSBCAQT}. However, we give a new proof for the case when $\psi$ and $\omega$ are in the same superselection sector. The formula \eqref{eq:norm_ray_product} is extremely useful.

\begin{thm}\label{thm:transition_probability}
Let $\fA$ be a $C^*$-algebra, let $\psi, \omega \in \pstate(\fA)$,  and let $(\hilbH, \pi)$ be a nondegenerate representation with unit vectors $\Psi, \Omega \in \hilbH$ representing $\psi$ and $\omega$, respectively. If $\psi$ and $\omega$ are in different superselection sectors, or if they are in the same superselection sector and $(\hilbH, \pi)$ is irreducible, then
\begin{equation}\label{eq:norm_ray_product}
\abs{\ev{\Psi, \Omega}}^2 = 1 - \frac{1}{4}\norm{\psi - \omega}^2.
\end{equation}
\end{thm}

\begin{proof}
If $\psi$ and $\omega$ are in different superselection sectors, then $\norm{\psi - \omega} = 2$ by Lemma \ref{lem:sector_is_open}, so the right hand side is zero, and $\ev{\Psi, \Omega} = 0$ by Proposition \ref{prop:inequivalent_superposition}.

We now consider the case where $\psi$ and $\omega$ are in the same superselection sector and $(\hilbH, \pi)$ is irreducible. If $\Psi$ and $\Omega$ are linearly dependent, then $\psi = \omega$ and $\abs{\ev{\Psi, \Omega}}^2 = 1$, so the identity holds. To prove the case where $\Psi$ and $\Omega$ are linearly independent, we will show inequality in both directions. 

Let $\lambda = 1 - \abs{\ev{\Psi, \Omega}}^2$ (which is strictly positive since $\Psi$ and $\Omega$ are linearly independent unit vectors) and define
\[
\Phi_\Psi = \lambda^{-1/2}\qty(\Omega - \ev{\Psi, \Omega}\Psi) \qqtext{and} \Phi_\Omega = \lambda^{-1/2}\qty(\Psi - \ev{\Omega, \Psi} \Omega).
\]
Then $\qty{\Psi, \Phi_\Psi}$ and $\qty{\Omega, \Phi_\Omega}$ are both orthonormal systems for $\vecspan\qty{\Psi, \Omega}$. We may construct a unitary $U \in \cB(\hilbH)$ such that $U\Psi = \Phi_\Omega$ and $U\Phi_\Psi = -\Omega$ by extending $U$ linearly on $\vecspan\qty{\Psi, \Omega}$ and having it act as the identity on the orthogonal complement. Of course, $\norm{U} = 1$ since $U$ is unitary. Since $\pi$ is irreducible, we know $U \in \pi(\fA)'' = \cB(\hilbH)$. Therefore, given $\varepsilon > 0$, the von Neumann bicommutant theorem and the Kaplansky density theorem yields $A \in \fA$ such that $\norm{\pi(A)} \leq 1$, 
\begin{align*}
\norm{\pi(A)\Psi - U\Psi} < \varepsilon \qqtext{and} \norm{\pi(A)\Omega - U\Omega} < \varepsilon.
\end{align*}
By Lemma \ref{lem:rep_almost_isometry} we may assume $\norm{A} < 1 + \varepsilon$. Note that
\[
U\Omega = U\qty(\lambda^{1/2}\Phi_\Psi + \ev{\Psi, \Omega}\Psi) = -\lambda^{1/2}\Omega + \ev{\Psi, \Omega}\Phi_\Omega.
\]
Then we can compute
\begin{align*}
\abs{(\psi - \omega)(A)} &= \abs{\ev{\Psi, \pi(A)\Psi} - \ev{\Omega, \pi(A)\Omega}}\\
&\geq \abs{\ev{\Psi, U\Psi} - \ev{\Omega, U\Omega}} - \abs{\ev{\Psi, \pi(A)\Psi - U\Psi}} - \abs{\ev{\Omega, \pi(A)\Omega - U\Omega}}\\
& \geq 2\lambda^{1/2} - 2\varepsilon.
\end{align*}
Thus,
\[
\norm{\psi - \omega} \geq \abs{(\psi - \omega)\qty(\frac{A}{1 + \varepsilon})} \geq \frac{2\lambda^{1/2} - 2\varepsilon}{1 + \varepsilon}.
\]
Since $\varepsilon > 0$ was arbitrary, this implies that $\norm{\psi - \omega} \geq 2\lambda^{1/2}$, which rearranges to
\[
\abs{\ev{\Psi, \Omega}}^2 \geq 1 - \frac{1}{4} \norm{\psi - \omega}^2.
\]

For the reverse inequality, let $A \in \frA$ with $\norm{A}\leq 1$. Given $\alpha, \beta, \gamma \in \bbC$ with $\abs{\alpha} = \abs{\beta} = \abs{\gamma} = 1$,  define $A' = \alpha A$, $\Psi' = \beta \Psi$, and $\Omega' = \gamma \Omega$, we note that $\Psi'$ and $\Omega'$ define the same states as $\Psi$ and $\Omega$, respectively, and $\abs{(\psi - \omega)(A)} = \abs{(\psi - \omega)(A')}$. We choose $\alpha$ so that $(\psi - \omega)(A') \geq 0$ and we choose $\beta$ and/or $\gamma$ such that $\ev{\Psi', \Omega'}$ is real. We compute
\begin{align*}
\ev{\Psi' \pm \Omega', \pi(A')(\Psi' \mp \Omega')} = (\psi  - \omega)(A')  \pm \qty[\ev{\Omega', \pi(A')\Psi'} - \ev{\Psi', \pi(A')\Omega'}].
\end{align*}
For one of the sign choices, the term in square brackets will have nonnegative real part, and therefore the magnitude of the left hand side will be at least $\abs{(\psi - \omega)(A)}$. For either sign choice, the Cauchy-Schwarz inequality gives
\begin{align*}
\abs{\ev{\Psi' \pm \Omega', \pi(A')\qty(\Psi' \mp \Omega')}} &\leq \norm{\Psi' + \Omega'} \norm{\Psi' - \Omega'} \\
&= 2\sqrt{1 - \qty(\Re \ev{\Psi', \Omega'})^2}\\
&= 2 \sqrt{1 - \abs{\ev{\Psi, \Omega}}^2}.
\end{align*}
Thus, we have
\[
\abs{(\psi - \omega)(A')} = \abs{(\psi - \omega)(A)} \leq 2\sqrt{1 - \abs{\ev{\Psi, \Omega}}^2}.
\]
Since $A$ was arbitrary, we have $\norm{\psi - \omega} \leq 2 \sqrt{1 - \abs{\ev{\Psi, \Omega}}^2}$, which rearranges to the desired inequality.
\end{proof}

\begin{corollary}\label{cor:PH_superselection_metric_equivalence}
Let $\fA$ be a $C^*$-algebra and let $(\hilbH, \pi)$ be a nonzero irreducible representation of $\fA$. The map $\bbP\hilbH \rightarrow \pstate_\pi(\fA)_\tn{n}$ which to each ray $\bbC \Psi$ with $\Psi \in \bbS \hilbH$ assigns the pure state $\psi$ represented by $\Psi$ is a homeomorphism. In particular, given rays $\bbC \Psi, \bbC \Omega \in \bbP \hilbH$ with $\Psi, \Omega \in \bbS \hilbH$ representing the pure states $\psi$ and $\omega$, we have
\[
2d_\tn{gap}(\bbC \Psi, \bbC \Omega) = \norm{\psi - \omega}.
\]
Thus, the open ball $\ball_1(\bbC \Omega)$ of radius one with respect to the gap metric maps homeomorphically onto $\ball_2(\omega)$.
\end{corollary}

\begin{proof}
The map is well-defined since two unit vectors $\Psi, \Psi' \in \bbS \hilbH$ which represent the same ray $\bbP \hilbH$ can only differ by a phase, and therefore define the same state. The map is bijective by Proposition \ref{prop:pure_folia_vector_state}. Finally, we observe that for any $\Psi, \Omega \in \bbS \hilbH$ representing pure states $\psi$ and $\omega$, Theorem \ref{thm:transition_probability} yields
\[
d_\tn{gap}(\bbC\Psi, \bbC\Omega)^2 = 1 - \ray{\bbC\Psi,\bbC\Omega}^2 = \frac{1}{4}\norm{\psi - \omega}^2
\]
This implies that our map is a homeomorphism.
\end{proof}

\begin{corollary}\label{cor:purestate_mapsto_SH}
Let $\fA$ be a $C^*$-algebra, let $\omega \in \pstate(\fA)$, and let $\ball_2(\omega) \subset \pstate(\fA)$ be the open ball of radius two around $\omega$. If $(\hilbH, \pi)$ is a nonzero irreducible representation of $\fA$ and $\Omega \in \sphere \hilbH$ represents $\omega$, then the map $\ball_2(\omega) \rightarrow \sphere \hilbH$ that associates to each pure state $\psi \in \ball_2(\omega)$ the unique unit vector $\Psi \in \sphere \hilbH$ representing $\psi$ and satisfying $\ev{\Psi,\Omega} > 0$ is continuous.
\end{corollary}

\begin{proof}
Combine Corollary \ref{cor:PH_superselection_metric_equivalence} and Corollary \ref{cor:PH_mapsto_SH}.
\end{proof}

\begin{corollary}\label{cor:superselection_sector_closed}
Let $\fA$ be a $C^*$-algebra. The union of an arbitrary collection of superselection sectors of $\fA$ is norm-closed in $\fA^*$.
\end{corollary}

\begin{proof}
As pointed out in Definition \ref{def:superselection_sector}, each superselection sector coincides with $\pstate_\pi(\fA)$ for some nonzero irreducible representation $(\hilbH, \pi)$. It follows from Theorem \ref{thm:metric_equivalences} and Corollary \ref{cor:PH_superselection_metric_equivalence} that each $\pstate_\pi(\fA)$ is complete, hence closed, in $\fA^*$ in the norm topology. By Lemma \ref{lem:sector_is_open}, each Cauchy sequence in $\pstate(\fA)$ is eventually in a single superselection sector. It follows that the union of an arbitrary collection of superselection sectors is complete, hence norm-closed.
\end{proof}

Corollary \ref{cor:superselection_sector_closed} is equivalent to \cite[Cor.~4.8]{KadisonLimitsofStates}. Kadison takes a different approach to the proof, but remarks that it may be proven with Theorem \ref{thm:transition_probability} as we have done in the proof of Corollary \ref{cor:PH_superselection_metric_equivalence}.

\begin{thm}\label{thm:superselection_sector_equivalences}
Let $\fA$ be a $C^*$-algebra and let $\psi, \omega \in \pstate(\fA)$ be pure states. The following are equivalent:
\begin{enumerate}
\item\label{ite:superselection-sector} $\psi$ and $\omega$ are in the same superselection sector,
\item\label{ite:irreducible-representation} there exists a nonzero irreducible representation $(\hilbH, \pi)$ such that $\psi, \omega \in \pstate_\pi(\fA)$,
\item\label{ite:quasilocal-perturbation} there exists $B \in \fA$ such that $\psi(B^*B)  = 1$ and $\omega = B \cdot \psi$,
\item\label{ite:normal-states} for every nondegenerate representation $(\hilbH, \pi)$, we have $\psi \in \pstate_\pi(\fA)$ if and only if $\omega \in \pstate_\pi(\fA)$,
\item\label{ite:path-component} $\psi$ and $\omega$ are in the same path component of
  $\pstate(\fA)_{\textup{n}}$,
\item\label{ite:superposable} $\psi$ and $\omega$ are  coherently superposable.
\end{enumerate}
If $\fA$ is unital, then the element $B \in \fA$ in \ref{ite:quasilocal-perturbation} may be chosen to be unitary. 
\end{thm}

The equivalence \ref{ite:superselection-sector} $\Leftrightarrow$ \ref{ite:irreducible-representation} $\Leftrightarrow$ \ref{ite:quasilocal-perturbation} $\Leftrightarrow$ \ref{ite:path-component} was stated by Roberts--Roepstorff in the case where $\fA$ is unital and the element $B \in \fA$ in \ref{ite:quasilocal-perturbation} is unitary \cite[Prop.~4.2 \& Thm.~4.5]{RobertsRoepstorffSBCAQT}. The equivalence \ref{ite:superselection-sector} $\Leftrightarrow$ \ref{ite:superposable} is implied by \cite[Thm.~6.1]{ArakiMTQF}, which gives a more general equivalence for states which are not necessarily pure. The equivalence of \ref{ite:normal-states} with the others is easily proved. We clarify that the result holds for nonunital $C^*$-algebras as well.

\begin{proof}
Denote by $(\hilbH_\psi, \pi_\psi,\Omega_\psi)$ and $(\hilbH_\omega, \pi_\omega,\Omega_\omega)$ the GNS representations of $\psi$ and $\omega$, respectively. 
Now verify the following.

\ref{ite:superselection-sector} $\Rightarrow$ \ref{ite:irreducible-representation}. This follows from our comments in Definition \ref{def:superselection_sector} that $\pstate_\omega(\fA) = \pstate_{\pi_\omega}(\fA)$, and the fact that $(\hilbH_{\omega}, \pi_{\omega})$ is a nonzero irreducible representation.

\ref{ite:irreducible-representation} $\Rightarrow$ \ref{ite:quasilocal-perturbation}. Let $\Psi, \Omega \in \hilbH$ be unit vectors representing $\psi$ and $\omega$. By the Kadison transitivity theorem, there exists $B \in \fA$ such that $\pi(B)\Psi = \Omega$. Then 
\[
\psi(B^*B) = \ev{\Psi, \pi(B^*B)\Psi} = \ev{\Omega, \Omega} = 1
\]
and
\[
(B \cdot \psi)(A) = \psi(B^*AB) = \ev{\Omega, \pi(A)\Omega} = \omega(A),
\]
as desired. Note that if $\fA$ is unital, then the Kadison transitivity theorem allows $B$ to be chosen to be unitary. 

\ref{ite:quasilocal-perturbation} $\Rightarrow$ \ref{ite:superselection-sector}. In the GNS representation of $\psi$ one has
\[
  \omega(A) = \ev{\pi_\psi(B)\Omega_\psi, \pi_\psi(A)\pi_\psi(B)\Omega_\psi} \quad\text{for all } A \in \fA \ .
\] 
Since $\pi_\psi(B)\Omega_\psi$ is a cyclic unit vector, uniqueness of the GNS representation up to unitary equivalence implies that $\omega$ and $\psi$ are in the same superselection sector.

\ref{ite:quasilocal-perturbation} $\Rightarrow$ \ref{ite:normal-states}. Given any nondegenerate representation $(\hilbH, \pi)$, $\psi \in \pstate_\pi(\fA)$ implies $\omega \in \pstate_\pi(\fA)$ by \ref{ite:folium-invariance}. Since \ref{ite:quasilocal-perturbation} $\Rightarrow$ \ref{ite:superselection-sector} and \ref{ite:superselection-sector} is symmetric in $\psi$ and $\omega$, we also have \ref{ite:quasilocal-perturbation} with $\psi$ and $\omega$ switched, hence $\omega \in \pstate_\pi(\fA)$ implies $\psi \in \pstate_\pi(\fA)$.

\ref{ite:normal-states} $\Rightarrow$ \ref{ite:path-component}. Let $(\hilbH_\psi, \pi_\psi, \Psi)$ be the GNS representation of $\psi$. By \ref{ite:normal-states}, there exists a unit vector $\Omega \in \hilbH_\psi$ representing $\omega$. Since the unit sphere of $\hilbH_\psi$ is path-connected, there exists a continuous path in the unit sphere from $\Psi$ to $\Omega$, hence there exists a continuous path in $\pstate(\fA)_{\textup{n}}$ from $\psi$ to $\omega$ by Lemma \ref{lem:vector_state}.

\ref{ite:path-component} $\Rightarrow$ \ref{ite:superselection-sector}. By Corollary \ref{cor:sector_is_open}, the superselection sector containing $\psi$ is open in $\pstate(\fA)$ with respect to the norm topology. By our comments in Definition \ref{def:superselection_sector}, we know $\pstate_\psi(\fA) = \pstate_{\pi_\psi}(\fA)$. But $\pstate_{\pi_\psi}(\fA) = \state_{\pi_\psi}(\fA) \cap \pstate(\fA)$ is norm-closed in $\pstate(\fA)$ since $\state_{\pi_\psi}(\fA)$ is norm-closed in $\state(\fA)$. Thus, the superselection sector containing $\psi$ is both an open and closed subset of $\pstate(\fA)$ in the norm
topology, so it contains the path component of $\pstate(\fA)_{\textup{n}}$ containing $\psi$.

\ref{ite:superselection-sector} $\Rightarrow$ \ref{ite:superposable}. 
  Assume that $\psi$ and $\omega$ are in the same superselection sector. Let $(\hilbH,\pi)$
  be an irreducible representation which is unitarily equivalent to $\pi_\omega$ and hence to $\pi_\psi$
  as well. Then there exist unit vectors $\Psi\in\hilbH$ and $\Omega \in \hilbH$ which induce
  the states $\psi$ and $\omega$, respectively. Let $\alpha,\beta \in \bbC$ such that
  $\Phi = \alpha\Psi + \beta\Omega$ is nonzero. The vector state $\varphi$
  corresponding to the unit vector $\Phi/\norm{\Phi}$ is pure by
  Proposition \ref{prop:pure_folia_vector_state} \ref{ite:irreducible-vector-state-pure},
  hence $\psi$ and $\omega$ are coherently superposable. 
  
\ref{ite:superposable} $\Rightarrow$ \ref{ite:superselection-sector}. This is contained in Proposition \ref{prop:inequivalent_superposition}.
\end{proof}

\begin{remark}
The equivalence of \ref{ite:superselection-sector} and \ref{ite:superposable} shows that the mathematical notion of superselection sector we have defined corresponds with the notion of superselection sector in physics. If $\fA$ is a quasi-local $C^*$-algebra, then the equivalence of \ref{ite:superselection-sector} and \ref{ite:quasilocal-perturbation} imply that the superselection sector of $\omega \in \pstate(\fA)$ is the set of all states that can be obtained from $\omega$ by acting on it with quasi-local operators.
\end{remark}

\begin{remark}
  For the quasi-local algebras of interest in quantum physics, it may not be the case that all superselection sectors are physically relevant. In their groundbreaking work \cite{DHRI,DHRII},
  Doplicher--Haag--Roberts (DHR) introduced a general theory of superselection sectors for algebraic
  quantum field theory as defined by Haag--Kastler \cite{HaagKastlerAAQFT,HaagLocalQP}. Roughly speaking, the relevant superselection sectors in the DHR analysis correspond to pure states that are unitarily equivalent to a vacuum state when restricted to observables localized in the causal complement of a bounded region of spacetime. In the $C^*$-algebraic formulation of quantum spin systems a similar approach to superselection sectors has been advocated in the work by Naaijkens,
  Cha, and Nachtergaele, see
  \cite{NaaijkensLEKTCP,NaaijkensKQDMLQPPV,ChaNaaijkensNachtergaeleSCIQSS}. There, the bounded region of spacetime from the DHR analysis is replaced by an infinite cone-like region of the lattice (in particular, the region is no longer bounded).  A similar approach appears also in the 
  work of Buchholz--Fredenhagen \cite{BuchholzFredenhagenLocalityParticleStates}
  on superselection sectors describing relativistic massive particles. There, the regions used
  are infinite cones in spacetime. 

  In general, one may also not be interested in all pure states in a given superselection sector either. For example, in work by Kapustin--Sopenko--Yang \cite{KapustinSopenkoYang,KSo_local_Noether}, one is interested in states which are obtained from product states by acting with so-called ``locally generated automorphisms.''

  In this dissertation, we will typically not worry about these subtleties and content ourselves to work with the full state space. At times, working with arbitrary states yields general results that apply no matter what special class of states one is interested in. In other cases, it becomes an interesting and nontrivial question to ask whether our results still hold (possibly in some modified form) when one restricts attention to a special class of states.
\end{remark}

\begin{corollary}\label{prop:decomposition-pure-state-space-superselection-sectors}
  Let $\fA$ be a $C^*$-algebra. The  space $\pstate(\frA)_{\textup{n}}$ of pure states endowed with the norm topology is locally path-connected.
  The superselection sectors are the path components of $\pstate(\frA)_{\textup{n}}$ and coincide with its connected components.
\end{corollary}

\begin{proof}
  We show every open ball $\ball_r(\omega) \subset \pstate(\fA)_{\textup{n}}$ with $r \leq 2$ is path-connected. If $\psi \in \ball_r(\omega)$, then Lemma \ref{lem:sector_is_open} implies $\psi$ and $\omega$ are in the same superselection sector, so by the equivalence of \ref{ite:superselection-sector} and \ref{ite:irreducible-representation} in  Theorem \ref{thm:superselection_sector_equivalences}, there exists a nonzero irreducible representation of $(\cH, \pi)$ such that $\psi, \omega \in \pstate_\pi(\fA)$. Let $\Psi, \Omega \in \bbS \cH$ represent $\psi$ and $\omega$. Since $\ev{\Psi, \Omega} \neq 0$ by Theorem \ref{thm:transition_probability}, we may choose $\Psi$ and $\Omega$ so that $\ev{\Psi, \Omega} > 0$. It follows that $\Phi(t) = t\Omega + (1 - t)\Psi$ is nonzero for all $t \in [0,1]$. Then
  \[
  \widehat \Phi(t) = \frac{\Phi(t)}{\norm{\Phi(t)}}
  \]
  is a path in $\bbS \cH$ starting at $\Psi$ and ending at $\Omega$. Moreover,
  \[
  \ev{\Omega, \widehat{\Phi}(t)} = \frac{t + (1 - t)\ev{\Omega, \Psi}}{\norm{\Phi(t)}} \geq \ev{\Omega, \Psi},
  \]
  which implies by Theorem \ref{thm:transition_probability} that $\norm{\varphi(t) - \omega} \leq  \norm{\psi - \omega}$, where $\varphi(t) \in \pstate(\fA)_\tn{n}$ is the pure state represented by $\widehat \Phi(t)$. In particular, $\varphi(t) \in \ball_r(\omega)$ for all $t$. By Lemma \ref{lem:vector_state} we know $\varphi(t)$ is continuous, so $\varphi(t)$ is a path in $\ball_r(\omega)$ starting at $\psi$ and ending at $\omega$. This proves that $\ball_r(\omega)$ is path-connected, and this implies that $\pstate(\fA)_\tn{n}$ is locally path-connected.

  The equivalence of \ref{ite:superselection-sector} and \ref{ite:path-component} in Theorem \ref{thm:superselection_sector_equivalences} implies
  that the superselection sectors are just the path components of $\pstate(\frA)_{\textup{n}}$. The connected components coincide with the path components since this is always true in a locally path-connected space.
\end{proof}

\section{Properties of the Weak* Topology}
\label{sec:P(A)_weak*}

The weak* topology on $\pstate(\fA)$ can be a strange creature. On the one hand, it is a good topology for discussing continuity of ground states of parametrized spin systems. It also often enjoys many nice elementary topological properties. The following theorem is an example.

\begin{theorem}[{\cite[Prop.~4.3.2]{PedersenCAlgAutomorphisms}}]\label{thm:P(A)_Polish}
If $\fA$ is a separable $C^*$-algebra, then the pure state space $\pstate(\fA)_\tn{w*}$ with the weak* topology is separable and completely metrizable, i.e., it is a Polish space.
\end{theorem}

On the other hand, the weak* topology can be difficult to work with. For example, a complete metric for $\pstate(\fA)_\tn{w*}$ as referred to in Theorem \ref{thm:P(A)_Polish} is somewhat artificial and difficult to describe. All open sets of the weak* topology are also ``very large'' in a certain sense. For a quasi-local $C^*$-algebra like $\fA = \bigotimes_{v \in \bbZ} M_2(\bbC)$, given any pure state $\omega \in \pstate(\fA)$ and any weak*-neighborhood $U$ of $\omega$, there will be pure states in $U$ that differ drastically from $\omega$ outside a finite region of the lattice. 

Let us further discuss the elementary topology of $\pstate(\fA)$, in particular its connectedness properties. For infinite-dimensional $C^*$-algebras of interest in the study of quantum lattice spin systems, $\pstate(\frA)_{\textup{n}}$ has uncountably many path components (this is true for any UHF algebra). In contrast, there is often only one component of $\pstate(\frA)_{\textup{w}^*}$, i.e.\ the pure state space with the weak$^*$ topology. This can be seen as a consequence of the theorem below. Following \cite{KadTSOTD} (in the version of \cite{AarFSSCSA}), we call a set $\folium$ of states \textdef{full}
if for all $A\in \fA$,  we have that $A$ is positive if and only if $\omega (A)\geq 0$ for all
$\omega \in \folium$. The theorem below gives a crucial characterization of when a set of states is full.

\begin{thm}[{\cite[Thm.~2.2]{KadTSOTD} \& \cite[Thm.~1 \& 2]{AarFSSCSA}}] \label{thm:full_set}
  For a set $\folium$ of states on a unital $C^*$-algebra $\fA$, the following are equivalent:
  \begin{enumerate}[label=\tn{(\alph*)}]
  \item\label{ite:full} $\folium$ is full,
  \item\label{ite:full_weak*_cch} the weak$^*$ closure of the convex hull of  $\folium$  contains  $\state(\fA)$,
  \item\label{ite:full_weak*_closure} the weak$^*$ closure of  $\folium$ contains $\pstate (\fA)$,
  \item\label{ite:full_norm} $\| A \| =\sup\limits_{\omega \in \folium} \omega (A)$ for every positive element $A \in \fA$. 
  \end{enumerate}
  If $\fA$ is nonunital and nonzero, then {\ref{ite:full_weak*_cch} $\Leftrightarrow$ \ref{ite:full_weak*_closure} $\Leftrightarrow$ \ref{ite:full_norm} $\Rightarrow$ \ref{ite:full}}.
\end{thm}

A nice proof of \ref{ite:full} $\Rightarrow$ \ref{ite:full_weak*_cch} and \ref{ite:full} $\Rightarrow$ \ref{ite:full_weak*_closure} can be found in \cite[Thm.~5.1.14]{Murphy}. We caution that \cite[Prop.~3.2.10]{BratteliRobinsonOAQSMII} falsely states \ref{ite:full} $\Rightarrow$ \ref{ite:full_weak*_cch} without requiring $\fA$ to be unital; a counterexample is provided in \cite{AarFSSCSA}. 

%This theorem is not new, however we provide a new elementary proof of (\ref{ite:full}) $\Rightarrow$ (\ref{ite:full_weak*_cch}) when $\fA$ is unital, in contrast to proofs in the literature which are more technical \cite[Thm.~2.2]{KadTSOTD} \& \cite[Prop.~3.2.10]{BratteliRobinsonOAQSMII} or lacking in details \cite[Lem.~3.4.1]{DixCA}. We caution that \cite[Prop.~3.2.10]{BratteliRobinsonOAQSMII} falsely states (\ref{ite:full}) $\Rightarrow$ (\ref{ite:full_weak*_cch}) without requiring $\fA$ to be unital; a counterexample is provided in \cite{AarFSSCSA}. 

%\begin{proof}
%(\ref{ite:full}) $\Rightarrow$ (\ref{ite:full_weak*_cch}). Let $\cch(\sF)$ be the weak$^*$ closure of the convex hull of $\sF$ and suppose there exists $\psi \in \sS(\fA) \setminus \cch(\sF)$. By the Hahn-Banach theorem (in the form of \cite[Thm.~3.4 (b)]{RudinFA}), there exists $A \in \frA$ and $\gamma_1, \gamma_2 \in \bbR$ such that
%\[
%\omega(B) = \Re \omega(A) < \gamma_1 < \gamma_2 < \Re \psi(A) = \psi(B)
%\]
%for all $\omega \in \cch(\sF)$, where $B = (A+A^*)/2$. Rearranging this, we see that
%\[
%\omega\qty(\psi(B) - \gamma_2 + \gamma_1 - B) \geq 0
%\]
%for all $\omega \in \sF$, so $\psi(B) - \gamma_2 + \gamma_1 - B \geq 0$ since $\sF$ is full. But applying $\psi$ yields
%\[
%\psi(\psi(B) - \gamma_2 + \gamma_1 - B) = -\gamma_2 + \gamma_1 < 0,
%\]
%which contradicts that $\psi$ is positive.
%\end{proof}

\begin{cor}
\label{cor:connectedness-cstaralgebra-faithful-irred-rep}  
If $\fA$ is a nonzero $C^*$-algebra with a faithful irreducible representation, then $\pstate(\fA)_{\textup{w}^*}$ is connected.
\end{cor}

\begin{proof}
Let $(\hilbH, \pi)$ be a faithful irreducible representation. Then for every $A \in \fA_+$,
\[
\norm{A} = \norm{\pi(A)} = \sup_{\Omega \in \hilbH} \ev{\Omega, \pi(A)\Omega} = \sup_{\omega \in \pstate_\pi(\fA)} \omega(A),
\]
so $\pstate_\pi(\fA)$ is weak$^*$-dense in $\pstate(\fA)$ by Theorem \ref{thm:full_set}. Since $\pstate_\pi(\fA)$ is connected in the norm topology, it is connected in the weak$^*$ topology, and density of $\pstate_\pi(\fA)$ now implies that $\pstate(\fA)$ is connected in the weak* topology.
\end{proof}

In particular, Corollary \ref{cor:connectedness-cstaralgebra-faithful-irred-rep} implies that $\pstate(\fA)_\tn{w*}$ is connected for every simple $C^*$-algebra $\fA$. In fact, for a simple $C^*$-algebra $\fA$, the pure state space $\pstate(\fA)_\tn{w*}$ is locally connected, as implied by Theorem 5.6 in \cite{EilCCCA}.

We now provide a few criteria entailing that $\pstate(\fA)_\tn{w*}$ is
path-connected or locally path-connected. The essential tool is the following.

\begin{thm}[{\cite[Prop.~5.9]{EilCCCA}}]\label{thm:locally_path_connected}
If $\fA$ is separable and $\pstate(\fA)_\tn{w*}$ is connected and locally connected, then $\pstate(\fA)_{\textup{w}^*}$ is path-connected and locally path-connected.
\end{thm}

This theorem is a trivial synthesis of Theorem \ref{thm:P(A)_Polish} and the fact that a locally connected complete metric space is locally path-connected. Unfortunately, it is surprisingly difficult to find a correct proof of the latter fact in the literature. One may find a roadmap of common errors along with corrections in \cite{PersistenceofErrors}, where several references to both correct and incorrect proofs are provided. Nonetheless, Theorem \ref{thm:locally_path_connected} applies in many cases of interest in physics.

\begin{thm}\label{thm:P(A)_path_connected_examples}
  Let $\fA$ be a $C^*$-algebra. 
  \begin{enumerate}[label=\tn{(\alph*)}]
  \item\label{ite:simple-separable}
    If $\fA$ is simple and separable, then $\pstate(\fA)_\tn{w*}$ is path-connected and locally path-connected.
  \item\label{ite:direct-sum-separables}
    If $\fA = \bigoplus_{n=1}^N \fA_n$, $N \in \mathbb{N} \cup \qty{\infty}$ is the direct sum of separable, simple $C^*$-algebras $\fA_n$, then $\pstate(\fA)_\tn{w*}$ is locally path-connected.
  \item\label{ite:colimit-injective-direct-system}
    If $\fA$ is the colimit of an injective directed system of countably many separable, simple $C^*$-algebras, then $\pstate(\fA)_\tn{w*}$ is path-connected and locally path-connected.
  \end{enumerate}  
\end{thm}

By the direct sum $\bigoplus_{n=1}^\infty \fA_n$, we mean the set of sequences $(A_n)_{n \in \bbN} \in \prod_{n =1}^\infty \fA_n$ such that $\norm{A_n} \rightarrow 0$, equipped with the max norm and componentwise algebraic operations. 

\begin{proof}
If $\fA$ is simple and separable, then $\pstate(\fA)_{\textup{w}^*}$ is connected, locally connected, and completely metrizable, hence path-connected and locally path-connected by the preceding remarks. The type of $C^*$-algebra described in \ref{ite:colimit-injective-direct-system} is itself simple and separable, and therefore path-connected and locally path-connected by \ref{ite:simple-separable}.

It is stated without proof in \cite{EilCCCA} that the pure state space of $\bigoplus_{n=1}^\infty \fA_n$ is homeomorphic to the disjoint union $\bigsqcup_{n=1}^\infty \pstate(\fA_n)$, from which it follows that \ref{ite:direct-sum-separables} holds since each $\pstate(\fA_n)$ is locally path-connected. We give a proof of this fact with $\infty$ replaced by $N \in \bbN \cup \qty{\infty}$. For each $n$ let $\pi_n:\fA \rightarrow \fA_n$ be the canonical projection and let $\iota_n :\fA_n \rightarrow \fA$ be the canonical inclusion $\iota_n(A) = (B_m)_{m \in \bbN}$ where $B_m = 0$ if $m \neq n$ and $B_n = A$. Given $\omega \in \pstate(\fA_n)$, the fact that $\pi_n$ is a surjective $*$-homomorphism implies $\omega \circ \pi_n \in \pstate(\fA)$. Thus, we have a map $\bigsqcup_{n=1}^N \pstate(\fA_n) \rightarrow \pstate(\fA)$. If $\psi \in \pstate(\fA_m)$ and $\omega \in \pstate(\fA_n)$ such that $\psi \circ \pi_m = \omega \circ \pi_n$, then $m = n$ must hold, hence surjectivity of the projections implies $\psi = \omega$, so this map is injective. For surjectivity, suppose $\omega \in \pstate(\fA)$. It cannot happen that $\omega \circ \iota_n = 0$ for all $n$ because the span of $\bigcup_{n} \iota_n\qty(\fA_n)$ is dense in $\fA$, therefore there exists $n$ such that $\omega \circ \iota_n \neq 0$. By definition of the algebraic operations on $\fA$, we see that $(\iota_n \circ \pi_n)(A) \leq A$ for all $A \in \fA_+$. Therefore $\omega \circ \iota_n \circ \pi_n$ is a positive linear functional on $\fA$ dominated by $\omega$. Since $\omega$ is pure, there exists $t \in [0,1]$ such that $\omega \circ \iota_n \circ \pi_n = t \omega$. Composing with $\iota_n$ on the right and using the fact that $\omega \circ \iota_n \neq 0$ implies $t = 1$. Purity of $\omega \circ \iota_n$ follows easily from purity of $\omega$.

We have established a bijective correspondence $\bigsqcup_{n=1}^N \pstate(\fA_n)_{\textup{w}^*} \rightarrow \pstate(\fA)_{\textup{w}^*}$. Continuity of this map follows from the universal property of the disjoint union and the fact that each map $f_n:\pstate(\fA_n)_{\textup{w}^*} \rightarrow \pstate(\fA)_{\textup{w}^*}$, $f_n(\omega) = \omega \circ \pi_n$ is weak$^*$ continuous. We show that each $f_n$ is open, from which it follows from the definition of the disjoint union topology that the map $\bigsqcup_{n=1}^N \pstate(\fA_n)_{\textup{w}^*} \rightarrow \pstate(\fA)_{\textup{w}^*}$ is open, and therefore a homeomorphism. Given $\omega \in \pstate(\fA_n)$, choose $A \in \fA_n$ such that $\abs{\omega(A)} > 1/2$. Then $U = \qty{\psi \in \pstate(\fA): \abs{\psi(\iota_n(A)) - \omega(A)} < 1/2}$ is a neighborhood of $\omega \circ \pi_n$ contained in $f_n(\pstate(\fA_n))$ since $\psi \in U$ implies $\psi \circ \iota_n \neq 0$. Thus, $f_n(\pstate(\fA_n))$ is open. For any basis neighborhood $U_{\varepsilon, \omega, A_1,\ldots, A_k} \subset \pstate(\fA_n)_{\textup{w}^*}$ around $\omega \in \pstate(\fA_n)_{\textup{w}^*}$, we have  
\[
f_n(U_{\varepsilon, \omega, A_1,\ldots, A_k}) = f_n(\pstate(\fA_n)) \cap U_{\varepsilon, \omega \circ \pi_n, \iota_n(A_1),\ldots, \iota_n(A_k)}.
\]
This exhibits $f_n(U_{\varepsilon, \omega, A_1,\ldots, A_k})$ as an open set in $\pstate(\fA)_{\textup{w}^*}$, so $f_n$ is open.
\end{proof}

Given that there may be unphysical superselection sectors for a particular physical system, it is reasonable to ask whether $\pstate(\fA)_{\textup{w}^*}$ remains path-connected after some superselection sectors are removed. For unital, separable, simple $C^*$-algebras, the answer is yes. In fact, one may find a path between arbitrary pure states $\psi, \omega \in \pstate(\fA)$ that remains in $\pstate_\psi(\fA)$ until the path reaches $\omega$ at the endpoint. This is a trivial consequence of Theorem \ref{thm:superselection_sector_equivalences} and the following remarkable theorem of Kishimoto, Ozawa, and Sakai.

\begin{thm}[{\cite[Thm.~1.1]{kishimoto_ozawa_sakai_2003}}]
Let $\fA$ be a separable $C^*$-algebra. If $\psi, \omega \in \pstate(\fA)$ and $\ker \pi_\psi = \ker \pi_\omega$, then there exists an automorphism $\alpha \in \Aut(\fA)$ and a continuous family of unitaries $U:[0,\infty) \rightarrow \Unitary(\fA)$, $t \mapsto U_t$ such that $U_0 = I$, $\psi \circ \alpha = \omega$, and
\[
\alpha(A) = \lim_{t \rightarrow \infty} U_t^*A U_t 
\]
for all $A \in \fA$.
\end{thm}

\section{The Pure State Space of a UHF Algebra is Simply Connected}
\label{sec:UHF_simply_connected}

This section is devoted to proving the titular fact. We begin by recalling the natural action of a $C^*$-algebra on its
state space and then show some crucial properties we later need. This is done in Section 
\ref{sec:prelimfund}.
Using this action we introduce in Section \ref{subsec:A-homotopies} a new notion of  homotopy in the state space which
essentially is
a kind of homotopy \emph{lifted} to the $C^*$-algebra. We then examine the space of density matrices
of a matrix algebra in Section \ref{subsec:fddensitymatrices} and finally use the results obtained
there to prove our main claim that the space of states of an UHF algebra is simply connected.

\subsection{Action of \texorpdfstring{$\mathfrak{A}$}{A} on Its State Space}\label{sec:prelimfund}
In this section, we study an action of $\mathfrak{A}$ on its state space.
\begin{definition}
Let $\fA$ be a $C^*$-algebra, let $\omega \in \state(\fA)$, and let 
\[
\fN_\omega = \qty{A \in \fA: \omega(A^*A) = 0}
\]
be the Gelfand ideal of $\omega$. Recall that if $A \notin \fN_\omega$, then we may define 
\begin{equation}\label{eq:action_on_state_def}
A \cdot \omega \colon \fA \to  \bbC, \quad (A \cdot \omega)(B) = \frac{\omega(A^*BA)}{\omega(A^*A)}.
\end{equation}
This is a state of $\fA$. In the GNS representation $(\cH_\omega, \pi_\omega, \Omega_\omega)$ of $\omega$, it is represented by the unit vector $\pi_\omega(A)\Omega_\omega/\sqrt{\omega(A^*A)}$. Thus, if $\omega$ is pure, then so is $A \cdot \omega$. Following the definition, we see that given $A, B \in \fA$, we have $AB\notin \fN_\omega$ if and only if $B \notin \fN_\omega$ and $A \notin \fN_{B \cdot \omega}$, in which case $(AB) \cdot \omega = A \cdot (B \cdot \omega)$. Of course, if $\fA$ is unital, then we have $\1 \cdot \omega = \omega$.
\end{definition}

Let us now prove  a few elementary facts of this action.

\begin{proposition}\label{prop:invariant_state}
Let $\fA$ be a $C^*$-algebra and let $\omega \in \state(\fA)$. If $A \in \fA$ is nonzero and satisfies $\abs{\omega(A)} = \norm{A}$, then $A \cdot \omega = \omega$. 
\end{proposition}

\begin{proof}
Let $(\cH_\omega, \pi_\omega, \Omega_\omega)$ be the GNS representation of $\omega$. By the Cauchy-Schwarz inequality, we have
\[
\abs{\omega(A)} = \abs{\ev{\Omega_\omega, \pi_\omega(A)\Omega_\omega}} \leq \norm{\Omega_\omega}\norm{\pi_\omega(A)\Omega_\omega} \leq \norm{A}.
\]
Since $\abs{\omega(A)} = \norm{A}$, we see that the Cauchy-Schwarz inequality is saturated, so $\pi_\omega(A)\Omega_\omega = \lambda \Omega_\omega$ for some $\lambda \in \bbC$. Observe that $\lambda \neq 0$, otherwise we find that $\abs{\omega(A)} = 0$, which contradicts the hypotheses. Note that $\sqrt{\omega(A^*A)} = \abs{\lambda}$, so that $A \cdot \omega$ is represented by the vector $\lambda \abs{\lambda}^{-1}\Omega_\omega$. Vectors that differ by a phase represent the same state, so we conclude that $A \cdot \omega = \omega$.
\end{proof}

\begin{proposition}\label{prop:linear_combo_action}
Let $\fA$ be a $C^*$-algebra and let $\omega \in \pstate(\fA)$. Suppose $A, B \in \fA \setminus \fN_\omega$ and $A \cdot \omega = B \cdot \omega$. If $\alpha, \beta \in \bbC$ and $\alpha A + \beta B \notin \fN_\omega$, then $(\alpha A + \beta B) \cdot \omega = A \cdot \omega = B \cdot \omega$. 
\end{proposition}

\begin{proof}
Let $(\cH_\omega, \pi_\omega, \Omega_\omega)$ be the GNS representation of $\omega$. Since $A \cdot \omega = B \cdot \omega$ and $\omega$ is pure, we know $\pi_\omega(A)\Omega_\omega$ and $\pi_\omega(B)\Omega_\omega$ are linearly independent, i.e., there exists $\lambda \in \bbC \setminus \qty{0}$ such that $\pi_\omega(B)\Omega_\omega = \lambda \pi_\omega(A)\Omega_\omega$. Then $(\alpha A + \beta B) \cdot \omega$ is represented by
\[
\frac{\pi_\omega(\alpha A + \beta B)\Omega_\omega}{\norm{\pi_\omega(\alpha A + \beta B)\Omega_\omega}} = \frac{(\alpha + \lambda \beta)\cdot \pi_\omega(A)\Omega_\omega}{\norm{\pi_\omega(\alpha A + \beta B)\Omega_\omega}} .
\]
This differs from $\pi_\omega(A)\Omega_\omega/\sqrt{\omega(A^*A)}$ by a phase, so $(\alpha A + \beta B) \cdot \omega = A \cdot \omega$.
\end{proof}

Next, let us examine the continuity properties of the action of $\fA$ on $\state(\fA)$.

\begin{proposition}\label{prop:evaluation_jointly_continuous}
Let $\fA$ be a $C^*$-algebra and equip $\state(\fA)$ with the weak* topology. The map
\begin{equation}\label{eq:evaluation_jointly_continuous}
\fA \times \state(\fA) \to  \bbC, \quad (A, \omega) \mapsto \omega(A)
\end{equation}
is continuous.
\end{proposition}

\begin{proof}
Fix $A_0 \in \fA$, $\omega_0 \in \state(\fA)$, and $\varepsilon > 0$. Given $A \in \fA$ such that $\norm{A - A_0} < \varepsilon/2$ and $\omega \in \state(\fA)$ such that $\abs{\omega(A_0) - \omega_0(A_0)} < \varepsilon/2$, we have
\begin{align*}
\abs{\omega(A) - \omega_0(A_0)} &\leq \abs{\omega(A) - \omega(A_0)} + \abs{\omega(A_0) - \omega_0(A_0)}  \\
&\leq \norm{A - A_0} + \abs{\omega(A_0) - \omega_0(A_0)} < \varepsilon.
\end{align*}
Thus, the map \eqref{eq:evaluation_jointly_continuous} is continuous at the arbitrary point $(A_0, \omega_0)$.
\end{proof}

\begin{proposition}\label{prop:action_on_state_continuous}
Let $\fA$ be a $C^*$-algebra and equip $\state(\fA)$ with the weak* topology. The map
\begin{equation}\label{eq:action_on_state}
\qty{(A, \omega) \in \fA \times \state(\fA)\colon  A \notin \fN_\omega} \to  \state(\fA), \quad (A, \omega) \to  A \cdot \omega
\end{equation}
is continuous.
\end{proposition}

\begin{proof}
Fix an arbitrary $B \in \fA$. Continuity of \eqref{eq:action_on_state} will follow from continuity of $(A, \omega) \mapsto (A \cdot \omega)(B)$ since $B$ was arbitrary. But since $A \mapsto A^*A$ and $A \mapsto A^*BA$ are continuous maps from $\fA$ to itself, it is easy to see from \eqref{eq:action_on_state_def} and Proposition \ref{prop:evaluation_jointly_continuous} that $(A, \omega) \mapsto (A \cdot \omega)(B)$ is continuous.
\end{proof}

\subsection{\texorpdfstring{$\mathfrak{A}$}{A}-Homotopies}\label{subsec:A-homotopies}
A key idea in our computation of the fundamental group is to study homotopies that can be factored through a special kind of lift from the state space to the $C^*$-algebra. We define what we mean here.

\begin{definition}\label{defn:Ahomotopy}
  Let $\fA$ be a unital $C^*$-algebra, equip $\state(\fA)$ with the weak* topology, and let $X$ be a topological space.
  We then say that a continuous map $\psi\colon X \to  \state(\fA)$ is $\fA$-\emph{homotopic} to a continuous map
  $\omega\colon X \to  \state(\fA)$ if there exists a continuous map  $A\colon X \times \interval \to  \fA$
  such that  
\begin{enumerate}
  \item\label{ite:notinideal} $A_{x,s} \notin \fN_{\psi_x}$ for all $(x,s) \in X \times I$,
  \item $A_{x,0} = \1$ for all $x \in X$,
  \item\label{ite:homotopy} $A_{x,1} \cdot \psi_x = \omega_x$ for all $x \in X$.
\end{enumerate}

In this case $(x,s) \mapsto A_{x,s} \cdot \psi_x$ is a homotopy from $\psi$ to $\omega$ in the standard sense.
We call the map $A\colon X \times \interval \to  \fA$ fulfilling conditions \eqref{ite:notinideal} to \eqref{ite:homotopy}
an $\fA$-\emph{homotopy} from $\psi$ to $\omega$. It can be understood as a lift of the homotopy from $\psi$ to $\omega$.
Let us write $\psi \htpy{\fA} \omega$ if $\psi$ is $\fA$-homotopic to $\omega$.

When $X$ is the unit interval $\interval$, then we say $\psi$ is $\fA$-\emph{homotopic} to $\omega$ \emph{relative endpoints}
if there exists an $\fA$-homotopy $A\colon \interval \times \interval \to  \fA$ from $\psi$ to $\omega$ such that
\begin{enumerate}
\setcounter{enumi}{3}
\item\label{ite:startpoint}
  $A_{0,s} \cdot \psi_0 = \omega_0$ for all $s\in \interval$, and 
\item\label{ite:endpoint}
   $A_{1,s} \cdot \psi_1 = \omega_1$ for all $s\in \interval$. 
\end{enumerate}
An $\fA$-homotopy fulfilling these properies  will be called an $\fA$-\emph{homotopy relative endpoints}. 
Let us write $\psi \pathhtpy{\fA} \omega$ if $\psi$ is $\fA$-homotopic to $\omega$ relative endpoints.
In case both $\psi$ and $\omega$ are loops we sometimes say that  $\psi$ is $\fA$-\emph{homotopic} to $\omega$
\emph{relative the basepoint} whenever  $\psi \pathhtpy{\fA} \omega$. 
\end{definition}

\begin{proposition}
The relations $\htpy{\fA}$ and $\pathhtpy{\fA}$ are transitive.
\end{proposition}

\begin{proof}
Suppose $\chi \htpy{\fA} \psi$ and $\psi \htpy{\fA} \omega$. Let $A\colon X \times \interval \to  \fA$ and $B\colon X \times \interval \to  \fA$ be respective $\fA$-homotopies. Define $C\colon X \times \interval \to  \fA$ by 
\[
  C_{x,s} = \begin{cases} A_{x,2s} &\text{for } s \in [0,1/2] \ , \\
  B_{x,2s-1}A_{x,1}  &\text{for } s \in [1/2,1] \ .\end{cases}
\]
This is a well-defined continuous map since $B_{x,0} = \1$ for all $x \in X$. By construction, $C$ then is 
an $\fA$-homotopy from $\chi$ to $\omega$. If $X = \interval$ and both $A$ and $B$ are $\fA$-homotopies relative
endpoints, then it is  easy to see that $C$ is also an $\fA$-homotopy relative endpoints. 
\end{proof}

We note that being $\fA$-homotopic is \textit{not} a symmetric relation. For example, in what follows, we will act on
non-pure states with projections to obtain pure states, but we cannot undo this operation by acting with an element of
$\fA$ because acting on a pure state with an element of $\fA$ always returns a pure state.

In our applications, $\fA$ will be unital (with unit denoted $\1$), we will have $X = \interval$, and the function
$A\colon \interval \times \interval \to  \fA$ will be of the form 
\[A_{t,s} = s\tilde A_t + (1 - s)\1\] for some continuous map
$\tilde A\colon  \interval \to  \fA$. Given a state $\psi$ and $A \in \fA \setminus \fN_\psi$, it is therefore important to examine when the linear interpolation $sA + (1 - s)\1$ falls outside the Gelfand ideal of $\psi$ for all $s$. We examine a few special cases of interest.

\begin{proposition}\label{prop:projection_linear_interp}
Let $\fA$ be a unital $C^*$-algebra and let $\omega \in \state(\fA)$. If $P \in \fA$ is a projection and $\omega(P) > 0$, then $sP + (1 - s)\1 \notin \fN_\omega$ for all $s \in \interval$.
\end{proposition}

\begin{proof}
For ease of notation, let $P_s = sP + (1-s)\1$. Then
\[
\omega\qty(P_s^*P_s) = s^2 \omega(P) + (1-s)^2 + 2s(1-s)\omega(P) > 0. \qedhere
\]
\end{proof}

\begin{proposition}\label{prop:unitary_linear_interp}
Let $\fA$ be a unital $C^*$-algebra and let $\omega \in \state(\fA)$. If $U \in \Unitary(\fA)$ and $s \in \interval$, then $sU + (1-s)\1 \in \fN_\omega$ if and only if $\omega(U) = -1$ and $s = 1/2$.
\end{proposition}

\begin{proof}
For ease of notation, let $U_s = sU + (1 - s)\1$. Observe that
\begin{align*}
\omega(U_s^*U_s) &= s^2 + (1-s)^2 +2s (1-s)\Re \omega(U)\\
&\geq s^2 + (1-s)^2 - 2s(1-s) = (1 - 2s)^2
\end{align*}
If $\omega(U) = -1$ and $s = 1/2$, then the inequality is an equality and we get $\omega(U_s^*U_s) = (1 - 2s)^2 = 0$. If $s \neq 1/2$, then we  have $\omega(U_s^*U_s) \geq (1 - 2s)^2 > 0$. If $s = 1/2$ and $\omega(U) \neq -1$, then the inequality is strict, so $\omega(U_s^*U_s) > (1-2s)^2 = 0$.
\end{proof}

Suppose we have a path $\omega\colon \interval \to  \state(\fA)$ of states and a path of unitaries $U\colon\interval \to  \Unitary(\fA)$, and we want to perform a homotopy of $\omega$ via a linear interpolation $(t,s) \mapsto sU_t + (1-s)\1$. We are in trouble if there exists $t \in \interval$ such that $\omega_t(U_t) = -1$. As the next lemma implies, we can avoid this difficulty by multiplying the path $U$ by a continuous phase $\lambda\colon \interval \to  S^1$.

\begin{lemma}\label{lem:phases_avoid_-1}
Let $\gamma\colon \interval \to  D^2$ be a path in the closed unit disk $D^2 \subset \bbC$. There exists a continuous map $\lambda\colon\interval \to  S^1 \subset \bbC$ such that for all $t \in \interval$ either $\lambda(t)\gamma(t) = 1$ or $\abs{\gamma(t)} < 1$. 
\end{lemma}
Intuitively, the lemma can be thought of as follows. Imagine $\gamma$ is the path of an ant walking on a round table and you are seated at the table. Any $\lambda = e^{i\theta} \in S^1$ corresponds to a rotation of the table by an angle $\theta$. The lemma says that as the ant is walking along the table, you can continuously rotate the table so that whenever the ant gets to the edge of the table, the ant will be directly in front of you.

\begin{proof}[Proof of Lemma \ref{lem:phases_avoid_-1}.]
The proof more or less follows the standard path-lifting argument for fiber bundles, although we are not looking at a fiber bundle here. 

Begin by covering $D^2$ with the open balls $B_1(0)$ and $B_1(z) \cap D^2$ for all $z \in S^1$. For any point $w_0 \in B_1(0)$ and any $\mu_0 \in S^1$, the constant function $\mu \colon  B_1(0) \to  S^1$ with value $\mu_0$ satisfies $\mu(w_0) = \mu_0$ and $\abs{w} < 1$ for all $w \in B_1(0)$. 

Now consider any $z \in S^1$, any point $w_0 \in B_1(z) \cap D^2$, and any $\mu_0 \in S^1$ such that either $\mu_0 w_0 = 1$ or $\abs{w_0} < 1$. If $\mu_0 w_0 = 1$, then the map 
\[
\mu\colon B_1(z) \cap D^2 \to  S^1, \quad \mu(w) = \frac{\abs{w}}{w}
\]
satisfies $\mu(w_0) = \mu_0$ and for all $w \in B_1(z) \cap D^2$ either $\mu(w)w  = 1$ or $\abs{w} < 1$. On the other hand, suppose $\abs{w_0} < 1$ and let $\theta \in [0,2\pi)$ such that $e^{i\theta} = \mu_0w_0/\abs{w_0}$. Then the function
\[
\mu\colon B_1(z) \cap D^2 \to  S^1, \quad \mu(w) = \mu_0 \cdot \frac{w_0}{\abs{w_0}} \cdot \frac{\abs{w}}{w} \cdot e^{-i\theta\qty(\abs{w} - \abs{w_0})/\qty(1 - \abs{w_0})}
\]
satisfies $\mu(w_0) = \mu_0$ and for all $w \in B_1(z) \cap D^2$, either $\mu(w) w = 1$ or $\abs{w} < 1$.

Now consider the path $\gamma$. Cover $\interval$ by the preimages $\gamma^{-1}(B_1(0))$ and $\gamma^{-1}(B_1(z) \cap D^2)$ for all $z \in S^1$. By the Lebesgue number lemma, there exists $N \in \bbN$ such that for all $k \in \qty{0,\ldots, N-1}$, the interval $[k/N, (k+1)/N]$ is contained in one of these preimages. 

Assume that for some $k \in \qty{0,\ldots, N-1}$ we have constructed a path $\lambda\colon [0,k/N] \to  S^1$ such that for all $t \in [0,k/N]$ either $\lambda(t) \gamma(t) = 1$ or $\abs{\gamma(t)} <1$. Choose one of the preimages above that contains $[k/N,(k+1)/N]$ and denote it $\gamma^{-1}(O)$. Setting $w_0 = \gamma(k/N)$ and $\mu_0 = \lambda(k/N)$, we may compose the restriction $\gamma \colon [k/N, (k+1)/N] \to  O$ with the appropriate map $\mu\colon O \to  S^1$ defined previously to obtain a map $\tilde \lambda\colon [k/N, (k+1)/N] \to  S^1$ such that $\tilde \lambda(k/N) = \mu_0 = \lambda(k/N)$ and for all $t \in [k/N, (k+1)/N]$ either $\tilde\lambda(t)\gamma(t) = 1$ or $\abs{\gamma(t)} < 1$. We may now glue $\lambda$ to $\tilde \lambda$ to get a continuous map $\lambda\colon [0,(k+1)/N] \to  S^1$ such that for all $t \in [0,(k+1)/N]$ either $\lambda(t)\gamma(t) = 1$ or $\abs{\gamma(t)} < 1$. Continuing in this manner constructs the desired path $\lambda \colon \interval \to  S^1$.
\end{proof}

\begin{cor}
Let $\fA$ be a unital $C^*$-algebra and let $\omega \colon \interval \to  \state(\fA)$ be weak*-continuous. Given a continuous map $U\colon \interval \to  \Unitary(\fA)$,  there exists a continuous $\lambda\colon \interval \to  S^1$ such that $s \lambda_tU_t + (1-s)\1 \notin \fN_{\omega_t}$ for all $t,s \in \interval$.
\end{cor}

\begin{proof}
This is immediate from Proposition \ref{prop:unitary_linear_interp} and Lemma \ref{lem:phases_avoid_-1} upon setting $\gamma(t) = \omega_t(U_t)$.
\end{proof}

\subsection{Finite Dimensional Density Matrices and Rectification of Paths}\label{subsec:fddensitymatrices}
Let us now fix $n \in \bbN$ and consider the specific example
$\fA = M_n(\bbC)$. In several steps we will show in this section that
every loop in the state space $\state( M_n(\bbC))$ is $M_n(\bbC)$-homotopic
relative the basepoint to the constant loop. In particular this implies that
the state space of a matrix algebra $M_n(\bbC)$ is simply connected. 
We start by identifying $\state (M_n(\bbC))$ with the set
$\denmat{n}$ of $n \times n$ density matrices, topologized as a subspace of
$M_n(\bbC)$. Recall that if $\varrho \in \denmat{n}$, then
$A \mapsto \tr(\varrho A)$ is a state on $M_n(\bbC)$.

\begin{proposition}
The map $\denmat{n} \to  \state(M_n(\bbC))$ that associates to a density matrix the corresponding state is a homeomorphism.
\end{proposition}

\begin{proof}
Suppose $\varrho_1, \varrho_2 \in \denmat{n}$ map to the same state. Then 
\[
\tr((\varrho_1 - \varrho_2)A) = 0
\] 
for all $A \in M_n(\bbC)$. Since $\varrho_1 - \varrho_2$ is self-adjoint, we can find an orthonormal basis of eigenvectors. Letting $A$ be the projection onto any eigenspace, the equation $\tr((\varrho_1 - \varrho_2)A) = 0$ implies that the corresponding eigenvalue is zero. Therefore $\varrho_1 - \varrho_2 = 0$. 

We show surjectivity. Since $M_n(\bbC)$ has only one superselection sector, we know $\pstate(M_n(\bbC)) \cong \bbP(\bbC^n)$, so $\pstate(M_n(\bbC))$ is compact. We know $\state(M_n(\bbC))$ is the closed convex hull of $\pstate(M_n(\bbC))$, but since $\pstate(M_n(\bbC))$ is a compact subset of a finite-dimensional vector space, the convex hull of $\pstate(M_n(\bbC))$ is already closed, so $\state(M_n(\bbC))$ is just the convex hull of $\pstate(M_n(\bbC))$. Any pure state can be written as $A \mapsto \tr(PA)$ for some rank-one projection $P$, and any convex combination of rank-one projections is a density matrix. Thus, any state can be represented by a density matrix.

Since $\denmat{n}$ and $\state(M_n(\bbC))$ are compact Hausdorff, all that remains to do is show continuity. Letting $\varrho_1, \varrho_2 \in \denmat{n}$ and letting $e_1,\ldots, e_n$ be an orthonormal basis of $\bbC^n$, we have
\[
\abs{\tr((\varrho_1 - \varrho_2)A)} \leq \sum_{i=1}^n \abs{\ev{e_i,(\varrho_1 - \varrho_2)A e_i}} \leq n \norm{\varrho_1 - \varrho_2}\norm{A}.
\]
Continuity follows.
\end{proof}

Note that if $\omega \in \state(M_n(\bbC))$ is represented by the density matrix $\varrho$ and $A \in M_n(\bbC) \setminus \fN_\omega$, then $A \cdot \omega$ is represented by the density matrix $A\varrho A^*/\tr(A\varrho A^*)$. We will use this correspondence between states and density matrices freely.

Our first goal is to show that any loop in $\state(M_n(\bbC))$ based at a pure state is $M_n(\bbC)$-homotopic to the constant map
relative the basepoint (in the sense of Definition \ref{defn:Ahomotopy}). Since all pure states on $M_n(\bbC)$ are unitarily equivalent, it suffices to take our loop to be based at the state represented by the first standard basis vector. Denote this state by $\omega_0^n$. Explicitly, $\omega_0^n$ is the state
\[
  \omega_0^n \colon M_n(\bbC) \to \C,\quad A \mapsto \langle e_0 ,Ae_0\rangle ,
\]
where $e_0 =(1,0,\ldots,0) \in \C^n$. 
The following notation will also be convenient. Given $n \in \bbN$ and $k \in \qty{0,\ldots, n-1}$, let 
\[
P^n_k = \diag(1,\ldots, 1, 0, \ldots, 0) \in M_n(\bbC)
\]
where there are $k$ zeros on the diagonal.

\begin{lemma}\label{lem:omega(P^n_1)=0}
  If $n > 1$ and $\omega \in \state(M_n(\bbC))$, then $\omega(P^n_1) = 0$ if and only if
  $\omega$ is represented by the last standard basis vector, that is, if and only if
  $\omega (A) =\langle e_n,A e_n \rangle$ for all $A\in M_n(\bbC)$, where
   $e_n =(0,\ldots,0,1) \in \C^n$.
\end{lemma}

\begin{proof}
If $\omega$ is represented by the last standard basis vector, then it is clear that $\omega(P^n_1) = 0$. Suppose $\omega(P^n_1) = 0$. Let $\varrho$ be the density matrix representing $\omega$. Then $\tr(P^n_1 \varrho P^n_1) = 0$, but $P^n_1 \varrho P^n_1$ is positive, so this implies that $P^n_1 \varrho P^n_1 = 0$. Note that $P^n_1 \varrho P^n_1$ is the matrix with the same entries as $\varrho$ in the upper left $(n-1) \times (n-1)$ submatrix, and with zeros everywhere else. Since this $(n-1) \times (n-1)$ submatrix is zero, it follows from the fact that $\tr(\varrho) = 1$ that the bottom right entry of $\varrho$ is $1$. Acting $\varrho$ on the last standard basis vector and using the fact that $\norm{\varrho} \leq 1$, we see that all other entries in the last column of $\varrho$ are zero. Since $\varrho$ is self-adjoint, all other entries in the last row of $\varrho$ are zero as well. Thus, the only nonzero entry of $\varrho$ is a one in the bottom right corner, and this is the density matrix corresponding to the last standard basis vector.
\end{proof}

Our next two lemmas follow the same train of thought as Lemma \ref{lem:phases_avoid_-1}. In  Lemma \ref{lem:phases_avoid_-1}, the situation was analogous to an ant walking on a round table $D^2$ while an observer rotated the table by multiplying with elements of $S^1$. In Lemma \ref{lem:state_neighborhood} and Lemma \ref{lem:unitary_avoid_0}, our aim is to make a similar statement for paths in $\state(M_n(\bbC))$, where we act on states with unitaries. 

For example, when $n = 2$ the state space $\state(M_n(\bbC))$ is the Bloch sphere and we can think of our unitaries acting on the Bloch sphere by rotations. Analogously, if an ant is eating through an apple, I can rotate the apple continuously so that whenever the ant emerges, it is facing straight up.

\begin{lemma}\label{lem:state_neighborhood}
For every $\psi \in \state(M_n(\bbC))$, there exists an open neighborhood 
\[\psi \in O \subset \state(M_n(\bbC))\] 
with the following property: 

For every $\psi_0 \in O$ and $U_0 \in \Unitary(n)$ such that either $U_0 \cdot \psi_0 = \omega_0^n$ or $U_0 \cdot \psi_0$ is not pure, there exists a continuous map $U\colon O \to  \Unitary(n)$ such that
\begin{enumerate}[label=\tn{(\alph*)}]
\item $U_{\psi_0} = U_0$, and 
\item for every $\phi \in O$, either $U_\phi \cdot \phi = \omega_0^n$ or $U_\phi \cdot \phi$ is not pure.
\end{enumerate}
\end{lemma}

\begin{proof}
Suppose $\psi$ is not pure. Then we may take 
\[O = \state(M_n(\bbC)) \setminus \pstate(M_n(\bbC)).\] 
Indeed, observe that for any $\psi_0 \in O$ and $U_0 \in \Unitary(n)$, the constant map at $U_0$ satisfies the desired properties, since a unitary cannot take a non-pure state to a pure state.

Suppose $\psi$ is pure, let $\Psi$ be a representing vector for $\psi$, and let $P = \ketbra{\Psi}$. Let 
\[
O = \qty{\phi \in \state(M_n(\bbC)): \phi(P) > 7/8}.
\]
Suppose $\phi \in O$ and let $\varrho_\phi \in \denmat{n}$ be the density matrix representing $\phi$. Write $\varrho_\phi$ using the spectral decomposition 
\[\varrho_\phi = \sum_{i=1}^n t_i \ketbra{\Phi_i},\] 
where $\Phi_1,\ldots, \Phi_n$ is an orthonormal basis of eigenvectors of $\varrho_\phi$. Observe that
\begin{align*}
\frac{7}{8} < \phi(P) = \tr(\varrho_\phi P) = \sum_{i=1}^n t_i \abs{\ev{\Phi_i, \Psi}}^2 \leq \qty(\max_i t_i) \sum_{i=1}^n \abs{\ev{\Phi_i, \Psi}}^2 = \max_i t_i.
\end{align*}
Let $k \in \qty{1,\ldots, n}$ be such that $t_k = \max_i t_i$. We have just shown that $t_k > 7/8$. Since $\sum_{i=1}^n t_i = 1$, we know that $t_k \leq 1$ and $t_i < 1/8$ for all $i \neq k$. 

Next, observe that
\begin{align*}
\frac{7}{8} < \phi(P) = t_k \abs{\ev{\Phi_k, \Psi}}^2 + \sum_{i \neq k} t_i \abs{\ev{\Phi_i, \Psi}}^2 \leq \abs{\ev{\Phi_k, \Psi}}^2 + \frac{1}{8}.
\end{align*}
Thus, $\abs{\ev{\Phi_k, \Psi}}^2 > 3/4$. 

Consider the map $O \ni \phi \mapsto \omega_\phi \in \pstate(M_n(\bbC))$, where $\omega_\phi$ is represented by an eigenvector of $\varrho_\phi$ with maximum eigenvalue. This is a continuous map that leaves $O \cap \pstate(M_n(\bbC))$ invariant. With $\Phi_k$ representing a maximum eigenvector of $\phi$ as above, we observe that
\[
\norm{\omega_\phi - \psi}^2 = 4 - 4\abs{\ev{\Phi_k, \Psi}}^2 < 4 - 4 \cdot \frac{3}{4} = 1.
\]
Thus,
\[
\norm{\omega_\phi - \omega_{\psi_0}} \leq \norm{\omega_\phi - \psi} + \norm{\psi - \omega_{\psi_0}} < 2. 
\]

There exists a continuous map $U\colon  B_2(\omega_{\psi_0}) \cap \pstate(M_n(\bbC)) \to  \Unitary(n)$ such that $U_{\omega_{\psi_0}} = I$ and $U_\omega \cdot \omega = \omega_{\psi_0}$ for every $\omega \in B_2(\omega_{\psi_0}) \cap \pstate(M_n(\bbC))$. If $\psi_0$ is pure, then $\psi_0 = \omega_{\psi_0}$, and the function
\[
\phi \mapsto U_0 U_{\omega_{\phi}}
\]
is the desired family of unitaries.

If $\psi_0$ is not pure, then choose a unitary $V$ such that $V \cdot \omega_{\psi_0} = \omega_0^n$. Write $U_0 = e^{iA}$ and $V = e^{iB}$ where $A,B \in M_n(\bbC)$ are self-adjoint. Then the function
\[
\phi \mapsto e^{iB\qty[S(\varrho_{\psi_0})-S(\varrho_\phi)]/S(\varrho_{\psi_0})} e^{iAS(\varrho_{\phi})/S(\varrho_{\psi_0})} U_{\omega_\phi}
\]
has the desired properties, where $S(\varrho)$ is the von Neumann entropy of $\varrho$.
\end{proof}

\begin{lemma}\label{lem:unitary_avoid_0}
Let $\omega\colon \interval \to  \state(M_n(\bbC))$ be a loop based at $\omega_0^n$. There exists a continuous family of unitaries $U\colon  \interval \to  \Unitary(n)$ such that 
\[U_0 \cdot \omega_0^n = U_1 \cdot \omega_0^n = \omega_0^n\] 
and $(U_t \cdot \omega_t)(P^n_{1}) > 0$ for all $t \in \interval$.
\end{lemma}

\begin{proof}
For each $\psi \in \state(M_n(\bbC))$, let $O_\psi$ be an open neighborhood of $\psi$ as provided by Lemma \ref{lem:state_neighborhood}. The neighborhoods $O_\psi$ cover $\state(M_n(\bbC))$, hence the neighborhoods $\omega^{-1}(O_\psi)$ cover $\interval$. Thus, there exists $N \in \bbN$ such that for each $k \in \qty{0,\ldots, N-1}$, the set $[k/N, (k+1)/N]$ lies in some $\omega^{-1}(O_\psi)$. 

We now follow the standard path-lifting argument. Assume that for some $k \in \qty{0,\ldots, N-1}$ we have defined a continuous map $U\colon [0,k/N] \to  \Unitary(n)$ such that $U_0 = I$ and for all $t \in [0,k/N]$ either $U_t \cdot \omega_t = \omega_0^n$ or $U_t \cdot \omega_t$ is not pure. Choose $\psi \in \state(M_n(\bbC))$ such that $[k/N, (k+1)/N] \subset \omega^{-1}(O_\psi)$. Composing $\omega|_{[k/N, (k+1)/N]}$ with such a continuous family of unitaries as provided by  Lemma \ref{lem:state_neighborhood}, we obtain a continuous map $V\colon [k/N, (k+1)/N] \to  \Unitary(n)$ such that $V_{k/N} = U_{k/N}$ and for every $t \in [k/N, (k+1)/N]$, either $V_t \cdot \omega_t = \omega_0^n$ or $V_t \cdot \omega_t$ is not pure. We may glue the maps $U$ and $V$ together to obtain a continuous map $W\colon [0,(k+1)/N] \to  \Unitary(n)$ such that $W_0 = I$ and for all $t$, either $W_t \cdot \omega_t = \omega_0^n$ or $W_t \cdot \omega_t$ is not pure. 

Proceeding in this fashion, we define a continuous map $U\colon \interval \to  \Unitary(n)$ such that $U_0 \cdot \omega_0^n = U_1 \cdot \omega_0^n = \omega_0^n$ and for all $t$ either $U_t \cdot \omega_t = \omega_0^n$ or $U_t \cdot \omega_t$ is not pure. The fact that $(U_t \cdot \omega_t)(P^n_1) > 0$ for all $t$ follows from Lemma \ref{lem:omega(P^n_1)=0}.
\end{proof}

\begin{lemma}\label{lem:P1}
  If $\omega\colon \interval \to  \state(M_n(\bbC))$ is a loop based at $\omega_0^n$, then $\omega$ is $M_n(\bbC)$-homotopic
  relative the basepoint to a loop $\psi\colon \interval \to  \state(M_n(\bbC))$ such that $\psi_t(P^n_1) = 1$
  for all $t \in \interval$.
\end{lemma}

\begin{proof}
Using Lemma \ref{lem:unitary_avoid_0} we find a continuous family of unitaries $U\colon \interval \to  \Unitary(n)$ such that $U_0 \cdot \omega_0^n = U_1 \cdot \omega_0^n = \omega_0^n$ and $(U_t \cdot \omega_t)(P_1^n) > 0$ for all $t \in \interval$. Using Lemma \ref{lem:phases_avoid_-1} we find a continuous path $\lambda\colon \interval \to  S^1$ such that $\lambda_t \omega_t(U_t) \neq -1$ for all $t \in \interval$. Then $(s,t) \mapsto s\lambda_t U_t + (1 - s)\1$ is continuous, equals $\1$ when $s = 0$, and satisfies $s\lambda_t U_t + (1 - s)\1 \notin \fN_{\omega_t}$ for all $t, s \in \interval$ by Proposition \ref{prop:unitary_linear_interp}. Furthermore, for all $s \in \interval$,  $s \lambda_0 U_0 + (1 - s)\1$ and $s\lambda_1 U_1 + (1- s)\1$ leave $\omega_0^n$ invariant by Proposition \ref{prop:linear_combo_action}. Therefore, $\omega$ is homotopic relative the base point to $\chi_t = \lambda_t U_t \cdot \omega_t = U_t \cdot \omega_t$.

Note that $\chi_t(P_1^n) > 0$ for all $t \in \interval$ and $\chi_0(P^n_1) = \chi_1(P^n_1) = 1$. Then by Proposition \ref{prop:projection_linear_interp} we have the homotopy $(t,s) \mapsto \qty[sP_1^n + (1-s)\1] \cdot \chi_t$. By  Proposition \ref{prop:invariant_state} and  Proposition \ref{prop:linear_combo_action}, this is a homotopy relative the basepoint. Finally, we note that $(P_1^n \cdot \chi_t)(P_1^n) = 1$ for all $t \in \interval$.
\end{proof}

We would now like to prove the result of Lemma \ref{lem:P1} but with $P^n_1$ replaced by $P^n_k$ for arbitrary $k \in \qty{1,\ldots, n-1}$. We note that $M_{n-k}(\bbC) \cong P^n_k M_n(\bbC) P^n_k$, where the $*$-isomorphism maps $A \in M_{n-k}(\bbC)$ to the $n \times n$ matrix with $A$ as the upper left $(n-k) \times (n-k)$ submatrix, and zeros everywhere else. If we have a state $\psi \in \state(M_n(\bbC))$ such that $\psi(P^n_k) = 1$, then $\psi$ is essentially a state on $M_{n-k}(\bbC) \cong P^n_k M_n(\bbC) P^n_k$. 

At this point it is worth taking a step back and considering this last statement through a more abstract lens. Note that if $\fA$ is a $C^*$-algebra and $P \in \fA$ is a projection, then $P \fA P$ is a unital $C^*$-algebra with unit $P$. In fact, it is a hereditary $C^*$-subalgebra.  Proposition \ref{prop:invariant_state} has the following consequence for states on $P \fA P$.

\begin{proposition}
Let $\fA$ be a $C^*$-algebra and let $P \in \fA$ be a projection.
\begin{enumerate}
  \item If $\psi \in \state(\fA)$, then $\psi|_{P \fA P} \in \state(P \fA P)$ if and only if $\psi(P) = 1$.
  \item Given a state $\omega \in \state(P\fA P)$ the function $\omega \circ \Ad(P)$ is the unique extension of $\omega$ to a state on $\fA$. 
  \item With respect to the weak* topologies on the domain and codomain, the map 
  \[
  \state(P\fA P) \ni \omega \mapsto \omega \circ \Ad(P) \in \state(\fA)
  \]
  is a closed embedding and has image $\qty{\psi \in \state(\fA): \psi(P) = 1}$.
\end{enumerate}
\end{proposition}

\begin{proof}
(1) This is immediate from the fact that $\psi|_{P\fA P}$ is a positive linear functional and $P$ is the unit of $P \fA P$.

(2) Since $P\fA P$ is a hereditary $C^*$-subalgebra of $\fA$ with unit $P$, this is provided by \cite[Thm.~3.3.9]{Murphy}.

(3) Using the characteristic mapping property of the weak* topology, we see that $\omega \mapsto \omega \circ \Ad(P)$ is weak*-continuous. It is clear that if $\omega_1, \omega_2 \in \state(P\fA P)$ and $\omega_1 \circ \Ad(P) = \omega_2 \circ \Ad(P)$, then $\omega_1 = \omega_2$. Since $\state(P\fA P)$ and $\state(\fA)$ are compact Hausdorff, this implies that $\omega \mapsto \omega \circ \Ad(P)$ is an embedding. We observe that $[\omega \circ \Ad(P)](P) = \omega(P) = 1$ for all $\omega \in \state(P\fA P)$. On the other hand, if $\psi(P) = 1$, then Proposition \ref{prop:invariant_state} implies that $P \cdot \psi = \psi$, so $\psi = \psi|_{P\fA P} \circ \Ad(P)$ and $\psi|_{P\fA P} \in \state(P\fA P)$. Thus, the image of the map $\omega \mapsto \omega \circ \Ad(P)$ is the set $\qty{\psi \in \state(\fA): \psi(P) = 1}$, and this set is manifestly weak* closed.
\end{proof}

\begin{lemma}\label{lem:path_homotopy_push_forward}
  Let $\fA$ and $\fB$ be unital $C^*$-algebras, let $P \in \fB$ be a projection, let $\pi\colon \fA \to  P\fB P$ be a $*$-isomorphism, and let $\psi\colon \interval \to  \state(\fB)$ be a weak*-continuous path such that $\psi_t(P) = 1$ for all $t \in \interval$. If $A\colon \interval \times \interval \to  \fA$ is an $\fA$-homotopy relative endpoints from
  the path $t \mapsto \psi_t|_{P\fB P} \circ \pi$ to $\omega \colon  \interval \to  \state(\fA)$,
  then $(t,s) \mapsto (\1 - P) + \pi(A_{t,s})$ is an $\fA$-homotopy relative endpoints 
  from $\psi$ to the path $t \mapsto \omega_t \circ \pi^{-1} \circ \Ad(P)$.
\end{lemma}

\begin{proof}
Note that $\pi(\1) = P$ since $\pi$ is a $*$-isomorphism. We check that $(\1 - P) + \pi(A_{t,s})$ has the desired properties. If $\pi(A_{t,s}) + \1 - P \in \fN_{\psi_t}$, then
\[
0 = \psi_t\qty[\qty(\qty(\1 - P) + \pi(A_{t,s}^*))\qty(\qty(\1 - P) + \pi(A_{t,s}))] = \psi_t(\pi(A_{t,s}^*A_{t,s})),
\] 
which implies $A_{t,s} \in \fN_{\psi_t|_{P\fB P} \circ \pi}$, which contradicts the definition of $A$. Thus, $(\1 - P) + \pi(A_{t,s}) \notin \fN_{\psi_t}$ for all $t$ and $s$.

Next, observe that for all $t$, we have 
\[
(\1 - P) + \pi(A_{t,0}) = (\1 - P) + \pi(\1) = \1 - P + P = \1.
\]

For arbitrary $t$ and $s$, we have
\begin{align*}
\qty[\qty[(\1 - P) + \pi(A_{t,s})] \cdot \psi_t](B) &= \frac{\psi_t(\pi(A_{t,s}^*)B\pi\qty(A_{t,s}))}{\psi_t(\pi(A_{t,s}^*A_{t,s}))} \\
&= \frac{\psi_t(\pi(A_{t,s}^* \pi^{-1}(PBP) A_{t,s}))}{\psi_t(\pi(A_{t,s}^*A_{t,s}))}\\
&= \qty[A_{t,s} \cdot \qty(\psi_t|_{P\fB P} \circ \pi)](\pi^{-1}(PBP))
\end{align*}
When $s = 1$, this yields
\[
\qty[\qty[(\1 - P) + \pi(A_{t,s})] \cdot \psi_t](B) = \omega_t(\pi^{-1}(PBP)),
\]
so $\qty[(\1 - P) + \pi(A_{t,s})] \cdot \psi_t = \omega_t \circ \pi^{-1} \circ \Ad(P)$. When $t = 0$ or $t = 1$, this yields
\[
\qty[\qty[(\1 - P) + \pi(A_{t,s})] \cdot \psi_t](B) = (\psi_t|_{P\fB P} \circ \pi)(\pi^{-1}(PBP)) =  \psi_t(B),
\]
so $\qty[(\1 - P) + \pi(A_{t,s})] \cdot \psi_t = \psi_t$ for all $s$ when $t = 0$ and $t = 1$. 
\end{proof}

\begin{theorem}\label{thm:M_n(C)_nulhomotopic}
  Fix $k \in \qty{1,\ldots, n-1}$. If $\omega\colon \interval \to  \state(M_n(\bbC))$ is a loop based at $\omega_0^n$, then $\omega$ is $M_n(\bbC)$-homotopic relative the basepoint to a loop $\psi\colon \interval \to  \state(M_n(\bbC))$ which satisfies
  $\psi_t(P^n_k) = 1$ for all $t \in \interval$.
\end{theorem}

\begin{proof}
  We prove the theorem by induction on $k$. The base case is given by Lemma \ref{lem:P1}. Suppose the theorem is true for some $k < n-1$. Then $\omega$ is $M_n(\bbC)$-homotopic relative the basepoint to a loop $\psi:\interval \to  \state(M_n(\bbC))$
  based at $\omega^n_0$ and which fulfills $\psi_t(P^n_{k}) = 1$ for all $t \in I$. 

Now consider the $*$-isomorphism $\pi\colon M_{n-k}(\bbC) \to  P^n_k M_n(\bbC) P^n_k$ that maps $A \in M_{n-k}(\bbC)$ to the matrix
\[
\pi(A) = \mqty(A & 0\\0 & 0).
\]
We have a loop $t \mapsto \psi_t|_{P^n_k M_n(\bbC) P^n_k} \circ \pi$ based at $\omega^{n-k}_0$. By Lemma \ref{lem:P1}, this loop is
$M_{n-k}(\bbC)$-homotopic relative the basepoint to a loop $\phi\colon \interval \to  \state(M_{n-k}(\bbC))$ such that $\phi_t(P^{n-k}_1) = 1$ for all $t \in \interval$. By Lemma \ref{lem:path_homotopy_push_forward}, $\psi$ is $M_n(\bbC)$-homotopic
relative the basepoint to $t \mapsto \phi_t \circ \pi^{-1} \circ \Ad(P^n_k)$. We observe that
\begin{align*}
(\phi_t \circ \pi^{-1} \circ \Ad(P^n_k))(P^n_{k+1}) &= (\phi_t \circ \pi^{-1} \circ \Ad(P^n_k))(\pi(P^{n-k}_1))\\
&= \phi_t(P^{n-k}_1) = 1
\end{align*}
for all $t \in \interval$. The result now follows by induction.
\end{proof}

We note that given $\omega \in \state(M_n(\bbC))$, the statement $\omega(P^n_{n-1}) = 1$ implies that $\omega = \omega^n_0$. Thus, setting $k = n-1$ in Theorem \ref{thm:M_n(C)_nulhomotopic}, the loop $\psi$ is the constant map at $\omega^n_0$.

\subsection{Triviality of the Fundamental Group for UHF Algebras}

We finally come to our goal to show that $\pstate(\fA)$ is simply connected for a  UHF algebra $\fA$.
So let us now consider the state space of a UHF algebra $\fA$. 

It is well-known (see \cite[\S12.1]{KadisonRingroseII})  that every UHF algebra is $*$-isomorphic to the directed colimit of a directed system of tensor products 
\[
\fA_n = \bigotimes_{i=1}^n M_{k_i}(\bbC)
\]
where we have assigned a dimension $k_i \in \bbN \setminus \qty{1}$ for each $i \in \bbN$. 
For $m < n$, the embedding $\fA_m \hookrightarrow \fA_n$ of the directed system is defined by tensoring on identity matrices in the factors $M_{k_i}(\bbC)$ for $i > m$. We may therefore assume $\fA$ is of this form.

 It is known that $\pstate(\fA)$ is path-connected in the weak$^*$ topology \cite{EilCCCA,Spiegel}, so it suffices to show that $\pi_1(\pstate(\fA), \omega_0) = 0$ for some choice of $\omega_0$. For convenience, we will choose the basepoint $\omega_0$ to be the product state represented by the first standard basis vector in each tensor factor. 

The $C^*$-algebras $M_{k_i}(\bbC)$ embed into any $\fA_n$ for $n > i$, and then embed into $\fA$. Let $\fB_i$ denote this copy of $M_{k_i}(\bbC)$ inside $\fA$. Let $P_i \in \fB_i$ be the copy of the projection $P^{k_i}_{k_i - 1}$ inside $\fB_i$. The following lemma will be useful.

\begin{lemma}[{\cite[Lem.~3.4.2]{BrownOzawa}}]\label{lem:restricted_state_factorized}
Let $\fA$ be a $C^*$-algebra and let $\fB$ be a $C^*$-subalgebra of $\fA$. If $\omega \in \state(\fA)$ and $\omega|_{\fB} \in \pstate(\fB)$, then
\[
\omega(AB) = \omega(A)\omega(B)
\]
for all $A \in \fB'$ and $B \in \fB$, where $\fB'$ is the commutant of $\fB$.
\end{lemma}

\begin{theorem}\label{thm:fundaUHFfinal}
  Let $\fA$ be a UHF algebra and let $\omega\colon \interval \to  \state(\fA)$ be a weak*-continuous loop based at the product state $\omega_0$. There exists a continuous map $A\colon \interval \times [0,1) \to  \fA$ and a homotopy $H\colon \interval \times \interval \to  \state(\fA)$ from $\omega$ to the constant loop at $\omega_0$ such that $H(t,s) = A_{t,s} \cdot \omega_t$ for all $s < 1$. It follows that $\pi_1(\pstate(\fA),\omega_0) = 0$.
\end{theorem}

The strategy is as follows. Beginning with the arbitrary loop $\omega$, we first focus on its restriction to the first ``lattice site'' $M_{k_1}(\bbC)$. Acting with unitaries and projections localized on ``site one'' (and their linear interpolations with the identity), we deform $\omega$ so that its restriction to site one becomes constant and pure, as in Theorem \ref{thm:M_n(C)_nulhomotopic}. Lemma \ref{lem:restricted_state_factorized} implies that site one is now disentangled from the rest of the sites. We then proceed to do the same thing to site two, site three, and so on. We can fit this countable infinity of homotopies in the half-open interval $[0,1)$, and then define the homotopy to be constant $\omega_0$ when the homotopy parameter reaches $s = 1$. This turns out to be weak*-continuous because for a fixed local operator $A$, the expectation values of $A$ become independent of the homotopy parameter after the support of $A$ has become disentangled from the rest of the sites.

\begin{proof}[Proof of Theorem \ref{thm:fundaUHFfinal}.]
  We will follow the notation in the discussion preceding the theorem statement.
  Moreover, for any $C^*$-subalgebra $\fB \subset \fA$ we will say that an $\fA$-homotopy
  $B\colon  \interval\times\interval \to\fA$ is \emph{supported} in $\fB$ if and only if it has
  image in $\fB$. As the last preparation we set $\psi^0 = \omega$ for convenience.
  Now we  construct recursively  for each positive integer $i$ a loop $\psi^i\colon \interval \to\state(\fA)$
  whose restriction to $\fB_i$ is a constant loop and an $\fA$-homotopy $B^i$ supported in  $\fB_i$.
    
  The restriction $t \mapsto \omega_t|_{\fB_1}$ of $\omega$ to the first tensor factor is a loop of states based at $\omega^{k_1}_0$, so it follows from Theorem \ref{thm:M_n(C)_nulhomotopic} that $\omega|_{\fB_1}$ is $\fB_1$-homotopic relative the basepoint to the constant loop at $\omega^{k_1}_0$. 
  Thus, there exists an $\fA$-homotopy $B^1\colon \interval \times \interval \to  \fA$ 
  relative the basepoint and supported on $\fB_1$ from $\psi^0 =\omega $ to a loop $\psi^1$
  such that $\psi^1_t(P_1) = 1$ for all $t \in \interval$. 

  Suppose that for some $n \in \bbN$ and all $i \in \qty{1,\ldots, n}$ we have constructed loops
  $\psi^i\colon \interval \to  \state(\fA)$ based at $\omega_0$ and continuous maps
  $B^i\colon \interval \times \interval \to  \fB_i \subset \fA$  such that 
  $B^i$ is   an $\fA$-homotopy relative basepoints from $\psi^{i-1}$ to $\psi^i$ and such that
  $\psi^i_t(P_j) = 1$ for all $j \leq i$ and $t \in \interval$.
  Restricting $\psi^n$ to $\fB_{n+1}$ and applying Theorem \ref{thm:M_n(C)_nulhomotopic}, we obtain
  an $\fA$-homotopy  $B^{n+1}\colon \interval \times \interval \to  \fB_{n+1} \subset \fA$ relative the basepoint
  from $\psi^n$ to a loop $\psi^{n+1}$ satisfying $\psi^{n+1}_t(P_{n+1}) = 1$ for all $t \in \interval$.
  Note that the $\fA$-homotopy $B^{n+1}$ is supported on $\fB_{n+1}$.
  We observe further that for $j < n+1$, the restricted state $\psi^n_t|_{\fB_j}$ is pure (and independent of $t$)
  since $\psi^n_t(P_j) = 1$, so Lemma \ref{lem:restricted_state_factorized} implies
\begin{align*}
\psi_t^{n+1}(P_j) &= (B_{t,1}^{n+1} \cdot \psi_t^n)(P_j) \\
&= \frac{\psi_t^n((B_{t,1}^{n+1})^*P_j B_{t,1}^{n+1})}{\psi_t^n((B_{t,1}^{n+1})^*B_{t,1}^{n+1})}\\
&= \frac{\psi_t^n((B_{t,1}^{n+1})^*B_{t,1}^{n+1}) \cdot \psi_t^n(P_j)}{\psi_t^n((B_{t,1}^{n+1})^*B_{t,1}^{n+1})} = 1.
\end{align*}

By recursion we therefore obtain for all positive integers $i$ a loop $\psi^i\colon \interval \to  \state(\fA)$ based at $\omega_0$
and an $\fA$-homotopy $B^i\colon \interval \times \interval \to  \fB_i \subset \fA$
from $\psi^{i-1}$ to $\psi^i$ such that $\psi^i_t(P_j) = 1$ for all $j \leq i$ and $t \in \interval$. 

Now define $A\colon \interval \times [0,1) \to  \fA$ as follows. Given $i \in \bbN$, define $A$ on $\interval \times [1 - 2^{-i+1}, 1-2^{-i}]$ as
\[
A_{t,s} = B_{t,r}^i B_{t,1}^{i-1} B_{t,1}^{i-2} \cdots B_{t,1}^1, \quad \tn{ where $r = \frac{s - 1 + 2^{-i+1}}{2^{-i+1} - 2^{-i}}$}.
\]
Since $B^i_{t,0} = \1$ for all $t$ and $i$, we see that 
\begin{align*}
A_{t,1-2^{-i+1}} = B_{t,1}^{i-1}B_{t,1}^{i-2}\cdots B_{t,1}^1.
\end{align*}
Note also that
\[
A_{t,1-2^{-i}} = B_{t,1}^iB_{t,1}^{i-1} \cdots B_{t,1}^1.
\]
Thus, the functions $A\colon \interval \times [1-2^{-i+1}, 1-2^{-i}] \to  \fA$ glue together to a continuous function $A\colon I \times [0,1) \to  \fA$. Note that $A_{t,0} = \1$ for all $t \in \interval$.

Finally, we define $H\colon \interval \times \interval \to  \state(\fA)$ as
\[
H(t,s) = \begin{cases} A_{t,s} \cdot \omega_t &: s < 1 \\ \omega_0 &: s = 1 \end{cases}.
\]
Since $A_{t,0} = \1$ for all $t$, we see that $H(t, 0) = \omega_t$. Since $B^i$ fixes the endpoints of $\psi^{i-1}$, we see that $H(0,s) = H(1,s) = \omega_0$ for all $s \in \interval$. Clearly $H(t,1)$ is constant at $\omega_0$. All that remains is to show weak*-continuity of $H$.

It suffices to show continuity of $H(t,s)(C)$ for fixed but arbitrary $C \in \bigcup_{n \in \bbN} \fA_n \subset \fA$. On $\interval \times [0,1)$, we have $H(t,s)(C) = (A_{t,s} \cdot \omega_t)(C)$, which is continuous since $A$ is continuous. Now, there exists $n \in \bbN$ such that $C \in \fA_n$. For $i > n+1$ and $s \in [1 - 2^{-i+1}, 1-2^{-i}]$, we have
\begin{align*}
H(t,s)(C) &= (A_{t,s} \cdot \omega_t)(C)\\
&= \qty[\qty(B_{t,r}^i B_{t,1}^{i-1} \cdots B_{t,1}^{n+1}) \cdot \qty(B_{t,1}^{n} \cdots B_{t,1}^1) \cdot \omega_t](C)\\
&= \qty[\qty(B_{t,r}^i  B_{t,1}^{i-1} \cdots B_{t,1}^{n+1}) \cdot \psi^n_t](C) = \psi^n_t(C) = \omega_0(C).
\end{align*}
The fact that $\psi^n_t(C) = \omega_0(C)$ follows from the fact that $\psi^n_t(P_i) = 1$ for all $i \leq n$, which implies $\psi^n_t|_{\fA_n} = \omega_0|_{\fA_n}$. The second to last step follows from the fact that $\psi^n_t|_{\fA_n}$ is pure, the fact that $B^i_{t,r} B_{t,1}^{i-1} \cdots B_{t,1}^{n+1} \in \fA_n'$, and an application of Lemma \ref{lem:restricted_state_factorized}. Thus, on $\interval \times [1-2^{-n-1}, 1]$, we see that $H(t,s)(C)$ is constant at $\omega_0(C)$. This proves that $H(t,s)$ is weak*-continuous on all of $\interval \times \interval$, completing the proof.
\end{proof}

%!TEX root = dissertation.tex

\newpage
\chapter{A Phase Invariant for One-Dimensional Parametrized Spin Systems}
\label{chp:cocycle_construction}

We come to our result of most immediate physical relevance. In this chapter we will construct an $\check{H}^3(X;\bbZ)$-valued invariant for one-dimensional parametrized spin systems. Section \ref{sec:construction} contains the mathematical details and proofs of our construction and in Section \ref{sec:1d_example} we give a nontrivial example of a one-dimensional parametrized system and construct its invariant explicitly.

\section{The Construction}
\label{sec:construction}

Before defining the invariant, we prove two original preliminary results. Theorem \ref{thm:cont_family_PU_from_aut} in Section \ref{subsec:projective_unitaries_from_aut} shows how to obtain a strongly continuous family of projective unitary operators from a strongly continuous family of inner automorphisms and a pure state, assuming the superselection sector of the pure state is invariant under the automorphisms. Theorem \ref{thm:equivalence_product_state} shows in a very general framework that two pure product states are equivalent if and only if their tensor factors are equivalent. These results are then used in the construction of the invariant, undertaken in Section \ref{subsec:the_construction}.

\subsection{Obtaining Projective Unitaries from Automorphisms}
\label{subsec:projective_unitaries_from_aut}

We begin with a point of notation. Given a nonzero complex Hilbert space $\hilbH$, observe that the homomorphism $\Ad:\Unitary(\hilbH) \rightarrow \Inn(\cB(\hilbH))$ contains $\Unitary(1)$ in its kernel, and therefore factors through $\PU(\hilbH)$. By slight abuse of notation we denote the factored map by $\Ad:\PU(\hilbH) \rightarrow \Inn(\cB(\hilbH))$ as well. The following is then a slight variation on a well-known result, see \cite[Cor.~2.5.8]{NaaijkensQSSIL}. Recall that we say two states are equivalent if they are in the same superselection sector, i.e., their GNS representations are unitarily equivalent.

\begin{proposition}\label{prop:PU_from_aut}
Let $\fA$ be a $C^*$-algebra, let $\omega \in \pstate(\fA)$, and let $(\hilbH, \pi, \Omega)$ be an irreducible representation with unit vector $\Omega \in \sphere \hilbH$ representing $\omega$. If $\alpha \in \Aut(\fA)$ and $\omega \circ \alpha$ is equivalent to $\omega$, then there exists a unique $\bbU \in \PU(\hilbH)$ such that
\begin{equation}\label{eq:projective_unitary_from_alpha}
\Ad(\bbU) \circ \pi = \pi \circ \alpha.
\end{equation}
\end{proposition}

\begin{proof}
We observe that $(\hilbH, \pi \circ \alpha, \Omega)$ represents $\omega \circ \alpha$. Since $\omega$ and $\omega \circ \alpha$ are in the same superselection sector, their GNS representations are unitarily equivalent. By uniqueness of the GNS representation up to unitary equivalence, we know $(\hilbH, \pi, \Omega)$ is unitarily equivalent the GNS representation of $\omega$, and we know $(\hilbH, \pi \circ \alpha, \Omega)$ is unitarily equivalent to the GNS representation of $\omega \circ \alpha$. By transitivity of unitary equivalence, we know $(\hilbH, \pi, \Omega)$ is unitarily equivalent to $(\hilbH, \pi \circ \alpha, \Omega)$, so there exists a unitary $U \in \Unitary(\hilbH)$ such that
\begin{equation}\label{eq:unitary_pi_alpha_commute}
\Ad(U) \circ \pi = \pi \circ \alpha.
\end{equation}
If $\bbU \in \PU(\hilbH)$ is the projectivization of $U$, then $\Ad(U) = \Ad(\bbU)$, so \eqref{eq:projective_unitary_from_alpha} holds. 

Suppose $\bbV \in \PU(\hilbH)$ also satisfies \eqref{eq:projective_unitary_from_alpha}. Lift $\bbV$ to a unitary $V \in \Unitary(\hilbH)$. Then \eqref{eq:unitary_pi_alpha_commute} holds with $U$ replaced by $V$, so
\[
\Ad(V^{-1}U) \circ \pi = \Ad(V)^{-1} \circ \Ad(U) \circ \pi = \Ad(V)^{-1} \circ \pi \circ \alpha = \pi.
\]
By Schur's lemma, we know $V^{-1}U \in \Unitary(1)$, so $\bbU = \bbV$.
\end{proof}

We now prove a new theorem that makes essential use of the understanding of the strong topology on $\PU(\hilbH)$ afforded by Theorem \ref{thm:topologies_PU(H)}.

\begin{theorem}\label{thm:cont_family_PU_from_aut}
Let $\fA$ be a $C^*$-algebra, let $\omega \in \pstate(\fA)$, and let $(\hilbH, \pi, \Omega)$ be an irreducible representation with $\Omega \in \sphere \hilbH$ representing $\omega$. Let $X$ be a topological space and let $\alpha:X \rightarrow \Aut(\fA)_\tn{s}$ be strongly continuous. If $X \rightarrow \pstate(\fA)_\tn{n}$, $x \mapsto \omega \circ \alpha_x$ is norm-continuous and $\omega \circ \alpha_x$ is equivalent to $\omega$ for all $x$, then there exists a unique function $\bbU:X \rightarrow \PU(\hilbH)$ such that 
\begin{equation}\label{eq:cont_family_PU_from_aut}
\Ad(\bbU_x) \circ \pi = \pi \circ \alpha_x
\end{equation}
for all $x \in X$. Moreover, $\bbU$ is strongly continuous.
\end{theorem}

\begin{proof}
By Proposition \ref{prop:PU_from_aut}, we know there exists a unique function $\bbU:X \rightarrow \PU(\hilbH)$ satisfying \eqref{eq:cont_family_PU_from_aut} for all $x \in X$. We must prove strong continuity of $\bbU$. Since $\PU(\hilbH)_\tn{s}$ is a topological group as a quotient of $\Unitary(\hilbH)_\tn{s}$, it suffices to prove strong continuity of $\bbU^{-1}$. Since by Theorem \ref{thm:topologies_PU(H)} the strong topology is also the initial topology generated by the evaluation maps $\widehat{\bbC \Psi}$, it suffices to prove that $\widehat{\bbC \Psi} \circ \bbU^{-1}$ is continuous for all $\Psi \in \sphere \hilbH$. 

For each $x \in X$, choose a representative $U_x \in \Unitary(\hilbH)$ of $\bbU_x$. Next, choose $A \in \fA$ such that $\pi(A)\Omega = \Psi$ and $\norm{A} \leq 1$, which we may do by the Kadison transitivity theorem. Then
\begin{equation}\label{eq:CPsiU_x}
\widehat{\bbC\Psi}(\bbU_x^{-1}) = \bbC U_x^* \Psi = \bbC U_x^* \pi(A)\Omega = \bbC \pi(\alpha^{-1}_x(A))U_x^*\Omega.
\end{equation}
Since $(\hilbH, \pi, U_x^*\Omega)$ represents $\omega \circ \alpha_x$ and $\omega \circ \alpha_x$ is continuous in $x$, we know $\bbC U_x^*\Omega$ is continuous in $x$ by Corollary \ref{cor:PH_superselection_metric_equivalence}.  Since $\alpha$ is strongly continuous and $\Aut(\fA)_\tn{s}$ is a topological group, we know $\alpha^{-1}$ is strongly continuous, so $\pi(\alpha_x^{-1}(A)) \in \cB(\hilbH)_{1,\tn{s}}$ is continuous in $x$. By Proposition \ref{prop:norm_unitary_PH_action} and \eqref{eq:CPsiU_x}, we know $\widehat{\bbC \Psi}\qty(\bbU_x^{-1})$ is continuous in $x$, as desired.
\end{proof}

\subsection{Equivalence of Product States}
\label{subsec:equivalence_of_product_states}

Our goal in this section is to prove that two product states are equivalent if and only if their factors are equivalent. This  appears to be a new result, although we note a related result by Wassermann \cite{Wassermann} stating that the tensor product of two $*$-automorphisms (with respect to the minimal tensor product) is inner if and only if the factors are inner. 

It is necessary for us to first review in greater detail some of the theory of tensor products of $C^*$-algebras. Let $\fA$ and $\fB$ be $C^*$-algebras and denote their algebraic tensor product by $\fA \odot \fB$. A \textdef{$C^*$-norm} on $\fA \odot \fB$ is a submultiplicative norm $\norm{\cdot}_\gamma:\fA \odot \fB \rightarrow [0,\infty)$ satisfying the $C^*$-property. The subscript $\gamma$ is merely a decoration to distinguish between norms when necessary. If $\norm{\cdot}_\gamma$ is $C^*$-norm on  $\fA \odot \fB$, then all properties of a  $C^*$-algebra are satisfied except possibly completeness. The completion of $\fA \odot \fB$ with respect to the norm $\norm{\cdot}_\gamma$ is then a $C^*$-algebra, denoted $\fA \otimes_\gamma \fB$. 

We give some key facts about tensor products of $C^*$-algebras that hold for arbitrary $C^*$-norms. Thus, let $\fA$ and $\fB$ be $C^*$-algebras and let $\norm{\cdot}_\gamma$ be a $C^*$-norm on $\fA \odot \fB$. For now, we need not assume that $\fA$ and $\fB$ are unital.
\begin{itemize}
\item The norm $\norm{\cdot}_\gamma$ is automatically a \textdef{cross norm}, meaning 
\[
\norm{A \otimes B}_\gamma = \norm{A} \norm{B}
\]
for all $A \in \fA$ and $B \in \fB$. \cite[Lem.~3.4.10]{BrownOzawa}

\item There exist approximate units $(E_i^{\fA})_{i \in I}$ and $(E_i^{\fB})_{i \in I}$ for $\fA$ and $\fB$, respectively, such that $(E_i^{\fA} \otimes E_i^{\fB})_{i \in I}$ is an approximate unit for $\fA \otimes_\gamma \fB$. \cite[Lem.~6.4.1]{Murphy} 

\item If $(\hilbH, \pi)$ and $(\hilbK, \rho)$ are representations of $\fA$ and $\fB$, respectively, then there exists a unique representation $\pi \otimes_\gamma \rho : \fA \otimes_\gamma \fB \rightarrow \cB(\hilbH \otimes \hilbK)$ such that
\[
(\pi \otimes_\gamma \rho)(A \otimes B) = \pi(A) \otimes \rho(B)
\]
for all $A \in \fA$ and $B \in \fB$. Here, $\hilbH \otimes \hilbK$ is the Hilbert space tensor product of $\hilbH$ and $\hilbK$. This is a fairly trivial synthesis of \cite[Thm.~6.4.4]{Murphy} and \cite[Thm.~6.4.18]{Murphy}. If $\pi$ and $\rho$ are nondegenerate, then $\pi \otimes_\gamma \rho$ is nondegenerate. This follows easily by using approximate units.

%\item If $(\hilbH, \pi)$ is a nondegenerate representation of $\fA \otimes_\gamma \fB$, then there exist unique $*$-homomorphisms $\pi_\fA:\fA \rightarrow \cB(\hilbH)$ and $\pi_\fB:\fB \rightarrow \cB(\hilbH)$ such that
%\[
%\pi(A \otimes B) = \pi_\fA(A) \pi_\fB(B) = \pi_{\fB}(B) \pi_\fA(A)
%\]
%for all $A \in \fA$ and $B \in \fB$. Moreover, $(\hilbH, \pi_\fA)$ and $(\hilbH, \pi_\fB)$ are nondegenerate representations of $\fA$ and $\fB$, respectively. \cite[Thm.~6.3.5]{Murphy}

\item If $\psi \in \state(\fA)$ and $\omega \in \state(\fB)$, then there exists a unique state $\psi \otimes_\gamma \omega \in \state(\fA \otimes_\gamma \fB)$ such that
\[
\qty(\psi \otimes_\gamma \omega)(A \otimes B) = \psi(A)\omega(B)
\]
for all $A \in \fA$ and $B \in \fB$ \cite[Prop.~3.4.7]{BrownOzawa}. A state of this form is called a \textdef{product state}.

%\item Suppose $\chi \in \state(\fA \otimes_\gamma \fB)$ and let $(\hilbH, \pi, \Omega)$ be the GNS representation of $\chi$. We may define restrictions $\chi|_{\fA} \in \state(\fA)$ and $\chi|_{\fB} \in \state(\fB)$ as 
%\[
%\chi|_{\fA}(A) = \ev{\Omega,\pi_{\fA}(A)\Omega} \qqtext{and} \chi|_{\fB}(B) = \ev{\Omega, \pi_{\fB}(B)\Omega}
%\]
%where $\pi_\fA:\fA \rightarrow \cB(\hilbH)$ and $\pi_\fB:\fB \rightarrow \cB(\hilbH)$ are as described above.
%\begin{itemize}
%	\item[$\diamond$] If $\psi \in \state(\fA)$ and $\omega \in \state(\fB)$, then $(\psi \otimes_\gamma \omega)|_{\fA} = \psi$ and $(\psi \otimes_\gamma \omega)|_{\fB} = \omega$.
%	\item[$\diamond$] If $\chi \in \state(\fA \otimes_\gamma \fB)$ and $\chi|_{\fA}$ or $\chi|_{\fB}$ is pure, then $\chi = \chi|_{\fA} \otimes_\gamma \chi|_{\fB}$. \cite[Cor.~3.4.3]{BrownOzawa}
%\end{itemize}
\end{itemize}

We are in particular interested in the tensor product of pure states and similarly the tensor product of irreducible representations. It seems to be folklore that the tensor product of two pure states (resp.\ two irreducible representations) is pure (resp.\ irreducible); I could not find a proof or even a sufficiently general statement of this in the literature. Therefore, I have provided full proofs below.

\begin{proposition}\label{prop:tensor_product_irreducible_reps}
Let $\fA$ and $\fB$ be $C^*$-algebras and let $\norm{\cdot}_\gamma$ be a $C^*$-norm on $\fA \odot \fB$. If $(\hilbH, \pi)$ and $(\hilbK, \rho)$ are nonzero irreducible representations of $\fA$ and $\fB$, respectively, then the only closed subspaces of $\hilbH \otimes \hilbK$ invariant under $\qty{\pi(A) \otimes \rho(B):A \in \fA,\, B \in \fB}$ are $\qty{0}$ and $\hilbH \otimes \hilbK$. In particular, $\pi \otimes_\gamma \rho$ is nonzero and irreducible.
\end{proposition}

\begin{proof}
It is easy to see that $\pi \otimes_\gamma \rho$ is nonzero. Suppose $X$ is a closed subspace invariant under $\qty{\pi(A) \otimes \rho(B):A \in \fA,\, B \in \fB}$ and suppose $X \neq \qty{0}$ and $X \neq \hilbH \otimes \hilbK$. We derive a contradiction. Let $(e_i)_{i \in I}$ and $(f_j)_{j \in J}$ be orthonormal bases for $\hilbH$ and $\hilbK$, respectively, so that $(e_i \otimes f_j)_{(i,j) \in I \times J}$ is an orthonormal basis for $\hilbH \otimes \hilbK$. Since $X \neq \hilbH \otimes \hilbK$, there exists $(i_0, j_0) \in I \times J$ such that $e_{i_0} \otimes f_{j_0} \notin X$. Let $P \in \cB(\hilbH \otimes \hilbK)$ be the projection onto $X$ and let 
\[
a = \norm{e_{i_0} \otimes f_{j_0} - P(e_{i_0} \otimes f_{j_0})} > 0.
\]

Since $X \neq \qty{0}$, there exists a unit vector $x \in X$. Since $x \neq 0$, there exists $(i_1, j_1) \in I \times J$ such $\ev{e_{i_1} \otimes f_{j_1}, x} \neq 0$. Set
\[
b = \abs{\ev{e_{i_1} \otimes f_{j_1}, x}} > 0.
\]
There exist finite subsets $I' \subset I$ and $J' \subset J$ such that $i_1 \in I'$, $j_1 \in J'$, and $\norm{x - Qx} < ab$, where $Q$ is the projection onto $\vecspan\qty{e_i \otimes f_j : (i,j) \in I' \times J'}$. By the Kadison transitivity theorem, there exists $A \in \fA$ and $B \in \fB$ such that $\norm{A} = \norm{B} = 1$ and
\[
\pi(A)e_i = \begin{cases} e_{i_0} &: i = i_1 \\ 0 &: i \in I' \setminus \qty{i_1} \end{cases} \qqtext{and} \rho(B)f_j = \begin{cases} f_{j_0} &: j = j_1 \\ 0 &: j \in J' \setminus \qty{j_1} \end{cases}
\]
We observe that
\begin{align*}
\qty[\pi(A) \otimes \rho(B)]x &= \qty[\pi(A) \otimes \rho(B)]Qx + \qty[\pi(A) \otimes \rho(B)](x - Qx)\\
&= \ev{e_{i_1} \otimes f_{j_1}, x}(e_{i_0} \otimes f_{j_0}) + \qty[\pi(A) \otimes \rho(B)](x - Qx)
\end{align*}
Applying $\1 - P$ to both sides sends the left side to zero since $X$ is invariant, so we obtain
\[
\ev{e_{i_1} \otimes f_{j_1}, x} \qty[e_{i_0} \otimes f_{j_0} - P(e_{i_0} \otimes f_{j_0})] = -(\1 - P)\qty[\pi(A) \otimes \rho(B)](x - Qx).
\]
Taking the norm of both sides yields
\[
ab = \norm{(\1 - P)\qty[\pi(A) \otimes \rho(B)](x - Qx)} \leq \norm{x - Qx} < ab,
\]
which is the desired contradiction.
\end{proof}

\begin{proposition}\label{prop:tensor_product_pure_is_pure}
Let $\fA$ and $\fB$ be $C^*$-algebras and let $\norm{\cdot}_\gamma$ be a $C^*$-norm on $\fA \odot \fB$. If $\psi \in \pstate(\fA)$ and $\omega \in \pstate(\fB)$, then $\psi \otimes_\gamma \omega \in \pstate(\fA \otimes_\gamma \fB)$. Moreover, if $(\hilbH, \pi, \Psi)$ and $(\hilbK, \rho, \Omega)$ are nonzero irreducible representations with unit vectors $\Psi$ and $\Omega$ representing $\psi$ and $\omega$, respectively, then $(\hilbH \otimes \hilbK, \pi \otimes_\gamma \rho, \Psi \otimes \Omega)$ represents $\psi \otimes_\gamma \omega$.
\end{proposition}

\begin{proof}
Let $(\hilbH, \pi, \Psi)$ and $(\hilbK, \rho, \Omega)$ be nonzero irreducible representations for $\psi$ and $\omega$, respectively. For any $A \in \fA$ and $B \in \fB$, we observe that
\begin{align*}
(\psi \otimes_\gamma \omega)(A \otimes B) &= \psi(A)\omega(B)\\
&= \ev{\Psi, \pi(A)\Psi}\ev{\Omega, \rho(B)\Omega}\\
&= \ev{\Psi \otimes \Omega, (\pi \otimes_\gamma \rho)(A \otimes B)(\Psi \otimes \Omega)}.
\end{align*}
By linearity, it follows that
\[
(\psi \otimes_\gamma \omega)(C) = \ev{\Psi \otimes \Omega, (\pi \otimes_\gamma \rho)(C)(\Psi \otimes \Omega)}
\]
for all $C \in \fA \odot \fB$. By density of $\fA \odot \fB$ in $\fA \otimes_\gamma \fB$ and continuity of the left and right hand sides, we see that the above equation holds for all $C \in \fA \otimes_\gamma \fB$. Thus, $\psi \otimes_\gamma \omega$ is represented by $(\hilbH \otimes \hilbK, \pi \otimes_\gamma \rho, \Psi \otimes \Omega)$. Since this is an irreducible representation by Proposition \ref{prop:tensor_product_irreducible_reps}, we see that $\psi \otimes_\gamma \omega$ is pure.
\end{proof}

We are almost ready to prove the main new result of this subsection. The key is the partial trace, a familiar tool in quantum information theory for finite-dimensional Hilbert spaces \cite{NielsenChuang}. For our purposes, however, we need the partial trace for the tensor product of two infinite-dimensional Hilbert spaces. This theory was developed by St\'ephane Attal in unpublished lecture notes available online \cite{AttalQuantumNoise}. The main result we need is the following. 

\begin{theorem}[{cf.~\cite[Thm.~2.28]{AttalQuantumNoise}}]\label{thm:partial_trace}
Let $\hilbH$ and $\hilbK$ be Hilbert spaces. If $T \in \cB(\hilbH \otimes \hilbK)$ is trace-class, then there exists a unique trace-class operator $\tr_{\hilbK}(T) \in \cB(\hilbH)$ such that
\[
\tr(\tr_{\hilbK}(T)A) = \tr(T(A \otimes \1))
\]
for all $A \in \cB(\hilbH)$. Likewise, there exists a unique trace-class operator $\tr_{\hilbH}(T) \in \cB(\hilbK)$ such that
\[
\tr(\tr_{\hilbH}(T)B) = \tr(T(\1 \otimes B))
\]
for all $B \in \cB(\hilbK)$.
\end{theorem}

Attal works with separable infinite-dimensional Hilbert spaces, but the above result holds for arbitrary Hilbert spaces as well, with essentially no modification to the proof. There is, however, a questionable step in Attal's proof in which two infinite sums are commuted. This uses Fubini's theorem, but Fubini's theorem requires that the summands be absolutely summable, and that is not clear in Attal's proof. Fortunately, this questionable step is not at all necessary to prove the theorem. In light of these subtleties, we will reprove Theorem \ref{thm:partial_trace} in Appendix \ref{app:partial_trace}. 
%\todo{Add this Appendix.}

With Theorem \ref{thm:partial_trace}, we can now prove the main new result of this subsection.

\begin{thm}\label{thm:equivalence_product_state}
Let $\fA$ and $\fB$ be $C^*$-algebras and let $\norm{\cdot}_\gamma$ be a $C^*$-norm on $\fA \odot \fB$. Consider pure states $\psi, \psi' \in \pstate(\fA)$ and $\omega, \omega' \in \pstate(\fB)$. Then $\psi \otimes_\gamma \omega$ is equivalent to $\psi' \otimes_\gamma \omega'$ if and only if $\psi$ is equivalent to $\psi'$ and $\omega$ is equivalent to $\omega'$.
\end{thm}

\begin{proof}
Suppose $\psi$ is equivalent to $\psi'$ and $\omega$ is equivalent to $\omega'$. Then by Theorem \ref{thm:superselection_sector_equivalences} there exists $A \in \fA$ such that $\psi(A^*A) = 1$ and $\psi(A^*CA) = \psi'(C)$ for all $C \in \fA$. Likewise, there exists $B \in \fB$ such that $\omega(B^*B) = 1$ and $\omega(B^*DB) = \omega'(D)$ for all $D \in \fB$. Then $A \otimes B \in \fA \otimes_\gamma \fB$ satisfies 
\[
(\psi \otimes_\gamma \omega)((A\otimes B)^*(A \otimes B)) = 1
\]
and
\begin{equation}\label{eq:equivalence_product_state_simple_tensor}
(\psi \otimes_\gamma \omega)((A \otimes B)^* (C \otimes D)(A \otimes B)) = (\psi' \otimes_\gamma \omega')(C\otimes D)
\end{equation}
for all $C \in \fA$ and $D \in \fB$. It follows by linearity that \eqref{eq:equivalence_product_state_simple_tensor} holds when $C \otimes D$ is replaced by an arbitrary element of $\fA \odot \fB$. It then follows by standard density and continuity arguments that \eqref{eq:equivalence_product_state_simple_tensor} holds when $C \otimes D$ is replaced by an arbitrary element of $\fA \otimes_\gamma \fB$. Thus, $\psi \otimes_\gamma \omega$ is equivalent to $\psi' \otimes_\gamma \omega'$ by Theorem \ref{thm:superselection_sector_equivalences}.

Now suppose $\psi \otimes_\gamma \omega$ is equivalent to $\psi' \otimes_\gamma \omega'$. Let $(\hilbH, \pi, \Psi)$ be the GNS representation of $\psi$ and let $(\hilbK, \rho, \Omega)$ be the GNS representation of $\omega$. Then $\pi \otimes_\gamma \rho :\fA \otimes_\gamma \fB \rightarrow \fB(\cH \otimes \cK)$ is nonzero and irreducible by Proposition \ref{prop:tensor_product_irreducible_reps} and $(\hilbH \otimes \hilbK, \pi \otimes_\gamma \rho, \Psi \otimes \Omega)$ represents $\psi \otimes_\gamma \omega$ by Proposition \ref{prop:tensor_product_pure_is_pure}. Equivalence of $\psi \otimes_\gamma \omega$ and $\psi' \otimes_\gamma \omega'$ implies that there exists a unit vector $\Phi \in \hilbH \otimes \hilbK$ representing $\psi' \otimes_\gamma \omega'$.

Consider the trace-class operator $\ketbra{\Phi} \in \cB(\hilbH \otimes \hilbK)$. Let $(E_i^\fA)_{i \in I}$ and $(E^\fB_i)_{i \in I}$ be approximate units for $\fA$ and $\fB$ such that $(E_i^\fA \otimes E_i^\fB)_{i\in I}$ is an approximate unit for $\fA \otimes_\gamma \fB$. Then for $A \in \fA$, we compute
\begin{align*}
\psi'(A) &= \lim_{i \in I} \psi'(E^\fA_i A)\omega'(E^\fB_i)\\
&= \lim_{i \in I} \ev{\Phi, (\pi \otimes_\gamma \rho)(E_i^\fA A \otimes E_i^\fB)\Phi}\\
&= \ev{\Phi, (\pi(A) \otimes \1)\Phi}\\
&= \tr\qty[(\pi(A) \otimes \1)\ketbra{\Phi}]\\
&= \tr(\pi(A)\tr_\hilbK(\ketbra{\Phi})),
\end{align*}
Note that $\tr_\hilbK(\ketbra{\Phi})$ is a positive trace-class operator of unit trace. Thus, $\psi' \in \pstate_\pi(\fA)$, so $\psi'$ and $\psi$ are equivalent by Theorem \ref{thm:superselection_sector_equivalences}.  A similar argument proves that $\omega'$ and $\omega$ are equivalent.
\end{proof}

\subsection{Definition of the Invariant}
\label{subsec:the_construction}

We will now present one of the main results of this dissertation: the $C^*$-algebraic construction of a $\check{H}^1(X;\PU(\hilbH))$ class from a sufficiently nice weak*-continuous family of pure states on a tensor product. Here, $\PU(\hilbH)$ is given the quotient of the strong operator topology.

We will restrict our attention to unital $C^*$-algebras $\fA$ and $\fB$ and will only work with the minimal and maximal $C^*$-norms on their algebraic tensor product. The reason for requiring unital $C^*$-algebras is because we will need to use inner automorphisms. In principle, we could pass to the unitization of $\fA$ and $\fB$, but we will stick to unital $C^*$-algebras for convenience. In the context of lattice spin systems, $\fA$ and $\fB$ are nuclear so the minimal and maximal tensor products are the same. In general, the reason for not using an arbitrary $C^*$-norm is more serious. In what follows, we will want to take the tensor product of two automorphisms $\alpha \in \Aut(\fA)$ and $\beta \in \Aut(\fB)$ to obtain an automorphism $\alpha \otimes_\gamma \beta \in \Aut(\fA \otimes_\gamma \fB)$. One can always take the algebraic tensor product to obtain a $*$-isomorphism $\alpha \odot \beta:\fA \odot \fB \rightarrow \fA \odot \fB$, but this function cannot always be extended continuously to $\fA \otimes_\gamma \fB$ \cite{Wassermann}. Fortunately, if $\norm{\cdot}_\gamma$ is the minimal or maximal $C^*$-norm on $\fA \odot \fB$,  then $\alpha \odot \beta$ does extend to an automorphism $\alpha \otimes_\gamma \beta \in \Aut(\fA \otimes_\gamma \fB)$ \cite[Thm.~3.5.3]{BrownOzawa}. Furthermore, we have the following proposition.

\begin{proposition}\label{prop:tensor_product_automorphism_continuous}
Let $\fA$ and $\fB$ be $C^*$-algebras and let $\norm{\cdot}_\gamma$ be either the minimal or maximal $C^*$-norm on $\fA \odot \fB$. The function
\[
\Aut(\fA)_\tn{s} \times \Aut(\fB)_\tn{s} \rightarrow \Aut(\fA \otimes_\gamma \fB)_\tn{s}, \quad (\alpha, \beta) \mapsto \alpha \otimes_\gamma \beta
\]
is strongly continuous.
\end{proposition}

\begin{proof}
By Proposition \ref{prop:dense_span_generates_strong_top} it suffices to show that 
\begin{equation}\label{eq:strong_continuity_tensor_prod}
(\alpha, \beta) \mapsto (\alpha \otimes_\gamma \beta)(A \otimes B) = \alpha(A) \otimes \beta(B)
\end{equation}
is continuous for all $A \in \fA$ and $B \in \fB$. The map $(\alpha, \beta) \mapsto (\alpha(A), \beta(B)) \in \fA \times \fB$ is manifestly continuous and the tensor product map $\otimes :\fA \times \fB \rightarrow \fA \otimes_\gamma \fB$ is continuous since $\norm{\cdot}_\gamma$ is a cross norm. This proves continuity of \eqref{eq:strong_continuity_tensor_prod}.
\end{proof}

We now begin the construction, starting with the data we assume to be given.

\begin{assumption}\label{asmpt:cocycle_data}
Let $\fA$ and $\fB$ be unital $C^*$-algebras and let $\norm{\cdot}_\gamma$ be either the minimal or maximal $C^*$-norm on $\fA \odot \fB$.  Let $X$ be a topological space and let $\omega:X \rightarrow \pstate(\fA \otimes_\gamma \fB)_{w*}$ be a weak*-continuous family of pure states. We assume to be given the following data:
\begin{itemize}
	\item fixed pure states $\omega_\fA \in \pstate(\fA)$ and $\omega_\fB \in \pstate(\fB)$ with GNS representations $(\hilbH_\fA, \pi_\fA, \Omega_\fA)$ and $(\hilbH_\fB, \pi_\fB, \Omega_\fB)$,
	\item an open cover $\cO = (O_i)_{i \in I}$ of $X$ indexed by a set $I$,
	\item for each index $i \in I$, two strongly continuous families of automorphisms $\alpha_i:O_i \rightarrow \Aut(\fA)_\tn{s}$ and $\beta_i:O_i \rightarrow \Aut(\fB)_\tn{s}$
	\item for each $i \in I$, a norm-continuous family of inner automorphisms $\eta_i:O_i \rightarrow \Inn(\fA \otimes_\gamma \fB)_\tn{n}$.
\end{itemize}
The states and automorphisms are assumed to be related by
\begin{equation}\label{eq:states_automorphisms_data_relation}
\omega_x = (\omega_\fA \otimes_\gamma \omega_\fB) \circ (\alpha_{i,x} \otimes_\gamma \beta_{i,x}) \circ \eta_{i,x}
\end{equation}
for all $i \in I$ and $x \in O_i$.
\end{assumption}

Moving the inner automorphism $\eta_{i,x}$ to the other side in \eqref{eq:states_automorphisms_data_relation}, we see that $\eta_{i,x}^{-1}$ splits $\omega_x$ into a product state 
\begin{equation}\label{eq:compose_tensor_prod_commute}
(\omega_\fA \otimes_\gamma \omega_\fB) \circ (\alpha_{i,x} \otimes_\gamma \beta_{i,x}) = (\omega_\fA \circ \alpha_{i,x}) \otimes_\gamma (\omega_\fB \circ \beta_{i,x}),
\end{equation}
while the automorphisms $\alpha_i$ and $\beta_i$ transform $\omega_\fA$ and $\omega_\fB$ into the restrictions of $\omega \circ \eta_{i}^{-1}$ to $\fA$ and $\fB$, respectively. We note that \eqref{eq:compose_tensor_prod_commute} holds since both sides evaluate to the same thing on all simple tensors $A \otimes B$.

We now come to our main result. Given the data in Assumption \ref{asmpt:cocycle_data}, we construct a \v{C}ech 1-cocycle over $X$ with values in $\PU(\hilbH_\fB)_\tn{s}$. The corresponding class in $\check{H}^1(X;\PU(\hilbH_\fB)_\tn{s})$ will be independent of all choices made for $\omega_\fA$, $\cO$, $\alpha$, $\beta$, and $\eta$. As $\PU(\hilbH_\fB)_\tn{s}$ is defined in terms of $\omega_\fB$, we cannot say literally that the cohomology class is independent of $\omega_\fB$, but for a different choice $\tilde \omega_\fB$ with GNS Hilbert space $\tilde \hilbH_\fB$, there exists an isomorphism $\PU(\hilbH_\fB)_\tn{s} \cong \PU(\tilde \hilbH_\fB)_\tn{s}$ such that the induced bijection on cohomology $\check{H}^1(X;\PU(\hilbH_\fB)_\tn{s}) \cong \check{H}^1(X;\PU(\tilde \hilbH_\fB)_\tn{s})$ identifies the two cohomology classes with each other. In Theorem \ref{thm:cocycle_construction1} we construct the cohomology class and prove independence of the class from the data, except for $\omega_\fB$. Independence from $\omega_\fB$ is proven in Theorem \ref{thm:independence_omega_B_same_O}. 

\begin{notation}
Given an indexed open cover $\cO = (O_i)_{i \in I}$ and $i_1,\ldots, i_n \in I$, we denote the $n$-fold overlap $O_{i_1\cdots i_n} = O_{i_1} \cap \cdots \cap O_{i_n}$.
\end{notation}

\begin{theorem}\label{thm:cocycle_construction1}
Under Assumption \ref{asmpt:cocycle_data}, for each $i, j \in I$ such that $O_{ij} \neq \varnothing$, there exists a unique map $\bbU_{ij}:O_{ij} \rightarrow \PU(\hilbH_\fB)$ such that
\begin{equation}\label{eq:PU(H)_cocycle_def}
\Ad(\bbU_{ij,x}) \circ \pi_\fB = \pi_\fB \circ \beta_{i,x} \circ \beta_{j,x}^{-1}
\end{equation}
for all $x \in O_{ij}$. Moreover, $\bbU_{ij}$ is strongly continuous and the collection of maps $\bbU_{ij}$ is a \v{C}ech 1-cocycle in $\check{Z}^1(\cO; \PU(\hilbH_\fB)_\tn{s})$ that does not depend on the choices for the fixed state $\omega_\fA$ and the automorphism families $\alpha_i$ and $\eta_i$. The associated cohomology class in $\check{H}^1(\cO; \PU(\hilbH_\fB)_\tn{s})$ is in addition independent of the choices of automorphism families $\beta_i$. The image of this cohomology class in $\check{H}^1(X;\PU(\hilbH_\fB)_\tn{s})$ is in addition independent of the choice of open cover $\cO$. 
\end{theorem}

\begin{proof}
Let $i,j \in I$ such that $O_{ij} \neq \varnothing$ and let $x \in O_{ij}$. Observe that
\[
(\omega_\fA \otimes_\gamma \omega_\fB) \circ (\alpha_{j,x} \otimes_\gamma \beta_{j,x}) \circ \eta_{j,x} = \omega_x = (\omega_\fA \otimes_\gamma \omega_\fB) \circ (\alpha_{i,x} \otimes_\gamma \beta_{i,x}) \circ \eta_{i,x}.
\]
Moving all the automorphisms to one side, we get
\begin{align}
\omega_\fA \otimes_\gamma \omega_\fB &= (\omega_\fA \otimes_\gamma \omega_\fB) \circ (\alpha_{i,x} \otimes_\gamma \beta_{i,x}) \circ \eta_{i,x} \circ \eta_{j,x}^{-1} \circ (\alpha_{j,x} \otimes_\gamma \beta_{j,x})^{-1}\\
&= (\omega_\fA \otimes_\gamma \omega_\fB) \circ (\alpha_{i,x} \otimes_\gamma \beta_{i,x}) \circ (\alpha_{j,x} \otimes_\gamma \beta_{j,x})^{-1} \circ \zeta_{ij,x}, \label{eq:cocycle_construction_step1}
\end{align}
where 
\[
\zeta_{ij,x} = (\alpha_{j,x} \otimes_\gamma \beta_{j,x})\circ \eta_{i,x} \circ \eta_{j,x}^{-1} \circ (\alpha_{j,x} \otimes_\gamma \beta_{j,x})^{-1}.
\]
If we also define
\[
\alpha_{ij,x} = \alpha_{i,x} \circ \alpha_{j,x}^{-1} \qqtext{and} \beta_{ij,x} = \beta_{i,x} \circ \beta_{j,x}^{-1},
\]
then we can more succinctly write \eqref{eq:cocycle_construction_step1} as
\begin{equation}\label{eq:cocycle_construction_step2}
(\omega_\fA \otimes_\gamma \omega_\fB) \circ \zeta_{ij,x}^{-1} = (\omega_\fA \circ \alpha_{ij,x}) \otimes_\gamma (\omega_\fB \circ \beta_{ij,x})
\end{equation}
where we have used the fact the tensor product respects both inversion of automorphisms and composition of automorphisms and states.

Since $\Inn(\fA \otimes_\gamma \fB)_\tn{n}$ is a topological group, we know the map $O_{ij} \rightarrow \Inn(\fA \otimes_\gamma \fB)_\tn{n}$, $x \mapsto \eta_{i,x} \circ \eta_{j,x}^{-1}$ is norm-continuous. Since the map $O_{ij} \rightarrow \Aut(\fA \otimes_\gamma \fB)_\tn{s}$, $x \mapsto \alpha_{j,x} \otimes_\gamma \beta_{j,x}$ is strongly continuous Proposition \ref{prop:tensor_product_automorphism_continuous}, Theorem \ref{thm:Aut(A)_Inn(A)_group_action} implies that $\zeta_{ij}:O_{ij} \rightarrow \Inn(\fA \otimes_\gamma \fB)_\tn{n}$ is norm-continuous. Again using the fact that $\Inn(\fA \otimes_\gamma \fB)_\tn{n}$ is a topological group, we see that $\zeta_{ij}^{-1}$ is norm-continuous.

Now we consider continuity of $\omega_\fB \circ \beta_{ij,x}$. Observe that for all $B \in \fB$, we have
\begin{align*}
\omega_\fB(\beta_{ij,x}(B)) &= \qty[(\omega_\fA \circ \alpha_{ij,x}) \otimes_\gamma (\omega_{\fB} \circ \beta_{ij,x})]\qty(\1 \otimes B)\\
&= (\omega_\fA \otimes_\gamma \omega_\fB)(\zeta_{ij,x}^{-1}(\1 \otimes B)).
\end{align*}
Therefore for $x, y \in O_{ij}$, we see that
\begin{align*}
\norm{\omega_\fB \circ \beta_{ij,y} - \omega_\fB \circ \beta_{ij,x}} \leq \norm{\zeta_{ij,y}^{-1} - \zeta_{ij,x}^{-1}}.
\end{align*}
Hence norm-continuity of $O_{ij} \rightarrow \pstate(\fB)$, $x \mapsto \omega_\fB \circ \beta_{ij,x}$ follows from norm continuity of $\zeta_{ij}^{-1}$. Furthermore, since $\zeta_{ij,x}^{-1}$ is inner, we know $\omega_\fA \otimes_\gamma \omega_\fB$ is equivalent to \eqref{eq:cocycle_construction_step2}. Therefore, by Theorem \ref{thm:equivalence_product_state} we know $\omega_\fB$ is equivalent to $\omega_\fB \circ \beta_{ij,x}$. By Theorem \ref{thm:cont_family_PU_from_aut}, we know there exists a unique function $\bbU_{ij}:O_{ij} \rightarrow \PU(\hilbH_\fB)$ satisfying \eqref{eq:PU(H)_cocycle_def} for all $x \in O_{ij}$, and $\bbU_{ij}$ is strongly continuous.

If $i,j,k \in I$ such that $O_{ijk} \neq \varnothing$, then for any $x \in O_{ijk}$ we have
\begin{align*}
\Ad(\bbU_{ij,x}\bbU_{jk,x}) \circ \pi_\fB &= \Ad(\bbU_{ij,x}) \circ \Ad(\bbU_{jk,x}) \circ \pi_\fB = \Ad(\bbU_{ij,x}) \circ \pi_\fB \circ \beta_{jk,x} \\ &= \pi_{\fB} \circ \beta_{ij,x} \circ \beta_{jk,x}= \pi_{\fB} \circ \beta_{ik,x} = \Ad(\bbU_{ik,x}) \circ \pi_\fB.
\end{align*}
It follows from irreducibility of $\pi_\fB$ and Schur's lemma that 
\[
\bbU_{ij,x}\bbU_{jk,x} = \bbU_{ik,x},
\]
so the collection $(\bbU_{ij})$ is a 1-cocycle in $\check{Z}^1(\cO, \PU(\hilbH_\fB)_\tn{s})$. It is clear from uniqueness of $\bbU_{ij}$ and the fundamental property \eqref{eq:PU(H)_cocycle_def} that the functions $\bbU_{ij}$ are independent of the choices for the pure state $\omega_\fA$ and the automorphisms $\alpha_i$ and $\eta_i$.

Suppose different choices are made for $\omega_\fA$, $\alpha_i$, $\beta_i$, and $\eta_i$, but that these different choices still satisfy Assumption \ref{asmpt:cocycle_data}. Denote these different choices by $\tilde \omega_\fA$, $\tilde \alpha_i$, $\tilde \beta_i$, and $\tilde \eta_i$. For each $i \in I$ and any $x \in I$, we have 
\[
(\tilde\omega_\fA \otimes_\gamma \omega_\fB) \circ (\tilde\alpha_{i,x} \otimes_\gamma \tilde \beta_{i,x}) \circ \tilde \eta_{i,x} = \omega_x = (\omega_\fA \otimes_\gamma \omega_\fB) \circ (\alpha_{i,x} \otimes_\gamma \beta_{i,x}) \circ \eta_{i,x}.
\]
By the exact same steps as before, we arrive at
\begin{equation}\label{eq:auto_equivalence_same_index}
(\tilde \omega_\fA \otimes_\gamma \omega_\fB) \circ \tilde \zeta_{i,x}^{-1} = (\omega_\fA \circ \alpha_{i,x} \circ \tilde \alpha_{i,x}^{-1}) \otimes_\gamma (\omega_\fB \circ \beta_{i,x} \circ \tilde \beta_{i,x}^{-1}),
\end{equation}
where
\begin{equation}\label{eq:tilde_zeta}
\tilde \zeta_{i,x} = (\tilde \alpha_{i,x} \otimes_\gamma \tilde \beta_{i,x}) \circ \eta_{i,x} \circ \tilde \eta_{i,x}^{-1} \circ (\tilde \alpha_{i,x} \otimes_\gamma \tilde \beta_{i,x})^{-1}.
\end{equation}
By the exact same reasons as before, we see that $O_i \rightarrow \pstate(\fB)_\tn{n}$, $x \mapsto \omega_\fB \circ \beta_{i,x} \circ \tilde \beta_{i,x}^{-1}$ is norm-continuous and equivalent to $\omega_\fB$ for all $x \in O_i$. Thus, we obtain a unique strongly continuous function $\bbV_i:O_i \rightarrow \PU(\hilbH_\fB)_\tn{s}$ such that
\[
\Ad(\bbV_{i,x}) \circ \pi_\fB = \pi_\fB \circ \beta_{i,x} \circ \tilde \beta_{i,x}^{-1}
\]
for all $x \in O_i$. Let $\tilde \bbU_{ij}$ denote the cocycle obtained from the $\tilde \beta_i$. Then for all $i,j \in I$ such that $O_{ij} \neq \varnothing$ and for all $x \in O_{ij}$ we have
\begin{align*}
\Ad(\bbV_{i,x}\tilde \bbU_{ij,x}\bbV_{j,x}^{-1}) \circ \pi_\fB &= \Ad(\bbV_{i,x}) \circ \Ad(\tilde \bbU_{ij,x}) \circ \pi_\fB \circ \tilde \beta_{j,x} \circ  \beta_{j,x}^{-1}\\
&= \Ad(\bbV_{i,x}) \circ \pi_\fB \circ \tilde \beta_{i,x} \circ \beta_{j,x}^{-1}\\
&= \pi_\fB \circ \beta_{i,x} \circ \beta_{j,x}^{-1}\\
&= \Ad(\bbU_{ij,x}) \circ \pi_\fB.
\end{align*}
By irreducibility of $\pi_\fB$ and Schur's lemma we conclude that
\[
\bbV_{i,x}\tilde \bbU_{ij,x}\bbV_{j,x}^{-1} = \bbU_{ij,x}.
\]
This proves that $(\tilde \bbU_{ij})$ and $(\bbU_{ij})$ are cohomologous, so the cohomology class $\class{\bbU} \in \check{H}^1(\cO; \PU(\hilbH_\fB)_\tn{s})$ is independent of the $\beta_i$ as well.

We show that the image of $\class{\bbU}$ under the canonical map $\check{H}^1(\cO; \PU(\hilbH_\fB)_\tn{s}) \rightarrow \check{H}^1(X; \PU(\hilbH_\fB)_\tn{s})$ is independent of the choice of open cover $\cO$. Suppose $\cO' = (O_{j}')_{j \in J}$ is a refinement of $\cO$ and choose a function $\kappa:J \rightarrow I$ such that $O_j' \subset O_{\kappa(j)}$ for all $j \in J$. For each $j \in J$, define $\alpha_j' = \alpha_{\kappa(j)}|_{O_j'}$, $\beta_j' = \beta_{\kappa(j)}|_{O_j'}$, and $\eta_j' = \eta_{\kappa(j)}|_{O_j}$. Evidently the data $(\omega, \omega_\fA, \omega_\fB, \cO', \alpha', \beta', \eta')$ satisfies Assumption \ref{asmpt:cocycle_data}. Let $(\bbU_{ij}')$ be the 1-cocycle in $Z^1(\cO;\PU(\hilbH_\fB)_\tn{s})$ induced by this data. Then for any $i,j \in J$ such that $O_{ij}' \neq \varnothing$, we have
\[
\Ad(\bbU_{ij,x}') \circ \pi_{\fB} = \pi_{\fB} \circ \beta_{i,x}' \circ \beta_{j,x}'^{-1} = \pi_\fB \circ \beta_{\kappa(i),x} \circ \beta_{\kappa(j),x} = \Ad(\bbU_{\kappa(i)\kappa(j),x}) \circ \pi_\fB.
\]
We see that $\bbU_{ij}' = \bbU_{\kappa(i)\kappa(j)}|_{O_{ij}'}$. Thus, the cohomology class $\class{\bbU'}$ is the image of $\class{\bbU}$ under the canonical map $\check{H}^1(\cO;\PU(\hilbH_\fB)_\tn{s}) \rightarrow \check{H}^1(\cO';\PU(\hilbH_\fB)_\tn{s})$. In particular, $\class{\bbU}$ and $\class{\bbU'}$ correspond to the same class in $\check{H}^1(X;\PU(\hilbH_\fB)_\tn{s})$. Thus, if $(\omega_\fA, \omega_\fB, \cO, \alpha, \beta, \eta)$ and $(\tilde\omega_\fA, \omega_\fB, \tilde\cO, \tilde\alpha, \tilde\beta, \tilde\eta)$ are two sets of data satisfying Assumption \ref{asmpt:cocycle_data} for the weak*-continuous family $\omega:X \rightarrow \pstate(\fA \otimes_\gamma \fB)_\tn{w*}$, then by choosing a common refinement of $\cO$ and $\tilde \cO$ and using the argument above, we see that the two sets of data induce the same cohomology class in $\check{H}^1(X; \PU(\hilbH_\fB)_\tn{s})$.
\end{proof}

Before showing that the cohomology class is independent from $\omega_\fB$, note the following. If $\hilbH$ and $\hilbK$ are complex Hilbert spaces and $W:\hilbH \rightarrow \hilbK$ is unitary (i.e., a bijective linear isometry), then the map $\Ad(W):\cB(\hilbH)_\tn{s} \rightarrow \cB(\hilbK)_\tn{s}$ defined as $\Ad(W)(A) = WAW^{-1}$ restricts to a group isomorphism and homeomorphism $\Ad(W):\Unitary(\hilbH)_\tn{s} \rightarrow \Unitary(\hilbK)_\tn{s}$ and this restriction factors through to a map on the projective unitary groups $\Ad(W):\PU(\hilbH)_\tn{s} \rightarrow \PU(\hilbK)_\tn{s}$ that is also a group isomorphism and  homeomorphism.

\begin{notation}
We will now consider what happens when we change $\omega_\fB$ to a different pure state $\tilde \omega_\fB$. We denote the GNS representation of $\tilde \omega_\fB$ by $(\tilde \hilbH_\fB, \tilde \pi_\fB, \tilde \Omega_\fB)$.
\end{notation}

\begin{theorem}\label{thm:independence_omega_B_same_O}
Let $(\omega_\fA, \omega_\fB, \cO, \alpha, \beta, \eta)$ and $(\tilde \omega_\fA, \tilde\omega_\fB, \cO, \tilde\alpha, \tilde\beta, \tilde\eta)$ be two sets of data satisfying Assumption \ref{asmpt:cocycle_data} for the same $\omega:X \rightarrow \pstate(\fA \otimes_\gamma \fB)_\tn{w*}$ and with the same open cover $\cO$. 
\begin{enumerate}
	\item\label{ite:auto_equiv_omega_B} There exists an automorphism $\xi \in \Aut(\fB)$ such that $\tilde \omega_\fB \circ \xi = \omega_\fB$.
	\item\label{ite:unitary_intertwiner_B} For any such $\xi$, there exists a unitary $W:\hilbH_\fB \rightarrow \tilde \hilbH_\fB$ such that
	\[
	\Ad(W) \circ \pi_\fB = \tilde \pi_\fB \circ \xi.
	\]
	\item\label{ite:omega_B_independence} The induced bijection $\Ad(W)_*:\check{H}^1(\cO;\PU(\hilbH_\fB)_\tn{s}) \rightarrow \check{H}^1(\cO;\PU(\tilde \hilbH_\fB)_\tn{s})$ maps the cohomology class $\class{\bbU}$ constructed from $(\omega_\fA, \omega_\fB, \cO, \alpha, \beta, \eta)$ to the cohomology class $\class{\tilde\bbU}$ constructed from $(\tilde \omega_\fA, \tilde\omega_\fB, \cO, \tilde\alpha, \tilde\beta, \tilde\eta)$.
\end{enumerate}
\end{theorem}

\begin{proof}
Given $i \in I$ and $x \in O_i$, we follow the exact same steps as in the proof of Theorem \ref{thm:cocycle_construction1} to arrive at
\begin{equation}\label{eq:auto_equivalence_same_index_tilde}
(\tilde \omega_\fA \otimes_\gamma \tilde \omega_\fB) \circ \tilde \zeta_{i,x}^{-1} = (\omega_\fA \circ \alpha_{i,x} \circ \tilde \alpha_{i,x}^{-1}) \otimes_\gamma (\omega_\fB \circ \beta_{i,x} \circ \tilde \beta_{i,x}^{-1}),
\end{equation}
analogous to \eqref{eq:auto_equivalence_same_index}, where $\tilde \zeta_{i,x}$ is defined by \eqref{eq:tilde_zeta}. Since $\tilde \zeta_{i,x}$ is inner, we see that $\tilde \omega_\fA \otimes_\gamma \tilde \omega_\fB$ is equivalent to the right hand side of \eqref{eq:auto_equivalence_same_index_tilde}, so $\tilde \omega_\fB$ is equivalent to $\omega_\fB \circ \beta_{i,x} \circ \tilde \beta_{i,x}^{-1}$ by Theorem \ref{thm:equivalence_product_state}.  This proves \ref{ite:auto_equiv_omega_B}  and \ref{ite:unitary_intertwiner_B} follows immediately from \ref{ite:auto_equiv_omega_B} and uniqueness of the GNS representation up to unitary equivalence.

Norm-continuity of $\tilde \zeta_{i}^{-1}$ implies that $\omega_\fB \circ \beta_{i,x} \circ \tilde \beta_{i,x}^{-1}$ is norm-continuous, so $\omega_\fB \circ \beta_{i,x} \circ \tilde \beta_{i,x}^{-1} \circ \xi$ is norm-continuous. Furthermore, $\omega_\fB \circ \beta_{i,x} \circ \tilde \beta_{i,x}^{-1} \circ \xi$ is equivalent to $\omega_\fB$, so by Theorem \ref{thm:cont_family_PU_from_aut} there exists a strongly continuous family $\bbV_i:O_i \rightarrow \PU(\hilbH_\fB)_\tn{s}$ such that
\[
\Ad(\bbV_{i,x}) \circ \pi_\fB = \pi_\fB \circ \beta_{i,x} \circ \tilde \beta_{i,x}^{-1} \circ \xi.
\]
For $i,j \in I$ such that $x \in O_{ij}$, we have
\begin{align*}
\Ad(W) \circ \Ad(\bbV_{i,x}^{-1}\bbU_{ij,x}\bbV_{j,x}) \circ \Ad(W^{-1}) \circ \tilde \pi_{\fB} &= \Ad(W) \circ \Ad(\bbV_{i,x}^{-1}\bbU_{ij,x}\bbV_{j,x}) \circ \pi_\fB \circ \xi^{-1}\\
&= \Ad(W) \circ \Ad(\bbV_{i,x}^{-1}\bbU_{ij,x}) \circ \pi_\fB \circ \beta_{j,x} \circ \tilde \beta_{j,x}^{-1}\\
&= \Ad(W) \circ \Ad(\bbV_{i,x}^{-1}) \circ \pi_\fB \circ \beta_{i,x} \circ \tilde \beta_{j,x}^{-1}\\
&= \Ad(W) \circ \pi_\fB \circ \xi^{-1} \circ \tilde \beta_{i,x} \circ \tilde \beta_{j,x}^{-1}\\
&= \tilde \pi_\fB \circ \tilde \beta_{i,x} \circ \tilde \beta_{j,x}^{-1}\\
&= \Ad(\tilde \bbU_{ij,x}) \circ \tilde \pi_{\fB}.
\end{align*}
By irreducibility of $\tilde \pi_\fB$ we see that 
\[
\Ad(W)\qty(\bbV_{i,x}^{-1}\bbU_{ij,x}\bbV_{j,x}) = \tilde \bbU_{ij,x}.
\]
On cohomology we have $\class{\bbV^{-1}\bbU\bbV} =\class{\bbU}$, this proves \ref{ite:omega_B_independence}.
\end{proof}

By taking refinements and restricting our automorphisms to open sets in the refinement, we can easily let the open cover vary as well. The following corollary is immediate from Theorem \ref{thm:cocycle_construction1} and Theorem \ref{thm:independence_omega_B_same_O}.

\begin{corollary}
Let $(\omega_\fA, \omega_\fB, \cO, \alpha, \beta, \eta)$ and $(\tilde \omega_\fA, \tilde\omega_\fB, \tilde \cO, \tilde\alpha, \tilde\beta, \tilde\eta)$ be two sets of data satisfying Assumption \ref{asmpt:cocycle_data} for the same $\omega:X \rightarrow \pstate(\fA \otimes_\gamma \fB)_\tn{w*}$.  There exists a unitary $W:\hilbH_\fB \rightarrow \tilde \hilbH_\fB$ such that the induced map $\Ad(W)_*:\check{H}^1(X;\PU(\hilbH_\fB)_\tn{s}) \rightarrow \check{H}^1(X; \PU(\tilde \hilbH_\fB)_\tn{s})$ sends the class $\class{\bbU}$ constructed from $(\omega_\fA, \omega_\fB, \cO, \alpha, \beta, \eta)$ to the  class $\class{\tilde\bbU}$ constructed from $(\tilde \omega_\fA, \tilde\omega_\fB, \cO, \tilde\alpha, \tilde\beta, \tilde\eta)$.
\end{corollary}

We will now use the formalism of Section \ref{subsec:exact_sequences} to map from $\check{H}^1(X;\PU(\hilbH_\fB)_\tn{s})$ to $\check{H}^1(X;\Unitary(1))$. We review this formalism briefly for the convience of the reader. Recall from Corollary \ref{cor:U(1)->U(H)->PU(H)}  that for any nonzero complex Hilbert space $\hilbH$ we have a fiber bundle $\Unitary(1) \rightarrow \Unitary(\hilbH)_\tn{s} \rightarrow \PU(\hilbH)_\tn{s}$. Since $\Unitary(1)$ is in the center of $\Unitary(\hilbH)_\tn{s}$, we obtain a ``medium-length'' exact sequence on \v{C}ech cohomology \cite[Thm.~4.1.4]{Brylinski}, as shown below, assuming that $X$ is paracompact Hausdorff.
\begin{align*}
1 &\longrightarrow \check{H}^0(X; \Unitary(1)) \longrightarrow \check{H}^0(X; \Unitary(\hilbH)_\tn{s}) \longrightarrow \check{H}^0(X; \PU(\hilbH)_\tn{s})\\
\phantom{1} &\longrightarrow \check{H}^1(X; \Unitary(1)) \longrightarrow \check{H}^1(X; \Unitary(\hilbH)_\tn{s}) \longrightarrow \check{H}^1(X; \PU(\hilbH)_\tn{s}) \overset{\delta_1}{\longrightarrow} \check{H}^2(X; \Unitary(1))
\end{align*}
We emphasize that this is an exact sequence of pointed sets and basepoint preserving functions, not an exact sequence of groups and homomorphisms. Brylinski's set-up is a bit more general, using an exact sequence  $1 \rightarrow A \rightarrow B \rightarrow C \rightarrow 1$ of sheaves of groups over $X$. The sheaves we use here are the sheaves of continuous functions into $\Unitary(1)$, $\Unitary(\hilbH)_\tn{s}$, and $\PU(\hilbH)_\tn{s}$. 

The connecting function $\delta_1:\check{H}^1(X;\PU(\hilbH)_\tn{s}) \rightarrow \check{H}^2(X;\Unitary(1))$ is defined as follows. Given a class $c \in \check{H}^1(X;\PU(\hilbH)_\tn{s})$, choose an open cover $\cO = \qty{O_i}_{i \in I}$ of $X$, a cocycle $\bbU \in \check{Z}^1(\cO;\PU(\hilbH)_\tn{s})$, and a cochain $U \in \check{C}^1(\cO;\Unitary(\hilbH)_\tn{s})$ such that:
\begin{itemize}
	\item $\class{\bbU}$ maps to $c$ under the canonical map $\check{H}^1(\cO;\PU(\hilbH)_\tn{s}) \hookrightarrow \check{H}^1(X;\PU(\hilbH)_\tn{s})$, and 
	\item $\bbU_{ij}$ is the projectivization of $U_{ij}$ for all $i,j \in I$ such that $O_{ij} \neq \varnothing$. 
\end{itemize}
That this is possible follows from the assumption that $X$ is paracompact Hausdorff (see the proof of \cite[Thm.~1.3.13]{Brylinski}). We then observe that for a nonempty triple overlap $O_{ijk} \neq \varnothing$ we have 
\[
U_{ik,x}^{-1}U_{ij,x}U_{jk,x}  = \lambda_{ijk,x}\1 \in \Unitary(1)\cdot \1
\]
Thus, we have a continuous function $\lambda_{ijk}:O_{ijk} \rightarrow \Unitary(1)$. It can be shown that the collection of maps $\lambda_{ijk}$ defines a 2-cocycle $\lambda \in \check{Z}^2(\cO;\Unitary(1))$. Furthermore, the image $\iota_\cO\class{\lambda}$ of the cohomology class $\class{\lambda}$ under the canonical inclusion $\iota_\cO: \check{H}^2(\cO;\Unitary(1)) \rightarrow \check{H}^2(X;\Unitary(1))$ is independent of the choice of open cover $\cO$, cocycle $\bbU$, and cochain $U$. We define $\delta_1c = \iota_\cO\class{\lambda}$.

\begin{corollary}\label{cor:data_independence_complete}
Let $(\omega_\fA, \omega_\fB, \cO, \alpha, \beta, \eta)$ and $(\tilde \omega_\fA, \tilde\omega_\fB, \tilde \cO, \tilde\alpha, \tilde\beta, \tilde\eta)$ be two sets of data satisfying Assumption \ref{asmpt:cocycle_data} for the same family $\omega:X \rightarrow \pstate(\fA \otimes_\gamma \fB)_\tn{w*}$ and assume $X$ is paracompact Hausdorff. Let $\bbU \in \check{Z}^1(\cO;\PU(\hilbH_\fB)_\tn{s})$ and $\tilde \bbU \in \check{Z}^1(\tilde \cO; \PU(\tilde \hilbH_\fB)_\tn{s})$ be the associated cocycles. Then
\begin{equation}\label{eq:H^2(X,U(1))_invariance}
\delta_1\iota_{\cO}\class{\bbU} = \tilde \delta_1\tilde \iota_{\tilde \cO}\class{\tilde \bbU},
\end{equation}
where $\iota_{\cO}:\check{H}^1(\cO;\PU(\hilbH_\fB)_\tn{s}) \rightarrow \check{H}^1(X;\PU(\hilbH_\fB)_\tn{s})$ and $\delta_1 : \check{H}^1(X;\PU(\hilbH_\fB)_\tn{s}) \to \check{H}^2(X;\Unitary(1))$ are the canonical inclusion and connecting map, respectively, and similarly for $\tilde \iota_{\tilde \cO}$ and $\tilde \delta_1$.  Thus, the class $\delta_1\iota_{\cO}\class{\bbU}$ is independent of all data $(\omega_\fA, \omega_\fB, \cO, \alpha, \beta, \eta)$ and is therefore intrinsic to $\omega$.
\end{corollary}

\begin{proof}
By passing to a common refinement we may assume $\cO = \tilde \cO = \qty{O_i}_{i \in I}$. Refining further if necessary, we may assume that each $\bbU_{ij}$ has image contained in a local trivialization of the fiber bundle $\Unitary(1) \rightarrow \Unitary(\hilbH_\fB)_\tn{s} \rightarrow \PU(\hilbH_\fB)_\tn{s}$, and therefore there exist lifts of the $\bbU_{ij}$ to strongly continuous families of unitaries $U_{ij}$. By Theorem \ref{thm:independence_omega_B_same_O} there exists a unitary $W:\hilbH_\fB \rightarrow \tilde \hilbH_\fB$ such that $\Ad(W)_*\class{\bbU} = \class{\tilde \bbU}$. This means that there exists $\bbV \in \check{C}^0(\cO;\PU(\tilde \hilbH_\fB)_\tn{s})$ such that
\[
\Ad(W)(\bbU_{ij}) = \bbV_i \tilde \bbU_{ij}\bbV_j^{-1}.
\]
Refining one more time if necessary, we may assume that each $\bbV_i$ has a lift to a strongly continuous family of unitaries $V_i \in \Unitary(\tilde \hilbH_\fB)_\tn{s}$. Then we observe that $\tilde \bbU_{ij}$ is the projectivization of $V_i^{-1}WU_{ij}W^{-1}V_j$. Moreover,
\begin{align*}
(V_k^{-1}WU_{ik}^{-1}W^{-1}V_i)(V_i^{-1}WU_{ij}W^{-1}V_j)(V_j^{-1}WU_{jk}W^{-1}V_k) &= V_k^{-1}WU_{ik}^{-1}U_{ij}U_{jk}W^{-1}V_k \\
&= U_{ik}^{-1}U_{ij}U_{jk},
\end{align*}
where in the second line we used the fact that $U_{ik}^{-1}U_{ij}U_{jk}$ is a phase, and therefore commutes with $W^{-1}V_k$. This proves \eqref{eq:H^2(X,U(1))_invariance} by definition of $\delta_1$ and $\tilde \delta_1$.
\end{proof}

Finally, we provide conditions under which the $\check{H}^1(X, \PU(\hilbH_\fB)_\tn{s})$ class is a homotopy invariant.

\begin{theorem}
Suppose $X$ is locally compact, normal, and Lindel\"of. Let Assumption \ref{asmpt:cocycle_data} be given for the topological space $X \times [0,1]$, so that $\omega:X \times [0,1] \rightarrow \pstate(\fA \otimes_\gamma \fB)_\tn{w*}$ is a homotopy from $\omega_0:X \rightarrow \pstate(\fA \otimes_\gamma \fB)_\tn{w*}$ to $\omega_1:X \rightarrow \pstate(\fA \otimes_\gamma \fB)_\tn{w*}$. Then Assumption \ref{asmpt:cocycle_data} can be fulfilled for $\omega_0$ and $\omega_1$ using the same fixed state $\omega_\fB \in \pstate(\fB)$ as used for $\omega$, and $\omega_0$ and $\omega_1$ generate the same class in $\check{H}^1(X;\PU(\hilbH_\fB)_\tn{s})$.
\end{theorem}

\begin{proof}
For $k \in \qty{0,1}$, define $f_k:X \rightarrow X \times [0,1]$ as $f_k(x) = (x, k)$. Let $(\omega_\fA, \omega_\fB, \cO, \alpha, \beta, \eta)$ satisfy Assumption \ref{asmpt:cocycle_data} for $\omega$, where $\cO = (O_i)_{i \in I}$. Then it is easy to see that $(\omega_\fA, \omega_\fB, f_k^{-1}(\cO), \alpha \circ f_k, \beta \circ f_k, \eta \circ f_k)$ satisfy Assumption \ref{asmpt:cocycle_data} for $\omega_k$, where $f_k^{-1}(\cO) = (f_k^{-1}(O_i))_{i \in I}$ and for example $(\alpha \circ f_k)_{ij} = \alpha_{ij} \circ f_k :f^{-1}(O_i \cap O_j) \rightarrow \Aut(\fA)_\tn{s}$. Thus, the $\check{H}^1(X;\PU(\hilbH_\fB)_\tn{s})$ class generated by $\omega_k$ for $k \in \qty{0,1}$ is represented by the cocycle $\bbU \circ f_k$ in $\check{Z}^1(f^{-1}(\cO);\PU(\hilbH_\fB)_\tn{s})$, where $\bbU \in \check{Z}^1(\cO;\PU(\hilbH_\fB)_\tn{s})$ is the cocycle generated by $\omega$. In other words, the cocycle generated by $\omega_k$ is $f_k^*\iota_\cO\class{\bbU}$, where $f_k^*:\check{H}^1(X \times [0,1]; \PU(\hilbH_\fB)_\tn{s}) \rightarrow \check{H}^1(X;\PU(\hilbH_\fB)_\tn{s})$ is map on \v{C}ech cohomology induced by $f_k$. The result now follows from Theorem \ref{thm:homotopy_invariance_Cech_1}.
\end{proof}

In the context of spin systems on the lattice $\bbZ$, one begins with the quasi-local algebra
\[
\bigotimes_{v \in \bbZ} M_{n_v}(\bbC)
\]
where to each $v \in \bbZ$ we have assigned a natural number $n_v \geq 2$. One assumes to be given a paracompact Hausdorff space $X$ and a weak*-continuous family $\omega:X \rightarrow \pstate(\bigotimes_{v \in \bbZ} M_{n_v}(\bbC))$. The $C^*$-algebras $\fA$ and $\fB$ above are defined by choosing a site $v_0 \in \bbZ$ and defining
\begin{equation}\label{eq:left_right_C*_alg}
\fA = \bigotimes_{v \leq v_0} M_{n_v}(\bbC) \qqtext{and} \fB = \bigotimes_{v > v_0} M_{n_v}(\bbC).
\end{equation}
We note that $\fA$ and $\fB$ are nuclear and $\fA \otimes \fB \cong \bigotimes_{v \in \bbZ} M_{n_v}(\bbC)$ (see, e.g., \cite{beaudry2023homotopical}). Suppose we are given data $(\omega_\fA, \omega_\fB, \cO, \alpha, \beta, \eta)$ satisfying Assumption \ref{asmpt:cocycle_data} for the family $\omega:X \rightarrow \pstate(\fA \otimes \fB)$. Let $\omega_0$ be a completely disentangled pure state, i.e., a pure state on $\bigotimes_{v \in \bbZ} M_{n_v}(\bbC)$ such that $\omega_0|_{M_{n_v}}(\bbC)$ is pure for all $v \in \bbZ$.  Since the automorphism group of a UHF algebra acts transitively on its pure state space, there exist automorphisms $\gamma_\fA \in \Aut(\fA)$ and $\gamma_\fB \in \Aut(\fB)$ such that $\omega_\fA = \omega_0|_{\fA} \circ \gamma_\fA$ and $\omega_\fB = \omega_0|_{\fB} \circ \gamma_\fB$. Thus $(\omega_0|_{\fA}, \omega_0|_{\fB}, \cO, \gamma_\fA \circ \alpha, \gamma_\fB \circ \beta, \eta)$ satisfies Assumption \ref{asmpt:cocycle_data} for $\omega$, where by $\gamma_\fA \circ \alpha$ we mean the family of functions $\gamma_\fA \circ \alpha_{i,x}$ for each $i \in I$ and $x \in O_i$ and similarly for $\gamma_\fB \circ \beta$. By Corollary \ref{cor:data_independence_complete}, $(\omega_\fA, \omega_\fB, \cO, \alpha, \beta, \eta)$ and $(\omega_0|_{\fA}, \omega_0|_{\fB}, \cO, \gamma_\fA \circ \alpha, \gamma_\fB \circ \beta, \eta)$ produce the same $\check{H}^2(X;\Unitary(1))$ class. Therefore we may without loss of generality take $\omega_\fA$ and $\omega_\fB$ to be of the form $\omega_0|_{\fA}$ and $\omega_0|_{\fB}$ for some completely disentangled pure state $\omega_0$.

Let us show that the invariant associated to $\omega$ is independent of the choice of $v_0$. Consider two choices $v_0, v_0' \in \bbZ$ and let $v_0' < v_0$ without loss of generality. Let $\fA'$ and $\fB'$ be the $C^*$-algebras defined using $v_0'$ instead of $v_0$ in \eqref{eq:left_right_C*_alg}. 

\begin{theorem}
Suppose $(\omega_\fA, \omega_\fB, \cO, \alpha, \beta, \eta)$ satisfy Assumption \ref{asmpt:cocycle_data} for $\omega$ with respect to the choice $v_0$ and $(\omega_{\fA'}, \omega_{\fB'}, \cO', \alpha', \beta', \eta')$ satisfy Assumption \ref{asmpt:cocycle_data} for $\omega$ with respect to the choice $v_0'$. These data define the same class in $\check{H}^2(X;\Unitary(1))$.
\end{theorem}

\begin{proof}
Let $\omega_0$ be a completely disentangled pure state on $\bigotimes_{v \in \bbZ} M_{n_v}(\bbC)$; by the previous paragraph we may assume that $\omega_{\fA} = \omega_0|_{\fA}$, $\omega_{\fB} = \omega_0|_{\fB}$, $\omega_{\fA'} = \omega_0|_{\fA'}$, and $\omega_{\fB'} = \omega_0|_{\fB'}$. Let $\fC = \bigotimes_{v_0' < v \leq v_0} M_{n_v}(\bbC)$ and let $\omega_\fC = \omega_0|_{\fC}$. Then $\omega_0 = \omega_{\fA'} \otimes \omega_\fC \otimes \omega_{\fB}$.

By passing to a common refinement of $\cO$ and $\cO'$, we may also assume that $\cO = \cO' = (O_i)_{i \in I}$. Given $i \in I$ and $x \in O_i$, we observe that 
\begin{align*}
\omega_0 \circ (\alpha_{i,x}'  \otimes \beta_{i,x}') \circ \eta_{i,x}' &= \omega_0 \circ (\alpha_{i,x}  \otimes \beta_{i,x}) \circ \eta_{i,x} \\
\Longrightarrow \quad \omega_0 \circ (\id_{\fA'} \otimes \beta_{i,x}') \circ (\id_{\fA' \otimes \fC} \otimes \beta_{i,x}^{-1}) &= \omega_0 \circ \zeta_{i,x} \circ (\alpha_{i,x} \otimes \id_{\fB}) \circ (\alpha_{i,x}'^{-1} \otimes \id_{\fC \otimes \fB}) ,
\end{align*}
where $\zeta_i$ is a norm-continuous family of inner automorphisms on $\bigotimes_{v \in \bbZ} M_{n_v}(\bbC)$. Restricting to $\fB'=\fC \otimes \fB$, we see that
\begin{equation}\label{eq:difference_tensor_identity}
\begin{split}
(\omega_\fC \otimes \omega_\fB) \circ \beta_{i,x}' \circ (\id_{\fC} \otimes \beta_{i,x}^{-1}) &= [\omega_0 \circ \zeta_{i,x} \circ (\alpha_{i,x} \otimes \id_\fB) \circ (\alpha_{i,x}'^{-1} \otimes \id_{\fC \otimes \fB})]|_{\fC \otimes \fB}\\
&= \omega_0 \circ \zeta_{i,x} \circ (\alpha_{i,x}|_{\fC} \otimes \id_\fB)
\end{split}
\end{equation}
where $\alpha_{i,x}|_\fC : \fC \rightarrow \fA$ is still a $*$-homomorphism but no longer an automorphism. Since $\alpha_i$ is strongly continuous and $\fC$ is finite-dimensional, it follows that $x \mapsto \alpha_{i,x}|_\fC$ is norm-continuous. Using the fact that $\fC$ is finite-dimensional and simple and therefore isomorphic to $M_n(\bbC)$ for some $n \in \bbN$, one can show that $x \mapsto \alpha_{i,x}|_{\fC} \otimes \id_\fB$ is a norm-continuous into the space of $*$-homomorphisms $\fC \otimes \fB \rightarrow \fA \otimes \fB$. 
%Take a family of matrix units for \fC and use estimates on the norm of elements of M_n(\bbC) \otimes \fB \cong M_n(\fB).
Since $\zeta_{i}$ is also norm-continuous, we may conclude that \eqref{eq:difference_tensor_identity} is norm-continuous into $\pstate(\fC \otimes \fB)$. Using \cite[Cor.~2.6.11]{BratteliRobinsonOAQSMI}, it follows from \eqref{eq:difference_tensor_identity} that $(\omega_\fC \otimes \omega_\fB) \circ \beta_{i,x}' \circ (\id_\fC \otimes \beta_{i,x}^{-1})$ is equivalent to $\omega_\fC \otimes \omega_\fB = \omega_{\fB'}$. 

Following the exact same steps in the proof of Theorem \ref{thm:cocycle_construction1} we see that the cocycles in $\check{Z}^1(\cO;\PU(\hilbH_{\fB'})_\tn{s})$ produced by $(\beta'_i)_{i \in I}$ and $(\id_\fC \otimes \beta_i)_{i \in I}$ are cohomologous. We conclude the proof by arguing that $(\id_\fC \otimes \beta_i)_{i \in I}$ and $(\beta_i)_{i \in I}$ produce the same invariant in $\check{H}^2(X;\Unitary(1))$. Since $(\hilbH_{\fB'}, \pi_{\fB'}, \Omega_{\fB'})$ and $(\hilbH_\fC \otimes \hilbH_\fB, \pi_\fC \otimes \pi_\fB, \Omega_{\fB} \otimes \Omega_\fB)$ both represent $\omega_{\fB'} = \omega_\fC \otimes \omega_\fB$, there exists a unitary $W:\hilbH_{\fB'} \rightarrow \hilbH_{\fC} \otimes \hilbH_{\fB}$ such that $\Ad(W) \circ \pi_{\fB'} = \pi_{\fC} \otimes \pi_{\fB}$. Let $\bbU \in \check{Z}^1(\cO;\PU(\hilbH_{\fB'})_\tn{s})$ be the cocycle satisfying
\[
\Ad(\bbU_{ij}) \circ \pi_{\fB'} = \pi_{\fB'} \circ (\id_{\fC} \otimes (\beta_{i} \circ \beta_{j}^{-1}))
\]
for all $i, j \in I$. We observe that
\[
\Ad(W\bbU_{ij}W^{-1}) \circ (\pi_\fC \otimes \pi_\fB) = (\pi_\fC \otimes \pi_\fB) \circ (\id_{\fC} \otimes (\beta_{i} \circ \beta_{j}^{-1})).
\]
A similar argument to that in Corollary \ref{cor:data_independence_complete} shows that $\bbU$ and $W\bbU W^{-1}$ map to the same class in $\check{H}^2(X;\Unitary(1))$. Furthermore, if $\bbV \in \check{Z}^1(\cO;\PU(\hilbH_\fB)_\tn{s})$ is the cocycle produced by $(\beta_i)_{i \in I}$, then we see that 
\[
\Ad(\1 \otimes \bbV_{ij}) \circ (\pi_\fC \otimes \pi_\fB) = (\pi_\fB \otimes \pi_\fB) \circ (\id_\fC \otimes (\beta_i \circ \beta_j^{-1})).
\]
Therefore $W\bbU_{ij} W^{-1} = \1 \otimes \bbV_{ij}$ for all $i,j \in I$. It is straightforward to verify from the definition of the connecting function $\check{H}^1(X;\PU(\hilbH)) \rightarrow \check{H}^2(X;\Unitary(1))$ that $\1 \otimes \bbV$ and $\bbV$ map to the same element of $\check{H}^2(X;\Unitary(1))$. Thus, $\bbU$ and $\bbV$ map to the same element of $\check{H}^2(X;\Unitary(1))$. This completes the proof.
\end{proof}

\section{An Example of a Nontrivial Parametrized 1d System}
\label{sec:1d_example}

We now present a one-dimensional spin system, invented by Michael Hermele and published in \cite{Xueda}, that has proven to be a useful toy model for the study of parametrized phases. 

We consider the lattice $\bbZ$ with a spin-1/2 particle on each lattice site. Our parameter space is the 3-sphere $X = \sphere^3$, whose elements we write as $w = (w_1, w_2, w_3, w_4) = (\w, w_4)$. The Hamiltonian is defined as
\begin{equation}\label{eq:1d_hamiltonian}
H(w) = \sum_{i \in \bbZ} H_i^1(w) + g_+(w) \sum_{\tn{$i$ even}} H_{i,i+1}^2 + g_{-}(w) \sum_{\tn{$i$ odd}} H^2_{i,i+1},
\end{equation}
where
\begin{align*}
H_i^1(w) = (-1)^i \w \cdot \bm{\sigma}_i \qqtext{and} H_{i,i+1}^2 &= \bm{\sigma}_i \cdot \bm{\sigma}_{i+1},
\end{align*}
and the functions $g_\pm : \bbS^3 \rightarrow [0,\infty)$ are continuous, satisfy $g_\pm(0,0,0,\pm1) \neq 0$, and admit an $\varepsilon \in (0,1)$ such that 
\[
\supp g_\pm \subset \qty{w \in \bbS^3: \pm w_4 \geq \varepsilon}.
\]
Here, $\bm{\sigma}_i = (\sigma_i^x, \sigma_i^y, \sigma_i^z)$ is a vector of Pauli matrices supported at site $i$. If we define
\[
O_\pm \defeq \qty{w \in \bbS^3: \pm w_4 > - \varepsilon},
\]
then on $O_+$ we have $g_-(w) = 0$ and on $O_-$ we have $g_+(w) = 0$. Thus, on $O_+$ the system decouples into pairs of even--odd dimers and on $O_-$ the system decouples into pairs of odd--even dimers.

\begin{figure}[H]
\begin{center}
\includegraphics{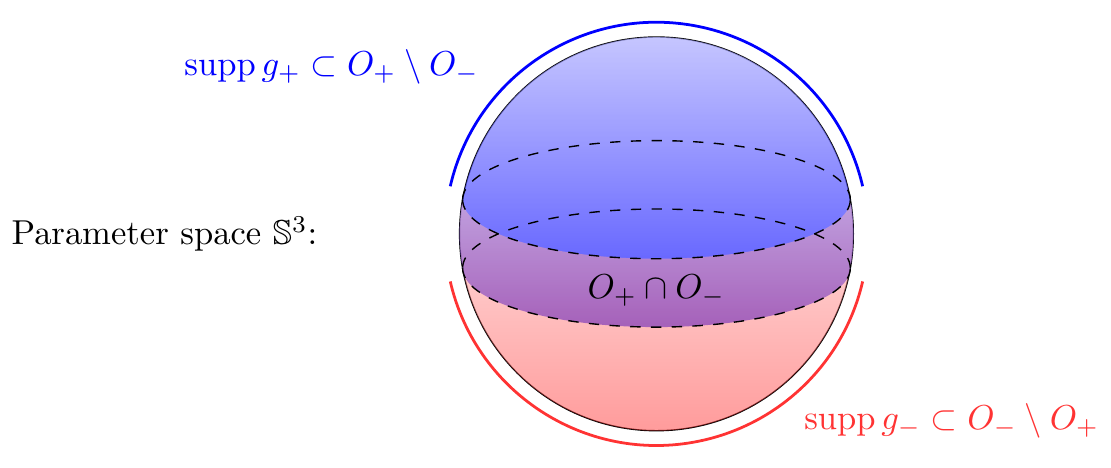}
\caption{The parameter space $\sphere^3$ with the open cover $\qty{O_-, O_+}$ and the supports of $g_-$ and $g_+$ depicted. The open sets $O_-$ and $O_+$ intersect in an $\varepsilon$-sized band around the equator.}
\end{center}
\end{figure}

\begin{figure}[H]
\begin{center}
\begin{tikzpicture}[scale=1.5]
\node[blue!80] at (-70pt,0pt){$\varepsilon<w_4\leq1$:};
\draw[blue!80] (-20pt,0pt) circle (4.5pt);
\node[blue!80] at (-20pt,0pt){\tiny $-$};
\draw[blue!80, thick] (-24.5pt, 0pt) -- (-30pt, 0pt);

\draw[blue!80] (0pt,0pt) circle (4.5pt);
\node[blue!80] at (0pt,0pt){\tiny$+$};
\draw[blue!80, thick](4.5pt,0pt)--(15.5pt,0pt);
\draw[blue!80] (20pt,0pt) circle (4.5pt);
\node[blue!80] at (20pt,0pt){\tiny$-$};

\draw[blue!80] (40pt,0pt) circle (4.5pt);
\node[blue!80] at (40pt,0pt){\tiny$+$};
\draw[blue!80, thick](44.5pt,0pt)--(55.5pt,0pt);
\draw[blue!80] (60pt,0pt) circle (4.5pt);
\node[blue!80] at (60pt,0pt){\tiny$-$};

\draw[blue!80] (80pt,0pt) circle (4.5pt);
\node[blue!80] at (80pt,0pt){\tiny$+$};
\draw[blue!80,thick](84.5pt,0pt)--(95.5pt,0pt);
\draw[blue!80] (100pt,0pt) circle (4.5pt);
\node[blue!80] at (100pt,0pt){\tiny$-$};

\draw[blue!80] (120pt,0pt) circle (4.5pt);
\node[blue!80] at (120pt,0pt){\tiny$+$};
\draw[blue!80,thick](124.5pt,0pt)--(135.5pt,0pt);
\draw[blue!80] (140pt,0pt) circle (4.5pt);
\node[blue!80] at (140pt,0pt){\tiny$-$};

\node[violet] at (-74pt,-20pt){$-\varepsilon \leq w_4 \leq \varepsilon$:};
\draw[violet] (-20pt,-20pt) circle (4.5pt);
\node[violet] at (-20pt,-20pt){\tiny $-$};

\draw[violet] (0pt,-20pt) circle (4.5pt);
\node[violet] at (0pt,-20pt){\tiny$+$};
\draw[violet] (20pt,-20pt) circle (4.5pt);
\node[violet] at (20pt,-20pt){\tiny$-$};

\draw[violet] (40pt,-20pt) circle (4.5pt);
\node[violet] at (40pt,-20pt){\tiny$+$};
\draw[violet] (60pt,-20pt) circle (4.5pt);
\node[violet] at (60pt,-20pt){\tiny$-$};

\draw[violet] (80pt,-20pt) circle (4.5pt);
\node[violet] at (80pt,-20pt){\tiny$+$};
\draw[violet] (100pt,-20pt) circle (4.5pt);
\node[violet] at (100pt,-20pt){\tiny$-$};

\draw[violet] (120pt,-20pt) circle (4.5pt);
\node[violet] at (120pt,-20pt){\tiny$+$};
\draw[violet] (140pt,-20pt) circle (4.5pt);
\node[violet] at (140pt,-20pt){\tiny$-$};

\node[red!80]  at (-78pt,-40pt){$-1 \leq w_4 < -\varepsilon$:};
\draw[red!80]  (-20pt,-40pt) circle (4.5pt);
\node[red!80]  at (-20pt,-40pt){\tiny $-$};

\draw[red!80]  (0pt,-40pt) circle (4.5pt);
\node[red!80]  at (0pt,-40pt){\tiny$+$};
\draw [thick,red!80](-15.5pt,-40pt)--(-4.5pt,-40pt);
\draw[red!80]  (20pt,-40pt) circle (4.5pt);
\node[red!80]  at (20pt,-40pt){\tiny$-$};

\draw[red!80] (40pt,-40pt) circle (4.5pt);
\node[red!80] at (40pt,-40pt){\tiny$+$};
\draw [thick,red!80](24.5pt,-40pt)--(35.5pt,-40pt);
\draw[red!80] (60pt,-40pt) circle (4.5pt);
\node[red!80] at (60pt,-40pt){\tiny$-$};

\draw[red!80] (80pt,-40pt) circle (4.5pt);
\node[red!80] at (80pt,-40pt){\tiny$+$};
\draw [red!80, thick](64.5pt,-40pt)--(75.5pt,-40pt);
\draw[red!80] (100pt,-40pt) circle (4.5pt);
\node[red!80] at (100pt,-40pt){\tiny$-$};

\draw[red!80] (120pt,-40pt) circle (4.5pt);
\node[red!80] at (120pt,-40pt){\tiny$+$};
\draw [red!80, thick](104.5pt,-40pt)--(115.5pt,-40pt);
\draw[red!80] (140pt,-40pt) circle (4.5pt);
\node[red!80] at (140pt,-40pt){\tiny$-$};
\draw [red!80, thick] (144.5pt, -40pt) -- (150pt, -40pt);
\end{tikzpicture}
\caption{Configuration of dimers as $w_4$ varies across the parameter space. Minus signs are on odd lattice sites and plus signs are on even lattice sites, consistent with the sign of $H_i^1(w)$. Figure adapted from \cite{Xueda}.}
\end{center}
\end{figure}
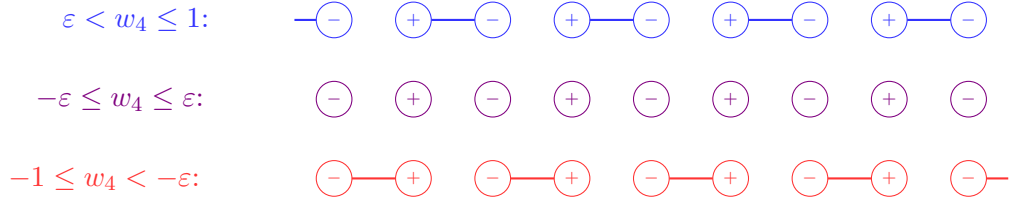

Strictly speaking, since we are working with an infinite lattice we should use the $C^*$-algebraic formalism. The quasi-local algebra is 
\[
\fA = \bigotimes_{i \in \bbZ} M_2(\bbC)
\]
and the Hamiltonian \eqref{eq:1d_hamiltonian} is a heuristic for the interaction
\[
\Phi_\Lambda(w) = \begin{cases} H_i^1(w) &: \Lambda = \qty{i} \\ g_+(w) H_{i,i+1}^2 &: \Lambda = \qty{i, i+1} \tn{ for even $i$}\\ g_-(w) H^2_{i,i+1} &: \Lambda = \qty{i, i+1} \tn{ for odd $i$}\\
0&: \tn{otherwise} \end{cases} 
\]

At any point of $\sphere^3$, the total ground state is thus the ground state on an individual dimer, tensored with itself across the whole lattice. It therefore suffices to understand the ground state on an individual dimer. This is a finite-dimensional problem, so the $C^*$-algebraic approach does not add much at this point, but it will come into play in the following section.  For $w \in O_+$, the Hamiltonian for the dimer on sites $0$ and $1$ is
\[
H_{01}(w) = g_+(w) \bm{\sigma}_0 \cdot \bm{\sigma}_1 + \w \cdot (\bm{\sigma}_0 - \bm{\sigma}_1).
\]
To solve the Hamiltonian we will once again employ the unitary from Section \ref{subsec:noninteracting}:
\begin{equation}\label{eq:U(theta,phi)_1d}
U(\theta, \phi) = \mqty(\cos(\theta/2)e^{i\phi/2} & \sin(\theta/2)e^{-i\phi/2} \\ -\sin(\theta/2)e^{i\phi/2} & \cos(\theta/2)e^{-i\phi/2})
\end{equation}
defined for $\theta, \phi \in \bbR$.
%For reference, the adjoint is
%\[
%U(\theta, \phi)^* = \mqty(\cos(\theta/2)e^{-i\phi/2} & -\sin(\theta/2)e^{-i\phi/2} \\ \sin(\theta/2)e^{i\phi/2} & \cos(\theta/2)e^{i\phi/2}).
%\]
Recall that
\begin{align}\label{eq:2x2_unitary_conjugate_H(r)_1d}
U(\theta, \phi)^*\sigma^z U(\theta, \phi) &= %\mqty(\cos(\theta/2)e^{-i\phi/2}&\sin(\theta/2)e^{-i\phi/2} \\ \sin(\theta/2)e^{i\phi/2}&-\cos(\theta/2)e^{i\phi/2})\mqty(\cos(\theta/2)e^{i\phi/2} & \sin(\theta/2)e^{-i\phi/2}  -\sin(\theta/2)e^{i\phi/2} & \cos(\theta/2)e^{-i\phi/2})\\
 \mqty(\cos \theta & e^{-i\phi}\sin \theta \\ e^{i\phi}\sin \theta & -\cos \theta) = \n(\theta, \phi) \cdot \bm{\sigma},
\end{align}
where 
\[
\n(\theta, \phi) = (\cos \phi \sin \theta, \sin \phi \sin \theta, \cos \theta).
\]

Given $i \in \bbZ$, let $U_i(\theta, \phi)$ denote the unitary $U(\theta, \phi)$ supported on site $i$, i.e., tensored with the identity on all other sites. Let $U_{i,i+1}(\theta, \phi) = U_{i}(\theta, \phi) U_{i+1}(\theta, \phi)$. One can verify by direct calculation that
\[
U_{i,i+1}(\theta, \phi)\qty(\bm{\sigma}_i \cdot \bm{\sigma}_{i+1}) U_{i,i+1}(\theta, \phi)^* = \bm{\sigma}_i \cdot \bm{\sigma}_{i+1}.
\]
Thus, choosing $\theta, \phi \in \bbR$ such that $\w = \norm{\w} \n(\theta, \phi)$, we obtain
\begin{equation}\label{eq:transformed_1d_hamiltonian}
U_{01}(\theta, \phi) H_{01}(w)U_{01}(\theta, \phi)^* = g_+(w) \bm{\sigma}_0 \cdot \bm{\sigma}_1 + \norm{\w}\qty(\sigma_0^z - \sigma_1^z).
\end{equation}

Eigenvalues and eigenvectors for $H_{01}(w)$ are now relatively straightforward to find. Using the spin raising and lowering operators
\[
\sigma^\pm_i = \frac{\sigma^x_i \pm i \sigma^y_i}{2},
\]
we can write
\[
\bm{\sigma}_0 \cdot \bm{\sigma}_1 = 2 \sigma_0^- \sigma_1^+ + 2 \sigma_0^+ \sigma_1^- + \sigma_0^z \sigma_1^z.
\]
Then we see that $\ket{\uparrow \uparrow}$ and $\ket{\downarrow \downarrow}$ are eigenvectors with eigenvalues $g_+(w)$. On the subspace spanned by the ordered basis $\ket{\uparrow \downarrow}$, $\ket{\downarrow \uparrow}$, the transformed Hamiltonian is represented by the matrix
\begin{equation}\label{eq:1d_hamiltonian_restricted}
\mqty(-g_+(w) + 2\norm{\w} & 2g_+(w)\\ 2g_+(w) & -g_+(w) - 2\norm{\w})
\end{equation}
It is economical to define a new function
\begin{align*}
f_+(w) = \sqrt{g_+(w)^2 + \norm{\w}^2}
\end{align*}
and note that $f_+$ is always strictly positive on $O_+$ by the assumption that $g_+(0,0,0,1) > 0$. The eigenvalues of \eqref{eq:1d_hamiltonian_restricted} are then easily found to be $-g_+(w) \pm 2f_+(w)$. Thus, the spectrum of $H_{01}(w)$ is 
\[
-g_+(w)-2f_+(w) < g_+(w) \leq -g_+(w) + 2f_+(w),
\]
with $g_+(w)$ doubly degenerate. We see that $H_{01}(w)$ has a unique ground state and a gap of $2g_+(w) + 2f_+(w)$. As a strictly positive continuous function on the compact set $\overline{O_+}$, we see that the gap has a positive lower bound over $O_+$.

To write the ground state, we introduce two more functions
\[
c_+(w) = \sqrt{\frac{f_+(w) +\norm{\w}}{2f_+(w)}} \qqtext{and} d_+(w) = \sqrt{\frac{f_+(w) - \norm{\w}}{2f_+(w)}}.
\]
It is straightforward to verify that an eigenvector of the transformed Hamiltonian \eqref{eq:transformed_1d_hamiltonian} with lowest eigenvalue is given by
\[
-g_+(w)\ket{\uparrow \downarrow} + (f_+(w) + \norm{\w})\ket{\downarrow \uparrow}
\] 
Normalizing this yields
\[
c_+(w)\ket{\downarrow \uparrow} - d_+(w)\ket{\uparrow \downarrow}.
\]
The ground state of $H_{01}(w)$ is therefore
\begin{align*}
\Omega_{01}(w) &= U_{01}(\theta, \phi)^*[c_+(w)\ket{\downarrow \uparrow} - d_+(w)\ket{\uparrow \downarrow}]\\
&= \frac{1}{2}(d_+(w) - c_+(w))\sin(\theta)e^{-i\phi}\ket{\uparrow \uparrow} - \frac{1}{2}(d_+(w) - c_+(w))\sin(\theta)e^{i\phi} \ket{\downarrow \downarrow}\\
&\qquad  - (d_+(w)\cos^2(\theta/2) + c_+(w)\sin^2(\theta/2))\ket{\uparrow \downarrow} + (d_+(w)\sin^2(\theta/2) + c_+(w)\cos^2(\theta/2))\ket{\downarrow \uparrow}\\
&= \frac{1}{2}(d_+(w) - c_+(w))\qty(w_1 - iw_2)\ket{\uparrow \uparrow} -  \frac{1}{2}(d_+(w) - c_+(w))(w_1 + iw_2) \ket{\downarrow \downarrow}\\
&\qquad - \frac{1}{2}\qty(d_+(w)(1 + w_3) + c_+(w)(1 - w_3))\ket{\uparrow \downarrow} + \frac{1}{2}\qty(d_+(w)(1 - w_3) + c_+(w)(1 +w_3))\ket{\downarrow \uparrow}.
\end{align*}
The point of expanding this out is to make it clear that the ground state function $\Omega_{01}:O_+ \rightarrow \bbC^2 \otimes \bbC^2$ is in fact continuous. This is not obvious from the first line above since $\theta$ and $\phi$ cannot be chosen so as to make $U_{01}(\theta, \phi)$ a continuous function on $O_+$. Finally, we note that
\[
\Omega_{01}(0,0,0,1) = \frac{1}{\sqrt{2}}\ket{\downarrow \uparrow} - \frac{1}{\sqrt{2}}\ket{\uparrow \downarrow} \qqtext{and} \Omega_{01}(0,0,1,0) = \ket{\downarrow \uparrow}.
\]
This suggests we have made a reasonable choice of phase for $\Omega_{01}$.

The analysis for $w \in O_-$ is much the same. If we consider the dimer on sites $-1$ and $0$, the Hamiltonian is
\[
H_{-10}(w) = g_-(w) \bm{\sigma}_{-1} \cdot \bm{\sigma}_{0} + \w \cdot (\bm{\sigma}_0 - \bm{\sigma}_{-1}).
\]
If we define
\[
f_-(w) = \sqrt{g_-(w)^2 + \norm{\w}^2}
\]
then the spectrum of $H_{-10}(w)$ is
\[
-g_-(w)-2f_-(w) < g_-(w) \leq -g_-(w) + 2f_-(w),
\]
with $g_-(w)$ doubly degenerate. Note that $f_-(w)$ is strictly positive on $O_-$ by the assumption that $g_-(0,0,0,-1) > 0$. We see that $H_{-10}(w)$ has a unique ground state and a gap of $2g_-(w) + 2f_-(w)$. As a strictly positive continuous function on the compact set $\overline{O_-}$, we see that the gap has a positive lower bound over $O_-$. We can now conclude that $H(w)$ has a unique gapped ground state for all $w \in \sphere^3$, with a uniform positive lower bound on the gap.

Defining
\[
c_-(w) = \sqrt{\frac{f_-(w) +\norm{\w}}{2f_-(w)}} \qqtext{and} d_-(w) = \sqrt{\frac{f_-(w) - \norm{\w}}{2f_-(w)}},
\]
the ground state $\Omega_{-10}(w)$ of $H_{-10}(w)$ is
\begin{align*}
\Omega_{-10}(w) &= U_{-10}(\theta, \phi)^*[c_-(w)\ket{\uparrow \downarrow} - d_-(w)\ket{\downarrow \uparrow}]\\
&= \frac{1}{2}(d_-(w) - c_-(w))\sin(\theta)e^{-i\phi}\ket{\uparrow \uparrow} - \frac{1}{2}(d_-(w) - c_-(w))\sin(\theta)e^{i\phi} \ket{\downarrow \downarrow}\\
&\qquad  - (d_-(w)\cos^2(\theta/2) + c_-(w)\sin^2(\theta/2))\ket{\downarrow \uparrow} + (d_-(w)\sin^2(\theta/2) + c_-(w)\cos^2(\theta/2))\ket{\uparrow \downarrow}\\
&= \frac{1}{2}(d_-(w) - c_-(w))\qty(w_1 - iw_2)\ket{\uparrow \uparrow} -  \frac{1}{2}(d_-(w) - c_-(w))(w_1 + iw_2) \ket{\downarrow \downarrow}\\
&\qquad - \frac{1}{2}\qty(d_-(w)(1 + w_3) + c_-(w)(1 - w_3))\ket{\downarrow \uparrow} + \frac{1}{2}\qty(d_-(w)(1 - w_3) + c_-(w)(1 +w_3))\ket{\uparrow \downarrow}.
\end{align*}
We see that $\Omega_{-10}:O_- \rightarrow \bbC^2 \otimes \bbC^2$ is again a continuous function. Note that in the kets above the first arrow pertains to site $-1$ and the second arrow pertains to site $0$. Finally, we have
\[
\Omega_{-10}(0,0,0,-1) = \frac{1}{\sqrt{2}}\ket{\uparrow \downarrow} - \frac{1}{\sqrt{2}}\ket{\downarrow \uparrow} \qqtext{and} \Omega_{-10}(0,0,1,0) = \ket{\uparrow \downarrow}.
\]

For all $w \in \sphere^3$, the system is invariant under shifts of the lattice by two sites. Therefore, given an even $i \in \bbZ$ and $w \in O_+$ we define $\Omega_{i,i+1}(w) = \Omega_{01}(w)$. Likewise, given an odd $i \in \bbZ$ and $w \in O_-$ we define $\Omega_{i,i+1}(w) = \Omega_{-10}(w)$. These represent the state of the dimer consisting of sites $i$ and $i+1$.  The full ground state may be heuristically written as
\[
\Omega(w) = \begin{cases} \bigotimes\limits_\tn{$i$ even} \Omega_{i,i+1}(w) &: w \in O_+ \\ \bigotimes\limits_\tn{$i$ odd} \Omega_{i,i+1}(w) &: w \in O_- \end{cases}
\]
As infinite tensor products, these are not strictly well-defined mathematical expressions. However, they are well-defined as $C^*$-algebraic states on the quasi-local algebra $\fA = \bigotimes_{i \in \bbZ} M_2(\bbC)$. In other words, the expectation values of the tensor products above are well-defined on all quasi-local operators. In $C^*$-algebraic notation, given $w \in O_+$, there exists a unique pure state $\omega_w \in \pstate(\fA)$ such that for any even $j \in \bbZ$ and $A \in \fA_{[-j,j+1]} = \bigotimes_{i=-j}^{j+1}M_2(\bbC)$ 
\begin{equation}\label{eq:ev_finite_tensor_product}
\omega_w(A) = \ev{\underset{-j \leq i \leq j}{\bigotimes_{\tn{$i$ even}}} \Omega_{i,i+1}(w), A \underset{-j \leq i \leq j}{\bigotimes_{\tn{$i$ even}}} \Omega_{i,i+1}(w)},
\end{equation}
and similarly for $w \in O_-$. These pure states agree for $w \in O_- \cap O_+$ and therefore define a family of states $\omega : \sphere^3 \rightarrow \pstate(\fA)$. Moreover, by continuity of $\Omega_{i,i+1}(w)$, we see from \eqref{eq:ev_finite_tensor_product} that the function $\omega:\sphere^3 \rightarrow \pstate(\fA)$ is weak*-continuous.

\subsection{Computing the Phase Invariant}

We now show how to compute our phase invariant for the model above. Let us outline the general procedure again. Suppose we are given a quantum spin system on the lattice $\bbZ$  parametrized by a topological space $X$. Let $n_v$ be the dimension of the on-site Hilbert space at site $v$, so that the quasi-local algebra is
\[
\fA = \bigotimes_{v \in \bbZ} M_{n_v}(\bbC).
\]
Assume the system has a unique ground state for all $x \in X$ and let $\omega:X \rightarrow \pstate(\fA)$ be the parametrized family ground states. The construction proceeds in the following steps:

\begin{enumerate}[label=\tn{(\arabic*)}]
  \item Choose a completely disentangled product state $\omega_0 \in \pstate(\fA)$.  
  \item Cut the lattice into left and right halves at a lattice site $v_0 \in \bbZ$. Define
  \[
  \fA_L = \bigotimes_{v \leq v_0} M_{n_v}(\bbC) \qqtext{and} \fA_R = \bigotimes_{v > v_0} M_{n_v}(\bbC).
  \]
  Note that there is a canonical $*$-isomorphism $\fA \cong \fA_L \otimes \fA_R$ \cite[Prop.~2.1]{beaudry2023homotopical}. Further define 
  \[
  \omega_L = \omega_0|_{\fA_L} \in \pstate(\fA_L) \qqtext{and} \omega_R = \omega_0|_{\fA_R} \in \pstate(\fA_R).
  \]
  Note that $\omega_L$ and $\omega_R$ are pure since $\omega_0$ is a product state.
  \item\label{step:automorphism1} Find an indexed open cover $\cO = \qty{O_i}_{i \in I}$ of $X$ and for each $i \in I$ a strongly continuous family of automorphisms $\alpha_i:O_i \rightarrow \Aut(\fA)$ such that
  \begin{equation}\label{eq:automorphic_equivalence_alpha_ix}
  \omega_x = \omega_0 \circ \alpha_{i,x}
  \end{equation}
  for all $x \in O_i$.

  \item\label{step:automorphism2} For each $i \in I$, find
  \begin{itemize}
    \item a strongly continuous family of automorphisms $\alpha_i^L:O_i \rightarrow \Aut(\fA_L)$,
    \item a strongly continuous family of automorphisms $\alpha_i^R:O_i \rightarrow \Aut(\fA_R)$,
    \item a norm-continuous family of inner automorphisms $\eta_i:O_i \rightarrow \Inn(\fA)$, 
  \end{itemize}
  satisfying
  \begin{equation}\label{eq:ch3_split_equation}
  \alpha_{i,x} = (\alpha_{i,x}^L \otimes \alpha_{i,x}^R) \circ \eta_{i,x}
  \end{equation}
  for all $x \in O_i$.
  \item Let $(\hilbH_R, \pi_R)$ be the GNS representation of $\omega_R$. If $i,j \in I$ and $x \in O_i \cap O_j$, then there exists a unitary $U_{ij,x} \in \Unitary(\hilbH_R)$ such that
  \[
  U_{ij,x} \pi_R(A)U^*_{ij,x} = (\pi_R \circ \beta_{i,x} \circ \beta_{j,x}^{-1})(A)
  \]
  for all $A \in \fA_R$. The unitary $U_{ij,x}$ is unique up to a phase. Let $\PU(\hilbH_R) = \Unitary(\hilbH_R)/\Unitary(1)$ be the projective unitary group of $\hilbH_R$ and equip $\PU(\hilbH_R)$ with the quotient of the strong operator topology on $\Unitary(\hilbH_R)$. Let $\bbU_{ij,x}$ be the image of $U_{ij,x}$ under the canonical projection $\Unitary(\hilbH_R) \rightarrow \PU(\hilbH_R)$. Then $\bbU_{ij}:O_i \cap O_j \rightarrow \PU(\hilbH_R)$ is strongly continuous and satisfies the cocycle condition
  \[
  \bbU_{ij}\bbU_{jk} = \bbU_{ik}
  \]
  for all $i,j,k \in I$ such that $O_i \cap O_j \cap O_k \neq \varnothing$. Thus, the collection of functions $\bbU_{ij}$ gives rise to a \v{C}ech cohomology class in $\check{H}^1(X;\PU(\hilbH_R))$.
\end{enumerate}

The \v{C}ech cohomology class in $\check{H}^1(X;\PU(\hilbH_R))$ is the desired invariant. There are bijective correspondences between the set $\check{H}^1(X;\PU(\hilbH_R))$, the set of isomorphism classes of principal $\PU(\hilbH_R)$-bundles over $X$, and the set of isomorphism classes of projective Hilbert bundles over $X$ (where the Hilbert space of the typical fiber is separable and infinite-dimensional) \cite{Lawson,SchottenloherUnitaryStrongTopology,Steenrod}. Using Dixmier-Douady theory \cite{Brylinski,DixmierDouady} and assuming $X$ is paracompact Hausdorff, there are also bijections
\[
\check{H}^1(X;\PU(\hilbH_R)) \longrightarrow \check{H}^2(X;\Unitary(1)) \longrightarrow \check{H}^3(X;\bbZ).
\]
We therefore obtain the desired $\check{H}^3(X;\bbZ)$ invariant through these maps. It is shown in Section \ref{subsec:the_construction} that the $\check{H}^2(X;\Unitary(1))$ class is independent of the all choices for $\omega_0$, $v_0$, $\cO$, $\alpha_i^L$, $\alpha_i^R$, and $\eta_i$.

It is a difficult problem to prove in generality that the automorphisms in steps \ref{step:automorphism1} and \ref{step:automorphism2} can be obtained with the desired continuity properties. It is known that for a fixed gapped local Hamiltonian (more precisely, gapped bounded finite-range interaction) of a 1d spin chain, and for any choices of $\omega_0$ and $i_0$, there exist automorphisms $\alpha$, $\alpha^L$, $\alpha^R$, and $\eta$ satisfying \eqref{eq:automorphic_equivalence_alpha_ix} and \eqref{eq:ch3_split_equation} \cite{Matsui_split_property,Matsui13,ogata_2d,OgataReview,Powers_UHF_Representations}. This is known as the \textit{split property} of gapped ground states. When the parameter space is the closed interval $[0,1]$ there are results on \textit{automorphic equivalence} (a.k.a.\ quasi-adiabatic evolution) of gapped ground states \cite{BMNS_automorphic_equivalence,Hastings_quasiadiabatic_evolution,MoonSplitProperty,OM_automorphic_equivalence} that allow one to obtain the desired continuity properties of the automorphisms in certain cases. Ultimately, however, there do not yet exist results that provide the required continuity properties in a context sufficiently general for the study of parametrized phases. Fortunately, for our exactly solvable model, we can find the desired automorphisms explicitly. We do this now.

It is natural to choose the completely factorized product state $\omega_0$ to be the state at parameter $w = (0,0,1,0)$. This state consists of an alternating pattern of up and down spins, with down spins on even lattice sites.  We then cut the lattice between sites $0$ and $1$ by choosing $i_0 = 0$. We will find the desired automorphisms using the open cover $\qty{O_-, O_+}$ of $\sphere^3$. 

The first step in constructing the automorphisms is to find a continuous family of unitaries $V_{+,01}: O_+ \rightarrow \Unitary(\bbC^2 \otimes \bbC^2)$  such that 
\[
\Omega_{01}(w) = V_{+,01}(w)\Omega_{01}(0,0,1,0) 
\]
for all $w \in O_+$. We will also need a continuous family $V_{-,-10}:O_- \rightarrow \Unitary(\bbC^2 \otimes \bbC^2)$ such that
\[
\Omega_{-10}(w) = V_{-,-10}(w)\Omega_{-10}(0,0,1,0).
\]
This requires a bit of care, but can be done using Lemma \ref{lem:unitary_rotator_continuous}, as we now explain. We will return to the discussion of automorphisms after $V_{+,01}(w)$ and $V_{-,-10}(w)$ have been computed.

Recall that Lemma \ref{lem:unitary_rotator_continuous} gives a unitary $V_{\Psi, \Omega} \in \Unitary(\hilbH)$, defined for unit vectors $\Psi$ and $\Omega$ in a Hilbert space $\hilbH$ satisfying $\ev{\Psi, \Omega} \neq 0$, such that $V_{\Psi,\Omega}$ is continuous in $\Psi$ and $\Omega$ and $V_{\Psi,\Omega}\Psi = \Omega$. Setting 
\[
\lambda(\Psi, \Omega) = \frac{\ev{\Omega,\Psi}}{\abs{\ev{\Omega,\Psi}}} \qqtext{and} \mu(\Psi, \Omega) = \frac{1}{1 + \abs{\ev{\Omega, \Psi}}},
\]
the proof of Lemma \ref{lem:unitary_rotator_continuous} gives an explicit expression
\begin{equation}\label{eq:unitary_rotator_example_explicit} 
V_{\Psi,\Omega} = \lambda \1 - \mu \ketbra{\Psi}{\Omega} - \lambda \mu \ketbra{\Psi} - \lambda \mu \ketbra{\Omega} + (1 + \lambda \mu \ev{\Omega, \Psi})\ketbra{\Omega}{\Psi}.
\end{equation}

To obtain $V_{+,01}(w)$, we might like to apply Lemma \ref{lem:unitary_rotator_continuous} by setting $\Psi = \Omega_{01}(0,0,1,0)$ and $\Omega = \Omega_{01}(w)$, but with this choice the condition $\ev{\Psi, \Omega} \neq 0$ does not hold for all $w \in O_+$. Fortunately, we can show that $\ev{\Omega_{01}(0,0,0,1), \Omega_{01}(w)} \neq 0$ for all $w \in O_+$. Observe that for any $\theta, \phi \in \bbR$,
\[
U(\theta, \phi) \otimes U(\theta, \phi)\qty(\frac{1}{\sqrt{2}}\ket{\downarrow \uparrow} - \frac{1}{\sqrt{2}}\ket{\uparrow \downarrow}) = \frac{1}{\sqrt{2}}\ket{\downarrow \uparrow} - \frac{1}{\sqrt{2}}\ket{\uparrow \downarrow},
\]
so in particular $U_{01}(\theta, \phi)\Omega_{01}(0,0,0,1) = \Omega_{01}(0,0,0,1)$. Then using the expression 
\[
\Omega_{01}(w) = U_{01}(\theta, \phi)^*\qty[c_+(w)\ket{\downarrow \uparrow} - d_+(w) \ket{\uparrow \downarrow}]
\]
we can compute
\begin{align*}
\ev{\Omega_{01}(0,0,0,1), \Omega_{01}(w)} &= \frac{c_+(w) + d_+(w)}{\sqrt{2}} > 0.
\end{align*}
Thus setting $\Psi = \Omega_{01}(0,0,0,1)$ and $\Omega = \Omega_{01}(w)$, Lemma \ref{lem:unitary_rotator_continuous} yields a continuous family of unitaries on $O_+$ that maps $\Omega_{01}(0,0,0,1)$ to $\Omega_{01}(w)$. Note that continuity of $\Omega_{01}(w)$, as pointed out earlier, is important for the continuous dependence on $w$ in this family of unitaries. To obtain $V_{+,01}(w)$, we then compose with one additional fixed unitary that sends $\Omega_{01}(0,0,1,0)$ to $\Omega_{01}(0,0,0,1)$. 

Better yet, we don't actually have to compute $V_{+,01}$ on all of $O_+$ (we do have to know that it exists and is continuous on all of $O_+$). Our parameter space $\bbS^3$ is covered by $O_+$ and $O_-$, and we only need to calculate things on the intersection 
\[
O \defeq O_+ \cap O_- = \qty{w \in \bbS^3: \abs{w_4} < \varepsilon}.
\]
For $w \in O$, we have 
\begin{align*}
g_+(w) = g_{-}(w) &= 0,\\
f_+(w) = f_-(w) &= \norm{\w},\\
c_+(w) = c_-(w) &= 1,\\
d_+(w)=d_-(w) &= 0,
\end{align*} 
so things simplify nicely. 

Turning the crank on the formula \eqref{eq:unitary_rotator_example_explicit} for $w \in O$ yields the continuous family of unitaries
\begin{equation}\label{eq:U*W*U}
V_{\Psi,\Omega} = \frac{1}{2}\qty(1 + \frac{1}{\sqrt{2}})\1 + \frac{1}{2}\qty(1 - \frac{1}{\sqrt{2}})\qty(\frac{\w}{\norm{\w}} \cdot \bm{\sigma}_0)\qty(\frac{\w}{\norm{\w}} \cdot \bm{\sigma}_1) - \frac{i}{2\sqrt{2}}\frac{\w}{\norm{\w}} \cdot (\bm{\sigma}_0 \times \bm{\sigma}_1)
\end{equation}
To better understand this formula, we introduce a new unitary. Given $i \in \bbZ$, we define
\begin{align*}
W_{i,i+1} = \frac{1}{2}\qty(1 + \frac{1}{\sqrt{2}})I + \frac{1}{2}\qty(1 - \frac{1}{\sqrt{2}})\sigma_i^z \sigma_{i+1}^z - \frac{1}{\sqrt{2}}(\sigma_i^+ \sigma_{i+1}^- - \sigma_{i}^- \sigma_{i+1}^+).
\end{align*}
This unitary is fixed; it does not depend on the parameter space. The above expression may look a little nasty, but the unitary acts nicely on the standard basis:
\begin{align*}
W_{i,i+1}\qty{\ket{\downarrow \uparrow}, \ket{\uparrow \downarrow}, \ket{\downarrow \downarrow}, \ket{\uparrow \uparrow}} = \qty{\frac{1}{\sqrt{2}}\ket{\downarrow \uparrow}-\frac{1}{\sqrt{2}}\ket{\uparrow \downarrow} ,  \frac{1}{\sqrt{2}}\ket{\downarrow \uparrow}+ \frac{1}{\sqrt{2}}\ket{\uparrow \downarrow}, \ket{\downarrow \downarrow}, \ket{\uparrow \uparrow} }
\end{align*}
In particular, 
\[
W_{01}\Omega_{01}(0,0,1,0) = \Omega_{01}(0,0,0,1).
\]
We now observe two additional identities. First:
\begin{align*}
 U_{i,i+1}(\theta, \phi)^* \qty[\z \cdot (\bm{\sigma}_i \times \bm{\sigma}_{i+1})]U_{i,i+1}(\theta, \phi) =  \n(\theta, \phi) \cdot (\bm{\sigma}_i \times \bm{\sigma}_{i+1}),
\end{align*}
where $\z = (0,0,1)$, and second:
\[
-\frac{i}{2}  \z \cdot (\bm{\sigma}_i \times \bm{\sigma}_{i+1}) = \sigma_i^+ \sigma_{i+1}^- - \sigma_{i}^- \sigma_{i+1}^+,
\]
Using these, we observe that
\[
V_{\Psi, \Omega} = U_{01}(\theta, \phi)^*W_{01}^* U_{01}(\theta, \phi).
\]
where $\theta, \phi \in \bbR$ such that $\w/\norm{\w} = \n(\theta, \phi)$. Note that the above operator is independent of the choice of $\theta$ and $\phi$. Finally, composing with $W_{01}$ to map $\Omega_{01}(0,0,1,0)$ to $\Omega_{01}(0,0,0,1)$ yields
\[
\boxed{V_{+,01}(w) = U_{01}(\theta, \phi)^*W_{01}^*U_{01}(\theta, \phi)W_{01}}
\]
for $w \in O$. A similar analysis holds on $O_-$, with the final result:
\[
\boxed{V_{-,-10}(w) = U_{-10}(\theta, \phi)^*W_{-10}U_{-10}(\theta, \phi) W_{-10}^*}
\] 
for $w \in O$.

We now return to the discussion of $C^*$-algebras and automorphisms. All the unitaries we found above compile into automorphisms of $\fA$ when tensored with themselves across all dimers. For example, while $\prod_{i \in \bbZ} U_i(\theta, \phi)$ does not represent a well-defined element of $\fA$, it does represent a well-defined automorphism $\gamma(\theta, \phi)$ of $\fA$, which we may heuristically write as
\[
\gamma(\theta, \phi) = \Ad\qty(\prod_{i \in \bbZ} U_i(\theta, \phi)).
\]
Likewise, we define automorphisms
\begin{align*}
\beta_+ = \Ad\qty(\prod_{\tn{$i$ even}} W_{i,i+1}) \qqtext{and} \beta_- = \Ad\qty(\prod_{\tn{$i$ odd}} W_{i,i+1})
\end{align*}

What we really want are automorphisms $\alpha_\pm:O_\pm \rightarrow \Aut(\fA)$ such that
\[
\omega_0 \circ \alpha_{\pm,w} = \omega_w 
\]
for all $w \in O_\pm$, where $\omega_0$ is the ground state at $(0,0,1,0)$. These are given by
\[
\alpha_{+,w} = \Ad\qty(\prod_{\tn{$i$ even}} V_{+,i,i+1}(w)^*) \qqtext{and}
\alpha_{-,w} = \Ad\qty(\prod_{\tn{$i$ odd}} V_{-,i,i+1}(w)^*).
\]
Note that $\alpha_+$ and $\alpha_-$ are strongly continuous on $O_+$ and $O_-$ since $V_{+,i,i+1}$ and $V_{-,i,i+1}$ are continuous for all $i$. 

As before, we only really need to work with $\alpha_+$ and $\alpha_-$ for $w \in O$. For $w \in O$, we have
\begin{equation}\label{eq:alphabetagamma}
\alpha_+ = \beta_+^{-1}\gamma^{-1}\beta_+^{\phantom{1}}\gamma \qqtext{and} \alpha_- = \beta_-^{\phantom{1}} \gamma^{-1}\beta_-^{-1}\gamma
\end{equation}
As you can see, we're going to start dropping the $w$ and $\theta$, $\phi$ arguments, as well as the composition symbol, when convenient. Suffice it to say that $\gamma$ depends on $\theta$ and $\phi$, $\alpha_+$ and $\alpha_-$ depend on the parameter $w$, and $\beta_+$ and $\beta_-$ are constant. 

We cut the chain between sites $0$ and $1$. Our first task is to decompose $\alpha_+$ and $\alpha_-$ as
\[
\alpha_\pm = (\alpha_{\pm,L} \otimes \alpha_{\pm,R}) \circ (\tn{inner}),
\]
where the inner automorphism should be a norm-continuous function of $w$. Notice that $\gamma$, $\beta_-$, and $\alpha_-$ are already naturally split up as $\gamma = \gamma_L \otimes \gamma_R$, $\beta_- = \beta_{-,L} \otimes \beta_{-,R}$, and
\[
\alpha_-^{\phantom{1}} = \alpha_{-,L}^{\phantom{1}} \otimes \alpha_{-,R}^{\phantom{1}} = \beta_{-,L}^{\phantom{1}}\gamma_{L}^{-1} \beta_{-,L}^{-1}\gamma_L^{\phantom{1}} \otimes \beta_{-,R}^{\phantom{1}}\gamma_{R}^{-1} \beta_{-,R}^{-1}\gamma_R^{\phantom{1}}.
\]
However, $\beta_+$ and $\alpha_+$ have terms that straddle the cut. We therefore define
\begin{align*}
\beta_{+,L} = \Ad\qty(\underset{i \leq -2}{\prod_{\tn{$i$ even}}} W_{i,i+1}) \qqtext{and} \beta_{+,R} = \Ad\qty(\underset{i \geq 2}{\prod_{\tn{$i$ even}}} W_{i,i+1}),
\end{align*}
whence
\[
\beta_{+} = (\beta_{+,L} \otimes \beta_{+,R}) \circ \Ad(W_{01})
\]
Note that $\Ad(W_{01})$ and $\Ad(\gamma^{-1}(W_{01}))$ commute with $(\beta_{+,L} \otimes \beta_{+,R})$. Substituting this into \eqref{eq:alphabetagamma} yields
\begin{align*}
\alpha_+ &= \Ad(W_{01}^*) \circ (\beta_{+,L}^{-1} \otimes \beta_{+,R}^{-1}) \circ \gamma^{-1} \circ \Ad(W_{01}) \circ (\beta_{+,L}^{\phantom{1}} \otimes \beta_{+,R}^{\phantom{1}}) \circ \gamma\\
&= \Ad(W_{01}^*) \circ (\beta_{+,L}^{-1} \otimes \beta_{+,R}^{-1}) \circ \Ad(\gamma^{-1}(W_{01})) \circ \gamma^{-1} \circ (\beta_{+,L}^{\phantom{1}} \otimes \beta_{+,R}^{\phantom{1}}) \circ \gamma\\
&= \Ad(W_{01}^* U_{01}^*W_{01}U_{01}) \circ (\beta_{+,L}^{-1} \gamma_{L}^{-1} \beta_{+,L}^{\phantom{1}} \gamma_L^{\phantom{1}} \otimes \beta_{+,R}^{-1} \gamma_{R}^{-1} \beta_{+,R}^{\phantom{1}} \gamma_R^{\phantom{1}}).
\end{align*}
We define
\[
\alpha_{+,L} = \beta_{+,L}^{-1} \gamma_{L}^{-1} \beta_{+,L}^{\phantom{1}} \gamma_L^{\phantom{1}} \qqtext{and} \alpha_{+,R} = \beta_{+,R}^{-1} \gamma_{R}^{-1} \beta_{+,R}^{\phantom{1}} \gamma_R^{\phantom{1}}.
\]

Note that $W_{01}^*U_{01}^*W_{01}U_{01}$ is indeed a norm-continuous function of $w$, as evidenced by \eqref{eq:U*W*U}. I have the inner automorphism on the ``wrong side,'' but this is not a big deal. I could have pushed the adjoint actions to the right instead and found that, in this case, the inner automorphism $\Ad(W_{01}^*U_{01}^*W_{01}U_{01})$ commutes with $\alpha_{+,L} \otimes \alpha_{+,R}$. I pushed the adjoint actions to the left because I now want to invert $\alpha_+$:
\begin{align*}
\alpha_+^{-1} = (\alpha_{+,L}^{-1} \otimes \alpha_{+,R}^{-1}) \circ \Ad(U_{01}^* W_{01}^* U_{01}W_{01}).
\end{align*}
Thus, for all $w \in O$, we have
\begin{equation}\label{eq:omega_0_invariance}
\omega_0 = \omega_0 \circ (\alpha_{-,L}^{\phantom{1}} \alpha_{+,L}^{-1} \otimes \alpha_{-,R}^{\phantom{1}}\alpha_{+,R}^{-1}) \circ \Ad(U_{01}^*W_{01}^*U_{01}W_{01}).
\end{equation}

We want to just be looking at the right hand side, but the inner automorphism is not a product between the left and right hand side, which looks like it might make things difficult to split up. Fortunately, this is easily fixed. We claim that
\begin{equation}\label{eq:claim}
\omega_0 \circ \Ad(W_{01}^*U_{01}^*(\theta, \phi)W_{01}) = \omega_0
\end{equation}
for all $\theta,\phi$. To prove this, first note that $\omega_0$ is a product state $\omega_0 = \omega_{0,01} \otimes \omega_{0, \bbZ \setminus \qty{0,1}}$, so it suffices to show that $\omega_{0,01} \circ \Ad(W_{01}^*U_{01}^*(\theta, \phi)W_{01}) = \omega_{0,01}$. Now $\omega_{0,01}$ is represented by $\ket{\downarrow \uparrow}$, so $\omega_{0,01} \circ \Ad(W_{01}^*U_{01}^*(\theta, \phi)W_{01})$ is represented by
\begin{align*}
W_{01}^* U_{01}(\theta, \phi)W_{01}\ket{\downarrow \uparrow} &= W_{01}^*U_{01}(\theta, \phi)\qty(\frac{1}{\sqrt{2}}\ket{\downarrow \uparrow} - \frac{1}{\sqrt{2}}\ket{\uparrow \downarrow}) = W_{01}^*\qty(\frac{1}{\sqrt{2}}\ket{\downarrow \uparrow} - \frac{1}{\sqrt{2}}\ket{\uparrow \downarrow}) = \ket{\downarrow \uparrow}.
\end{align*}
This proves \eqref{eq:claim}. Thus, precomposing with $\Ad(W_{01}^*U_{01}^*W_{01})$ on both sides of \eqref{eq:omega_0_invariance} yields
\begin{align*}
\omega_0 = \omega_0 \circ (\alpha_{-,L}^{\phantom{1}} \alpha_{+,L}^{-1} \otimes \alpha_{-,R}^{\phantom{1}}\alpha_{+,R}^{-1}) \circ \Ad(U_{01}^*)
\end{align*}
This is a tensor product on both sides, so we obtain
\begin{equation}\label{eq:omega_R}
\omega_{0,R} = \omega_{0,R} \circ \alpha_{-,R}^{\phantom{1}}\alpha_{+,R}^{-1} \circ \Ad(U_{1}(\theta, \phi)^*)
\end{equation}
for any choice of $\theta, \phi$ such that $\w = \norm{\w} \n(\theta, \phi)$.

Let $(\hilbH_R, \pi_R, \Omega_R)$ be the GNS representation of $\omega_{0,R}$, which is irreducible since $\omega_{0,R}$ is pure. Equation \eqref{eq:omega_R} implies that $(\hilbH_R, \pi_R \circ \alpha_{-,R}^{\phantom{1}}\alpha_{+,R}^{-1} \circ \Ad(U_1(\theta, \phi)^*), \Omega_R)$ represents $\omega_{0,R}$ as well, and therefore there exists a unique unitary $Y = Y(\theta, \phi) \in \Unitary(\hilbH_R)$ such that $Y(\theta, \phi)\Omega_R = \Omega_R$ and 
\[
\Ad(Y) \circ \pi_R = \pi_R \circ \alpha_{-,R}^{\phantom{1}}\alpha_{+,R}^{-1} \circ \Ad(U_1^*).
\]
Precomposing with $\Ad(U_1)$ on both sides yields
\[
\Ad(Y\pi_R(U_1)) \circ \pi_R = \Ad(Y)\circ \pi_R \circ \Ad(U_1) = \pi_R \circ \alpha_{-,R}^{\phantom{1}}\alpha_{+,R}.
\]
We define $Z = Z(\theta, \phi) = Y\pi_R(U_1) \in \Unitary(\hilbH_R)$. 

The function $Z:O \rightarrow \Unitary(\hilbH_R)$ is not guaranteed to be continuous, nor is it guaranteed to be independent of the choice of $\theta$ and $\phi$ to represent $\w/\norm{\w}$. However, it is guaranteed to be continuous and independent of the choice of $\theta$ and $\phi$ when we project $\Unitary(\hilbH_R) \rightarrow \PU(\hilbH_R)$ by Proposition \ref{prop:PU_from_aut}. Let $\bbZ:O \rightarrow \PU(\hilbH_R)$ be the projectivization of $Z$. 

Since we have only one overlap for the open cover $\cO = \qty{O_-, O_+}$, the function $\bbZ$ defines a \v{C}ech cocycle, hence a cohomology class in $\check{H}^1(\cO;\PU(\hilbH_R))$. If we can show that the cohomology class in $\check{H}^1(\cO;\PU(\hilbH_R))$ is nontrivial, then it will follow that its image in $\check{H}^1(X;\PU(\hilbH_R))$ is nontrivial since the canonical map $\check{H}^1(\cO;\PU(\hilbH_R)) \rightarrow \check{H}^1(X;\PU(\hilbH_R))$ is injective and maps the trivial class to the trivial class. We prove that the $\check{H}^1(\cO;\PU(\hilbH_R))$ class is nontrivial by contradiction.

Suppose the $\check{H}^1(\cO;\PU(\hilbH_R))$ class is trivial. Then there exist strongly continuous functions $\bbZ_{\pm}:O_\pm \rightarrow \PU(\hilbH_R)$ such that $\bbZ(w) = \bbZ_-(w)\bbZ_+(w)^{-1}$ for all $w \in O$. But since $O_-$ and $O_+$ are contractible and $\PU(\hilbH_R)$ is path-connected, we know $\bbZ_-$ and $\bbZ_+$ are nulhomotopic, hence $\bbZ$ is nulhomotopic. It follows that the composition 
\[
O \rightarrow \PU(\hilbH_R) \rightarrow \PU(\hilbH_R) \rightarrow  \bbP\hilbH_R, \quad w \mapsto \bbZ(w)\mapsto \bbZ(w)^{-1} \mapsto \bbC Z(\theta, \phi)^* \Omega_R
\]
is nulhomotopic. Observe that
\[
Z(\theta, \phi)^*\Omega_R = \pi_R(U_1(\theta, \phi)^*) Y(\theta, \phi)^* \Omega_R = \pi_R(U_1(\theta, \phi)^*)\Omega_R \in \pi_R(\fA_1)\Omega_R.
\]
Here $\fA_1 = M_2(\bbC) \subset \fA$ is the algebra at site 1. Define $\hilbK = \pi_R(\fA_1)\Omega_R$. One can show that $\hilbK$ is two-dimensional with an orthonormal basis given by $\Omega_R$ and $\pi_R(\sigma_1^x)\Omega_R$, which can be thought of in terms of spins as
\[
\Omega_R = \ket{\uparrow \downarrow \uparrow \downarrow \uparrow \downarrow \cdots } \qqtext{and} \pi_R(\sigma_1^x)\Omega_R = \ket{\downarrow \downarrow \uparrow \downarrow\uparrow \downarrow \cdots}.
\]
Using the equality condition of the Cauchy-Schwarz inequality one can show that $\pi_R(\sigma_1^z)\Omega_R = \Omega_R$ and $\pi_R(\sigma_1^y)\Omega_R = i\pi_R(\sigma^x_1)\Omega_R$. It follows that  we can expand $\pi_R(U_1(\theta, \phi)^*)\Omega_R$ in the basis $\Omega_R$, $\pi_R(\sigma^x_1)\Omega_R$ as
\[
\pi_R(U_1(\theta, \phi)^*)\Omega_R = \cos\qty(\frac{\theta}{2})e^{-i\phi/2} \Omega_R + \sin\qty(\frac{\theta}{2})e^{i\phi/2} \pi_R(\sigma^x_1)\Omega_R.
\]
This is exactly the ground state of the Hamiltonian $-\w \cdot \bm{\sigma}$. 

Let $P:\hilbH_R \rightarrow \hilbK$ be the orthogonal projection. This descends to a continuous map $\widebar P:\bbP\hilbH_R \rightarrow \bbP\hilbK$. We can conclude that the composition
\[
\sphere^2 \rightarrow O \rightarrow \bbP \hilbH_R \rightarrow \bbP \hilbK, \quad \w \mapsto (\w, 0) \mapsto \bbC Z(\theta, \phi)^*\Omega_R \mapsto \overline{P}\bbC Z(\theta, \phi)^*\Omega_R = \bbC Z(\theta, \phi)^*\Omega_R
\]
is nulhomotopic since the middle arrow is nulhomotopic. But this map is the homeomorphism $\sphere^2 \cong \bbP(\bbC^2)$ given by the ground state map of the single particle Hamiltonian $H(\w) = -\w \cdot \bm{\sigma}$ (Theorem \ref{thm:S^2_CP^1_homeomorphism}). Since $\sphere^2$ is not contractible, we have our contradiction. This proves that the invariant of our 1d parametrized model is nontrivial.

%!TEX root = dissertation.tex

\newpage
\chapter{Fiber Bundles over Pure State Space}
\label{chp:fiber_bundles}

In the Hilbert space setting, one encounters fiber bundles in many places. For example, $\bbS \hilbH$ is a principal $\Unitary(1)$-bundle over $\bbP \hilbH$ with fiber $\Unitary(1)$ and, if $\hilbH$ is infinite-dimensional, $\Unitary(\hilbH)$ is a fiber bundle over $\bbP \hilbH$ with fiber isomorphic to $\Unitary(1) \times \Unitary(\hilbH)$. In this chapter, we construct natural $C^*$-algebraic analogs of some such bundles. 

These constructions arose from our interest in understanding the continuous families of pure states that arise in the study of parametrized quantum systems. With a continuous family of pure states, it is natural to ask how to do classical $C^*$-algebraic maneuvers like the GNS construction and Kadison transitivity theorem in a way that depends continuously on their initial data. The answers to these questions were realized as the construction of the local trivializations of the aforementioned fiber bundles.

We will deal almost exclusively with the norm topology on pure state space in this chapter. Of course, the families of ground states are typically weak*-continuous, but we have seen that the norm topology also plays an important role in physics.

Finally, this chapter is based on the paper \textit{Continuous Dependence on the Initial Data in the Kadison Transitivity Theorem and GNS Construction}, authored by myself and others in our research group, and published in Reviews in Mathematical Physics \cite{Spiegel}. In that paper, several fiber bundles are constructed starting from a locally trivial $C^*$-algebra bundle. Here we will use a fixed $C^*$-algebra rather than a $C^*$-algebra bundle for the sake of simplicity and clarity.

\section{The Fiberwise GNS Construction}
\label{sec:fiberwise_GNS_construction}

Throughout this section, let $\fA$ be a nonzero $C^*$-algebra. Given a pure state $\omega \in \pstate(\fA)$, we let $(\hilbH_\omega, \pi_\omega,\Omega_\omega)$ denote its GNS representation. Throughout this section, endow $\pstate(\fA)$ with the norm topology. 

When we muse on what it means to perform the GNS construction in a way that depends norm-continuously on the state $\omega \in \pstate(\fA)$, our first idea is to give
\begin{equation}\label{eq:fiberwise_GNS_projection}
p:\bigsqcup_{\omega \in \pstate(\fA)} \hilbH_\omega \rightarrow \pstate(\fA)
\end{equation}
the structure of a fiber bundle, where $p$ is the canonical map that associates to each element $\Phi \in \hilbH_\omega$ the footpoint $\omega$. We will demonstrate how this can be done with a procedure we call the \textdef{fiberwise GNS construction}. 

When the GNS Hilbert spaces are infinite-dimensional, the topology of the bundle itself turns out to be of little interest. Kuiper's theorem tells us that the unitary group of an infinite-dimensional Hilbert space is contractible with respect to the norm topology \cite{Kuiper}, consequently all Hilbert bundles with infinite-dimensional fibers are trivial \cite{SchottenloherUnitaryStrongTopology}. Nonetheless, the construction of the local trivializations of this fiberwise GNS Hilbert bundle grant important insights that we will use to construct a different bundle. This second bundle will be nontrivial and serves as a generalization of the bundle $\Unitary(1) \rightarrow \sphere \hilbH \rightarrow \bbP \hilbH$ to the $C^*$-algebraic setting.

Let us establish some notation. Given $\omega \in \pstate(\fA)$ and $r > 0$, let $\ball_r(\omega) \subset \pstate(\fA)$ be the open ball of radius $r$ centered on $\omega$, with respect to the metric induced by the norm. Given $\scrl \in \bbP \hilbH_\omega$, let $\ball_r(\scrl)$ be the open ball of radius $r$ centered on $\scrl$, with respect to the gap metric. By Corollary \ref{cor:PH_superselection_metric_equivalence} we have a homeomorphism $\ball_2(\omega) \rightarrow \ball_1(\bbC \Omega_\omega)$ which assigns to $\psi \in \ball_2(\omega)$ the unique ray $\scrl \in \bbP \hilbH_\omega$ representing $\psi$. By Corollary \ref{cor:PH_mapsto_SH}, we have continuous map $\ball_1(\bbC \Omega_\omega) \rightarrow \sphere \hilbH_\omega$ which assigns to each ray $\scrl \in \ball_1(\bbC \Omega_\omega)$ the unique representative $\Psi \in \scrl \cap \sphere \hilbH_\omega$ such that $\ev{\Psi,\Omega_\omega} > 0$. Composing these maps, we have a continuous map 
\[
\Psi_\omega: \ball_2(\omega) \rightarrow \sphere \hilbH_\omega, \quad \psi \mapsto \Psi_\omega(\psi)
\]
where $\Psi_\omega(\psi)$ is the unique unit vector in $\hilbH_\omega$ satisfying 
\[
\ev{\Psi_\omega(\psi), \Omega_\omega} > 0 \qqtext{and} \psi(A) = \ev{\Psi_\omega(\psi), \pi_\omega(A)\Psi_\omega(\psi)}
\]
for all $A \in \fA$.  Note that $\Psi_\omega(\omega) = \Omega_\omega$. The subscripts here are meant to help one keep track of which Hilbert space a vector is in, for example $\Psi_\omega(\psi) \in \hilbH_\omega$ and similarly $\Omega_\omega \in \hilbH_\omega$.

If $\psi \in \ball_2(\omega)$, then by uniqueness of the GNS construction up to unitary equivalence, there exists a unique unitary $U_{\omega \psi}:\hilbH_\psi \rightarrow \hilbH_\omega$ such that
\begin{equation}\label{eq:unique_unitary_GNS_intertwiner}
U_{\omega \psi} \Omega_\psi = \Psi_\omega(\psi) \qqtext{and} U_{\omega \psi} \pi_\psi(A) = \pi_\omega(A) U_{\omega \psi}
\end{equation}
for all $A \in \fA$. Clearly $U_{\omega \omega} = \1$. The second equation says that $U_{\omega \psi}$ \textit{intertwines} the representations $\pi_\psi$ and $\pi_\omega$. At this point we make the following elementary observation.

\begin{proposition}
Let $(\hilbH_1, \pi_1)$ and $(\hilbH_2, \pi_2)$ be nonzero nondegenerate representations of $\fA$. If $U:\hilbH_1 \rightarrow \hilbH_2$ is a unitary intertwiner of these representations, then for any $\Phi \in \sphere \hilbH_1$, both $\Phi$ and $U\Phi$ represent the same state.
\end{proposition}

\begin{proof}
For any $A \in \fA$,
\[
\ev{U\Phi, \pi_2(A)U\Phi} = \ev{U\Phi, U\pi_1(A)\Phi} = \ev{\Phi, \pi_1(A)\Phi}. \qedhere
\]
\end{proof}

\begin{corollary}
If $\psi, \omega \in \pstate(\fA)$ and $\norm{\psi - \omega} < 2$, then $U_{\psi \omega}^{\phantom{1}} = U_{\omega \psi}^{-1}$.
\end{corollary}

\begin{proof}
Both $U_{\psi \omega}$ and $U_{\omega \psi}^{-1}$ are unitary intertwiners of the GNS representations of $\psi$ and $\omega$. Therefore $U_{\psi \omega}\Omega_\omega$ and $U_{\omega \psi}^{-1}\Omega_\omega$ both represent $\omega$.  Now we observe that
\[
\ev{U_{\omega \psi}^{-1}\Omega_\omega, \Omega_\psi} = \ev{\Omega_\omega, U_{\omega\psi} \Omega_\psi} = \ev{\Omega_\omega, \Psi_\omega(\psi)} > 0.
\]
We conclude that
\[
U_{\psi \omega}\Omega_\omega = U_{\omega \psi}^{-1}\Omega_\omega = \Psi_\psi(\omega),
\]
which is the unique unit vector in $\hilbH_\psi$ representing $\omega$ and having positive inner product with $\Omega_\psi$. By uniqueness of $U_{\psi \omega}$, we have $U_{\psi \omega} = U_{\omega \psi}^{-1}$.
\end{proof}

We observe that if we fix $\omega \in \pstate(\fA)$ and vary $\psi \in \ball_2(\omega)$, then $U_{\omega \psi}$ is beginning to look like a local trivialization with domain $\bigsqcup_{\psi \in \ball_2(\omega)} \hilbH_\psi$, in the sense that the diagram below commutes, albeit rather trivially, and the top arrow is a bijection.
\begin{equation}\label{eq:fiberwise_GNS_loc_triv}
\begin{tikzcd}
\bigsqcup\limits_{\psi \in \ball_2(\omega)} \hilbH_\psi \arrow[rr,"{(\psi,\Phi) \mapsto (\psi, U_{\omega \psi}\Phi)}"]\arrow[dr] && \ball_2(\omega) \times \hilbH_\omega \arrow[dl] \\
&\ball_2(\omega)&
\end{tikzcd}
\end{equation}
We now show that the transition functions are continuous.

\begin{proposition}\label{prop:unitary_transition_func_continuous}
Fix $\omega_1, \omega_2 \in \pstate(\fA)$ such that $\ball_2(\omega_1) \cap \ball_2(\omega_2) \neq \varnothing$. The function
\[
\ball_2(\omega_1) \cap \ball_2(\omega_2) \rightarrow \Unitary(\hilbH_1, \hilbH_2), \quad \psi \mapsto V_{\omega_2\omega_1}(\psi) \defeq U_{\omega_2\psi}U_{\psi \omega_1}
\]
is continuous with respect to the norm topologies on the domain and codomain.
\end{proposition}

\begin{proof}
Let $\chi, \psi \in \ball_2(\omega_1) \cap \ball_2(\omega_2)$ and let $\Phi \in \sphere \hilbH_{\omega_1}$ be arbitrary. By the Kadison transitivity theorem, there exists $A \in \fA$ with $\norm{A} \leq 1$ such that $\pi_{\omega_1}(A)\Psi_{\omega_1}(\psi) = \Phi$. Then
\begin{align*}
\norm{V_{\omega_2\omega_1}(\psi)\Phi - V_{\omega_2\omega_1}(\chi) \Phi} &= \norm{V_{\omega_2\omega_1}(\psi)\pi_{\omega_1}(A)\Psi_{\omega_1}(\psi) - V_{\omega_2\omega_1}(\chi)\pi_{\omega_1}(A)\Psi_{\omega_1}(\psi)}\\
&= \norm{\pi_{\omega_2}(A) V_{\omega_2\omega_1}(\psi)\Psi_{\omega_1}(\psi) - \pi_{\omega_2}(A)V_{\omega_2\omega_1}(\chi)\Psi_{\omega_1}(\psi) }\\
&\leq \norm{V_{\omega_2\omega_1}(\psi)\Psi_{\omega_1}(\psi) - V_{\omega_2\omega_1}(\chi)\Psi_{\omega_1}(\psi)},
\end{align*}
where we have used the fact that $\norm{\pi_{\omega_1}(A)} \leq \norm{A} \leq 1$. Our next step is to use the triangle inequality:
\begin{align*}
\norm{V_{\omega_2\omega_1}(\psi)\Phi - V_{\omega_2\omega_1}(\chi) \Phi} &\leq \norm{V_{\omega_2\omega_1}(\psi)\Psi_{\omega_1}(\psi) - V_{\omega_2\omega_1}(\chi)\Psi_{\omega_1}(\chi)} \\
&\qquad + \norm{V_{\omega_2\omega_1}(\chi)\Psi_{\omega_1}(\chi) - V_{\omega_2\omega_1}(\chi)\Psi_{\omega_1}(\psi)}\\
&= \norm{\Psi_{\omega_2}(\psi) - \Psi_{\omega_2}(\chi)} + \norm{V_{\omega_2\omega_1}(\chi)\Psi_{\omega_1}(\chi) - V_{\omega_2\omega_1}(\chi)\Psi_{\omega_1}(\psi)}\\
&\leq \norm{\Psi_{\omega_2}(\psi) - \Psi_{\omega_2}(\chi)} + \norm{\Psi_{\omega_1}(\psi) - \Psi_{\omega_1}(\chi)},
\end{align*}
where in the last line we have used the fact that $\norm{V_{\omega_2\omega_1}(\chi)} = 1$. Since $\Phi$ was arbitrary, we conclude that
\[
\norm{V_{\omega_2\omega_1}(\psi) - V_{\omega_2\omega_1}(\chi)} \leq  \norm{\Psi_{\omega_2}(\psi) - \Psi_{\omega_2}(\chi)} + \norm{\Psi_{\omega_1}(\psi) - \Psi_{\omega_1}(\chi)}.
\]
If we now fix $\psi$, then continuity of $\Psi_{\omega_1}$ and $\Psi_{\omega_2}$ implies that $\norm{V_{\omega_2\omega_1}(\psi) - V_{\omega_2\omega_1}(\chi)}$ can be made arbitrarily small if $\chi$ is closed enough to $\psi$. This is what we wanted to show.
\end{proof}

The following is now immediate from Proposition \ref{prop:unitary_transition_func_continuous} and Theorem \ref{thm:topologize_total_space}. 

\begin{corollary}
There exists a unique topology on $\bigsqcup_{\omega \in \pstate(\fA)} \hilbH_\omega$ such that $p:\bigsqcup_{\omega \in \pstate(\fA)} \hilbH_\omega \rightarrow \pstate(\fA)$ is continuous and the top arrow in the diagram \eqref{eq:fiberwise_GNS_loc_triv} is a homeomorphism for all $\omega \in \pstate(\fA)$.
\end{corollary}

Over each superselection sector $\sector \subset \pstate(\fA)$ the GNS Hilbert spaces are all isomorphic to each other. As mentioned previously, if the dimension of these spaces is infinite, then it is a consequence of Kuiper's theorem that the GNS Hilbert bundle is trivial over that superselection sector \cite{Kuiper,SchottenloherUnitaryStrongTopology}. Thus, in the infinite-dimensional case, we seem to have constructed a trivial bundle. 

However, by viewing the structure group of our bundle as the set of all unitaries, we miss an important point, addressed by the following proposition.

\begin{proposition}\label{prop:only_phase_changes}
Fix $\omega_1, \omega_2 \in \pstate(\fA)$ such that $\ball_2(\omega_1) \cap \ball_2(\omega_2) \neq \varnothing$. For any two pure states $\psi, \chi \in \ball_2(\omega_1) \cap \ball_2(\omega_2)$, there exists $\mu \in \Unitary(1)$ such that
\[
V_{\omega_2\omega_1}(\psi)= \mu V_{\omega_2\omega_1}(\chi).
\]
\end{proposition}

\begin{proof}
Observe that $V_{\omega_2\omega_1}(\chi)^{-1} V_{\omega_2\omega_1}(\psi)$ intertwines $\pi_{\omega_1}$ with itself. Since $\pi_{\omega_1}$ is irreducible, Schur's lemma implies $V_{\omega_2\omega_1}(\chi)^{-1} V_{\omega_2\omega_1}(\psi) = \mu\1$ for some $\mu \in \Unitary(1)$.
\end{proof}

Proposition \ref{prop:only_phase_changes} shows that the only thing that's changing in our transition functions $V_{\omega_2 \omega_1}$ is a phase. This suggests that we look for a $\Unitary(1)$-bundle. If we shrink the domains of our local trivializations from the radius-two balls $\ball_2(\omega)$ to the unit balls $\ball_1(\omega)$, then whenever we have a nonempty intersection $\ball_1(\omega_1) \cap \ball_1(\omega_2) \neq \varnothing$, we have $\norm{\omega_1 - \omega_2} < 2$, so the unitary $U_{\omega_2 \omega_1}$ is defined. This gives us a preferred unitary against which we can measure the phase of $V_{\omega_2 \omega_1}(\psi)$.

\begin{proposition}
Fix $\omega_1, \omega_2 \in \pstate(\fA)$ such that $\ball_1(\omega_1) \cap \ball_1(\omega_2) \neq \varnothing$. There exists a unique function
\[
\mu_{\omega_1 \omega_2}(\psi) :\ball_1(\omega_1) \cap \ball_1(\omega_2) \rightarrow \Unitary(1)
\]
such that
\begin{equation}\label{eq:phase_comparison_def}
V_{\omega_2\omega_1}(\psi) = \mu_{\omega_1 \omega_2}(\psi) U_{\omega_2\omega_1}
\end{equation}
for all $\psi \in \ball_1(\omega_1) \cap \ball_1(\omega_2)$. 
\end{proposition}

\begin{proof}
Setting $\chi = \omega_1$ in Proposition \ref{prop:only_phase_changes} and noting that $U_{\omega_1\omega_1} = \1$ reveals that for every $\psi \in \ball_1(\omega_1) \cap \ball_1(\omega_2)$, there exists $\mu_{\omega_1 \omega_2}(\psi) \in \Unitary(1)$ such that \eqref{eq:phase_comparison_def} holds. Uniqueness of $\mu_{\omega_1 \omega_2}$ is obvious. Continuity follows from Proposition \ref{prop:unitary_transition_func_continuous}.
\end{proof}

The functions $\mu_{\omega_1 \omega_2}$ form a \v{C}ech 1-cochain with respect to the open cover of $\pstate(\fA)$ by unit balls $\ball_1(\omega)$. If we take the differential $d\mu$ and examine the value on any element of a triple overlap $\psi \in \ball_1(\omega_1) \cap \ball_1(\omega_2) \cap \ball_1(\omega_3)$, we find
\begin{align*}
d\mu_{\omega_1 \omega_2 \omega_3}(\psi) \cdot U_{\omega_3 \omega_2} U_{\omega_2\omega_1} &= \mu_{\omega_2 \omega_3}(\psi)\mu_{\omega_1 \omega_3}(\psi)^{-1} \mu_{\omega_1\omega_2}(\psi) \cdot U_{\omega_3 \omega_2} U_{\omega_2\omega_1}\\
&= \mu_{\omega_1\omega_3}(\psi)^{-1} V_{\omega_3\omega_2}(\psi)V_{\omega_2\omega_1}(\psi)\\
&= \mu_{\omega_1\omega_3}(\psi)^{-1} V_{\omega_3\omega_1}(\psi)\\
&= U_{\omega_3\omega_1}\\
&= \mu_{\omega_1 \omega_3}(\omega_2)^{-1}V_{\omega_3\omega_1}(\omega_2)
\end{align*}
We conclude that $d\mu_{\omega_1\omega_2\omega_3}(\psi)$ is constant at $\mu_{\omega_1\omega_3}(\omega_2)^{-1}$ for all $\psi \in \ball_1(\omega_1)\cap \ball_1(\omega_2) \cap \ball_1(\omega_3)$. Now $d\mu$ is of course a $2$-coboundary, hence a $2$-cocycle $d\mu \in \check{Z}^2(\pstate(\fA);\Unitary(1))$, but the fact that $d\mu$ is constant for every triple overlap means that $d\mu$ is a $2$-cocycle with respect to the \textit{discrete} topology on $\Unitary(1)$, whereas $\mu$ was a $1$-cochain with respect to the standard topology on $\Unitary(1)$. It is unknown to the author if $d\mu$ is a $2$-coboundary with respect to the discrete topology on $\Unitary(1)$, so it is conceivable that we have produced a nontrivial class in $\check{H}^2(\pstate(\fA); \Unitary(1)_\tn{d})$, where $\Unitary(1)_\tn{d}$ is $\Unitary(1)$ with the discrete topology. We are unsure what the geometric implications of this would be. 

These observations are not immediately helpful for constructing fiber bundles, so we will not pursue this 2-cocycle further. To make progress on the construction of a $\Unitary(1)$-bundle, we first fix an arbitrary superselection sector $\sector \subset \pstate(\fA)$ and focus on constructing a $\Unitary(1)$-bundle over $\sector$. We choose a nonzero irreducible representation $(\hilbH, \pi)$ of $\fA$ such that $\sector = \pstate_\pi(\fA)$. We remind the reader that $\pstate_\pi(\fA)$ is the set of vector states of $\pi$. For every $\omega \in \sector$, we choose a unitary $U_\omega : \hilbH_\omega \rightarrow \hilbH$ that intertwines $\pi_\omega$ and $\pi$, i.e.,
\[
U_\omega\pi_\omega(A) = \pi(A)U_\omega
\]
for all $A \in \fA$. From this we will construct a $1$-cocycle $\lambda$ and show that the cohomology class is independent of the choice of $(\hilbH, \pi)$ and of the unitaries $U_\omega$. 

We cover $\sector$ by balls of radius two, i.e., we choose the open cover $\cO = \qty{\ball_2(\omega)}_{\omega \in \sector}$. Then for any $\omega_1, \omega_2 \in \sector$ such that $\ball_2(\omega_1) \cap \ball_2(\omega_2) \neq \varnothing$, we define
\[
W_{\omega_2\omega_1} : \ball_2(\omega_1) \cap \ball_2(\omega_2) \rightarrow \Unitary(\hilbH), \quad W_{\omega_2 \omega_1}(\psi) \defeq U_{\omega_2} U_{\omega_2 \psi} U_{\psi \omega_1} U_{\omega_1}^{-1}.
\]
By Proposition \ref{prop:unitary_transition_func_continuous} we know $W_{\omega_2 \omega_1}$ is norm-continuous. Since $W_{\omega_2 \omega_1}(\psi)$ intertwines $\pi$ with itself, Schur's lemma implies there exists a unique function $\lambda_{\omega_1 \omega_2}(\psi) :\ball_2(\omega_1)\cap \ball_2(\omega_2) \rightarrow \Unitary(1)$ such that 
\begin{equation}\label{eq:lambda_1-cocycle_def}
W_{\omega_2 \omega_1}(\psi) = \lambda_{\omega_1\omega_2}(\psi)\1
\end{equation}
for all $\psi \in \ball_2(\omega_1) \cap \ball_2(\omega_2)$. By continuity of $W_{\omega_2\omega_1}$, we know $\lambda_{\omega_1 \omega_2}$ is continuous. The reversal of the subscripts between $W_{\omega_2 \omega_1}$ and $\lambda_{\omega_1 \omega_2}$ has a purpose that will become apparent later.

\begin{theorem}
The assignment $(\omega_1, \omega_2) \mapsto \lambda_{\omega_1\omega_2}$ forms a \v{C}ech 1-cocycle $\lambda \in \check{Z}^1(\cO; \Unitary(1))$. The cohomology class of $\lambda$ is independent of the choices for the representation $(\hilbH, \pi)$ and unitaries $U_\omega$.
\end{theorem}

\begin{proof}
Given $\omega_1, \omega_2, \omega_3 \in \sector$ and $\psi\in  \ball_2(\omega_1) \cap \ball_2(\omega_2) \cap \ball_2(\omega_3)$, we have
\begin{align*}
\lambda_{\omega_1\omega_2}(\psi)\lambda_{\omega_2\omega_3}(\psi)\cdot\1 &= W_{\omega_3\omega_2}(\psi)W_{\omega_2\omega_1}(\psi)\\
&= U_{\omega_3}U_{\omega_3\psi}U_{\psi\omega_2}U_{\omega_2}^{-1}U_{\omega_2}U_{\omega_2\psi}U_{\psi \omega_1}U_{\omega_1}^{-1}\\
&= U_{\omega_3}U_{\omega_3\psi}U_{\psi \omega_1}U_{\omega_1}^{-1}\\
&= W_{\omega_3 \omega_1}(\psi)\\
&= \lambda_{\omega_1\omega_3}(\psi)\cdot\1.
\end{align*}
This proves that $\lambda \in \check{Z}^1(\cO; \Unitary(1))$.

Suppose we choose another nonzero irreducible representation $(\hilbK,\rho)$ such that $\sector = \pstate_\rho(\fA)$, and we choose other  unitaries $V_\omega:\hilbH_\omega \rightarrow \hilbK$ intertwining $\pi_\omega$ and $\rho$. Let $\mu \in \check{Z}^1(\cO;\Unitary(1))$ be the resulting 1-cocycle. 

Choose a unitary $T:\hilbH \rightarrow \hilbK$ intertwining $\pi$ and $\rho$. For each $\omega \in \sector$, observe that $U_\omega^{-1}T^{-1}V_\omega$ intertwines $\pi_\omega$ with itself, hence $U_\omega^{-1}T^{-1}V_\omega = \nu_\omega \1$ for some $\nu_\omega \in \Unitary(1)$. Finally, given $\omega_1, \omega_2 \in \sector$ and $\psi \in \ball_2(\omega_1) \cap \ball_2(\omega_2)$, we have
\begin{align*}
\lambda_{\omega_1 \omega_2}(\psi)\cdot \1 &= U_{\omega_2}U_{\omega_2\psi}U_{\psi \omega_1}U_{\omega_1}^{-1}\\
&= T^{-1} V_{\omega_2} \qty(V_{\omega_2}^{-1}T U_{\omega_2}) U_{\omega_2\psi}U_{\psi \omega_1}\qty(U_{\omega_1}^{-1}T^{-1}V_{\omega_1})V_{\omega_1}^{-1}T\\
&= \nu_{\omega_1} \cdot T^{-1}\qty(V_{\omega_2}U_{\omega_2 \psi}U_{\psi \omega_1} V_{\omega_1}^{-1})T \cdot \nu_{\omega_2}^{-1} \\
&= \nu_{\omega_1}\mu_{\omega_1\omega_2}(\psi) \nu_{\omega_2}^{-1} \cdot T^{-1}T = \nu_{\omega_1}\mu_{\omega_1\omega_2}(\psi)\nu_{\omega_2}^{-1} \cdot \1
\end{align*}
This proves that $\lambda$ and $\mu$ are cohomologous.
\end{proof}

We have thus constructed a ``representation-independent'' cohomology class $\class{\lambda} \in \check{H}^1(\cO; \Unitary(1))$. Applying the canonical inclusion $\iota_\cO:\check{H}^1(\cO;\Unitary(1)) \rightarrow \check{H}^1(\sector;\Unitary(1))$ gives a cohomology class $\iota_\cO \class{\lambda}$. Principal $\Unitary(1)$-bundles over $\sector$ are classified by $\check{H}^1(\sector;\Unitary(1))$, so $\iota_\cO\class{\lambda}$ corresponds to an isomorphism class of principal $G$-bundles. Thus, we have essentially achieved our goal! 

Let us show that this cohomology class is a generalization of the cohomology class arising from the bundle $\Unitary(1) \rightarrow \sphere \hilbH \rightarrow \bbP \hilbH$. 

\begin{theorem}\label{thm:SH->PH_generalization}
If $(\hilbH, \pi)$ is a nonzero irreducible representation of $\fA$ such that $\sector = \pstate_\pi(\fA)$ and $f:\bbP\hilbH \rightarrow \sector$ is the homeomorphism associating to each ray in $\bbP\hilbH$ the state in $\sector$ it represents, then $f^*\iota_\cO\class{\lambda} \in \check{H}^1(\bbP\hilbH; \Unitary(1))$ corresponds to the isomorphism class of the bundle $\Unitary(1) \rightarrow \sphere \hilbH \rightarrow \bbP\hilbH$.
\end{theorem}

This is the reason for the aforementioned subscript reversal between $W_{\omega_2 \omega_1}$ and $\lambda_{\omega_1\omega_2}$. With the opposite ordering of the subscripts on $\lambda$, we would have $f^*\iota_\cO\class{\lambda}$ corresponding to the inverse of the isomorphism class of $\Unitary(1) \rightarrow \sphere \hilbH \rightarrow \bbP \hilbH$.

\begin{proof}
For each $\omega \in \sector$, choose a representing unit vector $\Phi_\omega \in \sphere \hilbH$. Then by Corollary \ref{cor:PH_superselection_metric_equivalence} we have an open cover of $\bbP\hilbH$ defined as 
\[
f^{-1}(\cO) = \qty{f^{-1}(\ball_2(\omega)) : \omega \in \sector} =  \qty{\ball_1(\bbC \Phi_\omega)}_{\omega \in \sector},
\] 
where $\ball_1(\bbC\Phi_\omega)$ is the unit ball around $\bbC \Phi_\omega$ with respect to the gap metric. Recall from Theorem \ref{thm:local_triv_rho} that we an atlas of local trivializations of $\Unitary(1) \rightarrow \sphere \hilbH \rightarrow \bbP \hilbH$, where the local trivializations are defined as
\[
\rho_{\Phi_\omega} :\sphere \hilbH \setminus (\sphere \hilbH \cap \bbC \Phi_\omega^\perp) \rightarrow \ball_1(\bbC \Phi_\omega) \times \Unitary(1), \quad \rho_{\Phi_\omega}(\Psi) = \qty(\bbC \Psi, \frac{\ev{\Phi_\omega, \Psi}}{\abs{\ev{\Phi_\omega, \Psi}}}),
\]
where $\ball_1(\bbC \Phi_\omega)$ is the unit ball about $\bbC \Phi_\omega$ with respect to the gap metric. 

We compute the transition functions associated to these local trivializations. The inverse of $\rho_{\Phi_\omega}$ is given by
\[
\rho_{\Phi_\omega}^{-1}(\bbC \Psi, \mu) = \mu \cdot \frac{\abs{\ev{\Phi_\omega, \Psi}}}{\ev{\Phi_\omega, \Psi}}\cdot\Psi.
\]
Therefore, given $\omega_1, \omega_2 \in \sector$ such that $\ball_1(\bbC \Phi_{\omega_1}) \cap \ball_1(\bbC \Phi_{\omega_2}) \neq \varnothing$ the corresponding transition function $\ball_1(\bbC \Phi_{\omega_1}) \cap \ball_1(\bbC \Phi_{\omega_2}) \rightarrow \Unitary(1)$ is given by the second component of
\begin{equation}\label{eq:U(1)-SH-PH_transition_func}
\qty(\tn{proj}_{\Unitary(1)} \circ \rho_{\Phi_{\omega_1}}\circ\rho_{\Phi_{\omega_2}}^{-1})(\bbC \Psi, 1) = \frac{\abs{\ev{\Phi_{\omega_2}, \Psi}}}{\ev{\Phi_{\omega_2}, \Psi}} \frac{\ev{\Phi_{\omega_1}, \Psi}}{\abs{\ev{\Phi_{\omega_1}, \Psi}}},
\end{equation}
where $\tn{proj}_{\Unitary(1)}:\ball_1(\bbC \Phi_{\omega_1}) \times \Unitary(1) \rightarrow \Unitary(1)$ is the projection onto $\Unitary(1)$. 
This is a $1$-cocycle in $\check{Z}^1(f^{-1}(\cO);\Unitary(1))$ corresponding to the bundle $\Unitary(1) \rightarrow \sphere \hilbH \rightarrow \bbP \hilbH$. Notice that the right hand side is independent of the choice of representative $\Psi \in \bbC \Psi$.

We now compute $f^*\iota_\cO \class{\lambda} = \iota_{f^{-1}(\cO)} f^* \class{\lambda} = \iota_{f^{-1}(\cO)}\class{\lambda \circ f}$. We show that $\lambda_{\omega_1 \omega_2} \circ f$ is identically \eqref{eq:U(1)-SH-PH_transition_func}. First, to even define $\lambda \in \check{Z}^1(\cO;\Unitary(1))$, we need to choose the unitaries $U_{\omega} :\hilbH_\omega \rightarrow \hilbH$. Given $\omega \in \sector$, we choose $U_\omega$ to be the unique unitary intertwining $\pi_\omega$ and $\pi$ such that $U_\omega \Omega_\omega = \Phi_{\omega}$. This gives rise to $\lambda \in \check{Z}^1(\cO;\Unitary(1))$ as defined in \eqref{eq:lambda_1-cocycle_def}. 

Given $\omega_1, \omega_2 \in \sector$ and $\bbC\Psi \in \ball_1(\bbC\Phi_{\omega_1}) \cap \ball_1(\bbC\Phi_{\omega_2})$, we now show that $\lambda_{\omega_1\omega_2}(f(\bbC \Psi))$ is exactly \eqref{eq:U(1)-SH-PH_transition_func}. Denote $\psi = f(\bbC \Psi)$ and choose the representative $\Psi$ to be
\[
\Psi = U_{\omega_1}U_{\omega_1\psi}\Omega_\psi.
\]
Observe that
\[
\ev{\Phi_{\omega_1}, \Psi} = \ev{U_{\omega_1}\Omega_{\omega_1}, U_{\omega_1}U_{\omega_1\psi}\Omega_\psi} = \ev{\Omega_{\omega_1}, U_{\omega_1 \psi}\Omega_\psi} > 0,
\]
where positivity follows by definition of $U_{\omega_1\psi}$. Next, observe that
\begin{align*}
0 &< \ev{U_{\psi \omega_2} \Omega_{\omega_2}, \Omega_\psi} = \ev{U_{\psi \omega_2}U_{\omega_2}^{-1}\Phi_{\omega_2}, U_{\psi \omega_1}U_{\omega_1}^{-1}\Psi} = \ev{\Phi_{\omega_2}, W_{\omega_2\omega_1}(\psi)\Psi} = \lambda_{\omega_1\omega_2}(\psi) \ev{\Phi_{\omega_2}, \Psi}.
\end{align*}  
Taking the absolute value of the quantity on the right does nothing since it is already positive, so this yields $\lambda_{\omega_1 \omega_2}(\psi) \ev{\Phi_{\omega_2},\Psi} = \abs{\ev{\Phi_{\omega_2}, \Psi}}$. Thus,
\[
\lambda_{\omega_1\omega_2}(f(\bbC \Psi)) = \frac{\abs{\ev{\Phi_{\omega_2}, \Psi}}}{\ev{\Phi_{\omega_2}, \Psi}} = \frac{\abs{\ev{\Phi_{\omega_2}, \Psi}}}{\ev{\Phi_{\omega_2}, \Psi}}\frac{\ev{\Phi_{\omega_1}, \Psi}}{\abs{\ev{\Phi_{\omega_1}, \Psi}}}.
\]
This is what we wanted to show.
\end{proof}

The following corollary is immediate and concisely summarizes the main result of this section.

\begin{corollary}
Let $\sector$ be a superselection sector of $\fA$ and let $(\hilbH, \pi)$ be a nonzero irreducible representation such that $\sector = \pstate_\pi(\fA)$. Let $f:\bbP \hilbH \rightarrow \sector$ be the homeomorphism that associates to each ray in $\bbP\hilbH$ the state it represents. Consider the induced map on degree one \v{C}ech cohomology $(f^{-1})^*: \check{H}^1(\bbP \hilbH;\Unitary(1)) \rightarrow \check{H}^1(\pstate_\pi(\fA); \Unitary(1))$. The image under $(f^{-1})^*$ of the cohomology class corresponding to the bundle $\Unitary(1) \rightarrow \sphere \hilbH \rightarrow \bbP \hilbH$ is independent of the representation $(\hilbH,\pi)$.
\end{corollary}

\section{The Continuous Kadison Transitivity Theorem}
\label{sec:continuous_kadison}

In the previous section, we constructed a bundle over $\pstate(\fA)$ whose fiber over $\omega \in \pstate(\fA)$ was the GNS Hilbert space $\hilbH_\omega$. A natural follow-up question is to ask whether we can construct a bundle over $\pstate(\fA)$ whose fiber over $\omega$ is the Gelfand ideal $\fN_\omega$, in other words, whether the canonical map
\begin{equation}\label{eq:Gelfand_bundle}
\bigsqcup_{\omega \in \pstate(\fA)} \fN_\omega \rightarrow \pstate(\fA)
\end{equation}
can be endowed with a bundle structure. The Gelfand ideals $\fN_\omega$ are Banach algebras, so Kuiper's theorem does not apply in this case, in particular such a bundle need not be trivial. And even though the GNS Hilbert bundle turned out to be trivial in the infinite-dimensional case, valuable insights were nonetheless gained from its construction, leading to the construction of a different nontrivial bundle. The story will play out much the same way for the bundle of Gelfand ideals, with even more impressive results. 

In the fiberwise GNS construction, a local trivialization around a given $\omega \in \pstate(\fA)$ was built from a family of unitaries $U_{\omega \psi}:\hilbH_\psi \rightarrow \hilbH_\omega$ indexed by $\psi \in \ball_2(\omega)$. To construct a local trivialization for \eqref{eq:Gelfand_bundle}, we might start by looking for a family of Banach algebra isomorphisms $\alpha_{\omega \psi} : \fN_\psi \rightarrow \fN_\omega$. We note that if $\alpha_{\omega \psi} :\fA \rightarrow \fA$ is a $C^*$-algebra automorphism and 
\[
\psi = \omega \circ \alpha_{\omega \psi},
\]
then $\alpha_{\omega \psi}$ restricts to a Banach algebra isomorphism $\fN_\psi \rightarrow \fN_\omega$. In fact, if $\fA$ is unital then by Corollary \ref{cor:sector_is_open} and Theorem \ref{thm:superselection_sector_equivalences}, we know that for every $\psi \in \ball_2(\omega)$ there exists a unitary $U_{\omega \psi} \in \Unitary(\fA)$ such that
\[
\psi = \omega \circ \Ad(U_{\omega \psi}),
\]
where $\Ad(U_{\omega \psi}):\fA \rightarrow \fA$ is the inner automorphism $\Ad(U_{\omega \psi})(A) = U_{\omega \psi}A U_{\omega \psi}^*$. This is a consequence of the Kadison transitivity theorem. If we can choose the unitaries $U_{\omega \psi}$ to depend continuously on $\psi \in \ball_2(\omega)$, then our transition functions for $\omega_1, \omega_2 \in \pstate(\fA)$ with $\ball_2(\omega_1) \cap \ball_2(\omega_2) \neq \varnothing$
\[
\alpha_{\omega_2 \psi} \circ \alpha_{\psi \omega_1}^{-1} = \Ad(U_{\omega_2 \psi}) \circ \Ad(U_{\omega_1 \psi}^*)
\]
will be continuous, and our bundle structure will follow.

More precisely, recall that the Kadison transitivity theorem states that whenever a $C^*$-algebra $\fA$ acts
irreducibly on a Hilbert space $\hilbH$, there exists for every pair of $n$-tuples
of vectors $x_1,\ldots, x_n$ and $y_1,\ldots, y_n$ in $\hilbH$ such that $x_1,\ldots, x_n$
are linearly independent an element $A \in \fA$ such that $Ax_k = y_k$ for
$k=1,\ldots ,n$; see Section \ref{sec:Kadison_transitivity}. 
However, the solution to this problem---that is, the element $A \in \fA$---is in general not unique. The question of whether we can choose the unitaries $U_{\omega \psi}$ to depend continuously on $\psi$ can be generalized to the question of whether it is possible to choose the solutions $A \in \fA$ so as to depend continuously on the initial data $x_1,\ldots, x_n$ and $y_1,\ldots, y_n$. This is a problem amenable to the theory of selections developed by Ernest Michael in the 1950s, and indeed we use the Michael selection theorem to provide an affirmative answer to our question. In Section \ref{sec:main_results}, we recall the necessary terminology and results from Michael's original work on selections \cite{MichaelSelection}, then we prove our main results in Theorems \ref{thm:continuous_Kadison} and \ref{thm:continuous_Kadison_unitary}. We call these the \textdef{continuous Kadison transitivity theorems}. Theorem \ref{thm:continuous_Kadison_unitary} is by far the most difficult result of this dissertation. In Section \ref{sec:principal_fiber_bundles} we use Theorem \ref{thm:continuous_Kadison_unitary} to prove that $p_{\Unitary(\fA)}:\Unitary(\fA) \rightarrow \pstate_\omega(\fA)$, $p_{\Unitary(\fA)}(U) = U \cdot \omega$ has the structure of a principal $\Unitary_\omega(\fA)$-bundle, where $\Unitary_\omega(\fA) = \qty{U \in \Unitary(\fA): U \cdot \omega = \omega}$, for any unital $C^*$-algebra $\fA$ and pure state $\omega \in \pstate(\fA)$. We provide a few examples where this bundle is nontrivial. We conclude with a precise statement of the construction of the bundle of Gelfand ideals \eqref{eq:Gelfand_bundle}.

\subsection{Main results}\label{sec:main_results}

The key ingredient in proving our continuous Kadison transitivity theorems is the Michael selection theorem. We provide this result and the necessary definitions below.

\begin{defn}[{\cite{MichaelSelection}}]
Let $X$ and $Y$ be topological spaces and let $\npset(Y)$ be the set of nonempty subsets of $Y$. A \textdef{carrier} is a  function $\phi : X \rightarrow \npset(Y)$. A \textdef{selection} for $\phi$ is a continuous function $f:X \rightarrow Y$ such that $f(x) \in \phi(x)$ for all $x \in X$. A carrier $\phi$ is \textdef{lower semicontinuous} if for every open set $O_Y \subset Y$, the set
\[
\qty{x \in X: \phi(x) \cap O_Y \neq \varnothing}
\]
is open in $X$. Equivalently, $\phi$ is lower semicontinuous if for every $x_0 \in X$, $y_0 \in \phi(x_0)$, and neighborhood $O_Y$ of $y_0$, there exists a neighborhood $O_X$ of $x_0$ such that $\phi(x) \cap O_Y \neq \varnothing$ for all $x \in O_X$. 
\end{defn}

We will use the latter description of lower semicontinuity in our proofs. If $Y$ is metrizable, as it will be for our applications, the neighborhood $O_Y$ may be taken to be a ball of radius $\varepsilon > 0$ centered on $y_0$. The space $X$ will always be metrizable in our applications as well. We now state the Michael selection theorem for reference.

\begin{thm}[{\cite[Thm.\ 3.2$''$]{MichaelSelection}}]
Let $X$ be a paracompact Hausdorff space and let $Y$ be a real or complex Banach space. If $\phi:X \rightarrow \npset(Y)$ is a lower semicontinuous carrier such that $\phi(x)$ is closed and convex for all $x \in X$, then there exists a selection for $\phi$.
\end{thm}

Recall that metric spaces are paracompact Hausdorff. Since our spaces $X$ will be metrizable, the paracompact Hausdorff assumption will always be satisfied for us.

To apply the Michael selection theorem to the representation theory of $C^*$-algebras, we will use the following result. It is a lemma used in proving the Kadison transitivity theorem.

\begin{lem}[{\cite[Lem.\ 5.4.2]{KadisonRingroseI}}]\label{lem:transitivity_in_B(H)}
Let $\hilbH$ be a Hilbert space and let $e_1,\ldots, e_n \in \hilbH$ be an orthonormal system. For any vectors $z_1,\ldots, z_n \in \hilbH$ such that $\norm{z_i} \leq r$ for all $i$, there exists $T \in \cB(\hilbH)$ such that $\norm{T} \leq (2n)^{1/2}r$ and $Te_i = z_i$ for all $i$. If there exists a self-adjoint operator $S \in \cB(\hilbH)$ such that $Se_i = z_i$ for all $i$, then $T$ may be chosen to be self-adjoint.
\end{lem}

The norm bounds in the Kadison transitivity theorem as outlined in Theorem \ref{thm:Kadison_transitivity_theorem} will also be instrumental in proving lower semicontinuity of the carriers that we consider. For the convenience of the reader, we reproduce the essential statement, slightly rephrased.

\begin{thm}\label{thm:Pedersen}
Let $\frA$ be a $C^*$-algebra and let $(\hilbH, \pi)$ be a nonzero irreducible representation. If $x_1,\ldots, x_n \in \hilbH$ are linearly independent and $T \in \cB(\hilbH)$, then there exists $A \in \frA$ such that $\norm{A} \leq \norm{T}$ and $\pi(A)x_i = Tx_i$ for all $i$. If $T$ is self-adjoint, then $A$ may be chosen to be self-adjoint. 
\end{thm}

We are now ready to prove the ``continuous Kadison transitivity theorem'' in the general and self-adjoint cases. For notation, when $\hilbH$ is a Hilbert space we denote elements of the Hilbert space $\hilbH^n$ by bold letters $\x = (x_1,\ldots, x_n)$ and elements of $\hilbH^{2n}$ by pairs of bold letters $(\x, \y)$. Given an element $T \in \cB(\hilbH)$ and $n \in \bbN$, we denote $T^{\oplus n} = T\oplus \cdots \oplus T \in \cB(\hilbH^n)$.

\begin{thm}\label{thm:continuous_Kadison}
Let $\fA$ be a $C^*$-algebra, let $(\hilbH, \pi)$ be a nonzero irreducible representation, and let $n$ be a positive integer. Let
\[
X = \qty{(\x, \y) \in \hilbH^{2n}: x_1,\ldots, x_n \tn{ are linearly independent}},
\]
equipped with the subspace topology inherited from $\hilbH^{2n}$. There exists a continuous map $A : X \rightarrow \fA$ such that
\begin{equation}\label{eq:selection_criterion}
\pi(A(\x,\y))x_i = y_i \quad \tn{for all} \, \, i = 1,\ldots, n
\end{equation}
for all $(\x, \y) \in X$. Similarly, defining
\begin{align*}
X_\tn{sa} &= \qty{(\x, \y) \in X: \exists\, T \in \cB(\hilbH)_\tn{sa} \tn{ s.t.\ } Tx_i = y_i \tn{ for all }  i =1,\ldots, n},
\end{align*}
there exists a continuous map $A :X_\tn{sa} \rightarrow \fA_\tn{sa}$ satisfying \eqref{eq:selection_criterion} for all $(\x, \y) \in X_\tn{sa}$.
\end{thm}

\begin{proof}
Since $X$ and $X_\tn{sa}$ are subspaces of $\hilbH^{2n}$, they are metrizable, hence paracompact Hausdorff. We will use the Michael selection theorem for the carrier $\phi:X \rightarrow \npset(\fA)$ defined by
\[
\phi(\x, \y) = \qty{A \in \fA: \pi(A)x_i = y_i \tn{ for all $i = 1,\ldots, n$}}.
\]
For the self-adjoint  case, we define $\phi_\tn{sa}:X_\tn{sa} \rightarrow \npset(\fA)$ as
\begin{align*}
\phi_\tn{sa}(\x, \y) &= \fA_\tn{sa} \cap \phi(\x, \y)
\end{align*}

%note that $\fA_\tn{sa}$ is a real Banach space with the topology inherited from $\fA$, and we define $\phi_\tn{sa} :X_\tn{sa} \rightarrow \wp_+(\fA_\tn{sa})$ by  
%\[
%\phi_\tn{sa}(\x, \y) = \fA_\tn{sa} \cap \phi(\x, \y).
%\] 
By the Kadison transitivity theorem, $\phi(\x, \y)$ and $\phi_\tn{sa}(\x, \y)$ are nonempty for all $(\x, \y) \in X$ and$(\x, \y) \in X_\tn{sa}$, respectively.  Given $(\x, \y) \in X$, $t \in [0,1]$, and $A, B \in \phi(\x, \y)$, we have
\[
\pi(tA + (1 - t)B)x_i = t \pi(A) x_i + (1 - t)\pi(B)x_i = ty_i + (1 - t)y_i = y_i
\]
for all $i$, so $\phi(\x, \y)$ is convex. Since $\fA_\tn{sa}$ is convex, it follows that $\phi_\tn{sa}(\x, \y)$ is convex for all $(\x, \y) \in X_\tn{sa}$. Furthermore, if $\hat x_i : \cB(\hilbH) \rightarrow \hilbH$ denotes the evaluation map $\hat x_i(T) = Tx_i$, then we see that
\[
\phi(\x, \y) = \bigcap_{i=1}^n (\hat x_i \circ \pi)^{-1}(\qty{y_i}),
\]
so $\phi(\x, \y)$ is closed since $\hat x_i \circ \pi$ is continuous for each $i$. Since $\fA_\tn{sa}$ is closed in $\fA$, we see that $\phi_\tn{sa}(\x, \y)$ is closed for all $(\x, \y) \in X_\tn{sa}$.

All that remains to show is lower semicontinuity, then the result will follow immediately from the Michael selection theorem. Fix $(\x_0, \y_0) \in X$, $A_0 \in \phi(\x_0, \y_0)$, and let $\varepsilon > 0$; replace $X$ and $\phi$ by $X_\tn{sa}$ and $\phi_\tn{sa}$ for the self-adjoint case. Given $(\x, \y) \in X$, let $e_1(\x),\ldots, e_n(\x) \in \hilbH$ be the orthonormal basis obtained by applying the Gram-Schmidt method to $x_1,\ldots, x_n$, and let $\lambda_{ij}(\x) \in \bbC$ be such that $e_i(\x) = \sum_{j=1}^n \lambda_{ij}(\x)x_j$. Note that each $\lambda_{ij}(\x)$ is a continuous function $X \rightarrow \bbC$. Moreover, the matrix $\Lambda_\x = (\lambda_{ij}(\x))$ defines an invertible element $\Lambda_\x \in \cB(\hilbH^n)$. Observe that the map $X \rightarrow \cB(\hilbH^n)$, $(\x, \y) \mapsto \Lambda_\x$ is continuous, $\Lambda_\x \x = \be(\x)$, and $[\Lambda_\x, T^{\oplus n}] = 0$ for all $T \in \cB(\hilbH)$. Let $O_1$ be the preimage of the open ball of radius $\norm{\Lambda_{\x_0}}$ centered on $\Lambda_{\x_0}$ under the map $(\x, \y) \mapsto \Lambda_\x$. Let $O_2$ be the preimage of the open ball of radius $\varepsilon/(4n\norm{\Lambda_{\x_0}})$ centered at zero under the map $X\rightarrow \hilbH^n$, $(\x, \y) \mapsto \y - \pi(A_0)^{\oplus n}\x$, which is also continuous. Then $O = O_1 \cap O_2$ is a neighborhood of $(\x_0, \y_0)$ in $X$ and $(\x, \y) \in O$ implies 
\[
\norm{\Lambda_\x} < 2 \norm{\Lambda_{\x_0}} \qqtext{and} \norm{\y - \pi(A_0)^{\oplus n} \x} < \frac{\varepsilon}{4n\norm{\Lambda_{\x_0}}}.
\]
For the self-adjoint case we set $O_\tn{sa} = X_\tn{sa} \cap O$.

Given $(\x, \y) \in O$, set $\z(\x, \y) = \Lambda_\x \y$ and observe that $A \in \phi(\x, \y)$ if and only if $A \in \phi(\be(\x), \z(\x, \y))$ since $\Lambda_\x$ is invertible and commutes with $\pi(A)^{\oplus n}$ for all $A \in \fA$. For ease of notation we now suppress the arguments of $\be$ and $\z$. We estimate
\begin{align*}
\norm{z_i - \pi(A_0)e_i} &\leq \norm{\z - \pi(A_0)^{\oplus n}\be}\\
&<2\norm{\Lambda_{\x_0}}\norm{\y - \pi(A_0)^{\oplus n}\x} < \frac{\varepsilon}{2n}.
\end{align*}
By Lemma \ref{lem:transitivity_in_B(H)}, there exists $T \in \cB(\hilbH)$ such that $Te_i = z_i - \pi(A_0)e_i$ for all $i$ and $\norm{T} \leq \varepsilon/\sqrt{2n} < \varepsilon$.  In the self-adjoint case, we observe that we may choose $T$ to be self-adjoint since $(\be, \z) \in X_\tn{sa}$ and $\pi(A_0)$ is self-adjoint, so there exists a self-adjoint operator mapping $e_i$ to $z_i - \pi(A_0)e_i$ for all $i$. By Theorem \ref{thm:Pedersen}, there exists $A_1 \in \fA$ such that $\norm{A_1} \leq \norm{T} < \varepsilon$ and $\pi(A_1)e_i = z_i - \pi(A_0)e_i$. In the self-adjoint case, we may choose $A_1$ to be self-adjoint. Defining $A = A_0 + A_1$, we see that $\norm{A - A_0} < \varepsilon$ and
\[
\pi(A)e_i = \pi(A_0)e_i + \pi(A_1)e_i = z_i
\]
for all $i$, which implies $A \in \phi(\be, \z) = \phi(\x, \y)$. In the self-adjoint case, we have $A \in \fA_\tn{sa}$ by choice of $A_1$, so $A \in \phi_\tn{sa}(\x, \y)$. This proves lower semicontinuity, completing the proof.
\end{proof}

\begin{rem}
Suppose $(\hilbH, \pi, \Omega)$ is the GNS representation of $\omega \in \pstate(\fA)$. If we set $n = 1$ in the previous theorem and fix $x = \Omega$, then we get a continuous map $A: \hilbH \rightarrow \fA$ such that $\pi(A(y))\Omega = y = q(A(y))$ for all $y \in \hilbH$, where $q:\fA \rightarrow \hilbH$ is the quotient map. We see that $A$ is a right inverse for $q$. Since $q$ is a surjective bounded linear map between Banach spaces, the existence of a continuous right inverse is guaranteed by the Bartle-Graves theorem \cite{BartleGraves,MichaelSelection}. However, the existence of a continuous \textit{linear} right inverse is equivalent to $\ker q = \fN = \qty{A \in \fA: \omega(A^*A) = 0}$ having a closed complement in $\fA$. It is easy to see that this holds in certain special cases, such as when $\fA$ is commutative or finite-dimensional, or $\fA = \cB(\hilbK)$ for some Hilbert space $\hilbK$ and $\omega$ is a state corresponding to a unit vector in $\hilbK$. However, broader conditions under which $\fN$ is complemented (or whether this is true in general) are unknown to the authors.
\end{rem}

We now move towards a unitary version of Theorem \ref{thm:continuous_Kadison}.
%, which we call the ``continuous Kadison transitivity theorem.'' 
We do so through a series of lemmata.

\begin{lem}\label{lem:Stiefel_section}
Let $\hilbH$ be a Hilbert space, let $x_1,\ldots, x_n \in \hilbH$ be an orthonormal system. Given $\varepsilon > 0$, there exists $\delta > 0$ such that for any orthonormal system $y_1,\ldots, y_n \in \hilbH$ with $\norm{x_i - y_i} < \delta$ for all $i$, there exists a unitary $U \in \Unitary(\hilbH)$ such that 
\begin{enumerate}
\item[\tn{(i)}] $\hilbK_n = \vecspan\qty{x_1,\ldots, x_n, y_1,\ldots, y_n}$ is invariant under $U$, 
\item[\tn{(ii)}] $U$ acts as the identity on $\hilbK^\perp_n$, 
\item[\tn{(iii)}] $\norm{\1 - U} < \varepsilon$, 
\item[\tn{(iv)}] and $Ux_i = y_i$ for all $i$.
\end{enumerate}
\end{lem}

\begin{proof}
We prove the lemma by induction on $n$. Consider the case when $n = 1$. Given $x, y \in \sphere \hilbH$, we define $U_{x,y} \in \Unitary(\hilbH)$ by having $U_{x,y}$ act as the identity on the orthogonal complement of $\hilbK_1 = \vecspan\qty{x,y}$ and defining $U_{x,y}$ on $\hilbK_1$ by
\begin{equation}\label{eq:U_xy}
U_{x,y}z = \ev{y, x}z - \ev{y, z}x + \ev{x, z}y.
\end{equation}
One can check that this is indeed unitary and satisfies $U_{x,y}x = y$. When $\dim \hilbK_1 = 1$, we have $y = \ev{x, y} x$, so we see that $U_{x,y}|_{\hilbK_1}$ is multiplication by $\ev{x,y}$. When $\dim \hilbK_1 = 2$, the eigenvalues of $U_{x,y}|_{\hilbK_1}$ are
\[
\lambda_{x,y}^\pm = \Re \ev{x, y} \pm i \sqrt{1 - (\Re \ev{x, y})^2}.
\] 
Thus, $\sigma(U_{x,y}) \subset \qty{\lambda_{x,y}^+, \lambda_{x,y}^-, 1}$ for both possibilities of $\dim \hilbK_1$. Since $\1 - U_{x,y}$ is normal, its norm is given by its spectral radius, so
\begin{equation}\label{eq:U_xy_norm_identity}
\norm{\1 - U_{x,y}} = \abs{\lambda_{x,y}^+ - 1} = \abs{\lambda_{x,y}^- - 1} = \sqrt{2 - 2\Re \ev{x,y}} = \norm{x - y}.
\end{equation}
Therefore setting $\delta = \varepsilon$ and $U = U_{x,y}$ works for the base case.

Suppose the lemma is true for some $n$ and let $x_1,\ldots, x_{n+1}$ be an orthonormal system. Choose $\delta' > 0$ such that for any orthonormal system $y_1,\ldots, y_{n}$ with $\norm{x_i - y_i} < \delta'$ for all $i \leq n$, there exists a unitary $V \in \Unitary(\hilbH)$ satisfying (i), (ii), $\norm{\1 - V} < \varepsilon/3$, and $Vx_i = y_i$ for $i \leq n$. Let $\delta = \min(\delta', \varepsilon/3)$ and let $y_1,\ldots, y_{n+1}$ be an orthonormal system with $\norm{x_i - y_i} < \delta$ for all $i$. Since $V$ leaves $\hilbK_n$ invariant and acts as the identity on $\hilbK_n^\perp$, we see that
\[
Vx_{n+1} = VPx_{n+1} + (\1 - P)x_{n+1} \in \hilbK_{n+1},
\]
where $P$ is the projection onto $\hilbK_n$. Likewise $Vy_{n+1} \in \hilbK_{n+1}$, so $V$ leaves $\hilbK_{n+1}$ invariant. Since $\hilbK_{n+1}^\perp \subset \hilbK_n^\perp$, we also know that $V$ acts as the identity on $\hilbK_{n+1}^\perp$.

Set $z = Vx_{n+1}$ and note that $y_1,\ldots, y_n, z$ is an orthonormal system since it is the image of $x_1,\ldots, x_{n+1}$ under the unitary $V$. Furthermore,
\[
\norm{z - y_{n+1}} \leq \norm{(V-\1)x_{n+1}} + \norm{x_{n+1} - y_{n+1}} < \frac{\varepsilon}{3} + \delta \leq \frac{2\varepsilon}{3}
\]
Consider the unitary $W = U_{z, y_{n+1}}V$. Since $z \in \hilbK_{n+1}$, we see that $W$ leaves $\hilbK_{n+1}$ invariant and acts as the identity on $\hilbK_{n+1}^\perp$. Since $z$ and $y_{n+1}$ are orthogonal to all $y_i$ with $i \leq n$, it follows that $Wx_i = y_i$ for all $i \leq n +1$. Finally,
\begin{align*}
\norm{\1 - W} &\leq \norm{\1 - U_{z, y_{n+1}}} + \norm{U_{z, y_{n+1}} - U_{z, y_{n+1}}V}\\
&\leq \norm{z - y_{n+1}} + \norm{\1 - V} < \varepsilon,
\end{align*}
completing the proof. 
\end{proof}

\begin{rem}
We note that the $n = 1$ case of Lemma \ref{lem:Stiefel_section} can also be accomplished by polar decomposition. Assuming $x, y \in \sphere \hilbH$ and $\ev{x,y} \neq 0$, consider the operator
\[
A_{x,y} = \frac{\ev{x,y}}{\abs{\ev{x,y}}}P_yP_x + (\1 - P_y)(\1 - P_x),
\]
where $P_x = \ketbra{x}$ and $P_y = \ketbra{y}$ are the projections onto $\bbC x$ and $\bbC y$, respectively. The unitary $V_{x,y}$ obtained from the polar decomposition $A_{x,y} = V_{x,y}\abs{A_{x,y}}$ satisfies (i), (ii), and (iv). For fixed $x$, the map $y \mapsto V_{x,y}$ is norm-continuous on the domain $\qty{y \in \sphere \hilbH: \ev{x, y} \neq 0}$ and fulfills $V_{x,x} = I$. Therefore, $V_{x,y}$ satisfies (iii) for small enough $\delta$. 
\end{rem}

The following lemma contains the heart of the unitary version of the continuous Kadison transitivity theorem. 

\begin{lem}\label{lem:unitary_Kadison_lemma2}
Let $\fA$ be a unital $C^*$-algebra, let $(\hilbH, \pi)$ be a nonzero irreducible representation, and let $n$ be a non-negative integer. Define
\[
Y_+ = \qty{(\x, y) \in \hilbH^{n+1} \times \sphere \hilbH : \begin{array}{l} x_1,\ldots, x_{n+1} \tn{ are orthonormal, } \ev{x_{n+1}, y} > 0, \\ \tn{and } \ev{x_i, y} = 0 \tn{ for all }  i \leq n  \end{array}},
\]
equipped with the subspace topology. There exists a continuous map $U: Y_+ \rightarrow \Un(\fA)$ such that
\begin{equation}\label{eq:U_transitive_lemma}
\pi(U(\x, y))x_{n+1} = y \qqtext{and} \pi(U(\x, y))x_i = x_i
\end{equation}
for all $(\x, y) \in Y_+$ and $i \leq n$. 
\end{lem}

\begin{proof}
The function $\theta:Y_+ \rightarrow [0, \pi/2)$ defined as
\[
\theta(\x, y) = \cos^{-1} \ev{x_{n+1}, y}
\] 
is continuous on $Y_+$. If we set $Y'_+ = \qty{(\x, y) \in Y_+ : \ev{x_{n+1}, y} < 1}$, then we have another continuous map $w:Y'_+ \rightarrow \hilbH$ given by
\[
w(\x, y) = \frac{y - \ev{x_{n+1}, y}x_{n+1}}{\norm{y - \ev{x_{n+1}, y}x_{n+1}}}
\]
and $\qty{x_{n+1}, w(\x, y)}$ is a basis for $\vecspan\qty{x_{n+1}, y}$. In this basis, the unitary $U_{x_{n+1}, y}$ defined in Equation \eqref{eq:U_xy} is represented by the matrix
\[
\mqty(\cos \theta(\x, y) & - \sin \theta(\x, y) \\ \sin \theta(\x, y) & \cos \theta(\x, y))
\]
when restricted to $\vecspan\qty{x_{n+1}, y}$. Given $(\x, y) \in Y'_+$, we also define an operator $T_{\x, y} \in \cB(\hilbH)_\tn{sa}$ which acts as the zero operator on $\vecspan\qty{x_{n+1}, y}^\perp$ and, when restricted to $\vecspan\qty{x_{n+1}, y}$, is represented by the matrix
\[
\mqty(0 & i\theta(\x, y) \\ -i\theta(\x, y) & 0)
\]
with respect to the basis $\qty{x_{n+1}, w(\x, y)}$. Observe that $\norm{T_{\x,y}} = \theta(\x, y)$ and $U_{x_{n+1},y} = e^{iT_{\x,y}}$.

Note that $Y_+'$ is metrizable, hence paracompact Hausdorff. Given $(\x, y) \in Y_+$, define $\hilbK(\x, y) = \vecspan\qty{x_1,\ldots, x_{n+1}, y}$ and define a carrier $\phi:Y'_+ \rightarrow \npset(\fA_\tn{sa})$ by
\[
\phi(\x, y) = \qty{A \in \fA_\tn{sa} : \pi(A)|_{\hilbK(\x, y)} =T_{\x, y}|_{\hilbK(\x, y)} \tn{ and } \norm{A} \leq \theta(\x, y)}
\]
By the Kadison transitivity theorem (Theorem \ref{thm:Pedersen}), $\phi(\x, y)$ is nonempty for all $(\x, y) \in Y'_+$. We see that $\phi(\x, y)$ is closed and convex by the same arguments used in the proof of Theorem \ref{thm:continuous_Kadison}, along with the fact that the closed ball of radius $\theta(\x, y)$ is closed and convex. For lower semicontinuity, fix $(\x, y) \in Y'_+$, $A_0 \in \phi(\x, y)$, and $\varepsilon > 0$. Use continuity and positivity of $\theta$ on $Y_+'$ to choose a neighborhood $O$ of $(\x, y)$ such that for all $(\u, v) \in O$, we have 
\[
\abs{1 - \frac{\theta(\u, v)}{\theta(\x, y)}} < \frac{\varepsilon}{2\norm{A_0}}.
\]
Apply Lemma \ref{lem:Stiefel_section} to the orthonormal system $x_1,\ldots, x_{n+1}, w(\x, y)$  and the number
\[
\varepsilon' = \min\qty(2, \frac{\varepsilon}{4\norm{A_0}})
\]
to find a $\delta > 0$ with the properties described in Lemma \ref{lem:Stiefel_section}. By continuity of $w$ on $Y_+'$, we may shrink $O$ such that for all $(\u, v) \in O$, we have $\norm{x_i - u_i} < \delta$ and $\norm{w(\x, y) - w(\u, v)} < \delta$.

Now, given $(\u, v) \in O$, there exists a unitary $V \in \Unitary(\hilbH)$ such that $Vx_i = u_i$ for all $i$, $Vw(\x, y) = w(\u, v)$, and $\norm{\1 - V} < \varepsilon'$. The fact that $\norm{\1 - V} < 2$ implies that $-1 \notin \sigma(V)$, so we can use the continuous functional calculus to apply the principal branch of the logarithm and obtain a self-adjoint operator $S = -i \Log V$ with $\norm{S} \leq \pi$. Note that $S$ leaves the subspace 
\[
\hilbG = \vecspan\qty{x_1,\ldots, x_{n+1}, w_{x_{n+1}, y}, u_1,\ldots, u_{n+1}, w_{u_{n+1}, v}}
\] 
invariant since $V$ leaves $\hilbG$ invariant. By the Kadison transitivy  theorem (Theorem \ref{thm:Pedersen}), we can obtain a self-adjoint operator $B \in \fA$ such that $\pi(B)|_{\hilbG} = S|_{\hilbG}$ and $\norm{B} \leq \norm{S}$. Hence $W = e^{iB}$ acts as $V = e^{iS}$ on this subspace and by continuous functional calculus,
\begin{align*}
\norm{\1 - W} &= \sup_{\lambda \in \sigma(W)} \abs{\lambda - 1} = \abs{e^{i\norm{B}} - 1} \\
&\leq \abs{e^{i\norm{S}} - 1} = \sup_{\lambda \in \sigma(V)} \abs{\lambda - 1} = \norm{\1 - V}.
\end{align*}
It is easy to check that 
\[
A = \frac{\theta(\u, v)}{\theta(\x, y)}W A_0 W^{-1} \in \phi(\u, v)
\]
using the values of $V$ and $T_{\x, y}$  on $x_1,\ldots, x_{n+1}, w(\x, y)$, and the values of $V^{-1}$ and $T_{\u, v}$ on $u_1,\ldots, u_{n+1}, w(\u, v)$. Finally, observe that
\begin{align*}
\norm{A_0 - A} &\leq \norm{A_0 - WA_0W^{-1}} + \abs{1 - \frac{\theta(\u, v)}{\theta(\x, y)}} \norm{A_0}\\
&\leq \qty(2 \norm{I - W} + \abs{1 - \frac{\theta(\u, v)}{\theta(\x, y)}})\norm{A_0} < \varepsilon,
\end{align*}
as desired. This proves lower semicontinuity of $\phi$.

The Michael selection theorem now gives a continuous selection $A:Y'_+ \rightarrow \fA_\tn{sa}$ of $\phi$. We extend $A$ to $Y_+$ by defining $A(\x, y) = 0$ whenever $\ev{x_{n+1}, y} = 1$, equivalently, when $x_{n+1} = y$. Then $A:Y_+ \rightarrow \fA_\tn{sa}$ is continuous on $Y'_+$ since $Y'_+$ is open in $Y_+$ and $A$ is continuous on $Y_+ \setminus Y'_+$ by continuity of $\theta$ on $Y_+$ and the fact that $\norm{A(\x, y)} \leq \theta(\x, y)$ for all $(\x, y) \in Y_+$. Exponentiating $A$ yields a continuous map $U:Y_+ \rightarrow \Un(\fA)$, $U(\x, y) = e^{iA(\x, y)}$ that acts as $U_{x_{n+1}, y}$ on $\cK(\x, y)$, thereby satisfying \eqref{eq:U_transitive_lemma}.
\end{proof}

The purpose of our final lemma is to remove the condition that $\ev{x_{n+1}, y} > 0$ and replace it with the condition that $\ev{x_{n+1}, y} \notin \bbR_{\leq 0}$.

\begin{lem}\label{lem:unitary_Kadison_lemma}
Let $\fA$ be a unital $C^*$-algebra, let $(\hilbH, \pi)$ be a nonzero irreducible representation, and let $n$ be a non-negative integer. Define
\[
Y = \qty{(\x, y) \in \hilbH^{n+1} \times \bbS \hilbH : \begin{array}{l} x_1,\ldots, x_{n+1} \tn{ are orthonormal, } \ev{x_{n+1}, y} \notin  \bbR_{\leq 0}, \\ \tn{and } \ev{x_i, y} = 0 \tn{ for all }  i \leq n  \end{array}},
\]
equipped with the subspace topology. There exists a continuous map $U: Y \rightarrow \Unitary(\fA)$ such that
\begin{equation}\label{eq:unitary_Kadison_eq3}
\pi(U(\x, y))x_{n+1} = y \qqtext{and} \pi(U(\x, y))x_i = x_i
\end{equation}
for all $(\x, y) \in Y$ and $i \leq n$. 
\end{lem}

\begin{proof}
The angle $\alpha: Y \rightarrow (-\pi, \pi)$ defined by taking the principal branch of the logarithm:
\[
\alpha(\x, y) = \Im \Log \ev{x_{n+1}, y}
\]
is continuous on $Y$. The map $Y \rightarrow X_\tn{sa}$, $(\x, y) \mapsto (\x, 0, \ldots, 0, \alpha(\x, y)x_{n+1})$ is continuous, where $X_\tn{sa}$ is as in Theorem \ref{thm:continuous_Kadison}. We may therefore compose with the map $A:X_\tn{sa} \rightarrow \fA_\tn{sa}$ from Theorem \ref{thm:continuous_Kadison} and exponentiate to obtain a continuous map $V:Y \rightarrow \Unitary(\fA)$ such that
\begin{align*}
\pi(V(\x, y))x_{n+1} = e^{i\alpha(\x, y)}x_{n+1} = \frac{\ev{x_{n+1}, y}}{\abs{\ev{x_{n+1},y}}} x_{n+1}
\end{align*}
and $\pi(V(\x, y))x_i = x_i$ for all $i \leq n$. 

Note that
\[
\ev{\frac{\ev{x_{n+1}, y}}{\abs{\ev{x_{n+1}, y}}} x_{n+1}, y} = \abs{\ev{x_{n+1}, y}} > 0.
\]
Thus, we have a continuous map $Y \rightarrow Y_+$, $(\x, y) \mapsto (\pi(V(\x, y))^{\oplus n+1} \x, y)$ and we compose this with the continuous map from Lemma \ref{lem:unitary_Kadison_lemma2} to obtain a continuous map $W:Y \rightarrow \Unitary(\fA)$. Defining $U:Y \rightarrow \Unitary(\fA)$ by $U(\x, y) = W(\x, y)V(\x, y)$, we see that $U$ is continuous and satisfies \eqref{eq:unitary_Kadison_eq3}.
\end{proof}

We are finally ready to prove the unitary version of the continuous Kadison transitivity theorem.

\begin{thm}\label{thm:continuous_Kadison_unitary}
Let $\fA$ be a unital $C^*$-algebra, let $(\hilbH, \pi)$ be a nonzero irreducible representation, and let $n$ be a positive integer. Let 
\[
X_\tn{u} = \qty{(\x, \y) \in X: \exists\, T \in \Unitary(\hilbH) \tn{ s.t.\ } Tx_i = y_i \tn{ for all } i =1,\ldots, n},
\]
equipped with the subspace topology, where $X$ is as in Theorem \ref{thm:continuous_Kadison}. For every $(\x_0, \y_0) \in X_\tn{u}$, there exists a neighborhood $(\x_0, \y_0) \in O \subset X_\tn{u}$ and a continuous map $U:O \rightarrow \Unitary(\fA)$ such that
\begin{equation}\label{eq:unitary_selection}
\pi(U(\x, \y))x_i = y_i \quad \tn{ for all }  i = 1,\ldots, n
\end{equation}
for all $(\x, \y) \in O$.
\end{thm}

\begin{proof}
As in the proof of Theorem \ref{thm:continuous_Kadison}, let $\Lambda : X_\tn{u} \rightarrow \cB(\hilbH^n)^\times$ be the continuous map obtained by applying the Gram-Schmidt method to $x_1,\ldots, x_n$. Recall that for $T \in \cB(\hilbH)$, we have $T^{\oplus n}\x = \y$ if and only if $T^{\oplus n}\be(\x) = \z(\x, \y)$, where $\be(\x) = \Lambda_\x \x$ and $\z(\x, \y) = \Lambda_\x \y$. Since $(\x, \y) \mapsto (\be(\x), \z(\x, \y))$ is continuous, it suffices to prove the theorem with $X_\tn{u}$ replaced by 
\[
X_\tn{u}^\tn{on} = \qty{(\x, \y) \in X_\tn{u} : x_1,\ldots, x_n \tn{ are orthonormal}}.
\]
Therefore, suppose $(\x_0, \y_0) \in X_\tn{u}^\tn{on}$.

Suppose the theorem is true when $\x_0 = \y_0$. Then for arbitrary $(\x_0, \y_0) \in X_\tn{u}^\tn{on}$, we can find a neighborhood $(\x_0, \x_0) \in O \subset X_\tn{u}^\tn{on}$ and a continuous function $U:O \rightarrow \Unitary(\fA)$ for which \eqref{eq:unitary_selection} holds. By the Kadison transitivity theorem, there exists $V \in \Unitary(\fA)$ such that $\pi(V)^{\oplus n} \y_0 = \x_0$. Then $O' = (\1_{\hilbH^n} \oplus \pi(V)^{\oplus n})^{-1}(O) \subset X_\tn{u}^\tn{on}$ is a neighborhood of $(\x_0, \y_0)$ and $O' \ni (\x, \y) \mapsto V^{-1}U(\x, \pi(V)^{\oplus n}\y)$ is a continuous map satisfying \eqref{eq:unitary_selection}. Thus it suffices to prove the theorem for $(\x_0, \x_0) \in X_\tn{u}^\tn{on}$.

Let $Z_n = \qty{\x \in \hilbH^n : x_1,\ldots, x_n \tn{ are orthonormal}}$ with the subspace topology. Now suppose that for every $\x \in Z_n$, we have a neighborhood $\x \in O \subset Z_n$ and a continuous map $U: O \rightarrow \Unitary(\fA)$ such that 
\begin{equation}\label{eq:selection_unitary_simpler}
\pi(U(\y))^{\oplus n}\x = \y
\end{equation} 
for all $\y \in O$. Given $(\x_0, \x_0) \in X_\tn{u}^\tn{on}$ and such a neighborhood $\x_0 \in O \subset Z_n$, we see that $O' = O \times O \subset X_\tn{u}^\tn{on}$ is a neighborhood of $(\x_0, \x_0)$ and the map $O' \ni (\x, \y) \mapsto U(\y)U(\x)^{-1}$ satisfies \eqref{eq:unitary_selection}. 

We prove the hypothesis about $Z_n$ by induction on $n$. The $n = 1$ case follows from the $n = 0$ case of Lemma \ref{lem:unitary_Kadison_lemma}: one takes $O = \qty{y \in \sphere \hilbH: \ev{x, y} \notin \bbR_{\leq 0}}$ and the map $O \ni y \mapsto (x, y) \mapsto U(x, y)$ does the trick, where $U :Y \rightarrow \Unitary(\fA)$ is as in Lemma \ref{lem:unitary_Kadison_lemma}. Assume the hypothesis is true for some $n \geq 1$ and let $\x \in Z_{n+1}$. Let $P :Z_{n+1} \rightarrow Z_n$ be the projection onto the first $n$ components. We have a neighborhood $P\x \in O \subset Z_n$ and a continuous map $\tilde V: O \rightarrow \Unitary(\fA)$ such that \eqref{eq:selection_unitary_simpler} holds for $P\x$ and $\y \in O$. Define $V:O \rightarrow \Unitary(\fA)$ by $V(\y) = \tilde V(\y)\tilde V(P\x)^{-1}$ so that $V$ is continuous, satisfies \eqref{eq:selection_unitary_simpler} for $P\x$ and $\y \in O$, and has $V(P\x) = \1$. Define
\begin{align*}
O' = \qty{\y \in P^{-1}(O): \ev{y_{n+1}, \pi(V(P\y))x_{n+1}} \notin \bbR_{\leq 0}}
\end{align*}
Then $\x \in O' \subset Z_{n+1}$ and  $O' \rightarrow Y$, $\y \mapsto (\y, \pi(V(P\y))x_{n+1})$ is continuous. Composing this with the map from Lemma \ref{lem:unitary_Kadison_lemma} yields a continuous map $W:O' \rightarrow \Un(\fA)$, and one can easily check that $U:O' \rightarrow \Un(\fA)$, $U(\y) = W(\y)^{-1}V(P\y)$ satisfies \eqref{eq:selection_unitary_simpler}. This completes the proof.
\end{proof}

\subsection{Construction of Principal Fiber Bundles}\label{sec:principal_fiber_bundles}

Our first corollary uses the unitary version of the continuous Kadison transitivity theorem (Theorem \ref{thm:continuous_Kadison_unitary}) to make a statement about ``local automorphic equivalence'' of a norm-continuous family of pure states. By ``local automorphic equivalence,'' we mean that for any pure state $\omega$, there exists a neighborhood $\omega \in O \subset \pstate(\fA)$ and a norm-continuous family $\alpha:O \rightarrow \Aut(\fA)$ such that $\omega \circ \alpha_\psi = \psi$ for all $\psi \in O$. The name ``automorphic equivalence'' comes from seminal work by Bachmann et al.\ \cite{BMNS_automorphic_equivalence} in which a continuous path of ground states $\omega:[0,1] \rightarrow \pstate(\fA)$ corresponding to a  sufficiently nice path of gapped Hamiltonians is shown to admit a strongly continuous path $\alpha:[0,1] \rightarrow \Aut(\fA)$ such that $\omega_0 \circ \alpha_t = \omega_t$ for all $t \in [0,1]$. Bachmann et al. \cite{BMNS_automorphic_equivalence} made rigorous work by Hastings \& Wen \cite{Hastings_quasiadiabatic_evolution} and was later improved on by Moon \& Ogata \cite{OM_automorphic_equivalence}.

\begin{cor}\label{cor:continuous_Kadison_automorphism}
Let $\fA$ be a unital $C^*$-algebra and let $\omega \in \pstate(\fA)$. There exists a continuous map $U:\ball_2(\omega) \rightarrow \Unitary(\fA)$ such that 
\[
U_\psi \cdot \omega = \psi
\]
for all $\psi \in \ball_2(\omega)$. If $\fA$ is non-unital and $\omega \in \pstate(\fA)$, then there exists a norm-continuous map $\alpha:\ball_2(\omega) \rightarrow \Aut(\fA)_\tn{n}$ such that
\[
\omega \circ \alpha_\psi = \psi
\]
for all $\psi \in \ball_2(\omega)$.
\end{cor}

\begin{proof}
Let $(\hilbH, \pi, \Omega)$ be the GNS representation of $\omega$. 
Corollary \ref{cor:purestate_mapsto_SH} states continuity of the map $\bbB_2(\omega) \rightarrow \sphere \hilbH$, $\psi \mapsto \Psi$ where $\Psi$ represents $\psi$ and $\ev{\Psi, \Omega} > 0$. Hence $\bbB_2(\omega) \rightarrow Y$,  $\psi \mapsto (\Omega, \Psi)$ is well-defined, where $Y$ is as in Lemma \ref{lem:unitary_Kadison_lemma} with $n = 0$. Composing with the map from Lemma \ref{lem:unitary_Kadison_lemma} yields  a continuous map $U:\bbB_2(\omega) \rightarrow \Unitary(\fA)$, $\psi \mapsto U_\psi$ such that
\[
\pi(U_\psi)\Omega = \Psi
\]
for all $\psi \in \bbB_2(\omega)$. This implies that $U_\psi \cdot \omega = \psi$, as desired.

In the non-unital case, we consider the unitization $\tilde \fA$ and the isometry $\pstate(\fA) \rightarrow \pstate(\tilde \fA)$, $\psi \mapsto \tilde \psi$ where $\tilde \psi$ is the unique extension of $\psi$ to a pure state on $\tilde \fA$. This gives a continuous map $\ball_2(\omega) \rightarrow \ball_2(\tilde \omega)$, and we may apply the unital case to get a continuous map $U:\ball_2(\tilde \omega) \rightarrow \Unitary(\tilde \fA)$ such that $U_{\tilde\psi} \cdot \tilde \omega = \tilde\psi$ for all $\psi \in \ball_2(\omega)$. Since $\fA$ is a two-sided ideal in $\tilde \fA$, we have a continuous function $\Unitary(\tilde \fA) \rightarrow \Aut(\fA)_\tn{n}$, $U \mapsto (A \mapsto U^*AU)$ and composing with this function yields the desired map $\ball_2(\omega) \rightarrow \Aut(\fA)_\tn{n}$.
\end{proof}

With Corollary \ref{cor:continuous_Kadison_automorphism}, we can  immediately construct the following  principal fiber bundle. Recall that we denote the superselection sector of $\omega \in \pstate(\fA)$ by $\pstate_\omega(\fA)$.

\begin{corollary}
Let $\fA$ be a unital $C^*$-algebra, let $\omega \in \pstate(\fA)$, and let $p : \Unitary(\fA) \rightarrow \pstate_\omega(\fA)$ be the map $p(U) = U \cdot \omega$. Let
\[
\Unitary_\omega(\fA) = \qty{U \in \Unitary(\fA): U \cdot \omega = \omega}
\]
and let $\Unitary_\omega(\fA)$ act on $\Unitary(\fA)$ by right multiplication. Then $p$ is a principal $\Unitary_\omega(\fA)$-bundle.
\end{corollary}

\begin{proof}
The action $\Unitary(\fA) \times \pstate_\omega(\fA) \rightarrow \pstate_\omega(\fA)$ is transitive and continuous when all spaces are given their norm topologies. Corollary \ref{cor:continuous_Kadison_automorphism} provides a continuous section of $p$ in a neighborhood of $\omega$, so the result follows immediately. 
\end{proof}

Given that Kuiper's theorem implies the unitary group of an infinite-dimensional Hilbert space is contractible, we have a right to be skeptical when shown a new bundle built out of the unitary group of a possibly infinite-dimensional $C^*$-algebra. Nonetheless, we can show that the bundle $\Unitary_\omega(\fA) \rightarrow \Unitary(\fA) \rightarrow \pstate_\omega(\fA)$ is nontrivial in several cases with a simple strategy. By picking a particular unital $C^*$-algebra of interest and looking at the fundamental groups of the fiber, total space, and base space, one may be able to determine that the bundle is nontrivial by showing that $\pi_1(\Unitary(\fA)) \not\cong \pi_1(\Unitary_\omega(\fA) \times \pstate_\omega(\fA))$. We exhibit some examples below where these fundamental groups are indeed different.

If $\fA = M_n(\bbC)$ for some integer $n \geq 2$ and $\omega$ is any pure state, then $\pstate_\omega(\fA) \cong \bbC \bbP^{n-1}$, so $\pstate_\omega(\fA)$ is simply connected. The unitary group $\Unitary(\fA) = \Unitary(n)$ is path-connected and has fundamental group $\pi_1(\Unitary(n)) \cong \bbZ$. The stabilizer is $\Unitary_\omega(\fA) \cong \Unitary(1) \times \Unitary(n-1)$ which has fundamental group $\pi_1(\Unitary_\omega(\fA)) \cong \bbZ \times \bbZ$. Therefore the bundle is not trivial because
\[
\pi_1(\Unitary(\fA)) \cong \bbZ \not\cong \bbZ \times \bbZ \cong \pi_1(\Unitary_\omega(\fA) \times \pstate_\omega(\fA)).
\]

If $\fA = \cB(\hilbH)$ for a separable, infinite-dimensional Hilbert space $\hilbH$ and $\omega$ is a pure normal state, then $\pstate_\omega(\fA) \cong \bbP \hilbH$ is an Eilenberg-MacLane space of type $K(\bbZ, 2)$, 
%\todo{find a reference for this.}  
$\Unitary(\fA) = \Unitary(\hilbH)$ is contractible by Kuiper's theorem \cite{Kuiper}, and $\Unitary_\omega(\fA) \cong \Unitary(1) \times \Unitary(\hilbH)$. Therefore the bundle is not trivial because
\[
\pi_1(\Unitary(\fA)) \cong \qty{0} \not\cong \bbZ \cong \pi_1(\Unitary_\omega(\fA) \times \pstate_\omega(\fA)).
\]

We can also show that the bundle is nontrivial for any UHF algebra. If $\fA$ is a UHF algebra and
$\omega \in \pstate(\fA)$, then $\pstate_\omega(\fA) \cong \bbP \hilbH_\omega$ and $\hilbH_\omega$ is a separable,
infinite-dimensional Hilbert space \cite{Glimm_UHF_Algebras}, so $\pstate_\omega(\fA)$ is again a $K(\bbZ, 2)$.
In the following we will determine the homotopy groups of $\Unitary(\fA)$ and $\Unitary_\omega(\fA)$
which will then entail nontriviality of the bundle
$\Unitary_\omega(\fA) \rightarrow \Unitary(\fA) \rightarrow \pstate_\omega(\fA)$.
The computation relies on two major results, namely on Glimm's observation that the isomorphism
class of an UHF algebra can be encoded by its associated supernatural number \cite{Glimm_UHF_Algebras}
and on a theorem of Gl\"ockner \cite[Thm.~1.13]{glockner2010homotopy}.
Before we come to the computation of the homotopy groups $\pi_k(\Unitary(\fA))$ and
$\pi_k(\Unitary_\omega(\fA))$ we therefore first state these results in the form needed here and
provide a few preliminaries.

Recall from e.g.~\cite[Sec.~7.4]{RordamLarsenLaustsenKtheory}
that by a \textdef{supernatural number} one understands a sequence
$a= (a_i)_{i\in \bbN}$ of elements
$a_i \in \{ 0, 1,2\ldots ,\infty\}$.
By slight abuse of language one sometimes writes 
\[
  a = \prod_{i\in \bbN} p_i^{a_i} \ ,
\]
where $\{ p_1,p_2,\ldots\}$ is the set of primes listed in increasing order. One regards the right hand side of this formula as a formal prime
factorization of the supernatural number $a$. The product of two supernatural numbers $a,b$
is given by $ab = (a_i + b_i)_{i \in \bbN}$, i.e.,
\[
  ab = \prod_{i\in \bbN} p_i^{a_i+b_i}\ ,  
\]
but their sum is in general not defined. 
Associated to a supernatural number $a=(a_i)_{i\in \bbN}$ is the additive subgroup
$\UHFQ(a) \subset \bbQ$ consisting of all fractions $p/q$, where $p,q$ are integers
and $q$ has the prime decomposition
\[
  q = \prod_{i\in \bbN} p_i^{q_i} 
\]
such that $q_i \leq a_i$ for all $i\in \bbN$ and only finitely many of the $q_i$ are nonzero. 
By construction, $\UHFQ(a)$ contains $1$, and each additive subgroup $A\subset \bbQ$ containing $1$ equals
$\UHFQ(a)$ for some supernatural number $a$. Furthermore, two groups $\UHFQ(a)$ and  $\UHFQ(b)$ are isomorphic
if and only there are natural numbers $c,d \in \bbN$ such that $ac = bd$;
see \cite[Sec.~7.4]{RordamLarsenLaustsenKtheory} for details. 

To further clarify language let us remind the reader that by a UHF algebra one understands
a unital $C^*$-algebra $\fA$ with a strictly increasing sequence of finite-dimensional simple unital $C^*$-subalgebras:
\[
\1 \in \fA_1 \subsetneq \fA_2 \subsetneq \cdots \subsetneq \fA_j \subsetneq \cdots
\]
such that $\fA = \overline{\bigcup_{j=1}^\infty \fA_j}$. That each $\fA_j$ is finite-dimensional and simple implies that there exists $n_j \in \bbN$ such that $\fA_j$ is a \textdef{type $I_{n_j}$ factor}, meaning it is $*$-isomorphic to $M_{n_j}(\bbC)$. One says that $(\fA_j)_{j \in \bbN}$ is a \textdef{generating nest} for $\fA$ of type $(n_j)_{j \in \bbN}$ and that $\fA$ is a UHF algebra of type $(n_j)_{j \in \bbN}$. The fact that $\fA_j$ is a unital $C^*$-subalgebra of $\fA_{j'}$ for $j \leq j'$ implies that $n_j$ divides $n_{j'}$ whenever $j \leq j'$ \cite[Prop.~10.4.17]{KadisonRingroseII}.

\begin{comment}
which can be identified with the colimit of a countable strict inductive
system of type I factors of finite dimension and unital $*$-homomorphisms 
\[
  \fA_0 \lhook\joinrel\xrightarrow{\iota_{0,1}\:\:} \fA_1
  \lhook\joinrel\xrightarrow{\iota_{1,2}\:\:}\ldots  \lhook\joinrel\xrightarrow{\iota_{i-1,i}} \fA_i
  \lhook\joinrel\xrightarrow{\iota_{i,i+1}}\ldots \ . 
\]
Recall that strictness of the inductive system means that each of the unital $*$-homomorphisms
$\iota_i$ is injective and that by a type I factor of finite dimension
% or more pecisely a factor of type I$_n$ with $n\in \N$
one understands a von Neumann algebra which is $*$-isomorphic to the matrix algebra $M_n (\bbC)$
for some $n\in \bbN$.
Therefore,  each of the $C^*$-algebras $\fA_i$ is $*$-isomorphic to
a matrix algebra $M_{n_i} (\bbC)$ such that the sequence of ranks $(n_i)_{i\in \bbN}$ is increasing
and $n_i$ is a divisor of $n_j$ for all $i\leq j$. 
\end{comment}

\begin{comment}
Following Glimm \cite{Glimm_UHF_Algebras}, we say
that  $\fA$ is \emph{generated} by the inductive system $(\fA_i)_{i\in \N}$
of \emph{type} $(n_i)_{i\in \bbN}$. We always assume that the type is \emph{unbounded} meaning
that $\lim_{j\to \infty} n_j =\infty$. As in \cite{Glimm_UHF_Algebras}, a UHF algebra $\fA$ therefore has 
to be infinite dimensional. 
\end{comment}
Glimm \cite{Glimm_UHF_Algebras} associates to a UHF algebra $\fA$ of type $(n_j)_{j \in \bbN}$  a supernatural number
$a_\fA$ as follows. For each $j \in \bbN$ write 
\[
n_j = \prod_{i\in \bbN} p_i^{n_{ij}}
\]
where the $n_{ij} \in \qty{0} \cup \bbN$ are unique by prime decomposition.  Define 
\[
a_i = \sup_{j \in \bbN} n_{ij} \in \qty{0,1,2,\ldots, \infty}.
\] 
By construction, $a_\fA =(a_i)_{i\in \bbN}$ is a supernatural number.  As proven by Glimm \cite{Glimm_UHF_Algebras}, two UHF algebras $\fA$ and $\fB$ of type $(m_j)_{j \in \bbN}$ and $(n_j)_{j \in \bbN}$, respectively, are $*$-isomorphic if and only if their associated supernatural numbers are equal; see also
\cite[Thm.~7.4.5]{RordamLarsenLaustsenKtheory}. Note that the type of a UHF algebra $\fA$ is not uniquely determined, but the associated supernatural number is. We also note that by \cite[Lem.~7.4.4]{RordamLarsenLaustsenKtheory} the group $\UHFQ(a_\fA)$ associated to $a_\fA$ is
\begin{equation}\label{eq:relation-classifying-group-type}
\UHFQ(a_\fA) = \bigcup_{j=1}^\infty n_j^{-1}\bbZ.
\end{equation}

\begin{lem}\label{lem:UHF_unitary_closures}
  Let $\fA$ be a UHF algebra with a generating nest $(\fA_j)_{j \in \bbN}$ of type $(n_j)_{j \in \bbN}$. 
  If $\omega \in \pstate(\fA)$ and $\omega|_{\fA_j} \in \pstate(\fA_j)$
  for all $j$, then
\begin{equation}\label{eq:UHF_unitary_closures}
\Unitary_\omega(\fA) = \overline{\bigcup_{j\in \bbN} \Unitary(\fA_j) \cap \Unitary_\omega(\fA)}.
\end{equation}
and
\begin{equation}\Unitary(\fA_j) \cap \Unitary_\omega(\fA) = \Unitary_{\omega|_{\fA_j}}(\fA_j) \cong \Unitary(1) \times \Unitary(n_j - 1) \ .
  \end{equation}
  for all $j$.
\end{lem}

\begin{proof}
  Suppose $U \in \Unitary_\omega(\fA)$, fix $\varepsilon > 0$, and let $(\hilbH, \pi, \Omega)$ be the GNS representation of $\omega$. Lemma 3.1 in \cite{Glimm_UHF_Algebras} states that $\Unitary(\fA) = \overline{\bigcup_{j\in\bbN} \Unitary(\fA_j)}$, so there exists $j \in \bbN$ and $V \in \Unitary(\fA_j)$ such that $\norm{U - V} < \varepsilon/2$.  The fact that $U \in \Unitary_\omega(\fA)$ implies that $\pi(U)\Omega = \lambda \Omega$ for some $\lambda \in \Unitary(1)$, hence
\[
\norm{\pi(V)\Omega - \lambda \Omega} = \norm{\pi(V - U)\Omega} < \frac{\varepsilon}{2}.
\]
Now, $\hilbH_j \defeq \pi(\fA_j)\Omega$ is a finite-dimensional subspace of $\hilbH$ and $\pi$ restricts to a cyclic representation $\pi_j:\fA_j \rightarrow \cB(\hilbH_j)$ with cyclic unit vector $\Omega$ representing $\omega|_{\fA_j}$. Since $\omega|_{\fA_j}$ is pure by hypothesis, $\pi_j$ is an irreducible representation. In particular, since $\fA_j \cong M_{n_j}(\bbC)$, we know $\pi_j$ is a $*$-isomorphism. There exists a unitary $W \in \Unitary(\fA_j)$ such that $\pi_j(W) = U_{\pi(V)\Omega, \lambda\Omega}$, where $U_{\pi(V)\Omega, \lambda \Omega}$ is as defined in Lemma \ref{lem:Stiefel_section}. By \eqref{eq:U_xy_norm_identity} of Lemma \ref{lem:Stiefel_section}, $\norm{\1 - W} = \norm{\pi(V)\Omega - \lambda \Omega} < \varepsilon/2$ and $\pi(W)\pi(V)\Omega = \lambda \Omega$. Then $WV \in \Unitary(\fA_j) \cap \Unitary_\omega(\fA)$ and 
\[
\norm{U - WV} \leq \norm{U - V} + \norm{V - WV} < \varepsilon,
\]
as desired. 

Note that $U_0 \in \Unitary_{\omega|_{\fA_j}}(\fA_j)$ if and only if $U_0 \in \Unitary(\fA_j)$ and $\omega(U_0^*AU_0) = \omega(A)$ for all $A \in \fA_j$. Therefore it is clear that $\Unitary(\fA_j) \cap \Unitary_\omega(\fA)  \subset \Unitary_{\omega|_{\fA_j}}(\fA_j)$. Conversely, if $U_0 \in \Unitary_{\omega|_{\fA_j}}(\fA_j)$, then $\pi_j(U_0)\Omega = \pi(U_0)\Omega = \lambda_0 \Omega$ for some $\lambda_0 \in \Unitary(1)$, which implies that $U_0 \in \Unitary_\omega(\fA)$, i.e., $\omega(U_0^*AU_0) = \omega(A)$ for all $A \in \fA$. That $\Unitary_{\omega|_{\fA_j}}(\fA_j) \cong \Unitary(1) \times \Unitary(n_j - 1)$ is immediate from the fact that $\fA_j$ is $*$-isomorphic to $M_{n_j}(\bbC)$.
\end{proof}

\begin{lem}\label{lem:Cayley_transform}
Let $\fA$ be a UHF algebra with a generating nest $(\fA_j)_{j \in \bbN}$ of type $(n_j)_{j \in \bbN}$. 
% such that the sequence $(n_j)_{j\in \N}$ is strictly increasing.
Let $O = \qty{U \in \Unitary(\fA): \norm{\1 - U} < 2}$ and let $\phi:O \rightarrow \fA_\tn{sa}$ be defined by $\phi(U) = i(\1 - U)(\1 + U)^{-1}$. Then $\phi$ is a homeomorphism and $\phi(O \cap \Unitary_\omega(\fA))$ is a closed subspace of $\fA_\tn{sa}$ for all $\omega \in \pstate(\fA)$.
\end{lem}

\begin{proof}
If $U \in O$, then $-1 \notin \sigma(U)$ since this would contradict $\norm{\1 - U} < 2$. Therefore $\phi$ is well-defined. Note that $\phi$ is just multiplication by $-1$ composed with the inverse Cayley transform. In particular, $\phi$ is a homeomorphism  with inverse  $\phi^{-1}(A) = (i\1 - A)(i\1 + A)^{-1}$ by continuous functional calculus. 

Let $\omega \in \pstate(\fA)$. Then $\phi(O \cap \Unitary_\omega(\fA))$ is closed since $\Unitary_\omega(\fA)$ is closed in $\Unitary(\fA)$ and $\phi$ is a homeomorphism. Let $(\hilbH, \pi, \Omega)$ be the GNS representation of $\omega$. If $U \in O \cap \Unitary_\omega(\fA)$, then $\pi(U)\Omega = \lambda \Omega$ for some $\lambda \in \Unitary(1)$. Furthermore, there exists a sequence of polynomials $(p_n)$ such that $p_n(\lambda)$ converges to $\phi(\lambda)=i(1 - \lambda)(1 + \lambda)^{-1}$ uniformly on $\sigma(U)$, hence
\[
\pi(\phi(U))\Omega = \lim_{n \rightarrow \infty} \pi(p_n(U)) \Omega = \lim_{n \rightarrow \infty} p_n(\lambda)\Omega = \phi(\lambda)\Omega. 
\]
Now, let $U, V \in O \cap \Unitary_\omega(\fA)$, let $\alpha \in \bbR$, and set $A = \phi(U) + \alpha \phi(V)$. Then the argument above implies that $\pi(A)\Omega = \mu \Omega$ for some $\mu \in \bbR$, hence $\pi(\phi^{-1}(A))\Omega = \phi^{-1}(\mu)\Omega$ by the same argument. Thus, $\phi^{-1}(A) \in O \cap \Unitary_\omega(\fA)$, so $A \in \phi(O \cap \Unitary_\omega(\fA))$. Therefore, $\phi(O \cap \Unitary_\omega(\fA))$ is a subspace of $\fA_\tn{sa}$.
\end{proof}

The last tool we need for the computation of the homotopy groups is a theorem by Gl\"ockner \cite[Thm.~1.13]{glockner2010homotopy}, which says that under the existence of so-called well-filled charts, the homotopy groups of a space $X$ are the directed colimits of the homotopy groups of an ascending sequence of subspaces $X_1 \subset X_2 \subset \cdots$ whose union $\bigcup X_j$ is dense in $X$.  The notion of a well-filled chart is given by Definition 1.7 in the same article. The definition provided there is more general than we need; in fact, our well-filled charts are of a very simple form and the following more restrictive framework will suffice. Let $X$ be a Hausdorff topological group with a sequence of subgroups $(X_j)_{j \in \bbN}$ such that $X_j \subset X_{j+1}$ for all $j \in \bbN$ and $X = \overline{\bigcup_{j} X_j}$. Equip each $X_j$ with its subspace topology. Let $E$ be a
 locally convex Hausdorff topological vector space. If $O$ is an open subset of $X$ containing the identity of $X$ and $\phi:O \rightarrow E$ is a homeomorphism such that $\phi(O \cap X_j)$ is a closed
linear subspace of $E$ for all $j \in \bbN$, then $\phi$ is a \emph{well-filled chart} and
\[
\pi_k(X, x) = \colim_{j \in \bbN_x} \pi_k(X_j, x) \tn{ for all $k \in \bbN$ and $x \in \bigcup_{j\in \bbN} X_j$}\ ,
\]
where $\bbN_x = \qty{j \in \bbN: x \in X_j}$ and where the colimit is with respect to the homomorphisms induced by the inclusions $X_j \rightarrow X$ and $X_i \rightarrow X_{j}$ for $i \leq j$. Likewise,
\[
  \pi_0(X) = \colim_{j\in \bbN} \pi_0(X_j) \ .
\]
This distills what we need from Definition 1.7, Theorem 1.13, Corollary 1.14 and Lemma 8.1 in \cite{glockner2010homotopy}, although the full definition of a well-filled chart is more general.

\begin{thm}\label{thm:UHF_Uomega}
Let $\fA$ be a  UHF algebra.
Denote by $a_\fA$ the supernatural number associated to $\fA$. Then  
\begin{equation}\label{eq:homotopy_groups_U(A)_UHF}
\pi_k(\Unitary(\fA)) \cong \begin{cases} 0 &\text{for $k$ even},\\ \UHFQ(a_\fA) &\text{for $k$ odd}.  \end{cases}
\end{equation}
Furthermore, for every $\omega \in \pstate(\fA)$ the homotopy groups of the isotropy group $\Unitary_\omega (\fA)$ are  given by 
\begin{equation}\label{eq:homotopy_groups_Uomega(A)_UHF}
  \pi_k(\Unitary_\omega(\fA)) \cong \begin{cases} 0 &\text{for $k$ even}, \\
    \bbZ \times \UHFQ(a_\fA) &\text{for } k = 1, \\
    \UHFQ(a_\fA) &\text{for } k > 1 \text{ and } k \text{ odd}. \end{cases}
\end{equation}
\end{thm}

\begin{proof}
  %By
  % $\dim \fA =\infty$ the sequence $(n_j)_{j\in \N}$ is unbounded and therefore $\lim\limits_{j\to \infty} n_j =\infty$.
  %possibly passing to isomorphic $C^*$-algebras we can assume without loss of generality that
  %the inductive system defining $\fA$  is of the form
  %\[
  %  \bbC I \subset  \fA_0 \subset \fA_1 \subset \cdots  \ , 
  %\]
  %where $I$ is the unit of $\fA$. Then $\fA = \overline{\bigcup_{j\in \bbN} \fA_j}$
  Let $(\fA_j)_{j \in \bbN}$ be a generating nest for $\fA$ of type $(n_j)_{j \in \bbN}$. Recall that by \cite[Lem.\ 3.1]{Glimm_UHF_Algebras}, we know $\Unitary(\fA) = \overline{\bigcup_{j \in \bbN} \Unitary(\fA_j)}$. We want to prove a similar formula for the isotropy group $\Unitary_\omega(\fA)$. 
  To this end observe that by \cite[Cor.\ 3.8]{Powers_UHF_Representations} there exists for every pair
  of pure states $\psi, \omega \in \pstate(\fA)$ an automorphism $\alpha \in \Aut(\fA)$ such that $\psi = \omega \circ \alpha$ . Then $\alpha$ restricts to an isomorphism of topological groups $\Unitary_\psi(\fA) \rightarrow \Unitary_\omega(\fA)$. Therefore, for the purpose of computing the homotopy groups $\pi_k(\Unitary_\omega(\fA))$, we may choose $\omega$ to be any pure state we like; in particular, we may choose $\omega$ such that $\omega|_{\fA_j}$ is pure for all $j\in \bbN$. Then Lemma \ref{lem:UHF_unitary_closures} implies that $\Unitary_\omega(\fA) = \overline{\bigcup_{j\in\bbN} \Unitary(\fA_j) \cap \Unitary_\omega(\fA)}$.
  
\begin{comment}
  Formula \eqref{eq:relation-classifying-group-type} is an arithmetic result relating an abelian
  group obtained directly from the sequence $(n_j)_{j\in\bbN}$ with the abelian group constructed from
  the associated supernatural number $\delta_\fA$. The formula is proved in
  \cite[Lem.~7.4.4 (i)]{RordamLarsenLaustsenKtheory}.
\end{comment}

Now we will define a well-filled chart of $\Unitary(\fA)$, which restricts to a well-filled chart of $\Unitary_\omega(\fA)$, whose domain contains the identity $\1 \in \fA$. Then the remarks preceding the theorem will yield
\begin{equation}\label{eq:colimit_unitary_groups}
\pi_k(\Unitary(\fA)) = \colim_{j \in \bbN} \pi_k(\Unitary(\fA_j))
\end{equation}
and
\begin{equation}\label{eq:colimit_isotropy_groups}
\pi_k(\Unitary_\omega(\fA)) = \colim_{j\in \bbN} \pi_k(\Unitary(\fA_j) \cap \Unitary_\omega(\fA)).
\end{equation}
In particular, $\Unitary(\fA_j) \cong \Unitary(n_j)$ and $\Unitary(\fA_j) \cap \Unitary_\omega(\fA) \cong \Unitary(1) \times \Unitary(n_j - 1)$ by Lemma \ref{lem:UHF_unitary_closures}. These spaces are path-connected, so the homotopy groups are independent of the base point. We will define our well-filled chart, then analyze the colimit.

As in Lemma \ref{lem:Cayley_transform}, let $O = \qty{U \in \Unitary(\fA): \norm{\1 - U} < 2}$ and define $\phi:O \rightarrow \fA_\tn{sa}$ by $\phi(U) = i(\1 - U)(\1 + U)^{-1}$. Then $\phi$ is a homeomorphism and $\phi|_{O \cap \Unitary(\fA_j)}$ is a homeomorphism onto $(\fA_j)_\tn{sa}$ for all $j \in \bbN$, so $\phi$ is a well-filled chart for $\Unitary(\fA)$ and $\1 \in O$, as desired.

Since $\phi$ is a homeomorphism, the restriction
\[ \phi|_{O \cap \Unitary_\omega(\fA)}:O \cap \Unitary_\omega(\fA) \rightarrow \phi(O \cap \Unitary_\omega(\fA)) \]
is also a homeomorphism when $\phi(O \cap \Unitary_\omega(\fA))$ is given the subspace topology inherited from $\fA_\tn{sa}$. Lemma \ref{lem:Cayley_transform} entails that $\phi(O \cap \Unitary_\omega(\fA))$ is a linear subspace of $\fA_\tn{sa}$, hence a locally convex Hausdorff topological vector space. For each $j \in \bbN$, Lemma \ref{lem:UHF_unitary_closures} states that $\Unitary(\fA_j) \cap \Unitary_\omega(\fA) = \Unitary_{\omega|_{\fA_j}}(\fA_j)$, so Lemma \ref{lem:Cayley_transform} implies that $\phi(O \cap \Unitary_\omega(\fA) \cap \Unitary(\fA_j))$ is a linear subspace of
$(\fA_j)_\tn{sa}$. Since $(\fA_j)_\tn{sa}$ is finite-dimensional, we know $\phi(O \cap \Unitary_\omega(\fA) \cap \Unitary(\fA_j))$ is finite-dimensional as well, hence it is closed in $\phi(O \cap \Unitary_\omega(\fA))$. We see that $\phi|_{O \cap \Unitary_\omega(\fA)}$ is a well-filled chart with $\1 \in O \cap \Unitary_\omega(\fA)$, as desired.

We now analyze the colimits. Denote by $\iota_{ji}:\fA_i \rightarrow \fA_j$ the canonical inclusions
for $i\leq j$. Then there exist $*$-isomorphisms $\sigma_i:\fA_i \rightarrow M_{n_i}(\bbC)$ such that
\begin{equation}\label{eq:matrix_algebra_inclusion}
  \sigma_{ji}(A) = \mqty(A&&&\\&A&&\\&&\ddots &\\&&&A)
  \quad \text{for all } A \in M_{n_i}(\bbC)\ ,
\end{equation}
where $\sigma_{ji} = \sigma_j \circ \iota_{ji} \circ \sigma_i^{-1}: M_{n_i}(\bbC) \to M_{n_j}(\bbC)$ and
where there are $n_j/n_i \in \bbN$ copies of $A$ on the diagonal.
Choose $\omega$ to be the unique pure state determined by setting $\omega(A)$ to be the top left entry of the matrix $\sigma_i(A)$ for all $A \in \fA_i$. The colimit \eqref{eq:colimit_unitary_groups} is isomorphic to the colimit of the homomorphisms on homotopy groups $\pi_k(\Unitary(n_i))$ that are induced by the inclusions \eqref{eq:matrix_algebra_inclusion} restricted to the unitary groups $\Unitary(n_i)$. Furthermore, 
\[
\sigma_i(\Unitary_\omega(\fA) \cap \Unitary(\fA_i)) = \qty{\mqty(z & \\ & U) : z \in \Unitary(1), \, U \in \Unitary(n_i - 1)} \cong \Unitary(1) \times \Unitary(n_i - 1).
\]
Therefore, the colimit \eqref{eq:colimit_isotropy_groups} is isomorphic to the colimit of the homorphisms on homotopy groups $\pi_k(\Unitary(1) \times \Unitary(n_i - 1))$ that are induced by the continuous maps $g_{ji}:\Unitary(1) \times \Unitary(n_i - 1) \rightarrow \Unitary(1) \times \Unitary(n_j - 1)$ defined by
\begin{equation}
  \label{eq:pr1ofg}
  (\operatorname{pr}_1 \circ g_{ji})(z,U) = z
\end{equation}
and
\begin{equation}
  \label{eq:pr2ofg}
  (\operatorname{pr}_2 \circ g_{ji})(z, U) = \mqty(U&&&&&\\&z&&&&\\&&U&&&\\&&&z&&\\&&&&\ddots&&\\&&&&&U) \ ,
\end{equation}
where $\operatorname{pr}_1$ and $\operatorname{pr}_2$ are the projections.

We consider the former colimit. Recall that the map
\[ f_{ji}:\Unitary(n_i) \rightarrow \Unitary(n_j), \: U \mapsto \diag(U, \1)\]
  induces isomorphisms on homotopy groups $\pi_k$ for $k < 2n_i$.
  %, as seen by examining the long exact sequence emerging from the fiber bundle
  %\[ \Unitary(n_i) \rightarrow \Unitary(n_j) \rightarrow S^{2n_j-1} \]
  %defined by having $\Unitary(n_j)$ act on a fixed unit vector in $\bbC^{n_j}$.
  By the Bott periodicity
theorem \cite{BottStableHomotopyClassicalGroups}, the homotopy groups of the unitary groups
$\Unitary(n)$ are given for $k < 2n$ by 
\[
\pi_k(\Unitary(n)) = \begin{cases} 0 &\text{if } k \tn{ is even,} \\ \bbZ &\text{if } k \tn{ is odd.} \end{cases}
\]
Thus, when $k$ is even, the colimit is zero. Fix $k$ odd and let $i_0 \in \bbN$ be the smallest natural number such that $n_{i_0} > k/2$. Choosing a generator $x \in \pi_k(\Unitary(n_{i_0}))$ we obtain generators $(f_{ii_0})_*x \in \pi_k(\Unitary(n_i))$ for all $i \geq i_0$. Then for all $j\geq i \geq i_0 $,
\[
(\sigma_{ji})_*(f_{ii_0})_*x = \frac{n_j}{n_i}(f_{ji_0 })_*x,
\]
where we have restricted $\sigma_{ji}$ to the unitary groups. The homomorphisms $(\sigma_{ji})_*$ are thus multiplication by $n_{j}/n_i$.
% Set $Q_0 = \bigcup_{j \geq i \geq i_0} n_{ij}^{-1}\Z $ and $Q(\delta_\fA) = \bigcup_{j\in \N} n_j^{-1} \Z$.
Now define homomorphisms
\[
  \rho_i: \pi_k(\Unitary(n_i)) \rightarrow \UHFQ(a_\fA)=
  \bigcup_{j\in \bbN} n_j^{-1} \bbZ\]
as follows. If $i\geq i_0$, let $\rho_i$ by the unique group homomorphism
mapping the generator $(f_{ii_0})_*x$ to $1/n_{i}$. If $i< i_0$, put
$\rho_i = \rho_{i_0}(\sigma_{i_0i})_*$. By construction, the relation
$\rho_i = \rho_{j}(\sigma_{ji})_*$ then is fulfilled for all $j\geq i$.
Since the union of the images of
the homomorphisms $\rho_i$ coincides with  $\UHFQ(a_\fA)$ and since
$\rho_i$  is injective for $i\geq i_0$, $\UHFQ(a_\fA)$
together with the family $(\rho_j)_{j\in \bbN}$ is the directed colimit
we are looking for and formula \eqref{eq:homotopy_groups_U(A)_UHF} is proven.

 We now consider the direct system $g_{ji}:\Unitary(1) \times \Unitary(n_i - 1) \rightarrow \Unitary(1) \times \Unitary(n_j - 1)$.
 For the homotopy groups $\pi_k$ with $k > 1$, the analysis proceeds in an analogous way using the fact that
 the embedding $ \Unitary(n_j - 1) \hookrightarrow \Unitary(1) \times \Unitary(n_j - 1)$, $U \mapsto (1,U)$ 
 induces an isomorphism on homotopy $\pi_k(\Unitary(n_j - 1)) \to \pi_k(\Unitary(1) \times \Unitary(n_j - 1))$.
 Choose
 $i_0$ such that $n_{i_0}  > k/2+1$. If $k$ is even, then
 $\pi_k(\Unitary(1) \times \Unitary(n_j - 1)) = \pi_k(\Unitary(n_j - 1)) = 0$ for all $j \geq i_0$, hence the colimit is trivial.    
 If $k$ is odd, $k > 1$, and $j\geq i \geq i_0 $, the homomorphism $(g_{ji})_*$ maps a generator of
 $\pi_k(\Unitary(1) \times \Unitary(n_i - 1))$ to $n_{j}/n_i$ times a generator of $\pi_k(\Unitary(1) \times \Unitary(n_j - 1))$,
 hence the colimit coincides with $\UHFQ(a_\fA)$ as before. In case $k = 1$ we have $\pi_1(\Unitary(1) \times \Unitary(n_i - 1)) \cong \bbZ \times \bbZ$ for all $i\geq i_0$.
 Denote % by $f_j : \Unitary (1) \to \Unitary (n_j) $ the map $z \mapsto \diag (z,I)$,
 by $f_j : \Unitary (1) \to \Unitary (1) \times \Unitary (n_j-1) $ the map
 $z \mapsto (1,\diag (z,\1))$ and by
 $h_j : \Unitary (1) \to \Unitary (1) \times \Unitary (n_j-1) $ the map
 $z \mapsto (z,\1)$. After choosing a generator $x \in \pi_1(\Unitary (1))$
 the elements $x_j = (h_j)_* x$ and $y_j  = (f_j)_* x$ then are generators of
 $\pi_1(\Unitary (1) \times \Unitary (n_j-1))$.  Inspection of equations
 \eqref{eq:pr1ofg} and \eqref{eq:pr2ofg} then shows that for $j \geq i$ 
\begin{align*}
(g_{ji})_*x_i &= x_j + (n_jn_i^{-1} - 1)y_j \ ,\\
(g_{ji})_*y_i &= n_jn_i^{-1}y_j \ .
\end{align*}
In other words, $(g_{ji})_*$ is given by the matrix 
\[
\mqty(1 & 0 \\ n_jn_i^{-1} - 1 & n_jn_i^{-1}).
\]
Now let the homomorphisms
$g_j : \pi_1(\Unitary(1) \times \Unitary(n_j - 1)) \rightarrow \bbZ \times \UHFQ(a_\fA)$
be given by multiplication by the matrix
\[
\mqty(1 & 0 \\ 1 + n_j^{-1} & n_j^{-1}) \ .
\]
Then one checks easily that $ g_j (g_{ji})_* = g_i$ for all $j\geq i$. 
The union of the images of the maps $g_j$ covers  $\bbZ \times \UHFQ(a_\fA)$. Moreover, each
of the maps $g_j$ is injective, hence  $\bbZ \times \UHFQ(a_\fA)$ together with the family
of maps $(g_j)_{j\in \bbN}$ provides the colimit of the inductive system
of abelian groups $\left( \pi_1(\Unitary(1) \times \Unitary(n_j - 1)), (g_{ji})_*\right)_{i\leq j}$.
This finishes the proof of \eqref{eq:homotopy_groups_Uomega(A)_UHF}.
\end{proof}

\begin{rem}
  The proof of the theorem applies to more general situations. Namely, if $\fA$ is the colimit of an arbitrary (not necessarily countable) inductive system of unital $C^*$-algebras
  $(\fA_j)_{j \in J}$ and injective unital $*$-homomorphisms $\iota_{ji}:\fA_i \rightarrow \fA_j$ for $i \leq j$, then the same argument as above
  using \cite[Thm.\ 1.13]{glockner2010homotopy} yields
\[
\pi_k(\Unitary(\fA), U) = \colim_{j \in J_U} \pi_k(\Unitary(\fA_j), U) \tn{ for all $k \in \bbN$ and $U \in \bigcup_{j \in J} \Unitary(\fA_j)$},
\]
where $J_U = \qty{j \in J: U \in \Unitary(\fA_j)}$. This result was shown by Handelman in
\cite[Prop.\ 4.4]{HandelmanK0_AF_Algebras} through a different method of proof.
% Handelman also shows that $\pi_1(\Unitary(\fA)) \cong K_0(\fA)$ for any AF algebra $\fA$ \cite{HandelmanK0_AF_Algebras}.
Schr\"oder computed in \cite{SchroederUHFalgebras} the homotopy groups of the regular group of a UHF algebra which is
homotopy equivalent to its unitary group. Schr\"oder's result therefore entails ours.  
However, to our knowledge, the homotopy groups $\pi_k(\Unitary_\omega(\fA))$ for a UHF algebra $\fA$ and pure state $\omega$ have
not been computed before. 
\end{rem}

We now can show the claimed nontriviality of
the bundle $\Unitary(\fA) \rightarrow \pstate_\omega(\fA)$. 

\begin{cor}
  If $\fA$ is a UHF algebra and $\omega \in \pstate(\fA)$, then 
  the bundle  $\Unitary_\omega(\fA) \rightarrow \Unitary(\fA) \rightarrow \pstate_\omega(\fA)$ is nontrivial. 
\end{cor}

\begin{proof}
  As  a consequence of the preceding theorem, the   rationalized fundamental  groups
of $\Unitary(\fA)$
\[ \pi_1(\Unitary(\fA))\otimes_\bbZ \bbQ \cong \bbQ(a_\fA) \otimes_\bbZ \bbQ \cong \bbQ\] and of the trivial bundle
$\Unitary_\omega(\fA)\times \pstate_\omega(\fA)$ 
\[\pi_1(\Unitary_\omega(\fA)\times \pstate_\omega(\fA))\otimes_\bbZ \bbQ \cong (\bbZ \times \bbQ (a_\fA))\otimes_\bbZ \bbQ
\cong \bbQ^2\]
are not isomorphic, hence $\Unitary_\omega(\fA) \rightarrow \Unitary(\fA) \rightarrow \pstate_\omega(\fA)$ cannot be trivial.
\end{proof}

\subsection{A Selection Theorem for the Weak* Topology}

So far we have been considering the norm topology on $\pstate(\fA)$. Since families of ground states are typically only weak*-continuous, it is of great interest to prove selection theorems for the weak* topology on $\pstate(\fA)$. We make a small step in this direction.

\begin{theorem}\label{thm:weak*_selection_theorem}
Let $\fA$ be a unital separable $C^*$-algebra and let $P \in \fA$ be a  projection. Define
\[
X = \qty{\omega \in \pstate(\fA): \omega(P) > 0} \qqtext{and} Y = \qty{\omega \in \pstate(\fA): \omega(P) = 1}.
\]
Equip $X$ with the weak* topology. There exists a continuous map $U:X \rightarrow \Unitary(\fA)$ such that 
\[
U_\omega \cdot \omega \in Y
\]
for all $\omega \in X$ and 
\[
\norm{\1 - U_\omega} \leq \sqrt{2 - 2\sqrt{\omega(P)}}.
\]
\end{theorem}

\begin{proof}
First, let us establish some notation. Given $\omega \in X$, let $(\cH_\omega, \pi_\omega, \Omega_\omega)$ be its GNS representation. Let
\[
\Psi_\omega = \frac{\pi_\omega(P)\Omega_\omega}{\norm{\pi_\omega(P)\Omega_\omega}} = \frac{\pi_\omega(P)\Omega_\omega}{\sqrt{\omega(P)}}
\]
and let 
\[
\Phi_\omega = \Psi_\omega - \ev{\Omega_\omega, \Psi_\omega}\Omega_\omega.
\]
Let $\psi_\omega$ be the pure state induced by $\Psi_\omega$. We can easily compute:
\begin{align*}
\ev{\Omega_\omega, \Psi_\omega} &= \sqrt{\omega(P)}, \\
 \norm{\Omega_\omega - \Psi_\omega} &= \sqrt{2 - 2\sqrt{\omega(P)}}, \\
 \qqtext{and} \norm{\Phi_\omega} &= \sqrt{1 - \omega(P)}
\end{align*}
Let $\hilbK_\omega = \vecspan\qty{\Omega_\omega, \Psi_\omega}$ and let $V_\omega \in \cB(\hilbH)$ be our favorite unitary operator from Lemma \ref{lem:Stiefel_section} mapping $\Omega_\omega$ to $\Psi_\omega$, that is, $V_\omega \Omega_\omega = \Psi_\omega$, $V_\omega$ acts as identity on $\hilbK_\omega^\perp$,  and $\norm{\1 - V_\omega} = \norm{\Omega_\omega - \Psi_\omega} < \sqrt{2}$. Finally, let $T_\omega = -i\Log V_\omega$, where we use the principal branch of the logarithm. 

Recall from Lemma \ref{lem:Stiefel_section} that $\sigma(V_\omega) \subset \qty{\lambda_\omega, \lambda_\omega^*, 1}$, where
\begin{align*}
\lambda_\omega &= \Re \ev{\Psi_\omega, \Omega_\omega} + i \sqrt{1 - \qty(\Re\ev{\Psi_\omega, \Omega_\omega})^2}\\
&= \sqrt{\omega(P)} + i \sqrt{1 - \omega(P)} = e^{i\cos^{-1}\sqrt{\omega(P)}}
\end{align*}
It follows that $\sigma(T_\omega) \subset \qty{\cos^{-1}\sqrt{\omega(P)}, -\cos^{-1}\sqrt{\omega(P)}, 0}$, hence $\norm{T_\omega} = \cos^{-1}\sqrt{\omega(P)}$.

Let $Q_\omega \in \cB(\hilbH)$ be the projection onto $\hilbK_\omega$. Consider the carrier
\[
\phi:X \rightarrow \fA_\tn{sa}, \quad \phi(\omega) = \qty{A \in \fA_\tn{sa}: \pi_\omega(A)Q_\omega = T_\omega Q_\omega \tn{ and } \norm{A} \leq \norm{T_\omega}}
\]
By the Kadison transitivity theorem (with norm bounds), $\phi(\omega)$ is nonempty for all $\omega \in X$. It is obvious that $\phi(\omega)$ is closed and convex for all $\omega \in X$. Since $\fA$ is separable, $X$ is metrizable, hence paracompact Hausdorff, so the Michael selection theorem will apply if we can show that $\phi$ is lower semicontinuous. If there exists a selection $A:X \rightarrow \fA_\tn{sa}$ for $\phi$, then 
\[
e^{i\pi_\omega(A_\omega)}\Omega_\omega = e^{i\pi_\omega(A_\omega)Q_\omega}\Omega_\omega = e^{iT_\omega Q_\omega}\Omega_\omega = e^{iT_\omega}\Omega_\omega = V_\omega \Omega_\omega = \Psi_\omega,
\]
hence $U:X \rightarrow \Unitary(\fA)$, $U_\omega = e^{iA_\omega}$ is a continuous map such that
\[
U_\omega \cdot \omega = \psi_\omega \in Y
\]
and is therefore the desired continuous map.

All that remains to do is show that $\phi$ is lower semicontinuous. Fix $\omega_0 \in X$, $A_0 \in \phi(\omega_0)$, and $\varepsilon > 0$. If $\omega_0(P) = 1$, then 
\[
\norm{A_0} \leq \norm{T_\omega} = \cos^{-1}\sqrt{\omega(P)} = 0
\]
hence $A_0 = 0$. Defining 
\[
O = \qty{\omega \in X: \cos^{-1}\sqrt{\omega(P)} < \varepsilon},
\]
we see that $O$ is a neighborhood of $\omega_0$ in the weak* topology and $\phi(\omega) \subset B_\varepsilon(A_0)$ for all $\omega \in O$.

If $\omega_0(P) < 1$, then we first set $O = \qty{\omega \in X: \omega(P) < 1}$, which is a neighborhood of $\omega_0$ in the weak*-topology. On $O$, we know $\Psi_\omega$ and $\Omega_\omega$ are linearly independent, hence $\Phi_\omega \neq 0$. The functions
\[
\omega \mapsto \norm{T_\omega \Omega_\omega - \pi_\omega(A_0)\Omega_\omega} \qqtext{and} \omega \mapsto \norm{T_\omega \Phi_\omega - \pi_\omega(A_0)\Phi_\omega} \norm{\Phi_\omega}^{-1}
\]
are weak*-continuous on $O$ 
\begin{comment}
We determine the action of $T_\omega$ on $\Omega_\omega$ and $\Phi_\omega$. Recall from Lemma \ref{lem:Stiefel_section} that $V_\omega\Psi_\omega = 2\ev{\Psi_\omega, \Omega_\omega}\Psi_\omega - \Omega_\omega$. From this we see that
\begin{align*}
V_\omega \qty(\Omega_\omega - \lambda_\omega \Psi_\omega) &= \Psi_\omega - \lambda_\omega 2\ev{\Psi_\omega, \Omega_\omega}\Psi_\omega + \lambda_\omega \Omega_\omega\\
&= \lambda_\omega\qty(\Omega_\omega - \qty(2\ev{\Psi_\omega, \Omega_\omega} - \lambda_\omega^*)\Psi_\omega)\\
&= \lambda_\omega \qty(\Omega_\omega - \lambda_\omega \Psi_\omega).
\end{align*}
Likewise $V_\omega(\Omega_\omega - \lambda_\omega^*\Psi_\omega) = \lambda_\omega^*(\Omega_\omega - \lambda_\omega^*\Psi_\omega)$. It follows that
\begin{align*}
T_\omega (\Omega_\omega - \lambda_\omega \Psi_\omega) &= \cos^{-1}\sqrt{\omega(P)} \qty(\Omega_\omega - \lambda_\omega \Psi_\omega)\\
T_\omega\qty(\Omega_\omega - \lambda_\omega^*\Psi_\omega) &= -\cos^{-1}\sqrt{\omega(P)}\qty(\Omega_\omega - \lambda_\omega^*\Psi_\omega).
\end{align*}
Thus,
\begin{align*}
T_\omega\Omega_\omega = \frac{\lambda_\omega T_\omega(\Omega_\omega - \lambda_\omega^* \Psi_\omega) - \lambda_\omega^* T_\omega(\Omega_\omega - \lambda_\omega\Psi_\omega)}{\lambda_\omega - \lambda_\omega^*} =  
\end{align*}
\end{comment}
and evaluate to zero at $\omega_0$, and therefore we may shrink $O$ so that these norms are small on $O$. Then for any $\omega \in O$, there exists $A_1 \in \fA_\tn{sa}$ with $\norm{A_1} < \varepsilon$ such that
\[
\pi_\omega(A_1)\Omega_\omega = T_\omega \Omega_\omega - \pi_\omega(A_0)\Omega_\omega \qqtext{and} \pi_\omega(A_1)\Phi_\omega = T_\omega \Phi_\omega - \pi_\omega(A_0)\Phi_\omega
\]
Thus,
\[
A_0 + A_1 \in \phi(\omega) \cap B_\varepsilon(A_0).
\]
This proves lower semicontinuity.
\end{proof}

%%%%%%%%%   then the Bibliography, if any   %%%%%%%%%
\bibliographystyle{amsalpha}	% or "siam", or "alpha", etc.
%\nocite{*}		% list all refs in database, cited or not
\bibliography{refs}		% Bib database in "refs.bib"

%%%%%%%%%   then the Appendices, if any   %%%%%%%%%
\appendix
%!TEX root = dissertation.tex

\chapter{Directed Colimits}\label{app:directed_colimits}

We review the basics of directed colimits. None of this material is original; it can be found in \cite{EilenbergSteenrod}, for example. Nonetheless, given how much directed colimits are used in this dissertation, it is useful to have a catalog of elementary results close at hand.

\begin{appdefinition}
A \textdef{directed set} is a set $I$ with a preorder---a binary relation $\leq$ that is reflexive and transitive---such that for any $i,j \in I$ there exists $k \in I$ such that $i \leq k$ and $j \leq k$.
\end{appdefinition}

\begin{appdefinition}
Let $\sfC$ be a category. A \textdef{directed system} in $\sfC$ is a collection of objects $(A_i)_{i \in I}$ indexed by a directed set $I$, together with a collection of morphisms $\iota_{ji}:A_i \rightarrow A_j$ defined for all pairs $i, j \in I$ with $i \leq j$, that satisfy the following properties:
\begin{enumerate}[label={(DS\arabic*)}, leftmargin=1.5cm]
  \item $\iota_{ii} = \id_{A_i}$ for all $i \in I$,
  \item $\iota_{ki} = \iota_{kj} \circ \iota_{ji}$ for all $i,j,k \in I$ such that $i \leq j \leq k$.
\[
\begin{tikzcd}
A_i \rar["\iota_{ji}"] \arrow[rr,bend right = 45, "\iota_{ki}"]& A_j \rar["\iota_{kj}"]& A_k
\end{tikzcd}
\]
\end{enumerate}
We may denote a directed system by $\ev{A_i, \iota_{ji}}_{i,j \in I}$ or $\ev{A_i, \iota_{ji}}$ when the index set may be left implicit. 

A \textdef{target} of a directed system $\ev{A_i, \iota_{ji}}_{i,j \in I}$ consists of an object $A$ and morphisms $\iota_i:A_i \rightarrow A$ defined for all $i \in I$ such that $\iota_i = \iota_j \circ \iota_{ji}$ whenever $i \leq j$. We may denote a target as $\ev{A, \iota_i}_{i \in I}$ or $\ev{A, \iota_i}$. A \textdef{directed colimit} of a directed system $\ev{A_i, \iota_{ji}}_{i,j \in I}$ is a target $\ev{A, \iota_i}_{i \in I}$ such that for any other target $\ev{B, \kappa_i}_{i \in I}$, there exists a unique morphism $\phi:A \rightarrow B$ such that $\phi \circ \iota_i = \kappa_i$ for all $i \in I$. In summary, the diagram below commutes.
  \[
\begin{tikzcd}
&B&\\
&A \arrow[u, dashed, "\phi"]&\\
A_i \arrow[ur, "\iota_i"]\arrow[uur, bend left, "\kappa_i"] \arrow[rr,"\iota_{ji}"']& & A_j \arrow[ul, "\iota_j"'] \arrow[uul, "\kappa_j"', bend right]
\end{tikzcd}
\]
\end{appdefinition}

\begin{appproposition}\label{prop:directed_colim_function_identifier}
Let $\ev{A_i, \iota_{ji}}$ be a directed system in a category $\sfC$, let $\ev{A, \iota_i}$ be a directed colimit, and let $B$ be another object. If $f,g:A \rightarrow B$ are morphisms and $f \circ \iota_i = g \circ \iota_i$ for all $i$, then $f = g$.
\end{appproposition}

\begin{proof}
Define $f_i = f \circ \iota_i$ for all $i \in I$. For any $i$ and $j$ with $i \leq j$ we have $f_j \circ \iota_{ji} = f \circ \iota_i = f_i$. Therefore $\ev{B, f_i}$ is a target. By the universal property of the directed colimit, there exists a unique morphism $\phi:A \rightarrow B$ such that $\phi \circ \iota_i = f_i$ for all $i$. Since $f$ and $g$ satisfy $f \circ \iota_i = g \circ \iota_i = f_i$ for all $i$, we know $\phi = f = g$. 
\end{proof}

\begin{appproposition}\label{prop:inductive_limit_unique}
Let $\ev{A_i, \iota_{ji}}$ be a directed system and suppose a directed colimit $\ev{A, \iota_i}$ exists. If $\ev{B, \kappa_i}$ is any other directed colimit of $\ev{A_i, \iota_{ji}}$, then there exists a unique morphism $\phi:A \rightarrow B$ such that $\phi \circ \iota_i = \kappa_i$ for all $i \in I$. Moreover, $\phi$ is an isomorphism.
\end{appproposition}

\begin{proof}
By definition, there exist unique morphisms $\phi: A \rightarrow B$ and $\chi :B \rightarrow A$ such that $\phi \circ \iota_i = \kappa_i$ and $\chi \circ \kappa_i = \iota_i$ for all $i$. But then $\chi \circ \phi \circ \iota_i = \iota_i$ and $\phi \circ \chi \circ \kappa_i = \kappa_i$ for all $i$. By Proposition \ref{prop:directed_colim_function_identifier}, we know $\chi \circ \phi = \id_A$ and $\phi \circ \chi = \id_B$, so $\phi$ is an isomorphism.
\end{proof}

\begin{appproposition}
Let $\ev{A_i, \iota_{ji}}_{i,j \in I}$ and $\ev{B_i, \kappa_{ji}}_{i,j \in I}$ be directed systems indexed by the same directed set $I$, and let $\ev{A, \iota_i}$ and $\ev{B, \kappa_i}$ be colimits for these systems. If for each $i \in I$ we have a morphism $f_i:A_i \rightarrow B_i$ such that $f_j \circ \iota_{ji} = \kappa_{ji} \circ f_i$ whenever $i \leq j$, then there exists a unique morphism $f:A \rightarrow B$ such that $f \circ \iota_i = \kappa_i \circ f_i$ for all $i$. If each $f_i$ is an isomorphism, then $f$ is an isomorphism.
\[
\begin{tikzcd}
A\arrow[r,"f", dashed]&B\\
A_j \arrow[u,"\iota_j"] \arrow[r,"f_j"]& B_j\arrow[u,"\kappa_j"]  \\
A_i \arrow[uu,bend left = 60, "\iota_i"] \arrow[u, "\iota_{ji}"]\arrow[r,"f_i"]& B_i\arrow[uu,bend right = 60, "\kappa_i"] \arrow[u, "\kappa_{ji}"]
\end{tikzcd}
\]
\end{appproposition}

\begin{proof}
The system $\ev{B, \kappa_i \circ f_i}$ is a target for $\ev{A_i, \iota_{ji}}$, so there exists a unique morphism $f:A \rightarrow B$ such that $f \circ \iota_i = \kappa_i \circ f_i$ for all $i$. If each $f_i$ is an isomorphism, then we claim that $\ev{B, \kappa_i \circ f_i}$ is an inductive limit for $\ev{A_i, \iota_{ji}}$, from which it will follow that $f$ is an isomorphism by Proposition \ref{prop:inductive_limit_unique}.

Let $\ev{C, \lambda_i}$ be a target for $\ev{A_i, \iota_{ji}}$. Then $\ev{C, \lambda_i \circ f_i^{-1}}$ is a target for $\ev{B_i, \kappa_{ji}}$, so there exists a unique morphism $g:B \rightarrow C$ such that $g \circ \kappa_i = \lambda_i \circ f_i^{-1}$ for all $i$. Rearranging yields $g \circ \kappa_i \circ f_i = \lambda_i$, which shows that $\ev{B, \kappa_i \circ f_i}$ is an inductive limit for $\ev{A_i, \iota_{ji}}$.
\end{proof}

\begin{appproposition}
Let $\ev{A_i, \iota_{ji}}_{i,j \in I}$ be a directed system with directed colimit $\ev{A, \iota_i}$ and suppose there exists $i_0 \in I$ such that $\iota_{ji_0}$ is an isomorphism for all $j \geq i_0$. Then for all $j,k \in I$ with $i_0 \leq j \leq k$, the morphisms $\iota_{kj}$ and $\iota_j$ are isomorphisms.
\end{appproposition}

\begin{proof}
We observe that $\iota_{ki_0} = \iota_{kj} \circ \iota_{ji_0}$. Since $\iota_{ki_0}$ and $\iota_{ji_0}$ are isomorphisms, we know $\iota_{kj}$ is an isomorphism. 

If we can show that $\iota_{i_0}$ is an isomorphism, then it will follow that $\iota_j$ is an isomorphism for all $j \geq i$ since $\iota_j \circ \iota_{ji_0} = \iota_{i_0}$ and $\iota_{ji_0}$ is an isomorphism. For arbitrary $j \in I$, choose $k \in I$ such that ${i_0},j \leq k$. Define $\kappa_j = \iota_{ki_0}^{-1} \circ \iota_{kj}$. For any other $\ell \in I$ such that ${i_0},j \leq \ell$,  we can find $m \in I$ such that $k,\ell \leq m$, hence
\[
\iota_{\ell i_0}^{-1} \circ \iota_{\ell j} = (\iota_{m\ell}^{-1} \circ \iota_{mi_0})^{-1} \circ  \iota_{m\ell}^{-1} \circ \iota_{mj} = \iota_{mi_0}^{-1} \circ \iota_{mj}.
\]
The same argument applies when $\ell$ is switched with $k$, so we get
\[
\iota_{\ell i_0}^{-1} \circ \iota_{\ell j} = \iota_{mi_0}^{-1} \circ \iota_{mj} = \kappa_j.
\]
We see that the morphism $\kappa_j$ is independent of the choice of $k \in I$ satisfying ${i_0},j \leq k$. 

We claim that $\ev{A_{i_0}, \kappa_j}$ is a directed colimit for $\ev{A_i, \iota_{ji}}$. Given $j,k \in I$ with $j \leq k$, we choose $\ell \in I$ with $i_0, j,k \leq \ell$, hence
\[
\kappa_k \circ \iota_{kj}= \iota_{\ell i_0}^{-1} \circ \iota_{\ell k} \circ \iota_{kj} = \iota_{\ell i_0}^{-1} \circ \iota_{\ell j} = \kappa_j.
\]
Thus, $\ev{A_{i_0},\kappa_j}$ is a target. Suppose $\ev{B, \lambda_j}$ is any other target for $\ev{A_i, \iota_{ji}}$. Then the morphism $\lambda_{i_0}:A_{i_0} \rightarrow B$ satisfies, for any $j,k \in I$ with $i_0,j \leq k$,
\[
\lambda_{i_0} \circ \kappa_j = \lambda_{i_0} \circ \iota_{ki_0}^{-1} \circ \iota_{kj} = \lambda_k \circ \iota_{kj} = \lambda_j.
\]
If $\mu:A_{i_0} \rightarrow B$ is any other morphism satisfying $\mu \circ \kappa_j = \lambda_j$ for all $j \in I$, then
\[
\mu = \mu \circ \kappa_{i_0} = \lambda_{i_0}
\]
since $\kappa_{i_0} = \id_{A_{i_0}}$. This proves our claim.

By Proposition \ref{prop:inductive_limit_unique}, there exists a unique isomorphism $\phi:A \rightarrow A_{i_0}$ such that $\phi \circ \iota_j = \kappa_j$ for all $j \in I$. In particular, $\phi \circ \iota_{i_0} = \kappa_{i_0} = \id_{A_{i_0}}$. Since $\phi$ is an isomorphism, we see that $\iota_{i_0} = \phi^{-1}$ is an isomorphism.
\end{proof}

\begin{appdefinition}
Let $I$ be a directed set. A subset $J \subset I$ is \textdef{cofinal} in $I$ if for every $i \in I$, there exists $j \in J$ such that $i \leq j$.
\end{appdefinition}

\begin{appproposition}
Let $\ev{A_i, \iota_{ji}}_{i,j \in I}$ be a directed system indexed by $I$ and let $J$ be a subset of $I$ that is directed with respect to the preorder on $I$. If $\ev{A, \iota_i}$ is a directed colimit of $\ev{A_i, \iota_{ji}}_{i,j \in I}$ and $\ev{B, \kappa_j}$ is a directed colimit of $\ev{A_i, \iota_{ji}}_{i,j \in J}$, then there exists a unique morphism $\phi:B \rightarrow A$ such that $\phi \circ \kappa_j = \iota_j$ for all $j \in J$. If $J$ is cofinal in $I$, then $\phi$ is an isomorphism.
\end{appproposition}

\begin{proof}
The collection $\ev{A, \iota_j}_{j\in J}$ is a target for $\ev{A_i, \iota_{ji}}_{i,j \in J}$, so there exists a unique morphism $\phi:B \rightarrow A$ such that $\phi \circ \kappa_j = \iota_j$ for all $j \in J$. We show that $\phi$ is an isomorphism when $J$ is cofinal.

For $i \in I$, define $\lambda_i : A_i \rightarrow B$ by choosing $j \in J$ such that $i \leq j$ and setting $\lambda_i = \kappa_j \circ \iota_{ji}$. For any other choice $k \in J$ such that $i \leq k$, we may choose $\ell \in J$ such that $j,k \leq \ell$. Then
\[
\kappa_j \circ \iota_{ji} = \kappa_\ell \circ \iota_{\ell j} \circ \iota_{ji} = \kappa_\ell \circ \iota_{\ell i} = \kappa_\ell \circ \iota_{\ell k} \circ \iota_{ki} = \kappa_k \circ \iota_{ki}.
\]
Therefore the definition of $\lambda_i$ is independent of the choice of $j \in J$ such that $i \leq j$. In particular, $\lambda_j = \kappa_j$ for all $j \in J$. 

For $i,j \in I$ with $i \leq j$, we choose $k \in J$ with $i,j \leq k$ and observe
\[
\lambda_j \circ \iota_{ji} = \kappa_k \circ \iota_{kj} \circ \iota_{ji} = \kappa_k \circ \iota_{ki} = \lambda_i
\]
This proves that $\ev{B, \lambda_i}$ is a target for $\ev{A_i, \iota_{ji}}_{i,j \in I}$. Therefore there exists a unique morphism $\chi:A \rightarrow B$ such that $\chi \circ \iota_i = \lambda_i$ for all $i$. Observing that $\phi \circ \lambda_i = \iota_j \circ \iota_{ji} = \iota_i$ for all $i \in I$, we see that $\phi \circ \chi \circ \iota_i = \iota_i$ for all $i \in I$, so $\phi \circ \chi = \id_A$ by Proposition \ref{prop:directed_colim_function_identifier}. Furthermore, $\chi \circ \phi \circ \kappa_j = \chi \circ \iota_j = \lambda_j = \kappa_j$ for all $j \in J$, so $\chi \circ \phi = \id_B$ by Proposition \ref{prop:directed_colim_function_identifier}. This proves that $\phi$ is an isomorphism. 
\end{proof}

\chapter{Partial Traces for Infinite-Dimensional Hilbert Spaces}\label{app:partial_trace}

We review the construction of the partial trace provided by Attal for infinite-dimensional Hilbert spaces \cite{AttalQuantumNoise}. We will show that his method works for arbitrary Hilbert spaces (they need not be separable). Naturally, the theory of trace-class operators is essential and we will assume the reader is familiar with this theory.

We will need a form of Fubini's theorem for sums in Banach spaces; we prove this first. The sums below are meant as limits of the net of finite partial sums. In other words, a nonempty indexed collection $(x_i)_{i \in I}$ of elements of a Banach space $X$ is \textdef{summable} if the net $\qty(\sum_{i \in I'} x_i)_{I'}$ of finite partial sums converges, where $I'$ is a nonempty finite subset of $I$. If $(x_i)_{i \in I}$ is summable, then we write $\sum_{i \in I} x_i$ for the limit of the net of finite partial sums. We say that $(x_i)_{i \in I}$ is \textdef{absolutely summable} if $(\norm{x_i})_{i \in I}$ is summable in $\bbR$. This is equivalent to the set of finite partial sums $\sum_{i \in I'} \norm{x_i}$ having a finite upper bound. If this is the case, then $\sum_{i \in I} \norm{x_i}$ is the least upper bound of the finite partial sums. Note also that if $(x_i)_{i \in I}$ is absolutely summable, then it is summable and $\norm{\sum_{i \in I} x_i} \leq \sum_{i \in I} \norm{x_i}$.

\begin{apptheorem}\label{thm:Banach_Fubini}
Let $X$ be a Banach space. Let $I$ and $J$ be nonempty sets and let $(x_{ij})_{(i,j) \in I \times J}$ be a collection of elements of $X$. Suppose $(x_{ij})_{(i,j) \in I \times J}$ is absolutely summable. Then:
\begin{enumerate}
  \item for every $i \in I$, $(x_{ij})_{j \in J}$ is absolutely summable,
  \item for every $j \in J$, $(x_{ij})_{i \in I}$ is absolutely summable,
  \item $(\sum_{j \in J} x_{ij})_{i \in I}$ is absolutely summable,
  \item $(\sum_{i \in I} x_{ij})_{j \in J}$ is absolutely summable,
\end{enumerate}
and
\begin{equation}\label{eq:Banach_Fubini}
\sum_{i \in I} \qty(\sum_{j \in J} x_{ij}) = \sum_{j \in J}\qty(\sum_{i \in I} x_{ij}) = \sum_{(i,j) \in I \times J} x_{ij}.
\end{equation}
\end{apptheorem}

\begin{proof}
For (a) and (b), we observe that any finite partial sum of the collections $(\norm{x_{ij}})_{j \in J}$ and $(\norm{x_{ij}})_{i \in I}$ is a finite partial sum of the collection $(\norm{x_{ij}})_{(i,j) \in I \times J}$, and is therefore bounded above since $(x_{ij})_{(i,j) \in I \times J}$ is absolutely summable by assumption.

Consider (c). Let $I' \subset I$ be a nonempty finite subset. Given $\varepsilon > 0$, there exists a nonempty finite subset $J' \subset J$ such that
\[
\sum_{j \in J} \norm{x_{ij}} < \frac{\varepsilon}{\abs{I'}} + \sum_{j \in J'} \norm{x_{ij}}.
\]
for all $i \in I'$. Then
\begin{align*}
\sum_{i \in I'}\norm{\sum_{j \in J}x_{ij}} &\leq \sum_{i \in I'}\qty(\sum_{j \in J} \norm{x_{ij}}) \leq \sum_{i \in I'} \qty(\frac{\varepsilon}{\abs{I'}} + \sum_{j \in J'} \norm{x_{ij}}) \\
&= \varepsilon + \sum_{i \in I'} \qty(\sum_{j \in J'} \norm{x_{ij}}) \leq \varepsilon + \sum_{(i,j) \in I \times J} \norm{x_{ij}}.
\end{align*}
Since $\varepsilon$ was arbitrary, this shows that the partial sums of the collection $(\norm*{\sum_{j \in J} x_{ij}})_{i \in I}$ are bounded above by $\sum_{(i,j) \in I \times J} \norm{x_{ij}}$, proving (c). A similar argument proves (d). 

We must now prove \eqref{eq:Banach_Fubini}. Given $\varepsilon > 0$, we may find a nonempty finite subset $K \subset I \times J$ such that for any finite subset $K' \supset K$ we have
\[
\norm{\sum_{(i,j) \in I \times J} x_{ij} - \sum_{(i,j) \in  K'} x_{ij}} < \varepsilon.
\]
Let $\pi_I:I \times J \rightarrow I$ and $\pi_J:I \times J \rightarrow J$ be the projections. Let $ I' \subset I$ be a finite subset such that $\pi_I(K) \subset  I'$ and 
\[
\norm{\sum_{i \in I}\qty(\sum_{j \in J} x_{ij}) - \sum_{i \in  I'}\qty(\sum_{j \in J} x_{ij})} < \varepsilon.
\]
Then let $ J' \subset J$ be a finite subset such that $\pi_J(K) \subset J'$ and
\[
\norm{\sum_{j \in J} x_{ij}  - \sum_{j \in  J'} x_{ij}} < \frac{\varepsilon}{\abs*{ I'}}
\]
for all $i \in  I'$. Note that $ I' \times  J'$ contains $K$. Putting it all together yields
\begin{align*}
\norm{\sum_{i\in I}\qty(\sum_{j \in J} x_{ij}) - \sum_{(i,j) \in I \times J} x_{ij}} &\leq 2\varepsilon + \norm{\sum_{i \in  I'}\qty(\sum_{j \in J} x_{ij}) - \sum_{(i,j) \in  I' \times  J'} x_{ij}}\\
&\leq 2 \varepsilon + \sum_{i \in  I'} \norm{\sum_{j \in J} x_{ij} - \sum_{j \in  J'} x_{ij}} < 3\varepsilon.
\end{align*}
Since $\varepsilon$ was arbitrary, we conclude that
\[
\sum_{i \in I} \qty(\sum_{j \in J} x_{ij}) = \sum_{(i,j) \in I \times J} x_{ij}.
\]
A similar method argument shows that
\[
\sum_{j \in J} \qty(\sum_{i \in I} x_{ij}) = \sum_{(i,j) \in I \times J} x_{ij}. \qedhere
\]
\end{proof}

Let us return to the topic of partial traces, following \cite{AttalQuantumNoise}.

\begin{appdefinition}[{\cite[Def.~2.26]{AttalQuantumNoise}}]
Let $\hilbH$ and $\hilbK$ be (both real or both complex) Hilbert spaces and let $\hilbH \otimes \hilbK$ denote their Hilbert space tensor product. Let $y \in \hilbK$. We may use $y$ to define a linear map
\[
\tensor{\ket{y}}{_\hilbK} :\hilbH \rightarrow \hilbH \otimes \hilbK, \quad \tensor{\ket{y}}{_\hilbK}\qty(x) = x \otimes y,
\]
which we call a \textdef{partial ket}. Clearly $\tensor{\ket{y}}{_\hilbK}$ is linear and since $\norm{x \otimes y} = \norm{x}\norm{y}$ we see that $\tensor{\ket{y}}{_\hilbK}$ is also bounded with norm $\norm{y}$. We define the \textdef{partial bra} $\tensor[_\hilbK]{\bra{y}}{}:\hilbH \otimes \hilbK \rightarrow \hilbH$ to be the adjoint of $\tensor{\ket{y}}{_\hilbK}$. It is easy to check that
\[
\tensor[_\hilbK]{\bra{y}}{}(x \otimes z) = \ev{y, z}x
\]
for all $x \in \hilbH$ and $z \in \hilbK$. We note that the norm of $\tensor[_\hilbK]{\bra{y}}{}$ is also $\norm{y}$ since taking the adjoint preserves the norm.
\end{appdefinition}

Notice that if $T \in \cB(\hilbH \otimes \hilbK)$ and $y \in \hilbK$, then 
\[
\tensor[_\hilbK]{\langle y|T|y \rangle}{_\hilbK} \in \cB(\hilbH)
\]
just by composing operators.

\begin{applemma}[{\cite[Lem.~2.27]{AttalQuantumNoise}}]
If $T \in \cB(\hilbH \otimes \hilbK)$ is trace-class and $y \in \hilbK$, then $\tensor[_\hilbK]{\langle y|T|y \rangle}{_\hilbK}$ is trace-class.
\end{applemma}

\begin{proof}
Without loss of generality let $\norm{y} = 1$. Let $(u_i)_{i \in I}$ and $(v_i)_{i \in I}$ be orthonormal systems in $\hilbH$. Then
\begin{align*}
\sum_{i \in I} \abs{\langle u_i | \tensor[_\hilbK]{\langle y|T|y \rangle}{_\hilbK} | v_i \rangle} = \sum_{i \in I} \abs{\langle u_i \otimes y | T | v_i \otimes y \rangle}.
\end{align*}
Since $(u_i \otimes y)_{i \in I}$ and $(v_i \otimes y)_{i \in I}$ are orthonormal families in $\hilbH \otimes \hilbK$, we know the sum converges since $T$ is trace-class. This implies that $\tensor[_\hilbK]{\langle y|T|y \rangle}{_\hilbK}$ is trace-class.
\end{proof}

\begin{apptheorem}[{\cite[Thm.~2.28]{AttalQuantumNoise}}]
Let $T \in \cB(\hilbH \otimes \hilbK)$ be trace-class and let $(e_k)_{k \in K}$ be an orthonormal basis of $\hilbK$. The sum
\[
\tr_\hilbK(T) = \sum_{k \in K} \tensor[_\hilbK]{\langle e_k|T|e_k \rangle}{_\hilbK}
\]
is absolutely convergent with respect to the trace norm. The sum $\tr_\hilbK(T) \in \cB(\hilbH)$ is the unique trace-class operator satisfying
\[
\tr(\tr_\hilbK(T)A) = \tr(T(A \otimes \1))
\]
for all $A \in \cB(\hilbH)$. It follows that $\tr_\hilbK(T)$ is independent of the chosen basis for $\hilbK$.
\end{apptheorem}

\begin{proof}
Let $\norm{\cdot}_1$ denote the trace norm. For each $k \in K$, the operator $\tensor[_\hilbK]{\langle e_k|T|e_k \rangle}{_\hilbK}$ is trace-class on $\hilbH$, so there exist orthonormal families $(u_i^k)_{i \in I^k}$ and $(v^k_i)_{i \in I^k}$ such that
\begin{align*}
\norm{\tensor[_\hilbK]{\langle e_k|T|e_k \rangle}{_\hilbK}}_1 = \sum_{i \in I^k} \abs{\langle u_i^k | \tensor[_\hilbK]{\langle e_k|T|e_k \rangle}{_\hilbK} | v_i^k \rangle} = \sum_{i \in I^k} \abs{\langle u_i^k \otimes e_k| T|v_i^k \otimes e_k\rangle }.
\end{align*}

Let $\tilde K \subset K$ be nonempty and finite. For each $k \in \tilde K$, we may choose a finite subset $\tilde I^k \subset I^k$ such that
\[
\norm{\tensor[_\hilbK]{\langle e_k|T|e_k \rangle}{_\hilbK}}_1 < \frac{1}{\abs*{\tilde K}} + \sum_{i \in \tilde I^k} \abs{\langle u_i^k \otimes e_k | T|v_i^k \otimes e_k \rangle }.
\]
Then
\begin{align*}
\sum_{k \in \tilde K} \norm{\tensor[_\hilbK]{\langle e_k|T|e_k \rangle}{_\hilbK}}_1 < 1 + \sum_{k \in \tilde K} \sum_{i \in \tilde I^k}\abs{\langle u_i^k \otimes e_k | T|v_i^k \otimes e_k \rangle }.
\end{align*}
The families $(u_i^k \otimes e_k)_{k \in \tilde K, i \in \tilde I^k}$ and $(v_i^k \otimes e_k)_{k \in \tilde K, i \in \tilde I^k}$ are orthonormal, hence the double sum above is bounded above by $\norm{T}_1$, and we get
\[
\sum_{k \in \tilde K} \norm{\tensor[_\hilbK]{\langle e_k|T|e_k \rangle}{_\hilbK}}_1 < 1 + \norm{T}_1 < \infty,
\]
Since the finite partial sums are bounded above, the series converges absolutely.

Let $A \in \cB(\hilbH)$. We let $(d_j)_{j \in J}$ be an orthonormal basis for $\hilbH$ and we compute the trace
\begin{align*}
\tr(\tr_\hilbK(T)A) &= \sum_{j \in J} \langle d_j | \tr_\hilbK(T)A |d_j\rangle \\
&= \sum_{j \in J} \qty(\sum_{k \in K} \langle d_j | \tensor[_\hilbK]{\langle e_k|T|e_k \rangle}{_\hilbK} A | d_j \rangle )\\
&= \sum_{j \in J} \qty(\sum_{k \in K} \langle d_j \otimes e_k | T(A \otimes \1) | d_j \otimes e_k \rangle ).
\end{align*}
Since $T(A \otimes \1)$ is trace class and $(d_j \otimes e_k)_{(j,k) \in J \times K}$ is an orthonormal basis for $\hilbH \otimes \hilbK$, we know
\[
\sum_{(j,k) \in J \times K} \abs{\langle d_j \otimes e_k| T(A \otimes \1) | d_j \otimes e_k \rangle } < \infty
\] 
Therefore by Theorem \ref{thm:Banach_Fubini} we have
\begin{align*}
\tr(\tr_\hilbK(T)A) &= \sum_{(j,k) \in J \times K}  \ev{d_j \otimes e_k, T(A \otimes \1)(d_j \otimes e_k)} = \tr(T(A \otimes I)),
\end{align*}
as desired.

Finally, we show that $\tr_\hilbK(T)$ is the unique trace-class operator with this property. Suppose $S_1, S_2 \in \cB(\hilbH)$ are trace-class operators such that
\[
\tr(S_1A) = \tr(T(A \otimes I)) = \tr(S_2A)
\]
for all $A \in \cB(\hilbH)$. Then $\tr((S_1 - S_2)A) = 0$ for all $A \in \cB(\hilbH)$ by linearity of the trace. Let $x \in \hilbH$ and define $A = \ketbra{x}{(S_1 - S_2)x}$. Then
\[
0 = \tr((S_1 - S_2)A) = \tr\qty(\ketbra{(S_1 - S_2)x}{(S_1 - S_2)x}) = \norm{(S_1 - S_2)x}^2,
\]
so $(S_1 - S_2)x = 0$ for all $x \in \hilbH$. Thus, $S_1 = S_2$. This proves uniqueness of $\tr_\hilbK(T)$, hence independence from the basis $(e_k)_{k \in K}$.
\end{proof}

\begin{comment}
We conclude with some basic properties of the partial trace. Let $\cB_1(\hilbH)$ denote the set of trace-class operators on a Hilbert space $\hilbH$.

\begin{apptheorem}[{\cite[Thm.~2.29]{AttalQuantumNoise}}]
Let $\hilbH$ and $\hilbK$ be Hilbert spaces.
\begin{alignat*}{2}
\tr_\hilbK(A \otimes B) &= \tr(B)A, &&\qquad A \in \cB_1(\hilbH),\,\, B \in \cB_1(\hilbK)\\
\tr(\tr_\hilbK(T)) &= \tr(T), &&\qquad T \in \cB_1(\hilbK \otimes \hilbK)\\
\tr_\hilbK((A \otimes I)T(B \otimes I)) &= A\tr_\hilbK(T)B, &&\qquad A, B \in \cB(\hilbH),\,\, T \in \cB_1(\hilbH \otimes \hilbK)
\end{alignat*}
\end{apptheorem}
\end{comment}

\begin{comment}
\makeatletter
\newcommand{\leqnomode}{\tagsleft@true\let\veqno\@@leqno}
\newcommand{\reqnomode}{\tagsleft@false\let\veqno\@@eqno}
\makeatother
\leqnomode

\begin{proof}
Let $C \in \cB(\cH)$. 

\begin{equation}
 \hfill \tr(\tr(B)AC) = \tr(B)\tr(AC) = \tr(AC \otimes B) = \tr((A\otimes B)(C \otimes I)).  \tag{a}
\end{equation}

\begin{equation}
\tr(\tr_\cK(T)) = \tr(\tr_\cK(T)I) = \tr(T(I \otimes I)) = \tr(T). \tag{b}
\end{equation}

\begin{equation}
\tr(A\tr_\cK(T)BC) = \tr(\tr_\cK(T)BCA) = \tr(T(BCA \otimes I)) = \tr((A \otimes I)T(B \otimes I)(C \otimes I)). \tag{c}
\end{equation}
\end{proof}
\end{comment}

\end{document}